\documentclass[a4paper,11pt]{article}
\pdfoutput=1

\usepackage{jheppub}
\usepackage{amsmath,amssymb,amscd}
\usepackage{bbm}
\usepackage{graphicx,xcolor}
\usepackage{url}
\usepackage[export]{adjustbox}
\usepackage{accents}
\usepackage[]{algorithm2e}
\usepackage{verbatim}
\usepackage{slashed}
\usepackage{lmodern}

\graphicspath{ {fig/}{../fig} }

\makeatletter
\def\hlinewd#1{
  \noalign{\ifnum0=`}\fi\hrule \@height #1 \futurelet
   \reserved@a\@xhline}
\makeatother

\makeatletter
\renewcommand\@fpheader{}
\renewcommand\@journal{}
\makeatother

\newcommand{\pbar}[1]{\accentset{\textbf{\fontsize{3pt}{2pt}\selectfont(\fontsize{4pt}{2pt}\selectfont--\fontsize{3pt}{2pt}\selectfont)}}{#1}}

\definecolor{dgreen}{rgb}{0.,.3,0}
\definecolor{dblue}{rgb}{0.0,0.0,0.5}
\newcommand{\pole}[1]{\textcolor{dblue}{#1}}
\newcommand{\coupling}[1]{\textcolor{dgreen}{#1}}

\newcommand{\hide}[1]{}
\newcommand{\ep}{\epsilon}

\title{
\boldmath
Mixed EW-QCD two-loop amplitudes for $q\bar{q}\to \ell^+ \ell^-$
and $\gamma_5$ scheme independence of multi-loop corrections
}
\preprint{MITP/20-072, MSUHEP-20-020}

\author[a]{Matthias Heller,}
\author[b]{Andreas von Manteuffel,}
\author[b]{Robert M. Schabinger,}
\author[\,c]{and\\ Hubert Spiesberger}

\affiliation[a]{PRISMA$^+$ Cluster of Excellence, Institut f\"ur Kernphysik, Johannes Gutenberg Universit\"{a}t,\\
55099 Mainz, Deutschland}
\affiliation[b]{Department of Physics and Astronomy, Michigan State University, \\
East Lansing, Michigan 48824, USA}
\affiliation[c]{PRISMA$^+$ Cluster of Excellence, Institut f\"ur Physik, Johannes Gutenberg Universit\"{a}t,\\
55099 Mainz, Deutschland}

\emailAdd{maheller@students.uni-mainz.de}
\emailAdd{vmante@msu.edu}
\emailAdd{schabing@msu.edu}
\emailAdd{spiesber@uni-mainz.de}

\abstract{
We perform a dedicated study of the $q \bar{q}$-initiated two-loop electroweak-QCD Drell-Yan scattering amplitude in dimensional regularization schemes for vanishing light quark and lepton masses.
For the relative order $\alpha$ and $\alpha_s$ one-loop Standard Model corrections, details of our comparison to the original literature are given.
The infrared pole terms of the mixed two-loop amplitude are governed by a known generalization of the dipole formula and we show explicitly that exactly the same two-loop polarized hard scattering functions are obtained in both the standard 't\,Hooft-Veltman-Breitenlohner-Maison $\gamma_5$ scheme and Kreimer's anticommuting $\gamma_5$ scheme.
}

\begin{document}
\maketitle

\allowdisplaybreaks[4]
\newpage

\section{Introduction} 
\label{sec:intro}

A precise theoretical understanding of the neutral- and charged-current Drell-Yan processes~\cite{Drell:1970wh} in Standard Model perturbation theory is essential to the success of the Large Hadron Collider physics program. It is therefore unsurprising that significant resources have been allocated, over the course of a multi-generational effort, to bring the higher-order corrections to the Drell-Yan scattering processes under control. Very recently, the total cross section in pure Quantum Chromodynamics (QCD) was successfully calculated through to next-to-next-to-next-to-leading order in the strong coupling constant for both the neutral- and charged-current processes \cite{Hamberg:1990np,Harlander:2002wh,Anastasiou:2003ds,Melnikov:2006di,Duhr:2020seh,Duhr:2020sdp}. On the electroweak side, neutral gauge boson production and decay has been calculated through to next-to-next-to-leading order in pure Quantum Electrodynamics (QED) \cite{Berends:1987ab,Blumlein:2019srk}, but full electroweak (EW) corrections are known only at next-to-leading order \cite{Baur:2001ze,Dittmaier:2001ay,Baur:2004ig}. Due to the size of the next-to-leading order electroweak corrections, it has long been of interest to calculate the mixed EW-QCD corrections of relative order $\alpha \alpha_s$ as well. Indeed, many partial results in this direction have been put forth over the last decade \cite{Kilgore:2011pa,Dittmaier:2014qza,Dittmaier:2015rxo,Delto:2019ewv,Buccioni:2020cfi,Bonciani:2020tvf,Dittmaier:2020vra,Behring:2020cqi}, but a complete calculation in the regime of large lepton pair invariant mass is not available yet.

Despite the significant attention these two-loop EW-QCD calculations have received, it has long remained a challenge to include the non-factorizable two-loop virtual corrections. For many years, the primary technical problem was the evaluation of the massive two-loop box-type master integrals which emerge from the most complicated Feynman diagrams (see \cite{Bonciani:2016ypc,vonManteuffel:2017myy,Hasan:2020vwn} for various results). Recently, some of us successfully evaluated the complete set of two-loop box-type master integrals relevant to the non-factorizable corrections as linear combinations of standard multiple polylogarithms in the physical regions of phase space \cite{Heller:2019gkq}.
In this paper, we take an important step towards complete relative order $\alpha\alpha_s$ corrections by elucidating the treatment of $\gamma_5$ for the box-type diagrams and deriving the order $\alpha^2 \alpha_s$ two-loop helicity amplitudes for $q \bar{q} \to \ell^+\ell^-$.

As has been known since the earliest higher-order studies of the Drell-Yan process, it is important to have a solid understanding of the $\gamma_5$ problem of dimensional regularization \cite{Breitenlohner:1977hr,Breitenlohner:1975hg,Breitenlohner:1976te,Chanowitz:1979zu} if one wishes to calculate {\it all} Feynman diagrams at a given order, including those where a $W$ or $Z$ connect the initial- and final-state fermion lines.
As will be discussed in detail in what follows, care must be taken in practical calculations to consistently include scheme-dependent higher-order-in-$\epsilon$ terms, see also \cite{Gnendiger:2017pys} for a recent review.
In order to establish confidence in our results, we carried out two independent calculations, one where 't\,Hooft-Veltman-Breitenlohner-Maison's (HVBM's) $\gamma_5$ scheme \cite{tHooft:1972tcz,Breitenlohner:1977hr,Breitenlohner:1975hg,Breitenlohner:1976te} was employed and one where Kreimer's $\gamma_5$ scheme \cite{Kreimer:1989ke,Korner:1991sx,Kreimer:1992,Kreimer:1993bh} was employed. For our calculations, Kreimer's $\gamma_5$ scheme was practically superior because it does not require the introduction of finite counterterms to restore the chiral symmetry of the Standard Model.

This article is organized as follows. In Sections \ref{sec:dimreg} - \ref{sec:muterms}, a general discussion of the subtleties involved with the application of dimensional regularization in the presence of chiral couplings is given, both for HVBM's $\gamma_5$ scheme and Kreimer's $\gamma_5$ scheme. In Sections \ref{sec:couplingren} - \ref{sec:fermren} our renormalization scheme for the coupling constants, the wavefunctions, and the particle masses is described.
In Section \ref{sec:finiteren}, we discuss the finite renormalizations we perform in HVBM's $\gamma_5$ scheme. In Sections \ref{sec:process} - \ref{sec:Kreimercalcs} we define the scattering process we consider more precisely and relate further technical details of the numerator algebra scripts we used to carry out our Feynman diagram calculations. In Section \ref{sec:IRsing}, we elucidate the infrared structure of the one- and two-loop scattering amplitudes we calculate and define {\it hard scattering functions} which, at leading order in the parameter of dimensional regularization, encode physical information about the kernel of the virtual corrections which remains after infrared subtraction. In Section \ref{sec:helicities}, we provide a recipe to pass from our form factors to scattering amplitudes for states of definite helicity. In Sections \ref{sec:1Ldiags} - \ref{sec:1Lfinalres}, supplemented by Appendices \ref{sec:1Lasres} and \ref{sec:1Lares}, explicit results for the one-loop Standard Model corrections to the Drell-Yan process are given, together with a detailed comparison to the original literature. In Sections \ref{sec:2Ldiags} - \ref{sec:2Lassem}, we provide details specifically relevant to the two-loop mixed EW-QCD calculation and, in Section \ref{sec:2Lfinalres}, visualizations of our final results for the order $\alpha \alpha_s$, order $\alpha^2$, and order $\alpha^2 \alpha_s$ polarized hard scattering functions. Finally, in Section \ref{sec:outlook}, we summarize our findings.
\section{Regularization and renormalization}
\subsection{Universal features of dimensional regularization}
\label{sec:dimreg}
In this section, our aim is to review some background and motivation for the more technical, scheme-dependent discussions of chiral couplings in dimensional regularization which follow in Sections \ref{sec:HVBMdefs} and \ref{sec:Kreimerdefs}. The simultaneous dimensional regularization of ultraviolet and infrared divergences \cite{Speer:1971,Ashmore:1972uj,tHooft:1972tcz,Bollini:1972ui,Marciano:1974tv} has proven to be an essential tool for the calculation of higher-order perturbative scattering amplitudes in gauge theories. For a gauge theory Lagrange density defined in four spacetime dimensions, the method purports to provide a prescription for the continuation of all off-shell Feynman rules from four to $d$ dimensions. All higher-order correlation functions and scattering amplitudes in the model under consideration are then taken to be meromorphic functions of $d$ in a complex domain containing the point $d = 4$. Accordingly, the Fourier integrals which appear at higher orders in perturbation theory are continued from four dimensions to $d = 4 - 2\epsilon$ dimensions as
\begin{align}
    \int \frac{\mathrm{d}^4 k_i}{\left(2\pi\right)^4} \longrightarrow \left(\mu^2\right)^\ep \int \frac{\mathrm{d}^{4 - 2\ep} k_i}{\left(2\pi\right)^{4-2\ep}}\,,
\end{align}
where $\epsilon$ is the parameter of dimensional regularization and $\mu$ is the 't\,Hooft scale, a unit of mass introduced to maintain a dimensionless action. The ultraviolet and infrared singularities of higher-order correlation functions and scattering amplitudes manifest themselves as poles in $\epsilon$ which ultimately cancel only in higher-order corrections to physical observables.

While some of the prescribed continuations of the Lorentz and Dirac algebra such as
\begin{align}
    \label{eq:basicmetcontract}
    g^{\mu \nu}g_{\mu \nu} &= 4-2\epsilon\,,\\
    \label{eq:Diracalgebra}
    \left\{\gamma_\mu, \gamma_\nu\right\} &= 2 g_{\mu \nu} \,\mathbf{1}\,,\\
    \label{eq:basicDiraccontract}
   \mathrm{and}~~ \gamma^\mu \gamma_\mu &= \frac{1}{2} g^{\mu \nu} \left\{\gamma_\mu, \gamma_\nu\right\} = g^{\mu \nu}g_{\mu \nu} = 4-2\epsilon
\end{align}
feel very natural, other aspects of the program are far more subtle and require careful explanation. Before proceeding to a discussion of these subtleties, we state a universal identity for the $d$-dimensional contraction of two Dirac matrices with $n$ $d$-dimensional Dirac matrices interspersed which will be useful later on:
\begin{align}
\label{eq:genDiraccontract}
    \gamma^\mu \gamma_{\nu_1}\cdots\gamma_{\nu_n}\gamma_\mu =  2 \sum_{i = 1}^n (-1)^{i + n}  \gamma_{\nu_i} \gamma_{\nu_1}\cdots\gamma_{\nu_{i-1}}\gamma_{\nu_{i+1}}\cdots\gamma_{\nu_n}+(4-2\epsilon)(-1)^n \gamma_{\nu_1}\cdots\gamma_{\nu_n}.
\end{align}
This identity is simply obtained by repeatedly applying Eq. \eqref{eq:Diracalgebra} to move the $\gamma_\mu$ factor to the left until it is finally possible to apply Eq. \eqref{eq:basicDiraccontract}. The reader can easily check for the first few values of $n$ that the well-known textbook results (see {\it e.g.} \cite{Peskin:1995ev}) are all encoded in Eq. \eqref{eq:genDiraccontract}.

As has been clear from its inception \cite{Ashmore:1972uj,tHooft:1972tcz}, it is technically challenging to handle chiral couplings in dimensional regularization. This is primarily because there is no canonical way to continue the usual four-dimensional definition of $\gamma_5$,
\begin{align}
\label{eq:defg5}
    \gamma_5 = -i \,\gamma_0 \gamma_1 \gamma_2 \gamma_3\,,
\end{align}
into $d$ dimensions. It was emphasized in \cite{tHooft:1972tcz} by 't\,Hooft and Veltman that one {\it cannot} expect to maintain the usual anticommutation relation
\begin{align}
    \label{eq:anticommg5}
    \left\{\gamma_\mu,\gamma_5\right\} = 0
\end{align}
and, at the same time, work with traces of Dirac matrices which enjoy all of the usual mathematical properties. In particular, a cyclic Dirac trace was explicitly assumed alongside Eq. \eqref{eq:anticommg5} in the algebraic proof that the trace of $\gamma_\mu \gamma_\nu\gamma_\rho\gamma_\sigma\gamma_5$ must either vanish in dimensional regularization or be subject to a modified set of algebraic rules (see {\it e.g.} \cite{Collins:1984xc} for a review). 't\,Hooft and Veltman influenced most later authors to modify the $\gamma_5$ anticommutation relation,  but, as was pointed out by Kreimer \cite{Kreimer:1989ke}, retaining Eq. \eqref{eq:anticommg5} while abandoning the cyclicity of the Dirac trace is another path towards mathematical consistency. It would at this stage be reasonable to expect a concrete prescription for the $d$-dimensional continuation of classical trace formulae involving $\gamma_5$ such as
\begin{align}
\label{eq:basicoddtrace}
    \mathrm{tr}\left\{\gamma_\mu \gamma_\nu\gamma_\rho\gamma_\sigma\gamma_5\right\} = - 4 i \,\varepsilon_{\mu \nu \rho \sigma}
\end{align}
in some refined dimensional regularization scheme.

In fact, there {\it is} no $d$-dimensional replacement for $\gamma_5$ or the Levi-Civita tensor, $\varepsilon_{\mu \nu \rho \sigma}$, in the dimensional regularization schemes we consider; rather, one simply accepts that Eq. \eqref{eq:defg5}, Eq. \eqref{eq:basicoddtrace}, and the familiar properties
\begin{equation}
\gamma_5^2 = \mathbf{1}\qquad \mathrm{and} \qquad \gamma_5^\dagger = \gamma_5
\end{equation}
remain unchanged in $d$ dimensions. That $\gamma_5$ and $\varepsilon_{\mu \nu \rho \sigma}$ are intrinsically four-dimensional objects inevitably forces one to distinguish a four-dimensional subspace and sacrifice somewhat for what concerns the construction of a manifestly Lorentz-covariant formalism. In this work, it is frequently convenient to explicitly distinguish the four-dimensional and $(-2\epsilon)$-dimensional subspaces of $d$-dimensional Lorentz indices in perturbative calculations. Following \cite{Collins:1984xc}, we write
\begin{align}
\label{eq:splitindex1}
    k^\mu &= \bar{k}^\mu + \hat{k}^\mu\\
\label{eq:splitindex2}
    \mathrm{and} \qquad \gamma^\mu &= \bar{\gamma}^\mu + \hat{\gamma}^\mu\,,
\end{align}
where bars denote four-dimensional objects and hats denote $(-2\ep)$-dimensional objects. Similarly, we write the usual four-dimensional identities,
\begin{align}
    \label{eq:epscontract4}
    \varepsilon^{\mu \nu \rho \sigma} \,\varepsilon^{\vphantom{\mu}}_{\mu \nu \rho \sigma} &= -24\,, \\
    \label{eq:epscontract3}
    \varepsilon^{\mu \nu \rho}_{\hphantom{\mu \nu \rho} \alpha} \,\varepsilon^{\vphantom{\mu}}_{\mu \nu \rho \beta} &= -6 \,\bar{g}_{\alpha \beta}\,, \\
    \label{eq:epscontract2}
    \varepsilon^{\mu \nu}_{\hphantom{\mu \nu} \alpha \delta} \,\varepsilon^{\vphantom{\mu}}_{\mu \nu \beta \eta} &= -2\left(\bar{g}_{\alpha \beta}\bar{g}_{\delta \eta}-\bar{g}_{\alpha \eta}\bar{g}_{\beta \delta}\right)\,, \\
    \label{eq:epscontract1}
    \varepsilon^{\mu}_{\hphantom{\mu} \alpha \delta \kappa} \,\varepsilon^{\vphantom{\mu}}_{\mu \beta \eta \lambda} &=
    \bar{g}_{\alpha  \lambda } \bar{g}_{\beta  \kappa } \bar{g}_{\delta  \eta }-\bar{g}_{\alpha  \lambda } \bar{g}_{\beta  \delta } \bar{g}_{\eta  \kappa }-\bar{g}_{\alpha  \eta } \bar{g}_{\beta  \kappa } \bar{g}_{\delta  \lambda }
    \nonumber \\
    &+\bar{g}_{\alpha  \beta } \bar{g}_{\delta 
   \lambda } \bar{g}_{\eta  \kappa }+\bar{g}_{\alpha  \eta } \bar{g}_{\beta  \delta } \bar{g}_{\kappa  \lambda }-\bar{g}_{\alpha  \beta } \bar{g}_{\delta  \eta } \bar{g}_{\kappa  \lambda }\,, \\
   \label{eq:epscontract0}
   \mathrm{and}\qquad\varepsilon^{\vphantom{\mu}}_{\alpha \delta \kappa \tau} \,\varepsilon^{\vphantom{\mu}}_{\beta \eta \lambda \chi} &= -\bar{g}_{\alpha  \chi } \bar{g}_{\beta  \tau } \bar{g}_{\delta  \lambda } \bar{g}_{\eta  \kappa }+\bar{g}_{\alpha  \chi } \bar{g}_{\beta  \tau } \bar{g}_{\delta  \eta } \bar{g}_{\kappa  \lambda }+\bar{g}_{\alpha  \chi } \bar{g}_{\beta  \kappa }
   \bar{g}_{\delta  \lambda } \bar{g}_{\eta  \tau }-\bar{g}_{\alpha  \chi } \bar{g}_{\beta  \delta } \bar{g}_{\eta  \tau } \bar{g}_{\kappa  \lambda }
   \nonumber \\
   &\hspace{-1 cm}-\bar{g}_{\alpha  \chi } \bar{g}_{\beta  \kappa } \bar{g}_{\delta  \eta } \bar{g}_{\lambda  \tau }+\bar{g}_{\alpha  \chi }
   \bar{g}_{\beta  \delta } \bar{g}_{\eta  \kappa } \bar{g}_{\lambda  \tau }+\bar{g}_{\alpha  \lambda } \bar{g}_{\beta  \tau } \bar{g}_{\delta  \chi } \bar{g}_{\eta  \kappa }-\bar{g}_{\alpha  \eta } \bar{g}_{\beta  \tau } \bar{g}_{\delta  \chi } \bar{g}_{\kappa  \lambda
   }
   \nonumber \\
   &\hspace{-1 cm}-\bar{g}_{\alpha  \lambda } \bar{g}_{\beta  \kappa } \bar{g}_{\delta  \chi } \bar{g}_{\eta  \tau }+\bar{g}_{\alpha  \beta } \bar{g}_{\delta  \chi } \bar{g}_{\eta  \tau } \bar{g}_{\kappa  \lambda }+\bar{g}_{\alpha  \eta } \bar{g}_{\beta  \kappa } \bar{g}_{\delta  \chi }
   \bar{g}_{\lambda  \tau }-\bar{g}_{\alpha  \beta } \bar{g}_{\delta  \chi } \bar{g}_{\eta  \kappa } \bar{g}_{\lambda  \tau }
   \nonumber \\
   &\hspace{-1 cm}-\bar{g}_{\alpha  \lambda } \bar{g}_{\beta  \tau } \bar{g}_{\delta  \eta } \bar{g}_{\kappa  \chi }+\bar{g}_{\alpha  \eta } \bar{g}_{\beta  \tau }
   \bar{g}_{\delta  \lambda } \bar{g}_{\kappa  \chi }+\bar{g}_{\alpha  \lambda } \bar{g}_{\beta  \delta } \bar{g}_{\eta  \tau } \bar{g}_{\kappa  \chi }-\bar{g}_{\alpha  \beta } \bar{g}_{\delta  \lambda } \bar{g}_{\eta  \tau } \bar{g}_{\kappa  \chi }
   \nonumber \\
   &\hspace{-1 cm}-\bar{g}_{\alpha  \eta }
   \bar{g}_{\beta  \delta } \bar{g}_{\kappa  \chi } \bar{g}_{\lambda  \tau }+\bar{g}_{\alpha  \beta } \bar{g}_{\delta  \eta } \bar{g}_{\kappa  \chi } \bar{g}_{\lambda  \tau }+\bar{g}_{\alpha  \lambda } \bar{g}_{\beta  \kappa } \bar{g}_{\delta  \eta } \bar{g}_{\tau  \chi
   }-\bar{g}_{\alpha  \lambda } \bar{g}_{\beta  \delta } \bar{g}_{\eta  \kappa } \bar{g}_{\tau  \chi }
   \nonumber \\
   &\hspace{-1 cm}-\bar{g}_{\alpha  \eta } \bar{g}_{\beta  \kappa } \bar{g}_{\delta  \lambda } \bar{g}_{\tau  \chi }+\bar{g}_{\alpha  \beta } \bar{g}_{\delta  \lambda } \bar{g}_{\eta 
   \kappa } \bar{g}_{\tau  \chi }+\bar{g}_{\alpha  \eta } \bar{g}_{\beta  \delta } \bar{g}_{\kappa  \lambda } \bar{g}_{\tau  \chi }-\bar{g}_{\alpha  \beta } \bar{g}_{\delta  \eta } \bar{g}_{\kappa  \lambda } \bar{g}_{\tau  \chi }\,,
\end{align}
using $\bar{g}_{\mu  \nu}$ to denote the four-dimensional metric tensor.

In both schemes we consider, Eqs. \eqref{eq:epscontract4}-\eqref{eq:epscontract0} force us to apply Eq. \eqref{eq:splitindex1} to {\it split indices}, $\bar{g}_{\mu \nu} k_i^\mu k_j^\nu = k_i\cdot k_j - \hat{k}_i\cdot \hat{k}_j$.
Split loop momenta lead to well-known technical complications: in sufficiently-complicated Feynman diagrams, Dirac traces produce $d$-dimensional Feynman integrals with numerator insertions of $\hat{k}_i\cdot\hat{k}_j$. Fortunately, general approaches which involve the introduction of dimensionally-shifted Feynman integrals \cite{Bern:1992em,Bern:1993kr,Tarasov:1996br} have been worked out at one loop \cite{Bern:1995db} and at two loops \cite{Bern:2002tk}. For the convenience of the reader, we provide details of a modern formulation of these developments in Section \ref{sec:muterms} below.

As reviewed in \cite{Denner:2019vbn}, many authors have in the past adopted the so-called naive $\gamma_5$ scheme of \cite{Chanowitz:1979zu} (see also \cite{Trueman:1995ca}) for one-loop electroweak calculations. The naive scheme amounts to a set of {\it ad hoc} rules for $\gamma_5$ manipulations designed to lead to correct results while avoiding the technical complications which might arise in a more mathematically-rigorous handling of chiral couplings. Historically, this was justified at the one-loop level by checking {\it a posteriori} that the relevant Ward identities check out. Naturally, in order to calculate with confidence at higher orders in Standard Model perturbation theory, we feel that it is of considerable importance to adopt a mathematically consistent framework. It should be stressed that the naive $\gamma_5$ scheme is equivalent to Kreimer's $\gamma_5$ scheme (introduced below) for what concerns the calculation of the fermion self-energies and vertex form factors considered in this paper. As we shall see in what follows, the Feynman diagrams of box type are really the only ones in $2 \to 2$ scattering processes with sensitivity to the $\gamma_5$ problem.

HVBM's $\gamma_5$ scheme \cite{tHooft:1972tcz,Breitenlohner:1977hr,Breitenlohner:1975hg,Breitenlohner:1976te} and Kreimer's $\gamma_5$ scheme \cite{Kreimer:1989ke,Korner:1991sx,Kreimer:1992,Kreimer:1993bh} are the two variants of dimensional regularization we work with in this paper. The precise treatment of $\gamma_5$ in these schemes has a profound impact on both the Feynman rules and the handling of Dirac traces which we discuss at greater length in Sections \ref{sec:HVBMdefs} and \ref{sec:Kreimerdefs}.\footnote{It actually turns out that the {\it algebraic form} of the standard family of Dirac trace identities which generalizes Eq. \eqref{eq:basicoddtrace} is the same in both schemes, but this requires some explanation.}
Indeed, four-dimensional Dirac trace identities which rely on $\gamma_5$ manipulations must be, depending on the $\gamma_5$ scheme, either carefully reformulated or abandoned.
Consider, for example, the algebraic proof given in \cite{Peskin:1995ev} that the trace of an odd number of Dirac matrices must be zero. For $n$ odd, we have 
\begin{align}
    \mathrm{tr}\left\{\gamma_{\nu_1}\cdots \gamma_{\nu_n}\right\} & = \mathrm{tr}\left\{\gamma_5 \gamma_5 \gamma_{\nu_1}\cdots \gamma_{\nu_n}\right\} \qquad \qquad\qquad~\, (\gamma_5^2 = \mathbf{1})\nonumber \\
    & = (-1)^n \mathrm{tr}\left\{\gamma_5 \gamma_{\nu_1}\cdots \gamma_{\nu_n} \gamma_5\right\} \qquad \qquad (\mathrm{anticommutation~relation~ \eqref{eq:anticommg5}})
    \nonumber \\
    & = -\mathrm{tr}\left\{\gamma_5\gamma_5 \gamma_{\nu_1}\cdots \gamma_{\nu_n} \right\} \qquad \qquad \quad~~\, (n\mathrm{~odd,~trace~cyclicity})
    \nonumber \\
    & = -\mathrm{tr}\left\{\gamma_{\nu_1}\cdots \gamma_{\nu_n} \right\} \qquad \qquad \qquad\quad~\,\,(\gamma_5^2 = \mathbf{1})\,,
\end{align}
which implies $\mathrm{tr}\left\{\gamma_{\nu_1}\cdots \gamma_{\nu_n}\right\} = 0$ as desired. As should by now be clear, the above proof is valid {\it only} in four spacetime dimensions; different lines of reasoning must be adopted in $d = 4 - 2\epsilon$ dimensions. As we shall see, in HVBM's $\gamma_5$ scheme, Eq. \eqref{eq:genDiraccontract} facilitates a simple proof of the proposition by induction, whereas, in Kreimer's $\gamma_5$ scheme, the trace of an odd number of Dirac matrices is argued to be zero in $d$ dimensions by appealing to a mathematical property of the infinite-dimensional Dirac algebra \cite{Kreimer:1989ke}.
\subsection{HVBM's prescription for $\gamma_5$ in dimensional regularization}
\label{sec:HVBMdefs}
In HVBM's $\gamma_5$ scheme, chiral couplings are treated in a consistent way in $d$ dimensions by retaining the four-dimensional definition of $\gamma_5$, Eq. \eqref{eq:defg5}, at the expense of its anticommutation relation with the standard Dirac matrices. In place of Eq. \eqref{eq:anticommg5}, one can write
\begin{align}
\label{eq:modanticommg5pt1}
    \left\{\gamma_\mu,\gamma_5\right\} &= 0\qquad \mathrm{for}~~\mu = 0,1,2,\,\mathrm{and}~3 \\
\label{eq:modanticommg5pt2}
\mathrm{and}\qquad \left[\gamma_\mu,\gamma_5\right] &= 0\qquad \mathrm{in~the~ }(-2\ep)\mathrm{\mbox{-}dimensional~subspace,}
\end{align}
with the understanding that, as reviewed in {\it e.g.} \cite{Collins:1984xc}, the $(-2\ep)$-dimensional subspace is in fact infinite dimensional.
For our purposes, we find an equivalent formulation which utilizes Eq. \eqref{eq:splitindex2} to express the content of Eqs. \eqref{eq:modanticommg5pt1} and \eqref{eq:modanticommg5pt2} as
\begin{align}
\label{eq:finalanticommg5pt1}
    \left\{\bar{\gamma}_\mu,\gamma_5\right\} &= 0\\
\label{eq:finalanticommg5pt2}
    \mathrm{and}\qquad \left[\hat{\gamma}_\mu,\gamma_5\right] &= 0\,,
\end{align}
where
\begin{align}
\label{eq:mixedanticomm}
    \left\{\bar{\gamma}^\mu,\hat{\gamma}^\nu\right\} = 0\,,
\end{align}
to be much more transparent and practically useful.

As advertised above, this modification of the anticommutation relation has an immediate impact on the form of the Standard Model Feynman rules in dimensional regularization. As an example of particular relevance to this work, let us consider the axial part of the coupling of the $Z$ boson to the light quarks and leptons in four dimensions:
\begin{align}
\label{eq:Zint}
 - e\, a_f Z^\mu  \bar{\psi}_f\, \gamma_\mu \gamma_5\, \psi_f\,,
\end{align}
where $a_f$ is a function of the electroweak gauge boson masses and the isospin of fermion $f$.
Suppose now that we wish to consider interaction \eqref{eq:Zint} in $d$ dimensions. Due to the fact that we can no longer avail ourselves of Eq. \eqref{eq:anticommg5}, \eqref{eq:Zint} will {\it fail} to be Hermitian in $d$ dimensions. We can, however, preserve the Hermiticity of the interaction if we first rewrite \eqref{eq:Zint} as
\begin{align}
\label{eq:modZint}
 - e\, a_f Z^\mu  \bar{\psi}_f\, \frac{1}{2}\left[\gamma_\mu, \gamma_5\right] \psi_f
\end{align}
in four dimensions {\it before} giving up Eq. \eqref{eq:anticommg5} in favor of Eqs. \eqref{eq:modanticommg5pt1} and \eqref{eq:modanticommg5pt2}  \cite{Collins:1984xc}.

HVBM's $\gamma_5$ scheme allows for a very familiar treatment of the usual families of Dirac traces. As pointed out already in \cite{tHooft:1972tcz}, the trace of an even number of Dirac matrices with $d$ dimensional indices may be conveniently evaluated using the recursive formula
\begin{align}
\label{eq:eventracerecurse}    
    \mathrm{tr}\left\{\gamma_{\nu_1}\cdots \gamma_{\nu_n}\right\} = \sum_{i = 1}^{n-1} (-1)^{i+1} g_{\nu_1 \nu_{i+1}}\mathrm{tr}\left\{\gamma_{\nu_2}\cdots\gamma_{\nu_i} \gamma_{\nu_{i+2}}\cdots \gamma_{\nu_n}\right\},
\end{align}
valid for even $n$, with termination criterion
\begin{align}
    \label{eq:eventracetermcrit}
    \mathrm{tr}\left\{\mathbf{1}\right\} = 4\,.
\end{align}
Eq. \eqref{eq:eventracerecurse} is derived in a manner very similar to that of Eq. \eqref{eq:genDiraccontract}, but by repeatedly applying Eq. \eqref{eq:Diracalgebra} to move the $\gamma_{\nu_1}$ factor all the way to the right. Finally, the cyclic property of the trace is required to identify the final term, $-\mathrm{tr}\left\{\gamma_{\nu_2}\cdots \gamma_{\nu_n}\gamma_{\nu_1}\right\}$, with the original expression on the left-hand side.
 
The trace
\begin{align}
    \mathrm{tr}\left\{\gamma_{\nu_1}\cdots \gamma_{\nu_n}\gamma_5\right\}
\end{align}
for even $n$ may be evaluated in much the same way by replacing $\gamma_5$ by the covariant version of its definition,
\begin{align}
\label{eq:covdefg5}
    \gamma_5 = -\frac{i}{4!}\varepsilon^{\mu \nu \rho \sigma}\gamma_\mu \gamma_\nu\gamma_\rho\gamma_\sigma\,,
\end{align}
and then evaluating the resulting trace of $n + 4$ Dirac matrices using Eqs. \eqref{eq:eventracerecurse} and \eqref{eq:eventracetermcrit}.

As expected, the trace of an odd number of Dirac matrices is still zero in $d$ dimensions. As mentioned above, this can be proven by induction on the trace of Eq. \eqref{eq:genDiraccontract}. Let us consider the argument for $n = 1$. On the one hand, by trace cyclicity and Eq. \eqref{eq:basicDiraccontract}, we have
\begin{align}
\label{eq:oddproof1}
    \mathrm{tr}\left\{\gamma^\mu \gamma_{\nu_1}\gamma_\mu\right\} = (4 - 2\ep) \mathrm{tr}\left\{ \gamma_{\nu_1}\right\}.
\end{align}
On the other hand, applying Eq. \eqref{eq:Diracalgebra} before Eq. \eqref{eq:basicDiraccontract}, we have
\begin{align}
\label{eq:oddproof2}
    \mathrm{tr}\left\{\gamma^\mu \gamma_{\nu_1}\gamma_\mu\right\} = (-2 + 2\ep) \mathrm{tr}\left\{ \gamma_{\nu_1}\right\}.
\end{align}
Subtracting Eqs. \eqref{eq:oddproof1} and \eqref{eq:oddproof2}, we see that, indeed, $\mathrm{tr}\left\{ \gamma_{\nu_1}\right\}$ must be zero.

Now, trace cyclicity and Eq. \eqref{eq:basicDiraccontract} generically reduce the trace of the left-hand side of Eq. \eqref{eq:genDiraccontract}:
\begin{align}
\label{eq:oddproof3}
    \mathrm{tr}\left\{\gamma^\mu \gamma_{\nu_1}\cdots \gamma_{\nu_n}\gamma_\mu\right\} = (4 - 2\ep) \mathrm{tr}\left\{ \gamma_{\nu_1}\cdots \gamma_{\nu_n}\right\}.
\end{align}
However, we must simplify the trace of the summand of the first term on the right-hand side of Eq. \eqref{eq:genDiraccontract} using the induction hypothesis before we can finish the proof: it is easy to see that
\begin{align}
\label{eq:keystep}
    \mathrm{tr}\left\{\gamma_{\nu_i} \gamma_{\nu_1}\cdots\gamma_{\nu_{i-1}}\gamma_{\nu_{i+1}}\cdots\gamma_{\nu_n}\right\} = (-1)^{i-1}\mathrm{tr}\left\{\gamma_{\nu_1}\cdots\gamma_{\nu_n}\right\}
    \quad\text{(odd $n$)}
\end{align}
is a valid identity under the assumption $\mathrm{tr}\left\{\gamma_{\nu_1}\cdots \gamma_{\nu_{n-2}}\right\} = 0$. Eq. \eqref{eq:keystep} provides the desired reduction of the right-hand side of the trace of Eq. \eqref{eq:genDiraccontract},
\begin{align}
\label{eq:oddproof4}
    \mathrm{tr}\left\{\gamma^\mu \gamma_{\nu_1}\cdots \gamma_{\nu_n}\gamma_\mu\right\} &=  2 \sum_{i = 1}^n \mathrm{tr}\left\{ \gamma_{\nu_1}\cdots \gamma_{\nu_n}\right\}-(4-2\ep)\mathrm{tr}\left\{ \gamma_{\nu_1}\cdots \gamma_{\nu_n}\right\}\nonumber \\
    &= (2 n-4 + 2\ep)\mathrm{tr}\left\{ \gamma_{\nu_1}\cdots \gamma_{\nu_n}\right\} 
    \quad\text{(odd $n$)}. 
\end{align}
Finally, the difference of Eqs. \eqref{eq:oddproof3} and \eqref{eq:oddproof4} imply that
\begin{align}
\label{eq:oddDiraciszero}
    \mathrm{tr}\left\{\gamma_{\nu_1}\cdots \gamma_{\nu_n}\right\} = 0
    \quad\text{(odd $n$)}
\end{align}
by induction.

Similarly, Eq. \eqref{eq:covdefg5} and the line of reasoning used to establish Eq. \eqref{eq:oddDiraciszero} together reveal that
\begin{align}
\label{eq:oddoddDiraciszero}
    \mathrm{tr}\left\{\gamma_{\nu_1}\cdots \gamma_{\nu_n}\gamma_5\right\} = 0
    \quad\text{(odd $n$).}
\end{align}

HVBM's prescription for $\gamma_5$ has the very unfortunate drawback that the Standard Model's chiral symmetry is violated in all bare perturbative calculations involving chiral couplings. As was originally pointed out by Breitenlohner and Maison in \cite{Breitenlohner:1977hr,Breitenlohner:1975hg,Breitenlohner:1976te} and reviewed in \cite{Collins:1984xc}, it is possible to {\it restore} the chiral symmetry of the Standard Model as part of the renormalization program by introducing various finite counterterms on top of the usual divergent ones. Let us emphasize here that, in general, one expects {\it all} interactions in the Standard Model to require finite counterterms at some order in perturbation theory, not just those which explicitly involve $\gamma_5$. Furthermore, these finite counterterms must generically be calculated beyond $\mathcal{O}\left(\epsilon^0\right)$ to consistently restore chiral symmetry at still higher orders. We discuss some elementary examples relevant to this article in Section \ref{sec:finiteren}.

For what concerns the particular one- and two-loop calculations considered later on in this work, {\it Larin's principle} \cite{Larin:1993tq} allows us to bypass many finite counterterm computations for fermion self-energies and vertex form factors involving chiral couplings. Larin pointed out that, for these building blocks, chiral symmetry implies the finite renormalizations do nothing but turn results obtained in HVBM's $\gamma_5$ scheme into analogous results obtained in an anticommuting $\gamma_5$ scheme.
However, as we will discuss in greater detail in Section \ref{sec:finiteren}, the HVBM scheme requires the insertion of explicit finite counterterms and a careful treatment also of the higher-order-in-$\epsilon$ terms if box-type Feynman diagrams with chiral couplings contribute to the two-loop or higher-order corrections of interest (see also \cite{Ahmed:2020kme} for a related recent study). Overall, one comes away with the feeling that there must exist a pure dimensional regularization scheme better suited to the study of higher-order perturbative corrections in the Standard Model. In fact, the arduous and subtle character of perturbative calculations carried out in HVBM's $\gamma_5$ scheme is what led to the development of Kreimer's prescription for $\gamma_5$.
\subsection{Kreimer's prescription for $\gamma_5$ in dimensional regularization}
\label{sec:Kreimerdefs}
In Kreimer's $\gamma_5$ scheme, chiral couplings are treated in a consistent way in $d$ dimensions by retaining the four-dimensional definition of $\gamma_5$, Eq. \eqref{eq:defg5}, at the expense of Dirac trace cyclicity. Kreimer argued in \cite{Kreimer:1989ke} that Dirac trace cyclicity is {\it not} a natural property to retain in $d$ dimensions, due to the fact that the Dirac algebra necessarily becomes infinite dimensional. The standard trace of classical linear algebra, $\mathrm{tr}$, is replaced in Kreimer's $\gamma_5$ scheme with one of several possible exotic trace operations which differ only in their {\it reading point prescriptions}.
The reading point prescription effectively specifies with which Dirac matrix to start the non-cyclic trace.
In our case, the Lorentz projectors we define in Section \ref{sec:Kreimercalcs} provide a natural anchor to start all Dirac traces involving the external fermion lines. In order to clearly distinguish the trace operations in what follows, we will write Tr rather than tr when using a reading point prescription.
While not relevant to the specific higher-order calculations discussed in this paper,\footnote{As explained in Section \ref{sec:process}, we defer the calculation of contributions proportional to the number of light fermion flavors, as well as all top quark mass corrections, to future work in order to avoid unnecessary technical complications which have no bearing on the primary objective of our proof-of-concept study.} we would average over all possible choices of first entry for internal closed fermion loops.  Other reading point prescriptions, such as one which mandates an average over all possible choices of first entry in all cases, could have been chosen as well, but we find our choice particularly convenient. 

Kreimer's $\gamma_5$ scheme has practical advantages over HVBM's $\gamma_5$ scheme. Crucially, it does {\it not} force a change of algebraic form for any Standard Model Feynman rule when considering interactions in $d$ spacetime dimensions. Employing a consistent reading point prescription then {\it automatically} preserves the chiral symmetry of the Standard Model in bare perturbative calculations and the complete absence of finite counterterms makes renormalization straightforward. In addition, although box-type diagrams do ultimately force the introduction of split loop momenta also in Kreimer's $\gamma_5$ scheme, the bulk of the numerator algebra can be carried out {\it before} the split enters. This stands in stark contrast to HVBM's $\gamma_5$ scheme, where split loop momenta should be introduced immediately in order to progress with the necessary algebraic simplifications efficiently (this assertion will be clarified in Section \ref{sec:HVBMcalcs}). 

As alluded to above, the standard families of odd Dirac traces still vanish identically when $\mathrm{tr}$ is replaced by $\mathrm{Tr}$,
\begin{align}
    \mathrm{Tr}\left\{\gamma_{\nu_1}\cdots \gamma_{\nu_n}\right\} &= 0\\
   \mathrm{and}\qquad  \mathrm{Tr}\left\{\gamma_{\nu_1}\cdots \gamma_{\nu_n}\gamma_5\right\} &= 0\,.
\end{align}
For the standard families of even Dirac traces, \cite{Korner:1991sx} defines
\begin{align}
\label{eq:eventraceKreimer}
    \mathrm{Tr}\left\{\gamma_{\nu_1}\cdots \gamma_{\nu_{2 n}}\right\} &= 4 \sum_{\sigma} (-1)^{\mathrm{sgn}(\sigma)} g_{\nu_{i_1} \nu_{j_1}} \cdots g_{\nu_{i_n} \nu_{j_n}} ~~~~\mathrm{and}\\
    \label{eq:oddtraceKreimer}
    \mathrm{Tr}\left\{\gamma_{\nu_1}\cdots \gamma_{\nu_{2 n}}\gamma_5\right\} &=
- \frac{i}{3!} ~\varepsilon_{\nu_{2 n+1} \nu_{2 n+2} \nu_{2 n+3} \nu_{2 n+4}} \sum_{\sigma} (-1)^{\mathrm{sgn}(\sigma)} g_{\nu_{i_1} \nu_{j_1}} \cdots g_{\nu_{i_{n+2}} \nu_{j_{n+2}}}\,,
\end{align}
where $\sigma$ is the set of permutations of the $2 m$ indices which appear in the sums on the right-hand sides, subject to the restrictions
\begin{equation}
1 = i_1 < ... < i_m,\quad i_k < j_k,\quad\text{for all~}k \in \{1...m\}. \end{equation}
As is perhaps not immediate, these definitions are equivalent to the results one would obtain in HVBM's $\gamma_5$ scheme recursively using $\mathrm{tr}$ and Eqs. \eqref{eq:eventracerecurse}, \eqref{eq:eventracetermcrit}, and \eqref{eq:covdefg5}. Of course, Eqs. \eqref{eq:eventracerecurse} and \eqref{eq:eventracetermcrit} must ultimately yield the same answer as Eq. \eqref{eq:eventraceKreimer} above, as a string of plain Dirac matrices knows nothing about $\gamma_5$. As a corollary, due to Eq. \eqref{eq:covdefg5}, one immediately arrives at the same conclusion for the Kreimer trace of a string of Dirac matrices with a $\gamma_5$ factor appended (see also \cite{Kreimer:1993bh} for a supplementary discussion).

Despite the fact that Kreimer's $\gamma_5$ scheme was introduced a long time ago \cite{Kreimer:1989ke,Korner:1991sx,Kreimer:1992,Kreimer:1993bh}, it has not been universally accepted and embraced as a superior alternative to HVBM's $\gamma_5$ scheme; researchers continue to express doubts about the applicability of Kreimer's $\gamma_5$ scheme beyond one loop (see {\it e.g.} \cite{Denner:2019vbn} for a current review of the status of the $\gamma_5$ problem).
In this article, we present an explicit application of Kreimer's $\gamma_5$ scheme to a non-trivial calculation of two-loop box-type diagrams involving chiral couplings and show that it \emph{does} yield physically sensible results. Also worth mentioning is concurrent work by one of us and others, where the two-loop box contributions to $gg\to ZZ$ featuring a closed top-quark loop have been calculated in Kreimer's $\gamma_5$ scheme \cite{Agarwal:2020dye}. The results were compared against calculations in a more commonly used $\gamma_5$ scheme and found to agree after ultraviolet and infrared subtraction. While the exact top mass corrections to $gg\to ZZ$ in next-to-leading order QCD couple the $ZZ$ pair to the massive top quark line and involve anomalous double triangle contributions, the $\gamma_5$ dependence of the mixed two-loop EW-QCD corrections to the Drell-Yan process at the level of Feynman diagrams is significantly more complicated.

\subsection{$d$-dimensional Feynman integrals with four-dimensional scalar numerators}
\label{sec:muterms}
In this section, we describe how we deal with particularly awkward terms which often occur in the numerators of $d$-dimensional Feynman integrals in Standard Model perturbation theory: four-dimensional contractions of $d$-dimensional loop momenta, $\bar{g}_{\mu \nu} k_i^\mu k_j^\nu$. As mentioned above, we decompose our loop momenta into four-dimensional and $(-2\epsilon)$-dimensional components, $k_i = \bar{k}_i+\hat{k}_i$. This decomposition maps the awkward four-dimensional numerators to ordinary inverse propagators and ``$\mu$-term'' insertions of the form $\hat{k}_i \cdot \hat{k}_j$, as introduced long ago by Bern and Morgan \cite{Bern:1995db}. Below, we review algorithms which allow for a direct treatment of $\mu$-term insertions at one and two loops with dimensionally-shifted Feynman integrals. While the two-loop algorithm we review was, to our knowledge, the first to appear in the literature \cite{Bern:2002tk}, other approaches such as that of \cite{Boughezal:2017tdd} are certainly possible. It seems that some computational effort is simply unavoidable, as both approaches require the calculation of auxiliary integration by parts reductions \cite{Tkachov:1981wb,Chetyrkin:1981qh,Laporta:2001dd}.\footnote{Note that one can circumvent the discussion of this section altogether by performing a classical Passarino-Veltman tensor reduction \cite{Passarino:1978jh} on all one- and two-loop integrals. In fact, we found it very productive to carry out such an analysis as a comprehensive cross-check of our $\mu$-term implementation.} We found it very appealing to work with dimensionally-shifted Feynman integrals, as all of the necessary auxiliary integration by parts identities were already calculated by us some time ago for numerical cross-checks on our master integrals \cite{vonManteuffel:2017myy,Heller:2019gkq}, facilitated by passing to a finite basis of master integrals \cite{vonManteuffel:2014qoa,vonManteuffel:2015gxa} through Tarasov dimension shifts \cite{Tarasov:1996br}.

As has been clear at least since the appearance of \cite{Bern:1995db}, non-zero one-loop Feynman integrals with $\mu$-term insertions are of the form
\begin{align}
\label{eq:oneloopmu}
    \mathcal{I}^{(1)}_n\left[(\hat{k}_1 \cdot \hat{k}_1)^r\right] \equiv \int \frac{\mathrm{d}^d k_1}{i \pi^{d/2}} \frac{\left(\hat{k}_1 \cdot \hat{k}_1\right)^r}{D_1^{\nu_1}\cdots D_n^{\nu_n}}\,,
\end{align}
where the $D_i$ are the independent propagators of the integral family. Crucially, Eq. \eqref{eq:oneloopmu} can be rewritten in terms of standard dimensionally-shifted one-loop Feynman integrals with the same propagator structure. By absorbing the $(\hat{k}_1 \cdot \hat{k}_1)^r$ numerator insertion into the integration measure, it is shown in \cite{Bern:1995db} that 
\begin{align}
\label{eq:oneloopmusimp}
    \mathcal{I}^{(1)}_n\left[(\hat{k}_1 \cdot \hat{k}_1)^r\right] = (-1)^r\frac{\Gamma(r-\epsilon)}{\Gamma(-\epsilon)}\int \frac{\mathrm{d}^{d+2 r}k_1}{i \pi^{(d+2 r)/2}} \frac{1}{D_1^{\nu_1}\cdots D_n^{\nu_n}}\,.
\end{align}

Although Eq. \eqref{eq:oneloopmusimp} involves dimensionally-shifted integrals in an essential way, its form does not depend on the number of legs, the complexity of the kinematics, or the details of the $\nu_i$, observations which suggest a connection to the dimension shift of Tarasov \cite{Tarasov:1996br}. Since, to our knowledge, a manipulation of the Schwinger representation furnishes the simplest possible proof of Tarasov's relation, it is natural to expect the Schwinger representation to also allow for an efficient alternative proof of Eq. \eqref{eq:oneloopmusimp}. To proceed, it is useful to recall that the Tarasov dimension shift is derived in the Schwinger parametrization by simply multiplying and dividing by the first Symanzik polynomial. Furthermore, for an integral topology with $t$ propagator denominators, the first Symanzik polynomial at one loop has the especially simple, universal form
\begin{align}
   \mathcal{U}^{(1)}_t = \alpha_1 + \cdots + \alpha_t\,,
\end{align}
due to the uniqueness and triviality of the one-loop vacuum (tadpole) integral topology.

The general Schwinger representation is most simply expressed if one first orders the Feynman propagators such that the $\epsilon \to 0$ limits of their exponents form a monotonically decreasing sequence. We assume that, in this limit, the first $n^+$ propagators appear raised to positive powers and the remaining $n^-$ propagators appear raised to negative powers (we may assume without loss of generality that the final $n^-$ propagator exponents are in fact non-positive integers). In terms of the Symanzik polynomials of the integral family, $\overline{\mathcal{U}}$ and $\overline{\mathcal{F}}$, we have
\begin{align}
\label{eq:genSchwinger}
    \mathcal{I}^{(L)}_n &= e^{-i \pi \sum_{j = 1}^{n^+} \nu_j}\prod_{i = 1}^{~n^+} \left(\int_0^\infty \mathrm{d}\alpha_i \frac{\alpha_i^{\nu_i - 1}}{\Gamma(\nu_i)}\right)\times
    \nonumber \\
    &\quad\times \prod_{\ell = n^+ + 1}^{~n} \frac{\partial^{|\nu_\ell|}}{\partial \alpha_\ell^{|\nu_\ell|}}\left(\overline{\mathcal{U}}^{\,-d_0/2+\epsilon}e^{-\overline{\mathcal{F}}/\overline{\mathcal{U}}}\right)\Bigg|_{\alpha_{n^+ + 1} = \cdots = \alpha_{\scriptscriptstyle n^++n^-} = 0}
\end{align}
for $d = d_0 - 2\epsilon$, where $d_0$ is an even positive integer ($d_0 = 4$ is of course the main case of interest to us in this article). In the absence of inverse propagators, the usual result given in terms of the Symanzik polynomials of the integral topology under consideration, $\mathcal{U}$ and $\mathcal{F}$, is recovered.

Now, as emphasized in \cite{Bern:1995db}, generic $d$-dimensional Feynman propagators at one loop have the form $\bar{k}_1 \cdot \bar{k}_1 + 2 P_{ext} \cdot \bar{k}_1 + P_{ext}\cdot P_{ext} - m^2 + \hat{k}_1 \cdot \hat{k}_1$, where $P_{ext}$ denotes some sum of four-dimensional external momenta and $m$ some particle mass. This form motivates a split of the loop momentum integral in the derivation of the ordinary Schwinger parametrization into four-dimensional and $(-2\epsilon)$-dimensional pieces. Due to the fact that, by construction, $\mu$-term insertions are expected to lead to Feynman integrals explicitly suppressed by powers of $\epsilon$, the authors of \cite{Bern:2002tk} deduced that the $(-2\epsilon)$-dimensional part of the momentum integral must be proportional to the $\epsilon$-dependent part of Eq. \eqref{eq:genSchwinger},
\begin{align}
\label{eq:genfunconeloop}
    \int \mathrm{d}^{-2\epsilon}\hat{k}_1\, e^{\overline{\mathcal{U}}^{(1)}_n \,\hat{k}_1 \cdot \hat{k}_1} \propto \left(\overline{\mathcal{U}}^{(1)}_n\right)^\epsilon,
\end{align}
and therefore serve as a generating function for all $\mu$-term insertions via differentiation with respect to $\overline{\mathcal{U}}^{(1)}_n$. Indeed, by differentiating the right-hand side of the above relation $r$ times with respect to $\overline{\mathcal{U}}^{(1)}_n$, we find 
\begin{align}
    (-1)^r\frac{\Gamma(r-\epsilon)}{\Gamma(-\epsilon)}\left(\overline{\mathcal{U}}^{(1)}_n\right)^{\epsilon-r},
\end{align}
and, by recognizing that the shift $\epsilon \to \epsilon - r$ corresponds precisely to a shift $d_0 \to d_0 + 2 r$ in Eq. \eqref{eq:genSchwinger}, immediately recover the form of Eq. \eqref{eq:oneloopmusimp} as expected. 

\begin{figure}[!t]
\centering
\includegraphics[height=20mm]{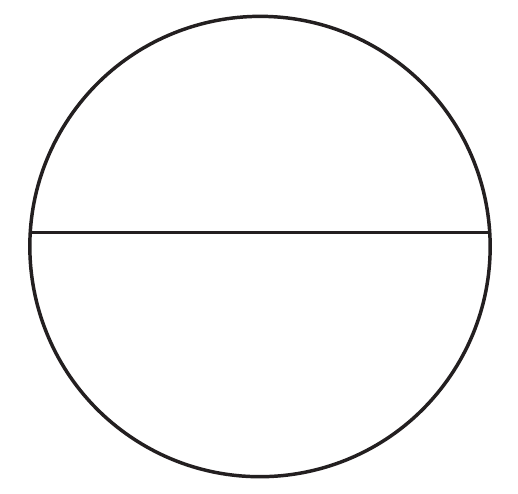} \qquad \qquad \includegraphics[height=20mm]{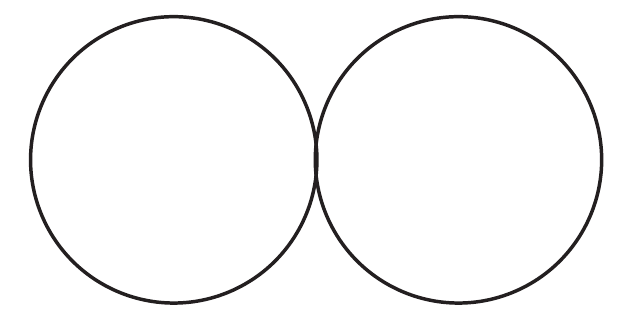}
\caption{The unique (up to internal mass assignment) non-factorizable (left) and factorizable (right) two-loop vacuum integral topologies. 
}
\label{fig:vactop}
\end{figure}

Fortunately, the approach described above may be fruitfully applied to the study of generic $\mu$ integrals at the two-loop level as well. This is primarily because there is a unique non-factorizable two-loop tadpole topology (see Figure \ref{fig:vactop}). The main new feature at the two-loop level is that shifts in the propagator exponents appear alongside the various dimension shifts. At two loops, generic non-zero $\mu$-term insertions can be constructed by taking products of $\hat{k}_1 \cdot \hat{k}_1$, $\hat{k}_2 \cdot \hat{k}_2$, and $(\hat{k}_1-\hat{k}_2)^2$. Due to the uniqueness of the non-factorizable two-loop tadpole topology, the first Symanzik polynomial of an arbitrary non-factorizable two-loop integral family is of the form
\begin{align}
    \overline{\mathcal{U}}^{(2)}_n = T_{k_1^2}T_{k_1\cdot k_2} + T_{k_2^2}T_{k_1\cdot k_2} + T_{k_1^2}T_{k_2^2},
\end{align}
where $T_{k_1\cdot k_2}$ is the sum of the Schwinger parameters associated to propagators which depend on the dot product $k_1\cdot k_2$, $T_{k_1^2}$ is the sum of the Schwinger parameters associated to propagators which depend on $k_1^2$ but not on $k_1 \cdot k_2$, and $T_{k_2^2}$ is the sum of the Schwinger parameters associated to propagators which depend on $k_2^2$ but not on $k_1 \cdot k_2$. It follows that the two-loop analog of Eq. \eqref{eq:genfunconeloop} ought to be
\begin{align}
\label{eq:genfunctwoloops}
    \int \mathrm{d}^{-2\epsilon}\hat{k}_1\, \mathrm{d}^{-2\epsilon}\hat{k}_2\, e^{T_{k_1^2} \,\hat{k}_1 \cdot \hat{k}_1+T_{k_2^2} \,\hat{k}_2 \cdot \hat{k}_2+T_{k_1\cdot k_2} \,(\hat{k}_1-\hat{k}_2)^2} \propto \left(\overline{\mathcal{U}}^{(2)}_n\right)^\epsilon,
\end{align}
a generating function via differentiation by $T_{k_1^2}$, $T_{k_2^2}$, and $T_{k_1\cdot k_2}$. As at one-loop, the above relation suggests that arbitrary two-loop $\mu$ integrals may be rewritten as a linear combination of ordinary Feynman integrals by simply comparing the result of each term of the prescribed differentiation of the right-hand side of \eqref{eq:genfunctwoloops} with the form of Eq. \eqref{eq:genSchwinger}. 

Starting at two loops, a subtle complication can arise in our formulation due to the fact that we prefer to write all numerator polynomials as linear combinations of power products of inverse propagators and particle masses. If a propagator depending on {\it e.g.} $k_1$ alone appears in the numerator of a Feynman integral alongside an explicit insertion of $\hat{k}_1 \cdot \hat{k}_1$, it will not work to use the above line of reasoning to eliminate the explicit $\hat{k}_1 \cdot \hat{k}_1$ factor. This is simply because the $k_1^2 = \bar{k}_1\cdot \bar{k}_1+\hat{k}_1\cdot \hat{k}_1$ term of the propagator itself depends on $\hat{k}_1\cdot \hat{k}_1$. In practice, we found it most convenient to temporarily delay rewriting all scalar products of the form $k_i \cdot p_j$ in terms of inverse propagators. Due to the fact that the $k_i \cdot p_j$ scalar products live in four dimensions, \eqref{eq:genfunctwoloops} may be applied to eliminate all explicit $\mu$ terms \textit{before} rewriting the remaining scalar products as linear combinations of propagators.

To clarify the procedure described above, we consider a non-trivial two-loop $\mu$ integral from integral family A of \cite{vonManteuffel:2017myy,Heller:2019gkq} which may be directly rewritten using the technology of \cite{Bern:2002tk}: 
\begin{align}
\label{eq:sampleintdef}
    &\includegraphics[valign=m,scale=0.4]{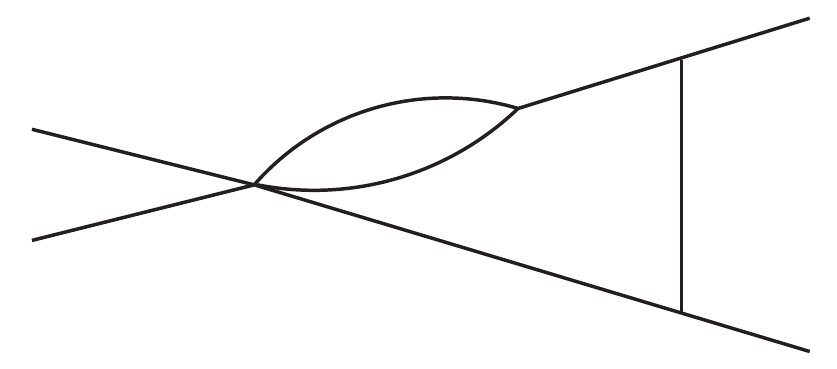}\left[D_7\, (\hat{k}_1 \cdot \hat{k}_1)^2\right] 
     \notag\\
    &\qquad \qquad \qquad= \int \frac{\mathrm{d}^d k_1\mathrm{d}^d k_2}{\left(i \pi^{d/2}\right)^2}\frac{(k_2-p_1-p_2)^2(\hat{k}_1 \cdot \hat{k}_1)^2}{k_1^2 k_2^2 (k_1-k_2)^2 (k_1-p_1)^2 (k_1-p_1-p_2)^2}\,,
\end{align}
where we have introduced the notation $D_7 \equiv (k_2-p_1-p_2)^2$ from \cite{vonManteuffel:2017myy,Heller:2019gkq} on the left-hand side for later convenience. Note that, in this example, there is no clash between the explicit factor of $(\hat{k}_1 \cdot \hat{k}_1)^2$ in Eq. \eqref{eq:sampleintdef} and the implicit factor of $\hat{k}_2\cdot \hat{k}_2$ present in
\begin{equation}
    D_7 = \bar{k}_2\cdot\bar{k}_2 - 2 (p_1+p_2)\cdot \bar{k}_2 + s + \hat{k}_2\cdot \hat{k}_2\,.
\end{equation}
The Schwinger parameter subsets of interest are
\begin{align}
   T_{k_1^2} = \alpha_1 + \alpha_4 + \alpha_6, \quad T_{k_2^2} = \alpha_2 + \alpha_7,~\mathrm{and}\quad T_{k_1 \cdot k_2} = \alpha_3\,,
\end{align}
and differentiating the right-hand side of Eq. \eqref{eq:genfunctwoloops} twice with respect to $T_{k_1^2}$ gives $\epsilon (\epsilon - 1)$ times an insertion of
\begin{align}
    \frac{(T_{k_2^2}+T_{k_1\cdot k_2})^2}{\left(\overline{\mathcal{U}}^{(2)}_6\right)^2} = \frac{\alpha_2^2+\alpha_3^2+\alpha_7^2+2\alpha_2\alpha_3+2\alpha_2 \alpha_7+2\alpha_3 \alpha_7}{\left(\overline{\mathcal{U}}^{(2)}_6\right)^2}
\end{align}
into Eq. \eqref{eq:genSchwinger} specialized to our two-loop example. 

The factor in the denominator induces the dimension shift $4 - 2\epsilon \to 8 - 2\epsilon$. Powers of Schwinger parameters associated to the propagator denominators induce positive shifts in the corresponding propagator exponents, and, for $h$ powers of parameter $i$, an overall factor of $(-1)^h \Gamma(\nu_i + h)/\Gamma(\nu_i)$. Powers of the Schwinger parameter associated to the numerator also induce positive shifts in the exponent, but only an overall factor of $h!$ from the action of the derivatives in Eq. \eqref{eq:genSchwinger}. Due to the fact that $\alpha_7$ is merely an auxiliary parameter set to zero at the end of the calculation, the insertion of $\alpha_7^2$ gives a vanishing contribution for the specific example considered. We find:
\begin{align}
\label{eq:sampleintrewrite}
    &\includegraphics[valign=m,scale=0.4]{samplemuint}\hspace{-.5 cm}\left[D_7\, (\hat{k}_1 \cdot \hat{k}_1)^2\right] =2\epsilon(\epsilon-1)\left[\includegraphics[valign=m,scale=0.4]{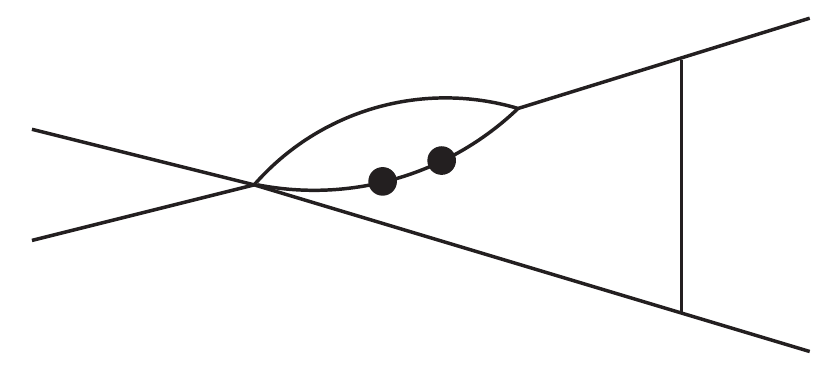}^{8-2\epsilon}\hspace{-.5 cm}
    \left[D_7\right]\right.
    \nonumber \\
    &\left.\qquad \qquad+\includegraphics[valign=m,scale=0.4]{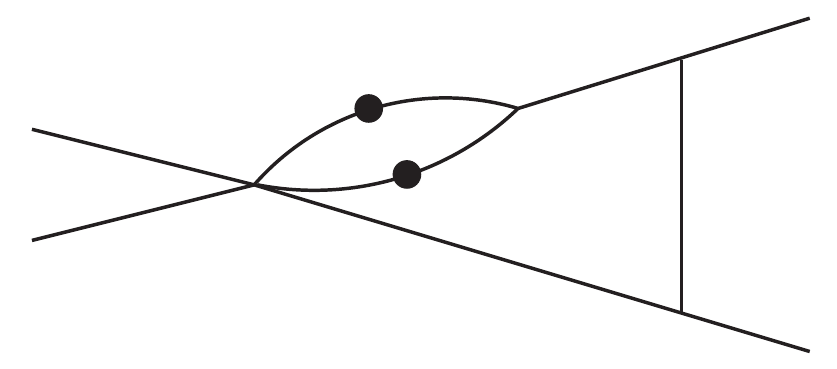}^{8-2\epsilon}\hspace{-.5 cm}\left[D_7\right]+\includegraphics[valign=m,scale=0.4]{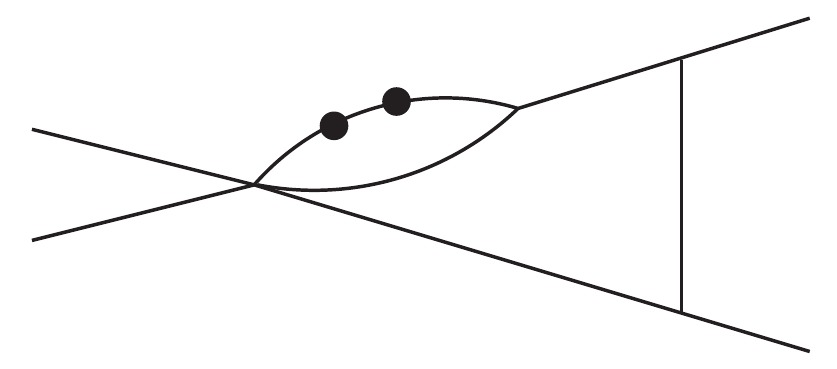}^{8-2\epsilon}\hspace{-.5 cm}\left[D_7\right]\nonumber\right.
     \\
    &\left.\qquad \qquad-\includegraphics[valign=m,scale=0.4]{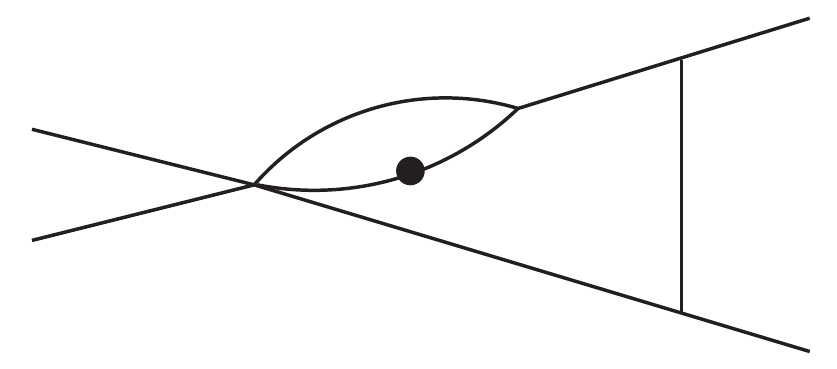}^{8-2\epsilon}-\includegraphics[valign=m,scale=0.4]{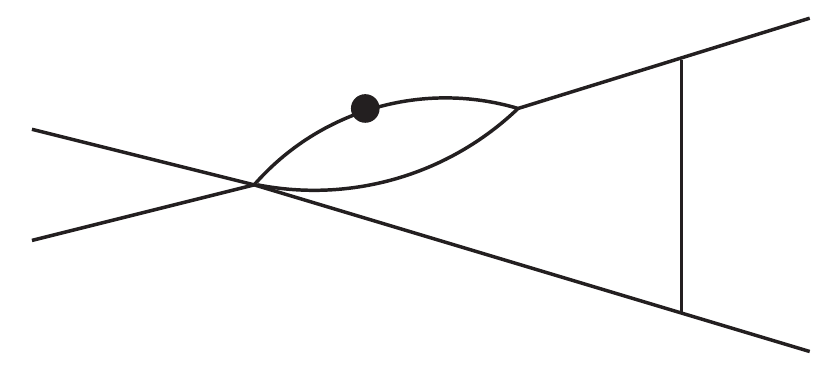}^{8-2\epsilon}\right].
\end{align}

The integrals on the right-hand side of Eq. \eqref{eq:sampleintrewrite} can be straightforwardly evaluated in Feynman parameters to all orders in $\epsilon$ due to the fact that they are iteratively one loop. After some simplifications, we obtain
\begin{align}
\label{eq:sampleintres}
    &\includegraphics[valign=m,scale=0.4]{samplemuint}\left[D_7\, (\hat{k}_1 \cdot \hat{k}_1)^2\right]
    \notag\\
    &\qquad \qquad \qquad=\frac{(5-2\epsilon)\Gamma^2(\epsilon)\Gamma^2(1-\epsilon)\Gamma(2\epsilon)\Gamma(1-2\epsilon)\Gamma(2-\epsilon)}{\Gamma(6-3\epsilon)\Gamma(3-2\epsilon)\Gamma^2(-1+\epsilon)}.
\end{align}
To check Eq. \eqref{eq:sampleintres} and our treatment of the two-loop $\mu$ terms, we invite the reader to first integrate out $k_2$ in integral \eqref{eq:sampleintdef} above and then treat the $\mu$-term insertion in the context of the remaining integration over $k_1$ ({\it i.e.} by applying Eq. \eqref{eq:oneloopmusimp}). Finally, let us stress again that a numerator insertion of $D_7 (\hat{k}_2 \cdot \hat{k}_2)^2$ would {\it not} be directly covered by our method.

To be clear, factorizable two-loop $\mu$ integrals also appear in our calculations. They could be evaluated as products of one-loop $\mu$ integrals, but we prefer to treat them along the lines described above. In the factorizable case, $(\hat{k}_1-\hat{k}_2)^2 \to \hat{k}_1 \cdot \hat{k}_1 + \hat{k}_2 \cdot \hat{k}_2$ and we have
\begin{align}
\label{eq:genfunctwoloopsfact}
    \int \mathrm{d}^{-2\epsilon}\hat{k}_1\, \mathrm{d}^{-2\epsilon}\hat{k}_2\, e^{T_{k_1^2} \,\hat{k}_1 \cdot \hat{k}_1+T_{k_2^2} \,\hat{k}_2 \cdot \hat{k}_2} \propto \left(T_{k_1^2}T_{k_2^2}\right)^\epsilon
\end{align}
for our generating function. This is nothing but a degenerate case of Eq. \eqref{eq:genfunctwoloops} above with $T_{k_1 \cdot k_2}$ set to zero. Note that, while we believe that it should be possible to make a similar construction to handle $\mu$-term insertions in $d$-dimensional Feynman integrals at higher loops, more effort would be required to identify the underlying vacuum topologies.
\subsection{$\overline{\mathrm{MS}}$ renormalization of $\alpha$ and $\alpha_s$}
\label{sec:couplingren}
In this work, we renormalize both $\alpha$ and $\alpha_s$ in the $\overline{\mathrm{MS}}$ scheme. Due to our neglect of all fermion masses, this choice is particularly convenient for our calculation of the two-loop mixed EW-QCD corrections to the neutral-current Drell-Yan process. As explained in greater detail in Section \ref{sec:process}, we neglect the gauge-invariant subset of Feynman diagrams with internal closed fermion loops throughout this work for simplicity. This includes in particular the two-loop Standard Model vacuum polarization diagrams featuring a single quark loop and a single gluon exchange. As a corollary, the form of the contributing two-loop Feynman diagrams (see Section \ref{sec:2Ldiags}) immediately implies that the order $\alpha \alpha_s$ corrections to the electric charge renormalization constant must vanish identically. 

Throughout this work, we follow the convention of \cite{Kilgore:2011pa} for fixed-order results. Therefore, quantities in Standard Model perturbation theory are written in the form
\begin{align}
\label{eq:relativenotation}
  \mathbf{Z} =  \sum_{m,n = 0}^\infty \mathbf{Z}^{(m,n)} \left(\frac{\alpha}{4\pi}\right)^m\left(\frac{\alpha_s}{4\pi}\right)^n
\end{align}
to all orders in the coupling constants.
In this notation, neglecting contributions with closed fermion loops which are proportional to the number of light leptons ($n_l$) or light quarks ($n_q$) or involve the mass of the top quark ($m_t$), we have
\begin{align}
\label{eq:2Lchargerenorm}
    \delta Z_e^{(1,1)}\Bigg|_{n_\ell, n_q \rightarrow 0;\, m_t \rightarrow \infty} = 0\,,
\end{align}
for the order $\alpha \alpha_s$ corrections to the electric charge renormalization constant.

Since the tree-level Drell-Yan amplitude is of order $\alpha$, the trivial renormalization of the strong coupling constant
\begin{align}
    (4 \pi)^\ep e^{-\gamma_E \ep}\alpha_s^{\rm bare} = \alpha_s
\end{align}
is all that is required for our order $\alpha^2 \alpha_s$ two-loop calculation. On the other hand, the one-loop electric charge renormalization constant does make an appearance. It has been known since the work of \cite{Bohm:1986rj} in the standard on-shell scheme where the charge renormalization constant is fixed order-by-order in the Thomsen limit (see {\it e.g.} \cite{Denner:1991kt} for a review). In the $\overline{\mathrm{MS}}$ scheme it is given by,
\begin{align}
\label{eq:1Lchargerenorm}
    \delta Z_e^{(1,0)}\Bigg|_{n_\ell, n_q \rightarrow 0;\, m_t \rightarrow \infty} = -\frac{7}{2 \ep}\,,
\end{align}
the pole part of Eq. (5.40) of \cite{Bohm:1986rj} once one discards the first, fermionic term inside the bracket. For notational convenience, we shall suppress the qualifiers provided on the left-hand sides of Eqs. \eqref{eq:2Lchargerenorm} and \eqref{eq:1Lchargerenorm} throughout the rest of this work. When necessary, appropriate caveats will be provided in the text.
\subsection{On-shell electroweak gauge boson wavefunction and mass renormalization}
\label{sec:EWren}
In this section, we review the on-shell renormalization of the electroweak gauge bosons in a generalized 't\,Hooft-Feynman gauge. Our conventions for the electroweak sector of the Standard Model Lagrange density and Feynman rules are identical to those of \cite{Bohm:1986rj}, upon which the useful resources \cite{Denner:1991kt,Bohm:2001yx,Denner:2019vbn} were modeled. Our electroweak gauge boson renormalization constants are also defined with exactly the same renormalization conditions, but we have chosen to factor the squared electroweak gauge boson mass, $m_v^2$, out of the mass renormalization constants, $Z_{m_v^2}$, to render them dimensionless.

As usual, a particularly convenient choice for the gauge parameter renormalization constants is
\begin{align}
\label{eq:gaugeparamren}
    Z_{\xi_\gamma} = Z_{\gamma \gamma},\quad Z_{\xi_z} = Z_{ZZ}, \quad 
    \mathrm{and}\quad Z_{\xi_w} = Z_{W^\pm}
\end{align}
because it simplifies the all-orders form of the electroweak gauge boson kinetic terms in the renormalized Lagrange density. Subsequently, our generalized 't\,Hooft-Feynman gauge is defined by setting all of the renormalized gauge parameters to one at the outset,
\begin{align}
    \xi_\gamma = \xi_z = \xi_w = 1\, .
\end{align}
This is justified because, due to Eqs. \eqref{eq:gaugeparamren}, the gauge parameters in this scheme receive no radiative corrections.

Apart from the exactly-transverse photon-photon self-energy, 
\begin{align}
    - i\, \Sigma_{\gamma \gamma}^{\mu \nu}(q) = -i \left(g^{\mu \nu} - \frac{q^\mu q^\nu}{q^2}\right) \bar{\Sigma}_{\gamma \gamma}\left(q^2\right)\,,
\end{align}
the electroweak gauge boson self-energies naturally decompose into a sum of transverse and longitudinal components:
\begin{align}
    - i\, \Sigma_{\gamma Z}^{\mu \nu}(q) &= -i \left(g^{\mu \nu} - \frac{q^\mu q^\nu}{q^2}\right) \bar{\Sigma}_{\gamma Z}\left(q^2\right) - i \frac{q^\mu q^\nu}{q^2}\Sigma_{\gamma Z}^L\left(q^2\right)\,,\\
    - i\, \Sigma_{Z Z}^{\mu \nu}(q) &= -i \left(g^{\mu \nu} - \frac{q^\mu q^\nu}{q^2}\right) \bar{\Sigma}_{Z Z}\left(q^2\right) - i \frac{q^\mu q^\nu}{q^2}\Sigma_{Z Z}^L\left(q^2\right)\,,\\
  \mathrm{and}\quad - i\, \Sigma_{W^+ W^-}^{\mu \nu}(q) &= -i \left(g^{\mu \nu} - \frac{q^\mu q^\nu}{q^2}\right) \bar{\Sigma}_{W^+ W^-}\left(q^2\right) - i \frac{q^\mu q^\nu}{q^2}\Sigma_{W^+ W^-}^L\left(q^2\right)\,.
\end{align}
The longitudinal components of the self-energies, $\Sigma_{V V^\prime}^L$, are unphysical and need not be calculated explicitly.

Working in $d$ dimensions, we see that the relevant Lorentz projectors for the transverse gauge boson self-energies can be cleanly summarized as 
\begin{align}
    \Big(\mathbb{P}_{V V^\prime}^T\Big)_{\mu \nu} = \frac{i}{3-2\ep}\left(g_{\mu \nu} - \frac{q_\mu q_\nu}{q^2}\right).
\end{align}
For all of the Lorentz projection operators considered in this work, a sum over the available open Lorentz and spin indices is always implied for their action on Feynman diagrams. We find the following results for the transverse projections of the Fourier-transformed electroweak gauge boson kinetic terms, $\tilde{\mathcal{L}}_{V V^\prime}^{\mu \nu}$, of order $\alpha$:
\begin{align}
\label{eq:gamgamFT}
    \Big(\mathbb{P}_{V V^\prime}^T\Big)_{\mu \nu} \tilde{\mathcal{L}}^{\mu \nu}_{\gamma \gamma} &= \Big(\mathbb{P}_{V V^\prime}^T\Big)_{\mu \nu} \left(-i\, \delta Z_{\gamma \gamma}^{(1,0)} \left(q^2 g^{\mu \nu} - q^\mu q^\nu\right)\right) = \delta Z_{\gamma \gamma}^{(1,0)} q^2\,,\\
\label{eq:gamZFT}
    \Big(\mathbb{P}_{V V^\prime}^T\Big)_{\mu \nu} \tilde{\mathcal{L}}^{\mu \nu}_{\gamma Z} &= \Big(\mathbb{P}_{V V^\prime}^T\Big)_{\mu \nu} \left(-\frac{i}{2} \left(\left(\delta Z_{\gamma Z}^{(1,0)} + \delta Z_{Z \gamma}^{(1,0)}\right)q^2 - m_z^2 \delta Z_{Z \gamma}^{(1,0)}\right)g^{\mu \nu}\right) 
    \nonumber \\
    &= \frac{1}{2} \left(\left(\delta Z_{\gamma Z}^{(1,0)} + \delta Z_{Z \gamma}^{(1,0)}\right)q^2 - m_z^2 \delta Z_{Z \gamma}^{(1,0)}\right),\\
\label{eq:ZZFT}
    \Big(\mathbb{P}_{V V^\prime}^T\Big)_{\mu \nu} \tilde{\mathcal{L}}^{\mu \nu}_{Z Z} &= \Big(\mathbb{P}_{V V^\prime}^T\Big)_{\mu \nu} \left(-i\left( \delta Z_{Z Z}^{(1,0)} \left(q^2 g^{\mu \nu} - q^\mu q^\nu\right)- m_z^2 \left(\delta Z_{Z Z}^{(1,0)}+ \delta Z_{m_z^2}^{(1,0)}\right)g^{\mu \nu}\right)\right) \nonumber \\
    &= \delta Z_{Z Z}^{(1,0)} q^2 - m_z^2 \left(\delta Z_{Z Z}^{(1,0)} + \delta Z_{m_z^2}^{(1,0)}\right),\quad \mathrm{and} \\
\label{eq:WWFT}
    \Big(\mathbb{P}_{V V^\prime}^T\Big)_{\mu \nu} \tilde{\mathcal{L}}^{\mu \nu}_{W^\pm} &= \Big(\mathbb{P}_{V V^\prime}^T\Big)_{\mu \nu} \left(-i\left( \delta Z_{W^\pm}^{(1,0)} \left(q^2 g^{\mu \nu} - q^\mu q^\nu\right)- m_w^2 \left(\delta Z_{W^\pm}^{(1,0)}+ \delta Z_{m_w^2}^{(1,0)}\right)g^{\mu \nu}\right)\right) \nonumber \\
    &= \delta Z_{W^\pm}^{(1,0)} q^2 - m_w^2 \left(\delta Z_{W^\pm}^{(1,0)} + \delta Z_{m_w^2}^{(1,0)}\right).
\end{align}

Analyzing Eqs. \eqref{eq:gamgamFT} - \eqref{eq:WWFT} and taking into account the renormalization conditions of \cite{Bohm:1986rj}, we see immediately that
\begin{align}
    \delta Z_{\gamma \gamma}^{(1,0)} &= - \frac{\mathrm{d}\bar{\Sigma}_{\gamma \gamma}^{(1,0)}\left(q^2\right)}{\mathrm{d}q^2}\Bigg|_{q^2 = 0}\,,
    &
    \delta Z_{Z \gamma}^{(1,0)} &= \frac{2\bar{\Sigma}_{\gamma Z}^{(1,0)}\left(q^2\right)}{m_z^2}\Bigg|_{q^2 = 0}\,, \nonumber \\
    \delta Z_{\gamma Z}^{(1,0)} &= - \frac{2 \bar{\Sigma}_{\gamma Z}^{(1,0)}\left(q^2\right)}{m_z^2}\Bigg|_{q^2 = m_z^2}\,,
&
    \delta Z_{Z Z}^{(1,0)} &= - \frac{\mathrm{d}\bar{\Sigma}_{Z Z}^{(1,0)}\left(q^2\right)}{\mathrm{d}q^2}\Bigg|_{q^2 = m_z^2}\,, \nonumber \\
\label{eq:countertermdefs}
    \delta Z_{m_z^2}^{(1,0)} &= \frac{\bar{\Sigma}_{Z Z}^{(1,0)}\left(q^2\right)}{m_z^2}\Bigg|_{q^2 = m_z^2}\,,
&    
     \delta Z_{m_w^2}^{(1,0)} &= \frac{\bar{\Sigma}_{W^+ W^-}^{(1,0)}\left(q^2\right)}{m_w^2}\Bigg|_{q^2 = m_w^2}\,.
\end{align}
Apart from light fermion and top quark contributions which we consistently neglect throughout this work, explicit expressions for the one-loop self-energies and counterterms relevant to the higher-order corrections of interest to us will be provided to all orders in $\ep$ in Section \ref{sec:1Lares}. Note that we do not provide an explicit expression for $\delta Z^{(1,0)}_{W^\pm}$ because it plays no role in the renormalization of any of the Feynman diagrams we calculate.
\subsection{On-shell wavefunction renormalization for massless fermion fields}
\label{sec:fermren}
In this section, we review the systematics of on-shell renormalization for massless fermion fields in the Standard Model. The on-shell scheme for massless fermions is widely used in QCD because, due to the scalelessness of the contributing gluon-exchange diagram, the one-loop quark wavefunction counterterm vanishes identically in dimensional regularization. Although we will ultimately take the $q^2 \rightarrow 0$ limit, it is nevertheless instructive to begin with the off-shell setup of \cite{Bohm:1986rj}, where the momentum transfer $q^2$ in the Lorentz decomposition of the fermion self energy
\begin{align}
\label{eq:KreimerSEdecomp}
    i \Big(\bar{\Sigma}_{f}(q)\Big)_{\alpha \beta} = i\, \bar{\Sigma}_{\mathrm{V},\, f}\left(q^2\right) \left(\slashed{q}\right)_{\alpha \beta} + i\, \bar{\Sigma}_{\mathrm{A},\, f}\left(q^2\right)\left(\slashed{q}\gamma_5\right)_{\alpha \beta}
\end{align} 
is taken to be different from zero.\footnote{Throughout this section, we suppress the color indices of the quarks to streamline our discussion and allow for a unified description of propagator corrections at one loop. The two-loop mixed EW-QCD scattering amplitudes of primary interest to us in this work always come with a color factor of $C_F$, due to the fact that all contributing Feynman diagrams feature exactly one gluon exchange between quark lines.} In writing Eq. \eqref{eq:KreimerSEdecomp}, we have implicitly assumed Kreimer's $\gamma_5$ scheme; analogous calculations in HVBM's $\gamma_5$ scheme will be discussed in Section \ref{sec:finiteren} and, accordingly, we have written $\bar{\Sigma}_{f}(q)$ instead of $\Sigma_{f}(q)$ in order to allow for a side-by-side comparison in the next section.

We employ the notation of \cite{Bohm:1986rj} for the interactions of the electroweak gauge bosons with matter. The $Z$ interaction is parametrized by flavor-dependent axial vector and vector coupling coefficients,
\begin{align}
\label{eq:zalias}
    a_f = \frac{m_z^2\, I_f^3}{2 m_w \sqrt{m_z^2 - m_w^2}}\quad \mathrm{and} \quad v_f = \frac{m_z^2}{2 m_w \sqrt{m_z^2 - m_w^2}}\left(I^3_f - 2 \frac{m_z^2 - m_w^2}{m_z^2}Q_f\right)\,.
\end{align}
The $W$ interaction, on the other hand, is parametrized by a universal coupling coefficient
\begin{align}
\label{eq:walias}
    a_w = v_w = \frac{m_z}{2\sqrt{2}\sqrt{m_z^2 - m_w^2}}\,.
\end{align}
We will make extensive use of these aliases throughout the rest of this paper.

\begin{figure}[!t]
\centering
\begin{align}
&\hspace{5.1 cm} \includegraphics[scale=0.6]{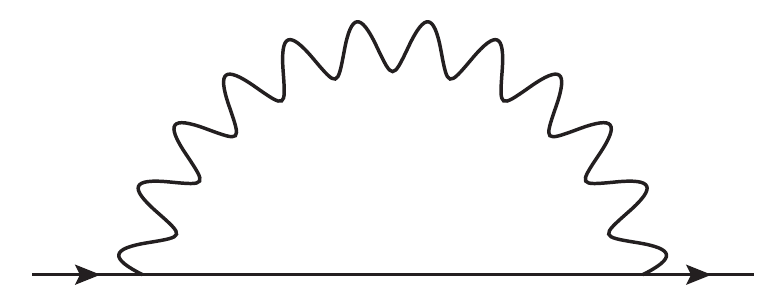} \nonumber \\ &\includegraphics[scale=0.6]{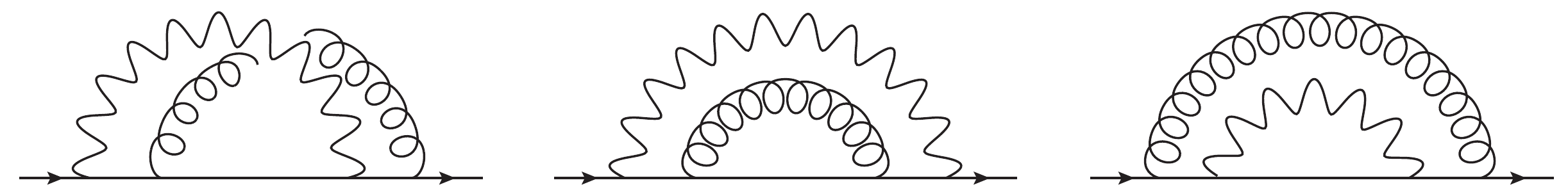}\nonumber
\end{align}
\caption{Independent $Z$ exchange diagrams at one and two loops. Due to the absence of fermion masses, diagrams which feature only gluon and/or photon exchanges vanish identically.
}
\label{fig:fermSE}
\end{figure}

At first sight, it might seem counterproductive not to employ Eq. \eqref{eq:walias} to eliminate $a_w$ and $v_w$, as they are equal and have no dependence on the fermion flavor. However, as will become clear later on in this section, it is useful to retain the dependence on $a_w$ and $v_w$ at intermediate stages of fermion self-energy and vertex calculations because it generically allows for a determination of axial vector components from vector components through chiral symmetry. Actually, the fermion self-energies considered in this section are particularly simple and have no independent $W$-exchange Feynman diagrams; all $W$-exchange diagrams can be obtained from the $Z$-exchange diagrams for free by making the replacements $a_f \rightarrow a_w$ and $v_f \rightarrow v_w$ (see Figure \ref{fig:fermSE}).

Thus, for the relatively low perturbative orders of interest to us, it suffices to consider the calculation of the $Z$-exchange contributions to the vector component of the fermion self-energy. At this stage, one is faced with a conceptual hurdle. Taken at face value, the self-energy in the on-shell scheme for massless fermions would seem to be most appropriately defined as the $q^2 \rightarrow 0$ limit of the fermion self-energy calculated off-shell.\footnote{Here, the $q^2 \rightarrow 0$ limit should be taken {\it before} expanding in $\epsilon$.} Of course, it also seems that one ought to be able to simply set $q^2 = 0$ in all Feynman diagrams at the very beginning, significantly simplifying all subsequent calculations. In fact, as we now demonstrate at the one-loop level, both approaches are possible and yield the same answer.

If we begin with $q^2$ generic, one can immediately verify that the Lorentz projector of interest is
\begin{align}
\label{eq:projSEoffshell}
    \Big(\mathbb{P}_f^\mathrm{V}\Big)_{\beta \alpha}~\Big|_{\mathrm{off-shell}} = - \frac{i}{4\, q^2} \left(\slashed{q}\right)_{\beta \alpha}
\end{align}
and our Kreimer reading point prescription instructs us to begin reading our Dirac traces from the projector insertion. Therefore, using the Feynman rules of \cite{Bohm:1986rj} and the anticommutation relation satisfied by Kreimer's $\gamma_5$, Eq. \eqref{eq:anticommg5}, the contribution of the one-loop $Z$-exchange diagram to the vector component of the fermion self-energy is
\begin{align}
\label{eq:offshelldemoKstart}
\bar{\Sigma}_{\mathrm{V},\, f}^{(1,0)}(q^2)\Big|_{Z~\mathrm{exchange}} =&  -\frac{c_1\left(\epsilon,\mu^2\right) \left(a_f^2+v_f^2\right)}{4\,q^2} \int \mathrm{d}^d k_1\frac{\mathrm{Tr}\Big\{\slashed{q}\gamma^\nu (\slashed{q}+\slashed{k}_1)\gamma_\nu\Big\}}{(q+k_1)^2(k_1^2 - m^2_z)}\,,
\end{align}
for\footnote{As explained in Section \ref{sec:couplingren}, $\alpha$ is renormalized in the $\overline{\mathrm{MS}}$ scheme.}
\begin{align}
    c_1\left(\epsilon,\mu^2\right) = \frac{e^{\gamma_E \epsilon} \left(\mu^2\right)^{\epsilon}}{i \pi^{2-\epsilon}}\,.
\end{align}

As Eq. \eqref{eq:offshelldemoKstart} does not depend on $\gamma_5$, it can be trivially simplified using the contraction identity and even trace formulae of Section \ref{sec:HVBMdefs}. After carrying out the numerator algebra, we obtain:
\begin{align}
    \bar{\Sigma}_{\mathrm{V},\, f}^{(1,0)}(q^2)\Big|_{Z~\mathrm{exchange}} =  \frac{2 \,c_1\left(\epsilon,\mu^2\right) \left(a_f^2+v_f^2\right)(1-\epsilon)}{q^2} \int\mathrm{d}^d k_1 \frac{q^2 + q\cdot k_1}{(q+k_1)^2(k_1^2 - m^2_z)},
\end{align}
which can immediately be rewritten as a linear combination of standard scalar integrals using Passarino-Veltman reduction \cite{Passarino:1978jh}; the result is
\begin{align}
\label{eq:offshellwvfuncinte}
    \bar{\Sigma}_{\mathrm{V},\, f}^{(1,0)}(q^2)\Big|_{Z~\mathrm{exchange}} &= \frac{c_1\left(\epsilon,\mu^2\right)\left(a_f^2+v_f^2\right)(1-\epsilon)}{q^2}\left((q^2-m^2_z)\int \frac{\mathrm{d}^d k_1}{(q+k_1)^2(k_1^2 - m^2_z)}
    \right.\nonumber \\
    &\quad\left.+\int \frac{\mathrm{d}^d k_1}{k_1^2 - m^2_z}\right),
\end{align}
where, using Feynman/Schwinger parameters,
\begin{align}
\label{eq:onemasstadpoleres}
 \int \frac{\mathrm{d}^d k_1}{k_1^2 - m^2_z} &= -\frac{e^{\gamma_E \epsilon}}{c_1\left(\epsilon,\mu^2\right)}\left(\frac{\mu^2}{m^2_z}\right)^{\epsilon} m^2_z\, \Gamma(-1+\epsilon) \quad \mathrm{and} \\
\label{eq:onemassbubbleres}
\int \frac{\mathrm{d}^d k_1}{(q+k_1)^2(k_1^2 - m^2_z)} &= 2\frac{e^{\gamma_E \epsilon}}{c_1\left(\epsilon,\mu^2\right)}\left(\frac{\mu^2}{m^2_z}\right)^{\epsilon}\frac{\Gamma(\epsilon)\Gamma(2-2\epsilon)}{\Gamma(3-2 \epsilon)} {}_2 F_1\left(1,\epsilon; 2-\epsilon; \frac{q^2}{m^2_z}\right) .
\end{align}
In writing Eq. \eqref{eq:onemassbubbleres}, $q^2$ was assumed to be less than $m^2_z$. Modulo minor typos,\footnote{While comparing to \cite{Bohm:1986rj}, we noticed a typo in the expression for $\Sigma_A^{\hphantom{A}i \sigma}$ on the 3rd line of Eq. (5.27): the first term inside the bracket should actually read $-2\, v_{i \sigma} a_{i \sigma}\left(2\,B_1\left(k^2; m_{i \sigma}, M_z\right)+1\right)$. Additional typos were discovered in Eq. (B.2); the tensor reduction formula needed for $B_1\left(k^2; m_{i \sigma}, M_z\right)$ should read
\begin{align}
    2\, k^2 B_1\left(k^2; M_1, M_2\right) = A\left(M_1\right) - A\left(M_2\right) + \left(M_2^2 - M_1^2 - k^2\right)B_0 \left(k^2; M_1, M_2\right). \nonumber
\end{align}} our results agree as expected with those of \cite{Bohm:1986rj} through to $\mathcal{O}\left(\epsilon^0\right)$ in the limit of vanishing fermion mass.

 From Eqs. \eqref{eq:onemasstadpoleres} and \eqref{eq:onemassbubbleres} it is a simple matter to work out the $q^2 \rightarrow 0$ limit of Eq. \eqref{eq:offshellwvfuncinte}. From l'H{\^o}pital's rule, we find:
\begin{align}
    \lim_{q^2 \to 0} \left( \bar{\Sigma}_{\mathrm{V},\, f}^{(1,0)}(q^2)\Big|_{Z~\mathrm{exchange}}\right) =&  \frac{2 \left(a_f^2+v_f^2\right)(1-\epsilon)\Gamma(\epsilon)e^{\gamma_E \epsilon}}{2-\epsilon} \left(\frac{\mu^2}{m^2_z}\right)^{\epsilon} .
\end{align}
 The above derivation is certainly valid, but let us stress that attempting to calculate the $q^2 \rightarrow 0$ limit of $\bar{\Sigma}_{\mathrm{V},\,f}(q^2)$ in this way is not recommended at higher orders in perturbation theory; while we were able to confirm numerically that the correct $q^2 \rightarrow 0$ limit exists at order $\alpha \alpha_s$ for fixed, finite $\epsilon$, it is not amenable to analytic evaluation.

In general, our opinion is that it is far simpler to implement the on-shell condition $q^2 = 0$ from the get-go. Of course, Eq. \eqref{eq:projSEoffshell} is ill-defined at $q^2 = 0$; a valid projector when $q^2 = 0$ is given by
\begin{align}
\label{eq:projSEonshell}
\Big(\mathbb{P}_f^\mathrm{V}\Big)_{\beta \alpha}~\Big|_{\mathrm{on-shell}} = - \frac{i}{4\, \eta \cdot q} \left(\slashed{\eta}\right)_{\beta \alpha},
\end{align}
where $\eta$ is an arbitrary four-vector such that $\eta \cdot q \neq 0$.

Setting $q^2 = 0$ from the beginning of the calculation, we have
\begin{align}
\label{eq:onshelldemoKstart}
\bar{\Sigma}_{\mathrm{V},\, f}^{(1,0)}(0)\Big|_{Z~\mathrm{exchange}} &=  -\frac{c_1\left(\epsilon,\mu^2\right) \left(a_f^2+v_f^2\right)}{4\,\eta \cdot q} \int \mathrm{d}^d k_1\frac{\mathrm{Tr}\Big\{\slashed{\eta}\gamma^\nu (\slashed{q}+\slashed{k}_1)\gamma_\nu\Big\}}{(q+k_1)^2(k_1^2 - m^2_z)} \nonumber \\
&= \frac{2\,c_1\left(\epsilon,\mu^2\right) \left(a_f^2+v_f^2\right)(1-\epsilon)}{\eta \cdot q} \int\mathrm{d}^d k_1 \frac{\eta \cdot q + \eta \cdot k_1}{(q+k_1)^2(k_1^2 - m^2_z)}.
\end{align}
Proceeding directly from Feynman/Schwinger parameters for the integral evaluations, we find
\begin{align}
\int \mathrm{d}^d k_1 \frac{1}{(q+k_1)^2(k_1^2 - m^2_z)}\Big|_{q^2 = 0} &= -\frac{e^{\gamma_E \epsilon}}{c_1\left(\epsilon,\mu^2\right)}\left(\frac{\mu^2}{m^2_z}\right)^{\epsilon} \Gamma(-1+\epsilon) \\
\mathrm{and}\quad \int \mathrm{d}^d k_1 \frac{\eta\cdot k_1}{(q+k_1)^2(k_1^2 - m^2_z)}\Big|_{q^2 = 0} &= \frac{\eta\cdot q \, e^{\gamma_E \epsilon}}{c_1\left(\epsilon,\mu^2\right)}\left(\frac{\mu^2}{m^2_z}\right)^{\epsilon}  \frac{\Gamma(-1+\epsilon)}{2-\epsilon}.
\end{align}
Note that, in the degenerate $q^2 = 0$ kinematics under consideration, {\it all} contributing one-loop Feynman integrals must ultimately turn out to be proportional to the one-loop tadpole.

Finally, we obtain
\begin{align}
\label{eq:finalZVSE}
  \bar{\Sigma}_{\mathrm{V},\, f}^{(1,0)}(0)\Big|_{Z~\mathrm{exchange}} =&  \frac{2\,\left(a_f^2+v_f^2\right)(1-\epsilon)\Gamma(\epsilon)e^{\gamma_E \epsilon}}{2-\epsilon} \left(\frac{\mu^2}{m^2_z}\right)^{\epsilon}
\end{align}
and see immediately that 
\begin{align}
    \lim_{q^2 \to 0} \left( \bar{\Sigma}_{\mathrm{V},\, f}^{(1,0)}(q^2)\Big|_{Z~\mathrm{exchange}}\right) = \bar{\Sigma}_{\mathrm{V},\, f}^{(1,0)}(0)\Big|_{Z~\mathrm{exchange}}
\end{align}
as claimed. 

We now consider the ramifications of chiral symmetry. In the Standard Model, the right- and left-handed components of the $Z$-exchange contribution to the one-loop fermion self-energy are exchanged under $\gamma_5 \rightarrow -\gamma_5$. This implies that the axial vector component of the $Z$-exchange contribution to the fermion self-energy ought to be obtained by a replacement of the form $v_f^2+a_f^2 \rightarrow \pm 2\,a_f v_f$ in Eq. \eqref{eq:finalZVSE}. In order to fix the overall sign, recall that the coupling of the $W$ to matter is exactly left-handed, that the $W$-exchange contribution is obtained by making the replacements $a_f \rightarrow a_w$ and $v_f \rightarrow v_w$, and that $a_w = v_w$. Taken together, these considerations allow us to conclude that
\begin{align}
  \bar{\Sigma}_{\mathrm{A},\, f}^{(1,0)}(0)\Big|_{Z~\mathrm{exchange}}
  =&  -\frac{4\,a_f v_f (1-\epsilon)\Gamma(\epsilon)e^{\gamma_E \epsilon}}{2-\epsilon} \left(\frac{\mu^2}{m^2_z}\right)^{\epsilon}
\end{align}
with no additional calculation. 

In the on-shell scheme, massless fermion wavefunction counterterms are defined to exactly cancel the perturbative corrections to the corresponding fermion self-energies order-by-order. Therefore, we find
\begin{align}
\label{eq:wavefuncaV}
\delta Z_{\mathrm{V},\,f}^{(1,0)} &=  -\bar{\Sigma}_{\mathrm{V},\, f}^{(1,0)}(0) 
\\
&= -\frac{2\left(a_f^2+v_f^2\right)(1-\epsilon)\Gamma(\epsilon)e^{\gamma_E \epsilon}}{2-\epsilon} \left(\frac{\mu^2}{m^2_z}\right)^{\epsilon}- \frac{2\left(a_w^2+v_w^2\right)(1-\epsilon)\Gamma(\epsilon)e^{\gamma_E \epsilon}}{2-\epsilon} \left(\frac{\mu^2}{m^2_w}\right)^{\epsilon}\nonumber\\
\label{eq:wavefuncaAV}
\delta Z_{\mathrm{A},\,f}^{(1,0)} &=  -\bar{\Sigma}_{\mathrm{A},\, f}^{(1,0)}(0) 
\\
&= \frac{4\,a_f v_f (1-\epsilon)\Gamma(\epsilon)e^{\gamma_E \epsilon}}{2-\epsilon} \left(\frac{\mu^2}{m^2_z}\right)^{\epsilon}+\frac{4\,a_w v_w (1-\epsilon)\Gamma(\epsilon)e^{\gamma_E \epsilon}}{2-\epsilon} \left(\frac{\mu^2}{m^2_w}\right)^{\epsilon}\nonumber
\end{align}
for the vector and axial vector components of the one-loop fermion wavefunction counterterm.

We now turn to the vector and axial vector components of the two-loop quark self-energy of order $\alpha \alpha_s$. To our knowledge, these mixed two-loop corrections to the self-energy were first considered in \cite{Franzkowski:1997zz}, though not in the on-shell scheme with massless fermions. It is straightforward to calculate the desired quantities by acting with the appropriate Lorentz projector in Kreimer's $\gamma_5$ scheme, Eq. \eqref{eq:projSEonshell}, on the eight non-zero two-loop Feynman diagrams summarized by Figure \ref{fig:fermSE}. Remarkable simplifications are achieved by setting $q^2 = 0$ at the beginning of the calculation.
 In the end, we find that the vector component of the quark wavefunction counterterm is given by
\begin{align}
\label{eq:wavefuncaasV}
\delta Z_{\mathrm{V},\,q}^{(1,1)} &=  -\bar{\Sigma}_{\mathrm{V},\, q}^{(1,1)}(0) \\
&= -\frac{2(a_q^2 + v_q^2) C_F (1 - \epsilon) (1 - 3 \epsilon) (3 - 2 \epsilon)\Gamma^2(1-\epsilon)\Gamma(1+\epsilon)\Gamma(-1+2\epsilon)e^{2\gamma_E \ep}}{\Gamma(3-\epsilon)}\left(\frac{\mu^2}{m^2_z}\right)^{2\epsilon}\nonumber\\
&-\frac{2(a_w^2 + v_w^2) C_F (1 - \epsilon) (1 - 3 \epsilon) (3 - 2 \epsilon)\Gamma^2(1-\epsilon)\Gamma(1+\epsilon)\Gamma(-1+2\epsilon)e^{2\gamma_E \ep}}{\Gamma (3-\epsilon)}\left(\frac{\mu^2}{m^2_w}\right)^{2\epsilon}\nonumber
\end{align}
and the axial vector component of the quark wavefunction counterterm is given by
\begin{align}
\label{eq:wavefuncaasAV}
\delta Z_{\mathrm{A},\,q}^{(1,1)} &= -\bar{\Sigma}_{\mathrm{A},\, q}^{(1,1)}(0) \\
&= \frac{4\,a_q v_q C_F (1 - \epsilon) (1 - 3 \epsilon) (3 - 2 \epsilon)\Gamma^2(1-\epsilon)\Gamma(1+\epsilon)\Gamma(-1+2\epsilon)e^{2\gamma_E \ep}}{\Gamma(3-\epsilon)}\left(\frac{\mu^2}{m^2_z}\right)^{2\epsilon}
\nonumber \\
&+\frac{4\,a_w v_w C_F (1 - \epsilon) (1 - 3 \epsilon) (3 - 2 \epsilon)\Gamma^2(1-\epsilon)\Gamma(1+\epsilon)\Gamma(-1+2\epsilon)e^{2\gamma_E \ep}}{\Gamma(3-\epsilon)} \left(\frac{\mu^2}{m^2_w}\right)^{2\epsilon}.\nonumber
\end{align}
As at one loop, the simple form of Eqs. \eqref{eq:wavefuncaasV} and \eqref{eq:wavefuncaasAV} results from the fact that only the one-mass two-loop tadpole integral can survive in the $q^2 \rightarrow 0$ limit.
\subsection{Chiral symmetry restoration in HVBM's $\gamma_5$ scheme and Larin's principle}
\label{sec:finiteren}
\begin{figure}[!t]
\centering
\begin{align}
\includegraphics[scale=0.6]{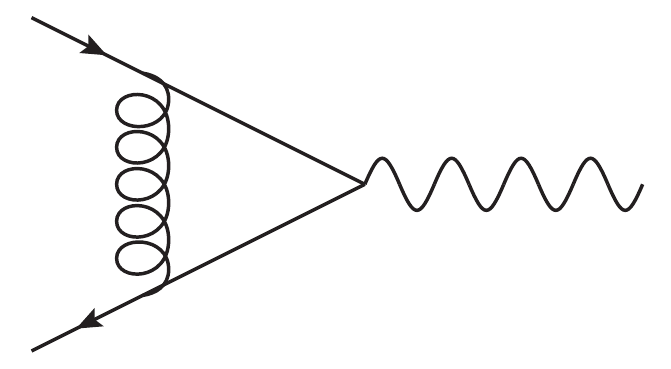}\nonumber
\end{align}
\caption{The order $\alpha_s$ gluon-exchange correction to the $Z\bar{q} q$ vertex form factor. 
}
\label{fig:gluonFF}
\end{figure}

As mentioned in Section \ref{sec:HVBMdefs}, HVBM's $\gamma_5$ scheme is especially challenging to work with in practice due to the fact that, in general, a plethora of finite counterterms must be introduced to restore the chiral symmetry of the Standard Model. We shall see that Larin's principle \cite{Larin:1993tq} provides a practical recipe which helps tremendously to minimize the number of finite counterterms which must be computed explicitly for a given application. While we will be most interested in the standard one-loop example of the order $\alpha_s$ gluon-exchange correction to the $Z\bar{q} q$ vertex shown in Figure \ref{fig:gluonFF} \cite{Collins:1984xc}, let us first revisit the one-loop fermion self-energy calculation of Section \ref{sec:fermren} in HVBM's $\gamma_5$ scheme to illustrate the utility of Larin's principle with minimal computational effort.

In HVBM's $\gamma_5$ scheme, the Lorentz decomposition for the massless fermion self-energy reads
\begin{align}
\label{eq:HVBMSEdecomp}
    i \Big(\Sigma_{f}(q)\Big)_{\alpha \beta} = i\, \Sigma_{\mathrm{V},\, f}\left(0\right) \left(\slashed{q}\right)_{\alpha \beta} + i\, \Sigma_{\mathrm{A},\, f}\left(0\right)\frac{1}{2}\left[\slashed{q}, \gamma_5\right]_{\alpha \beta}\,,
\end{align} 
where we have assumed the $q^2 = 0$ Lorentz projector, Eq. \eqref{eq:projSEonshell}, will be employed to compute $\Sigma_{\mathrm{V},\, f}\left(0\right)$ and, as in Section \ref{sec:fermren}, the color indices are suppressed. It is once again sufficient to consider the $Z$-exchange contribution:
\begin{align}
\label{eq:onshelldemoHVBMstart}
\Sigma_{\mathrm{V},\, f}^{(1,0)}(0)\Big|_{Z~\mathrm{exchange}} &=  -\frac{c_1\left(\epsilon,\mu^2\right) v_f^2}{4\,\eta \cdot q} \int \frac{\mathrm{d}^d k_1}{(q+k_1)^2(k_1^2 - m^2_z)}\mathrm{tr}\Big\{\slashed{\eta}\gamma^\nu (\slashed{q}+\slashed{k}_1)\gamma_\nu\Big\} \nonumber \\
&\quad-\frac{c_1\left(\epsilon,\mu^2\right) a_f^2}{16\,\eta \cdot q} \int \frac{\mathrm{d}^d k_1}{(q+k_1)^2(k_1^2 - m^2_z)}\mathrm{tr}\Big\{\slashed{\eta}\left[\gamma^\nu, \gamma_5\right] (\slashed{q}+\slashed{k}_1)\left[\gamma_\nu,\gamma_5\right]\Big\}.
\end{align}
In order to simplify the term proportional to $a_f^2$, it is useful to resolve the $d$-dimensional momenta and Dirac matrices into four-dimensional and $(-2\ep)$-dimensional components using Eqs. \eqref{eq:splitindex1} and \eqref{eq:splitindex2}. 
Crucially, the linear dependence on $k_1$ in the numerator of Eq. \eqref{eq:onshelldemoHVBMstart} implies that the Dirac traces {\it cannot} produce non-zero $\mu$ integrals. This observation allows for the immediate neglect of all terms involving $\hat{k}_1$.

Using the HVBM $\gamma_5$ anticommutation relations of Section \ref{sec:HVBMdefs}, Eq. \eqref{eq:onshelldemoHVBMstart} simplifies to \begin{align}
\label{eq:onshelldemoHVBMsimp}
\Sigma_{\mathrm{V},\, f}^{(1,0)}(0)\Big|_{Z~\mathrm{exchange}} &=  -\frac{c_1\left(\epsilon,\mu^2\right) v_f^2}{4\,\eta \cdot q} \int \frac{\mathrm{d}^d k_1}{(q+k_1)^2(k_1^2 - m^2_z)}\mathrm{tr}\Big\{\slashed{\eta}\gamma^\nu (\slashed{q}+\slashed{k}_1)\gamma_\nu\Big\} \nonumber \\
&\quad-\frac{c_1\left(\epsilon,\mu^2\right) a_f^2}{4\,\eta \cdot q} \int \frac{\mathrm{d}^d k_1}{(q+k_1)^2(k_1^2 - m^2_z)}\mathrm{tr}\Big\{\slashed{\eta}\bar{\gamma}^\nu (\slashed{q}+\slashed{\bar{k}}_1)\bar{\gamma}_\nu\Big\}\,.
\end{align}
Finally, the trace on the second line of Eq. \eqref{eq:onshelldemoHVBMsimp} is nothing but the $\ep \rightarrow 0$ limit of the trace on the first line of Eq. \eqref{eq:onshelldemoHVBMsimp}; from the analysis of Section \ref{sec:fermren}, we see that
\begin{align}
\label{eq:onshelldemoHVBMfinal}
&\Sigma_{\mathrm{V},\, f}^{(1,0)}(0)\Big|_{Z~\mathrm{exchange}} =  \frac{2\,\left(v_f^2 (1-\ep)+a_f^2\right)\Gamma(\ep)e^{\gamma_E \ep}}{2-\ep}\left(\frac{\mu^2}{m_z^2}\right)^\ep.
\end{align}

Starting at $\mathcal{O}\left(\ep^0\right)$, Eq. \eqref{eq:onshelldemoHVBMfinal} clearly exhibits a {\it chiral mismatch}, indicating that the vector components of the fermion kinetic terms of the Standard Model Lagrange density receive finite counterterm corrections already at order $\alpha$. In other words, a correction must be made because the coefficient of $a_f^2$ is not equal to the coefficient of $v_f^2$ in Eq. \eqref{eq:onshelldemoHVBMfinal} at $\mathcal{O}\left(\ep^0\right)$. As Larin emphasized in \cite{Larin:1993tq}, the form of the finite counterterm may be deduced by simply {\it demanding} that the $a_f^2$ and $v_f^2$ coefficients be equal. For the present calculation, Larin's principle dictates that the finite counterterm is given to all orders in $\ep$ by the difference
\begin{align}
\label{eq:SEdiff}
\bar{\Sigma}_{\mathrm{V},\, f}^{(1,0)}(0) - \Sigma_{\mathrm{V},\, f}^{(1,0)}(0) = - \frac{2\,a_f^2 \Gamma(1+\ep) e^{\gamma_E \ep}}{2-\ep}\left(\frac{\mu^2}{m_z^2}\right)^\ep - \frac{2\,a_w^2 \Gamma(1+\ep) e^{\gamma_E \ep}}{2-\ep}\left(\frac{\mu^2}{m_w^2}\right)^\ep.
\end{align}
In fact, we could have simply skipped the calculation of the bare self-energy in HVBM's $\gamma_5$ scheme {\it altogether}; requiring that our result respect the chiral symmetry of the Standard Model to all orders in $\ep$ {\it uniquely} leads to the result obtained in Kreimer's $\gamma_5$ scheme in Section \ref{sec:fermren} after finite renormalization.\footnote{In \cite{Larin:1993tq}, it is suggested that one ought to perform additional renormalizations for the purpose of chiral symmetry restoration only after the removal of all ultraviolet divergences with ordinary counterterms. In our view, this is not necessarily the most transparent approach. However, at higher orders in perturbation theory, it may be that restoring the chiral symmetry first in the way we propose leads to divergent ``finite'' counterterms. Therefore, in our approach, ``finite'' may not be the best moniker to generally describe the counterterms responsible for the restoration of the Standard Model chiral symmetry in HVBM's $\gamma_5$ scheme.}

\begin{figure}[!t]
\centering
\begin{align}
\includegraphics[scale=0.6]{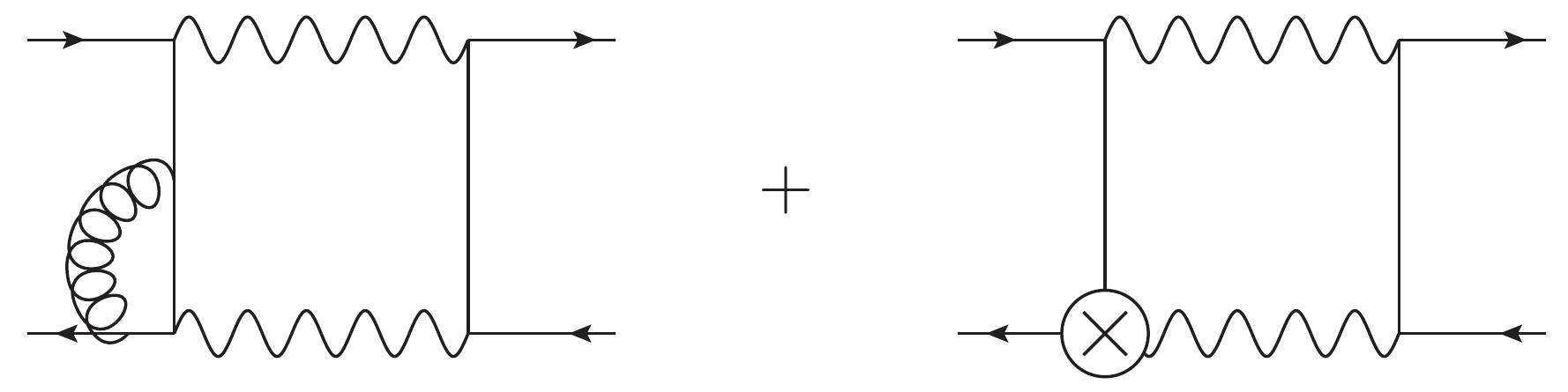}\nonumber
\end{align}
\caption{A two-loop $\gamma Z$ box-type diagram for the mixed EW-QCD corrections of relative order $\alpha \alpha_s$ to the neutral-current Drell-Yan process which receives a correction from the insertion of a finite one-loop counterterm ($\delta Z_{Z\bar{q} q}^{(0,1)}$ of Eq. \eqref{eq:qbarZqfiniteCT}) into a corresponding one-loop box diagram. 
}
\label{fig:finiteren}
\end{figure}

Of course, Larin's principle can be successfully applied to the fermion self-energies and vertex form factors because they have a direct correspondence to terms in the renormalized Lagrange density. For more complicated contributions to higher-multiplicity scattering amplitudes at two loops and beyond, it is expected that non-trivial corrections will arise from the finite counterterms, beginning with insertions into lower-loop box-type diagrams. Indeed, as we shall see later on in our discussion of the two-loop mixed EW-QCD corrections to the neutral-current Drell-Yan process, certain two-loop box-type diagrams which involve axial vector couplings to quarks receive corrections from finite counterterm insertions into corresponding one-loop box diagrams (see Figure \ref{fig:finiteren}). The bare calculation in HVBM's $\gamma_5$ scheme needed to derive the finite counterterm featured in Figure \ref{fig:finiteren} is a familiar one \cite{Collins:1984xc,Larin:1993tq} and, fortunately, the {\it only} one relevant to this work which we cannot bypass.

In HVBM's $\gamma_5$ scheme, the $Z \bar{q} q$ vertex has the Lorentz decomposition
\begin{align}
\label{eq:LorentzVFFHVBM}
    i e\, \Big(\mathcal{F}_\mu^{Z \bar{q} q}(p_1,p_2)\Big)_{kj} &= i e\, \mathcal{V}_{Z \bar{q} q}(s)\,\bar{v}_k (p_2)\gamma_\mu u_j (p_1)
     \\
    &+ i e\, \mathcal{A}_{Z \bar{q} q}(s)\,\bar{v}_k (p_2)\frac{1}{2}\left[\gamma_\mu,\gamma_5\right] u_j (p_1)\,,\nonumber
\end{align}
where $s = (p_1 + p_2)^2$ denotes the virtuality of the $Z$ and color indices are suppressed for brevity. Here, we have assumed an incoming massless quark of momentum $p_1$ and an incoming massless antiquark of momentum $p_2$ for later convenience. The Lorentz projectors
\begin{align}
\label{eq:projVFFHVBMV}
    \Big(\mathbb{P}^{\hspace{.025 cm}\mu}_{\mathcal{V}_q}\Big)_{j k} &= \frac{i}{4 e N_c s(1-\ep)} \bar{u}_j (p_1)\gamma^\mu v_k (p_2) \\
\label{eq:projVFFHVBMAV}
    \mathrm{and}\qquad\Big(\mathbb{P}^{\hspace{.025 cm} \mu}_{\mathcal{A}_q}\Big)_{jk} &= \frac{i}{4 e N_c s} \bar{u}_j (p_1)\frac{1}{2}\left[\gamma^\mu,\gamma_5\right] v_k (p_2)
\end{align}
deliver $\mathcal{V}_{Z \bar{q} q}(s)$ and $\mathcal{A}_{Z \bar{q} q}(s)$ respectively when acting on the sum of all $Z \bar{q} q$ vertex Feynman diagrams with an implicit sum on the open Lorentz and spin indices.

For the axial vector component of the order $\alpha_s$ correction to the $Z \bar{q} q$ vertex calculated in HVBM's $\gamma_5$ scheme, Larin's principle is applied by demanding equality to all orders in $\epsilon$ between its $a_q$ coefficient and minus the $v_q$ coefficient of the {\it vector} component of the order $\alpha_s$ correction to the $Z \bar{q} q$ vertex form factor, a condition which would hold automatically in Kreimer's $\gamma_5$ scheme. In other words, in the notation of \cite{Bohm:1986rj}, chiral symmetry guarantees that
\begin{align}
\label{eq:LarinVFF}
    \bar{\mathcal{A}}^{(0,1)}_{Z\bar{q} q}(s) = -\frac{a_q}{v_q}\, \bar{\mathcal{V}}^{(0,1)}_{Z\bar{q} q}(s)\,
\end{align}
where $\bar{\mathcal{V}}^{(0,1)}_{Z\bar{q} q}(s) = \mathcal{V}^{(0,1)}_{Z\bar{q} q}(s)$ is nothing but $v_q$ times the bare time-like one-loop quark vertex form factor of massless QCD:
\begin{align}
\label{eq:oneloopQCDquarkFF}
    \bar{\mathcal{V}}^{(0,1)}_{Z \bar{q} q}(s) &= v_q\bar{\mathcal{F}}_1^{q}(\epsilon) e^{i \pi \ep} \left(\frac{\mu^2}{s}\right)^\ep \nonumber \\
    &= -v_q \frac{\left(2-\ep+2\ep^2\right)\Gamma^2(1-\ep)\Gamma(\ep)e^{\gamma_E\ep}}{\ep\Gamma(2-2\ep)} C_F e^{i \pi \ep}\left(\frac{\mu^2}{s}\right)^\ep.
\end{align}
The derivation of the analytic expression for $\bar{\mathcal{F}}_1^{q}(\epsilon)$ given above is reviewed in many places in the literature, see {\it e.g.} \cite{Gehrmann:2005pd}. Thus, $\bar{\mathcal{A}}^{(0,1)}_{Z \bar{q} q}(s)$ is determined through Eq. \eqref{eq:LarinVFF} and, to fix the finite counterterm of interest, it remains only to directly calculate $\mathcal{A}^{(0,1)}_{Z \bar{q} q}(s)$.

Applying Eq. \eqref{eq:projVFFHVBMAV}, the axial vector component of the order $\alpha_s$ correction to the $Z \bar{q} q$ vertex form factor calculated in HVBM's $\gamma_5$ scheme is
\begin{align}
\label{eq:qbarZqdemoHVBMstart}
    \mathcal{A}^{(0,1)}_{Z \bar{q} q}(s) &= -\frac{a_q c_1\left(\epsilon,\mu^2\right) C_F}{16\,s} \int \frac{\mathrm{d}^d k_1}{(p_1 - k_1)^2 k_1^2(p_2 + k_1)^2}\notag\\
    &\quad \mathrm{tr}\Big\{\left[\gamma^\rho,\gamma_5\right] \slashed{p}_2 \gamma^\sigma (\slashed{p}_2+\slashed{k}_1)\left[\gamma_\rho,\gamma_5\right](\slashed{p}_1-\slashed{k}_1)\gamma_\sigma \slashed{p}_1\Big\}.
\end{align}
Splitting $d$-dimensional indices with Eqs. \eqref{eq:splitindex1} and \eqref{eq:splitindex2} and eliminating all $\gamma_5$ matrices using the HVBM anticommutation relations, \eqref{eq:finalanticommg5pt1} and \eqref{eq:finalanticommg5pt2}, we find that Eq. \eqref{eq:qbarZqdemoHVBMstart} simplifies to
\begin{align}
\label{eq:qbarZqdemoHVBMsimp}
    \mathcal{A}^{(0,1)}_{Z \bar{q} q}(s) &= -\frac{a_q c_1\left(\epsilon,\mu^2\right) C_F}{4\,s}\int \frac{\mathrm{d}^d k_1}{(p_1 - k_1)^2 k_1^2(p_2 + k_1)^2}\notag\\
&\quad\bigg[\mathrm{tr}\Big\{\bar{\gamma}^\rho \slashed{p}_2 \bar{\gamma}^\sigma (\slashed{p}_2+\slashed{\bar{k}}_1)\bar{\gamma}_\rho(\slashed{p}_1-\slashed{\bar{k}}_1)\bar{\gamma}_\sigma \slashed{p}_1\Big\}-
\mathrm{tr}\Big\{\bar{\gamma}^\rho \slashed{p}_2 \hat{\gamma}^\sigma (\slashed{p}_2+\slashed{\bar{k}}_1)\bar{\gamma}_\rho(\slashed{p}_1-\slashed{\bar{k}}_1)\hat{\gamma}_\sigma \slashed{p}_1\Big\}\notag \\
&\quad\; +\mathrm{tr}\Big\{\bar{\gamma}^\rho \slashed{p}_2 \bar{\gamma}^\sigma \slashed{\hat{k}}_1\bar{\gamma}_\rho \slashed{\hat{k}}_1\bar{\gamma}_\sigma \slashed{p}_1\Big\}-
\mathrm{tr}\Big\{\bar{\gamma}^\rho \slashed{p}_2 \hat{\gamma}^\sigma \slashed{\hat{k}}_1\bar{\gamma}_\rho \slashed{\hat{k}}_1\hat{\gamma}_\sigma \slashed{p}_1\Big\}\bigg].
\end{align}

Using Eq. \eqref{eq:mixedanticomm} and the relation
\begin{align}
\label{eq:HVBMepDiraccontract}
    \hat{\gamma}^\mu\hat{\gamma}_\mu = -2\ep\,,
\end{align}
we can rewrite Eq. \eqref{eq:qbarZqdemoHVBMsimp} solely in terms of four-dimensional Dirac traces:
\begin{align}
\label{eq:qbarZqdemoHVBMsimp2}
&\mathcal{A}^{(0,1)}_{Z \bar{q} q}(s) =-\frac{a_q c_1\left(\epsilon,\mu^2\right) C_F}{4\,s}\int \frac{\mathrm{d}^d k_1}{(p_1 - k_1)^2 k_1^2(p_2 + k_1)^2}\\
&\bigg[\mathrm{tr}\Big\{\bar{\gamma}^\rho \slashed{p}_2 \bar{\gamma}^\sigma (\slashed{p}_2+\slashed{\bar{k}}_1)\bar{\gamma}_\rho(\slashed{p}_1-\slashed{\bar{k}}_1)\bar{\gamma}_\sigma \slashed{p}_1\Big\}-2\ep\,\mathrm{tr}\Big\{\bar{\gamma}^\rho \slashed{p}_2 (\slashed{p}_2+\slashed{\bar{k}}_1)\bar{\gamma}_\rho(\slashed{p}_1-\slashed{\bar{k}}_1) \slashed{p}_1\Big\}\nonumber \\
&\qquad \qquad -\hat{k}_1\cdot \hat{k}_1\mathrm{tr}\Big\{\bar{\gamma}^\rho \slashed{p}_2 \bar{\gamma}^\sigma \bar{\gamma}_\rho \bar{\gamma}_\sigma \slashed{p}_1\Big\}+2\ep\, \hat{k}_1\cdot \hat{k}_1
\mathrm{tr}\Big\{\bar{\gamma}^\rho \slashed{p}_2 \bar{\gamma}_\rho  \slashed{p}_1\Big\}\bigg].\nonumber
\end{align}
Clearly, the remaining Dirac traces in Eq. \eqref{eq:qbarZqdemoHVBMsimp2} may be evaluated in a straightforward manner by applying Eqs. \eqref{eq:genDiraccontract}, \eqref{eq:eventracerecurse}, and \eqref{eq:eventracetermcrit} with $\ep$ set to zero and the $d$-dimensional metric tensor replaced by the four-dimensional metric tensor throughout.
Discarding all contributions which vanish due to scalelessness and setting $\bar{k}_1 \cdot \bar{k}_1 = k_1^2 - \hat{k}_1 \cdot \hat{k}_1$, we find
\begin{align}
\label{eq:qbarZqdemoHVBMint}
\mathcal{A}^{(0,1)}_{Z \bar{q} q}(s) =\frac{2\,a_q c_1\left(\epsilon,\mu^2\right) C_F}{s}\int\mathrm{d}^d k_1 \frac{2 s\, k_1^2 + (1+\ep)k_1^4 + s^2 + (1+\ep) s\, \hat{k}_1\cdot \hat{k}_1}{(p_1 - k_1)^2 k_1^2(p_2 + k_1)^2}\,.
\end{align}

From Eq. \eqref{eq:oneloopmusimp}, we have
\begin{align}
\label{eq:qbarZqdemoHVBMmuterm}
    \int\frac{\mathrm{d}^d k_1}{i \pi^{d/2}} \frac{ \hat{k}_1\cdot \hat{k}_1}{(p_1 - k_1)^2 k_1^2(p_2 + k_1)^2} = \ep \int \frac{\mathrm{d}^{d+2}k_1}{i \pi^{(d+2)/2}} \frac{1}{(p_1 - k_1)^2 k_1^2(p_2 + k_1)^2}\,.
\end{align}
It is worth pointing out that, by construction, such explicit $\mu$ terms could {\it not} occur in a direct calculation of $\bar{\mathcal{A}}^{(0,1)}_{Z \bar{q} q}(s)$ in Kreimer's $\gamma_5$ scheme. Finally, taking into account the algorithms of \cite{vonManteuffel:2014qoa} and their implementation in {\tt Reduze 2} \cite{Bauer:2000cp,Studerus:2009ye,vonManteuffel:2012np} to treat the dimensionally-shifted integral on the right-hand side of Eq. \eqref{eq:qbarZqdemoHVBMmuterm}, it follows that
\begin{align}
\label{eq:qbarZqdemoHVBMredint}
\mathcal{A}^{(0,1)}_{Z \bar{q} q}(s) =\frac{a_q c_1\left(\epsilon,\mu^2\right) C_F (2-3\ep+\ep^2)}{(1-\ep)\ep}\int \frac{\mathrm{d}^d k_1}{(p_1 - k_1)^2 (p_2 + k_1)^2}
\end{align}
after a straightforward integration by parts reduction. As is well-known (see {\it e.g.} \cite{Smirnov:2004ym}),
\begin{align}
\label{eq:zeromassbubbleres}
\int \frac{\mathrm{d}^d k_1}{(p_1 - k_1)^2 (p_2 + k_1)^2} = \frac{e^{\gamma_E \epsilon}}{c_1\left(\epsilon,\mu^2\right)}e^{i\pi \ep}\left(\frac{\mu^2}{s}\right)^\ep \frac{\Gamma^2(1-\ep)\Gamma(\ep)}{\Gamma(2-2\ep)}
\end{align}
and therefore
\begin{align}
\label{eq:qbarZqdemoHVBMfinal}
\mathcal{A}^{(0,1)}_{Z \bar{q} q}(s) = a_q\frac{ (2-3\ep+\ep^2)\Gamma^2(1-\ep)\Gamma(\ep)e^{\gamma_E \epsilon}}{(1-\ep)\ep\Gamma(2-2\ep)}C_F  e^{i\pi \ep}\left(\frac{\mu^2}{s}\right)^\ep.
\end{align}

Now, as explained above, the finite counterterm of interest is obtained as the difference between the Kreimer's $\gamma_5$ scheme expression deduced above and Eq. \eqref{eq:qbarZqdemoHVBMfinal}:
\begin{align}
\label{eq:qbarZqfiniteCT}
\delta Z_{Z \bar{q} q}^{(0,1)} &= \bar{\mathcal{A}}^{(0,1)}_{Z \bar{q} q}(s) - \mathcal{A}^{(0,1)}_{Z \bar{q} q}(s)\nonumber \\
&= 2\,a_q\frac{ (2-\ep)\Gamma^2(1-\ep)\Gamma(1+\ep)e^{\gamma_E \epsilon}}{(1-\ep)\Gamma(2-2\ep)}C_F  e^{i\pi \ep}\left(\frac{\mu^2}{s}\right)^\ep.
\end{align}
Eq. \eqref{eq:qbarZqfiniteCT} is manifestly finite at $\ep = 0$; as expected, we see that $\delta Z_{Z \bar{q} q}^{(0,1)} = 4 a_q C_F + \mathcal{O}(\ep)$, in complete agreement with Eq. (9) of \cite{Larin:1993tq}.\footnote{In making this comparison, note that Eq. (3) of \cite{Larin:1993tq} was not written down for the $Z \bar{q} q$ interaction verbatim and therefore does not have the overall factor of $-a_q$ present in interaction \eqref{eq:modZint}. Note also that the cited equation numbers refer to the journal version of \cite{Larin:1993tq}.} 

Before leaving this section, let us stress that we do {\it not} employ in this paper what is commonly referred to as ``Larin's scheme'' in the QCD literature. It is clear that the goal of \cite{Larin:1993tq} was to provide a recipe in {\tt FORM} for the consistent treatment of multi-loop QCD corrections to the axial vector components of Standard Model vertex form factors in HVBM's $\gamma_5$ scheme; Larin did not discuss a prescription for the treatment of multi-loop box-type diagrams, for example.
Indeed, in contrast to the prescription of \cite{Larin:1993tq} for the treatment of multi-loop QCD corrections to Standard Model vertex form factors, we found it essential for the restoration of chiral symmetry in the two-loop box-type diagrams treated below to consistently keep also higher-order terms of the $\ep$ expansions of our finite counterterms in order to avoid any premature truncation.

\section{Calculational details}
\subsection{Process definition}
\label{sec:process}
In this work, we study lepton pair production in quark-antiquark annihilation,
\begin{align}
\label{eq:process}
    q(p_1) + \bar{q}(p_2) \to \ell^-(p_3) + \ell^+(p_4)\,,
\end{align}
at one and two loops. We obtain complete results for scattering amplitudes at order $\alpha \alpha_s$, order $\alpha^2$, and order $\alpha^2 \alpha_s$ in the approximation of vanishing light quark and lepton masses, {\it i.e.} $p_1^2 = p_2^2 = p_3^2 = p_4^2 = 0$. As the present paper represents only a first decisive step towards the assembly of phenomenological results for Drell-Yan lepton production including complete two-loop EW-QCD effects in all channels, we defer the calculation of the gauge-invariant contributions proportional to the number of light quark or lepton flavors and all top quark mass corrections to future work. Our expectation is that the most important two-loop corrections we omit, coming from the light fermion contributions to the electroweak gauge boson self-energies, are already known (see {\it e.g.} \cite{Djouadi:1993ss}).

As usual, the $2 \rightarrow 2$ kinematics of \eqref{eq:process} is conveniently described by the Mandelstam variables
\begin{align}
    s = (p_1 + p_2)^2 \quad \mathrm{and} \quad t = (p_1 - p_3)^2; \quad u = (p_2 - p_3)^2 = - s - t\,.
\end{align}
In total, five independent kinematic scales, $s$, $t$, $m_w$, $m_z$, and $m_h$, appear in our results. Here, $m_w$ is the mass of the $W$ boson, $m_z$ is the mass of the $Z$ boson, and $m_h$ is the mass of the Higgs boson.
For simplicity, we present our final results in a physical phase-space region where $t$ is allowed to take arbitrary physical values while $s$ is constrained to be above all two-particle thresholds present in the contributing Feynman diagrams. 
This amounts to the condition
\begin{align}
\label{eq:kinrestriction}
    s > \left(m_h + m_z\right)^2 \qquad\qquad (\,-s < t < 0\,)
\end{align}
imposed by the $Z$-$Z$ self-energy. 
Working in the kinematic region defined by \eqref{eq:kinrestriction} allows us to do without the complex mass scheme, thereby avoiding further technical complications for our proof-of-concept study. As discussed in \cite{Heller:2019gkq}, this restriction also simplifies the explicit expressions for the two-loop master integrals.

\subsection{Form factor decomposition}
\label{sec:ffdecomp}

We decompose the one- and two-loop scattering amplitudes for our process into basic Lorentz structures (``standard matrix elements'') multiplied by form factors, which depend only on the kinematic invariants.
Since we are considering massless external fermions in the Standard Model, the amplitude in conventional dimensional regularization can be written up to two loops in terms of the Lorentz structures~\cite{Glover:2004si}
\begin{align}
\bar{\mathcal{T}}_1 &= \bar{v}(p_2)\gamma^\mu u(p_1)\; \bar{u} (p_3)\gamma_\mu v(p_4),\\
\bar{\mathcal{T}}_2 &= \bar{v}(p_2)\slashed{p}_3 u(p_1)\; \bar{u}(p_3) \slashed{p}_{1} v(p_4),\\
\bar{\mathcal{T}}_3 &= \bar{v}(p_2)\gamma^\mu \gamma^\nu \gamma^\rho u(p_1)\; \bar{u}(p_3) \gamma_\mu \gamma_\nu \gamma_\rho v(p_4),\\
\bar{\mathcal{T}}_4 &= \bar{v}(p_2)\gamma^\mu \slashed{p}_3 \gamma^\nu u(p_1)\; \bar{u}(p_3) \gamma_\mu  \slashed{p}_{1} \gamma_\nu v(p_4),\\
\bar{\mathcal{T}}_5 &= \bar{v}(p_2)\gamma^\mu \gamma^\nu \gamma^\rho \gamma^\sigma \gamma^\tau u (p_1)\; \bar{u}(p_3) \gamma_\mu \gamma_\nu \gamma_\rho \gamma_\sigma \gamma_\tau v(p_4),
\\
\mathrm{and}\qquad\bar{\mathcal{T}}_6 &= \bar{v}(p_2)\gamma^\mu \gamma^\nu \slashed{p}_3 \gamma^\rho \gamma^\sigma u(p_1)\; \bar{u}(p_3) \gamma_\mu \gamma_\nu \slashed{p}_{1} \gamma_\rho \gamma_\sigma v(p_4),
\end{align}
where we suppressed color and spin indices and ignored insertions of $\gamma_5$.
For our EW-QCD corrections, structures $\mathcal{T}_5$ and $\mathcal{T}_6$ can be omitted even in a general $R_\xi$ gauge because the gluon couples only to the initial-state quark line.

Allowing also for insertions of $\gamma_5$ at the end of each Dirac chain, we obtain sixteen different Lorentz structures.
For the sake of our argument, let us consider an anticommuting $\gamma_5$ to illustrate the general principle of our projector method.
The described form of the amplitude can be obtained either using Passarino-Veltman reduction of the tensor integrals or using projectors.
In $d=4$ dimensions, only four out of the sixteen Lorentz structures are independent, as can be seen using Schouten and Fierz identities.
In order to facilitate a smooth $d \to 4$ limit, we can define a $d$-dimensional basis of the four structures derived from $\bar{\mathcal{T}}_1$ by inserting $\gamma_5$ matrices:
\begin{align}
\label{eq:4dLorentzGenericVV}
\bar{\mathcal{T}}_{\mathrm{VV}} &= \bar{v}(p_2)\gamma^\mu u(p_1)\;
 \bar{u}(p_3)\gamma_\mu v(p_4),\\
\label{eq:4dLorentzGenericAA}
\bar{\mathcal{T}}_{\mathrm{AA}} &= \bar{v}(p_2)\gamma^\mu \gamma_5 u(p_1)\;  
 \bar{u}(p_3)\gamma_\mu \gamma_5 v(p_4), \\
\label{eq:4dLorentzGenericVA}
\bar{\mathcal{T}}_{\mathrm{VA}} &= \bar{v}(p_2)\gamma^\mu u(p_1)\;
 \bar{u}(p_3)\gamma_\mu \gamma_5 v(p_4),\\
\label{eq:4dLorentzGenericAV}
\bar{\mathcal{T}}_{\mathrm{AV}} &= \bar{v}(p_2)\gamma^\mu \gamma_5 u(p_1)\; 
 \bar{u}(p_3)\gamma_\mu v(p_4),
\end{align}
and twelve further structures (derived from $\bar{\mathcal{T}}_2$, $\bar{\mathcal{T}}_3$, and $\bar{\mathcal{T}}_4$ in a similar way), which are strictly ``orthogonal'' in $d$ dimensions and smoothly vanish for $d\to 4$.
The $d$ dimensional projection onto the structures \eqref{eq:4dLorentzGenericVV}-\eqref{eq:4dLorentzGenericAV} obtained in this way are identical to those calculated by ignoring the evanescent directions at the outset. This procedure may be justified by considering the fact that, even in $d$ dimensions, helicity is conserved and, as is clear from Eq. \eqref{eq:eventracetermcrit}, the spin space retains its usual (integer) number of degrees of freedom, see also~\cite{Peraro:2019cjj,Peraro:2020sfm,Caola:2020dfu}.
We note that the four independent structures in $d=4$ are of course directly related to the four independent helicity amplitudes;
alternatively, for a given $\gamma_5$ scheme, we could have written polarized Lorentz structures by
replacing {\it e.g.} $u(p_i)$ with
    $\frac{1}{2}\left(1\pm\gamma_5\right) u(p_i)$.

For finite, renormalized and infrared-subtracted scattering amplitudes, $\mathcal{O}(\ep)$ deviations from the four-dimensional projectors will be irrelevant for the $\mathcal{O}\left(\ep^0\right)$ result.
For infrared divergent amplitudes, the $\mathcal{O}\left(\ep^0\right)$ term depends on the exact choice of the projector.
However, as long as we use the exact same prescription for the projector for all divergent contributions, those ambiguities will exactly cancel for a finite sum.
This argument assumes that we use an analytic continuation which gives a well-defined meaning to the higher-order-in-$\epsilon$ terms of the divergent quantities and that we do not truncate their Laurent expansions in $\epsilon$ prematurely.
In the schemes we consider here, for contributions involving Levi-Civita tensors, we expect that consistency is ensured because Levi-Civita tensors effectively truncate the higher-order-in-$\ep$ terms of finite quantities. 
We will discuss the details of how we construct our projectors for the $\gamma_5$ schemes we consider in the following.

\subsection{Calculations in HVBM's $\gamma_5$ scheme}
\label{sec:HVBMcalcs}
The Lorentz decomposition of the Drell-Yan scattering amplitude in HVBM's $\gamma_5$ scheme is
\begin{align}
\label{eq:LorentzstructureHVBMamp}
   i \mathcal{A}_{\mathrm{DY}} &= 
    i \left({\bf C}_{\mathrm{VV}}\,\mathcal{T}_{\mathrm{VV}}
  + {\bf C}_{\mathrm{AA}}\,\mathcal{T}_{\mathrm{AA}}
  + {\bf C}_{\mathrm{VA}}\,\mathcal{T}_{\mathrm{VA}}
  + {\bf C}_{\mathrm{AV}}\,\mathcal{T}_{\mathrm{AV}}\right)
\end{align}
with the Lorentz structures
\begin{align}
\mathcal{T}_{\mathrm{VV}} &= \bar{v}_k (p_2)\gamma^\mu u_j (p_1)\bar{u}_m (p_3)\gamma_\mu v_n (p_4),\\
\mathcal{T}_{\mathrm{AA}} &= \bar{v}_k (p_2)\frac{1}{2}\left[\gamma^\mu, \gamma_5\right] u_j (p_1)\bar{u}_m (p_3)\frac{1}{2}\left[\gamma_\mu, \gamma_5\right] v_n (p_4),\\
\mathcal{T}_{\mathrm{VA}} &= \bar{v}_k (p_2)\gamma^\mu u_j (p_1)\bar{u}_m (p_3)\frac{1}{2}\left[\gamma_\mu, \gamma_5\right] v_n (p_4),\\
\mathcal{T}_{\mathrm{AV}} &= \bar{v}_k (p_2)\frac{1}{2}\left[\gamma^\mu, \gamma_5\right] u_j (p_1)\bar{u}_m (p_3)\gamma_\mu v_n (p_4).
\end{align}
In the manner described in Section \ref{sec:ffdecomp}, we employ Lorentz projectors to calculate the form factors ${{\bf C}_{\mathrm{VV}}}$, ${{\bf C}_{\mathrm{AA}}}$, ${{\bf C}_{\mathrm{VA}}}$, and ${{\bf C}_{\mathrm{AV}}}$ in $d$ dimensions.
We request
\begin{equation}
    {\bf C} = \sum_{\text{spin,color}} \mathbb{P}_{{\bf C}}\, i \mathcal{A}_{\mathrm{DY}},\qquad\text{for~}{\bf C} ={{\bf C}_{\mathrm{VV}}},{{\bf C}_{\mathrm{AA}}},{{\bf C}_{\mathrm{VA}}},{{\bf C}_{\mathrm{AV}}},
\end{equation}
and obtain
\begin{align}
   \mathbb{P}_{{\bf C}_{\mathrm{VV}}} &= \frac{i\left(s^2 - 2 t u\right)}{8 N_c \left(\ep s^2\left(s^2 - 2 t u\right)-4 t^2 u^2\right)} \mathcal{T}^\dagger_{\mathrm{VV}} 
  -\frac{i s \left(t - u\right)}{8 N_c \left(\ep s^2\left(s^2 - 2 t u\right)-4 t^2 u^2\right)} \mathcal{T}^\dagger_{\mathrm{AA}}\,,
\\
   \mathbb{P}_{{\bf C}_{\mathrm{AA}}} &= -\frac{i s\left(t - u\right)}{8 N_c \left(\ep s^2\left(s^2 - 2 t u\right)-4 t^2 u^2\right)} \mathcal{T}^\dagger_{\mathrm{VV}} 
    +\frac{i\left((1-\ep)s^2 - 2 t u\right)}{8 N_c \left(\ep s^2\left(s^2 - 2 t u\right)-4 t^2 u^2\right)} \mathcal{T}^\dagger_{\mathrm{AA}}\,,
\\
    \mathbb{P}_{{\bf C}_{\mathrm{VA}}} &= -\frac{i\left(s^2 - 2 t u\right)}{32 N_c t^2 u^2} \mathcal{T}^\dagger_{\mathrm{VA}}
    +\frac{i s \left(u-t\right)}{32 N_c t^2 u^2} \mathcal{T}^\dagger_{\mathrm{AV}}\,,
\\
    \mathbb{P}_{{\bf C}_{\mathrm{AV}}} &= \frac{i s \left(u-t\right)}{32 N_c t^2 u^2} \mathcal{T}^\dagger_{\mathrm{VA}}
    -\frac{i \left(s^2 - 2 t u\right)}{32 N_c t^2 u^2} \mathcal{T}^\dagger_{\mathrm{AV}}.
\end{align}
In practice, we use these projectors only for box-type diagrams; for diagrams with self-energy or vertex insertions, we choose to sew effective propagators and vertices into appropriate tree-level diagrams. Note that, as we plan to perform a calculation of all Feynman diagrams in Kreimer's $\gamma_5$ scheme anyway, Larin's principle allows us to {\it completely bypass} the direct calculation of almost all vertex Feynman diagrams in HVBM's $\gamma_5$ scheme; as explained in Section \ref{sec:finiteren}, the vertex form factor projector defined in Eq. \eqref{eq:projVFFHVBMAV} above was employed for the calculation of the finite counterterm $\delta Z_{Z \bar{q} q}^{(0,1)}$ of Eq. \eqref{eq:qbarZqfiniteCT}, but that is all.\footnote{Strictly speaking, there is another finite counterterm which is needed as well, $\delta Z_{W^\pm \bar{q} q}^{(0,1)}$, but it may be trivially obtained from Eq. \eqref{eq:qbarZqfiniteCT} via the replacement $a_q \rightarrow a_w$.}

After carrying out the renormalization program outlined in Sections \ref{sec:couplingren}-\ref{sec:finiteren}, the loop-level amplitude form factor coefficients still diverge in the infrared. These singularities in the virtual amplitudes manifest as poles in $\epsilon$ which are canceled by taking real radiative corrections into account and, in the end, considering perturbative corrections to physical observables. Fortunately, we have {\it a priori} knowledge of the infrared singularity structure of the scattering amplitudes we calculate in this work due to the well-known dipole formula \cite{Becher:2009cu,Gardi:2009qi,Becher:2009qa} and a simple generalization thereof. In Section \ref{sec:IRsing}, we review the two-loop mixed EW-QCD neutral-current Drell-Yan dipole singularity structure derived in \cite{Kilgore:2011pa} and show how subtraction functions may be defined which allow for the comparison of finite hard scattering functions in different $\gamma_5$ schemes.

We now turn to a technical discussion of our approach to Dirac trace manipulation and numerator algebra in {\tt Mathematica}. First of all, note that all terms which involve an odd number of $\gamma_5$ matrices {\it in total} must vanish due to the fact that we consider a $2 \rightarrow 2$ scattering process with purely massless external momenta which cannot support a contraction with a Levi-Civita tensor. One possible approach to calculate the non-vanishing terms in HVBM's $\gamma_5$ scheme, popularized by \cite{Larin:1993tq}, would be to make use of the identity
\begin{align}
    \frac{1}{2} [\gamma_\mu, \gamma_5] = - \frac{i}{3!} ~\varepsilon_{\mu \nu \rho \sigma} \gamma^\nu \gamma^\rho \gamma^\sigma,
\end{align}
derived from Eqs. \eqref{eq:Diracalgebra} and \eqref{eq:covdefg5}, at the level of the $d$-dimensional Feynman rules. This prescription, however, has the very unpleasant side-effect that the lengths of {\it all} Dirac traces involving $\gamma_5$ are artificially inflated at intermediate stages. 

While the associated performance penalty could be ameliorated to some extent by exploiting the total antisymmetry of the Levi-Civita tensor, we prefer instead to follow \cite{Bern:1995db} and split indices from the very beginning. We use the anticommutation relations, Eqs. \eqref{eq:finalanticommg5pt1} and \eqref{eq:finalanticommg5pt2}, to write\footnote{To be clear, we distilled Eq. \eqref{eq:smartHVBMsimp} from \cite{Collins:1984xc}; axial vector couplings were not considered in \cite{Bern:1995db}.}
\begin{align}
\label{eq:smartHVBMsimp}
    \frac{1}{2} [\gamma_\mu, \gamma_5] = \bar{\gamma}_\mu \gamma_5\,.
\end{align}
Once all explicit $\gamma_5$ matrices have been anticommuted away and traces with an odd number of hatted variables have been discarded, we apply Eq. \eqref{eq:genDiraccontract} and its companion identity,
\begin{align}
\label{eq:genslashedmomcontract}
    \slashed{q} \gamma_{\nu_1}\cdots\gamma_{\nu_n}\slashed{q} =  2 \sum_{i = 1}^n (-1)^{i + n}  q_{\nu_i}\,\slashed{q} \gamma_{\nu_1}\cdots\gamma_{\nu_{i-1}}\gamma_{\nu_{i+1}}\cdots\gamma_{\nu_n}+q^2 (-1)^n \gamma_{\nu_1}\cdots\gamma_{\nu_n}\,,
\end{align}
to shorten all Dirac traces at an early stage to the maximal extent possible.  

In our view, this strategy is most sensible because, due to the presence of box-type diagrams, it is anyway inevitable that we must split indices at {\it some} stage. As a bonus, Eq. \eqref{eq:genslashedmomcontract} also allows us to exploit the on-shell kinematics to discard vanishing terms at an early stage. While it may not be obvious at first sight, Eqs. \eqref{eq:genDiraccontract}, \eqref{eq:eventracerecurse}, and \eqref{eq:genslashedmomcontract} can be applied verbatim to traces which contain both barred and hatted variables, provided that mixed contractions of indices are discarded as soon as they are produced by the formulae. Although it was not really needed for the two-loop calculations considered in this work, it is worth mentioning that an aggressive use of memoization for the Dirac traces appearing in our calculations could have significantly decreased the runtime required by our code to process all Feynman diagrams. 

Once all Dirac traces have been evaluated and all compatible Lorentz indices contracted, contractions and products of Levi-Civita tensors from the box-type diagrams must be dealt with using Eqs. \eqref{eq:epscontract4} - \eqref{eq:epscontract0}. 
Subsequently, all scalar products of the form $\bar{g}_{\mu \nu} k_i^\mu k_j^\nu$ must be processed using the methods reviewed in Section \ref{sec:muterms}. Finally, taking into account the algorithms of \cite{vonManteuffel:2014qoa} and their implementation in {\tt Reduze 2}, we obtain the amplitude form factor coefficients via integration by parts reduction as a linear combination of master integrals. At both one and two loops, we find it particularly convenient to rotate to {\it normal form bases} of Feynman integrals, characterized by the simple, $\ep$-decoupled differential equations they satisfy \cite{Kotikov:2010gf,Henn:2013pwa}. Our one- and two-loop normal form bases are defined in Sections \ref{sec:1Lints} and \ref{sec:2Lints}.

\subsection{Calculations in Kreimer's $\gamma_5$ scheme}
\label{sec:Kreimercalcs}
The Lorentz decomposition of the Drell-Yan scattering amplitude in Kreimer's $\gamma_5$ scheme is essentially
Eq. (B.10) of \cite{Bohm:1986rj}. Suppressing the color indices, we have
\begin{align}
\label{eq:LorentzstructureKreimeramp}
   i \bar{\mathcal{A}}_{\mathrm{DY}}
= i\left(
  {\bf \bar{C}}_{\mathrm{VV}} \bar{\mathcal{T}}_{\mathrm{VV}}
 +{\bf \bar{C}}_{\mathrm{AA}} \bar{\mathcal{T}}_{\mathrm{AA}}
 +{\bf \bar{C}}_{\mathrm{VA}} \bar{\mathcal{T}}_{\mathrm{VA}}
 +{\bf \bar{C}}_{\mathrm{AV}} \bar{\mathcal{T}}_{\mathrm{AV}}
\right)
\end{align}
where the Lorentz structures are defined in \eqref{eq:4dLorentzGenericVV}-\eqref{eq:4dLorentzGenericAV}.
The relevant Lorentz projectors in Kreimer's $\gamma_5$ scheme fulfill
\begin{equation}
    {\bf \bar{C}} = \sum_{\text{spin,color}} \mathbb{P}_{{\bf \bar{C}}}\, i \bar{\mathcal{A}}_{\mathrm{DY}},\qquad\text{for~}{\bf \bar{C}} ={{\bf \bar{C}}_{\mathrm{VV}}},{{\bf \bar{C}}_{\mathrm{AA}}},{{\bf \bar{C}}_{\mathrm{VA}}},{{\bf \bar{C}}_{\mathrm{AV}}},
\end{equation}
and are given by
\begin{align}
\label{eq:boxprojKVV}
  \mathbb{P}_{{\bf \bar{C}}_{\mathrm{VV}}} &=
  -\frac{i\left((1-\ep)s^2 + 2 t (s+t)\right) }{8 N_c \left(2 t^2-\ep s^2\right)D} \bar{\mathcal{T}}^\dagger_{\mathrm{VV}}
  +\frac{i s \left(s + 2 t\right)}{8 N_c \left(2 t^2-\ep s^2\right)D} \bar{\mathcal{T}}^\dagger_{\mathrm{AA}}\,,
    \\
\label{eq:boxprojKAVAV}
    \mathbb{P}_{{\bf \bar{C}}_{\mathrm{AA}}} &= \frac{i s \left(s + 2 t\right)}{8 N_c \left(2 t^2-\ep s^2\right)D} \bar{\mathcal{T}}^\dagger_{\mathrm{VV}}
  -\frac{i\left((1-\ep)s^2 + 2 t (s+t)\right)}{8 N_c \left(2 t^2-\ep s^2\right)D} \bar{\mathcal{T}}^\dagger_{\mathrm{AA}}\,,
    \\
\label{eq:boxprojKVAV}
    \mathbb{P}_{{\bf \bar{C}}_{\mathrm{VA}}} &= -\frac{i\left((1-\ep)s^2 + 2 t (s+t)\right)}{8 N_c \left(2 t^2-\ep s^2\right)D} \bar{\mathcal{T}}^\dagger_{\mathrm{VA}}
    +\frac{i s \left(s + 2 t\right)}{8 N_c \left(2 t^2-\ep s^2\right)D} \bar{\mathcal{T}}^\dagger_{\mathrm{AV}}\,,
    \\
\label{eq:boxprojKAVV}
    \mathbb{P}_{{\bf \bar{C}}_{\mathrm{AV}}} &= \frac{i s \left(s + 2 t\right)}{8 N_c \left(2 t^2-\ep s^2\right)D} \bar{\mathcal{T}}^\dagger_{\mathrm{VA}}
    -\frac{i\left((1-\ep)s^2 + 2 t (s+t)\right)}{8 N_c \left(2 t^2-\ep s^2\right)D} \bar{\mathcal{T}}^\dagger_{\mathrm{AV}}\,,
\end{align}
with $D \equiv (2-\ep)s^2 + 2 t (2 s + t)$.

Again, we employ Eqs. \eqref{eq:boxprojKVV} - \eqref{eq:boxprojKAVV} only for the box contributions. For the vertex corrections, the Kreimer's $\gamma_5$ scheme analogs of Eqs. \eqref{eq:LorentzVFFHVBM}, \eqref{eq:projVFFHVBMV}, and \eqref{eq:projVFFHVBMAV},
\begin{align}
\label{eq:LorentzVFFKquark}
    i e\, \Big(\bar{\mathcal{F}}_\mu^{V\bar{q} q}(p_1,p_2)\Big)_{kj} &= i e\, \bar{\mathcal{V}}_{V \bar{q} q}(s)\,\bar{v}_k (p_2)\gamma_\mu u_j (p_1)
     \\
    &+ i e\, \bar{\mathcal{A}}_{V \bar{q} q}(s)\,\bar{v}_k (p_2)\gamma_\mu\gamma_5 u_j (p_1)\,,\nonumber
    \\
\label{eq:projVFFKqV}
    \Big(\mathbb{P}^{\hspace{.025 cm}\mu}_{\bar{\mathcal{V}}_q}\Big)_{j k} &= \frac{i}{4 e N_c s(1-\ep)} \bar{u}_j (p_1)\gamma^\mu v_k (p_2)\,,
    \\
\label{eq:projVFFKqAV}
    \Big(\mathbb{P}^{\hspace{.025 cm} \mu}_{\bar{\mathcal{A}}_q}\Big)_{jk} &= \frac{i}{4 e N_c s (1-\ep)} \bar{u}_j (p_1)\gamma^\mu \gamma_5 v_k (p_2)\,,
\end{align}
are useful.\footnote{In Eqs. \eqref{eq:LorentzVFFKquark} and \eqref{eq:LorentzVFFKlepton}, $V$ may be either a photon or a $Z$ boson.} Of course, a decomposition completely analogous to \eqref{eq:LorentzVFFKquark} exists for the lepton vertex form factors,
\begin{align}
\label{eq:LorentzVFFKlepton}
  i e\, \Big(\bar{\mathcal{F}}_\mu^{V\bar{\ell} \ell}(p_3,p_4)\Big)_{m n} &= i e\, \bar{\mathcal{V}}_{V \bar{\ell} \ell}(s)\,\bar{u}_m (p_3)\gamma_\mu v_n (p_4)
    \\
    &+ i e\, \bar{\mathcal{A}}_{V \bar{\ell} \ell}(s)\,\bar{u}_m (p_3)\gamma_\mu \gamma_5 v_n (p_4)\,.\nonumber
\end{align}
$\Big(\mathbb{P}^{\hspace{.025 cm}\mu}_{\bar{\mathcal{V}}_\ell}\Big)_{n m}$ and $\Big(\mathbb{P}^{\hspace{.025 cm}\mu}_{\bar{\mathcal{A}}_\ell}\Big)_{n m}$ are obtained by setting $N_c = 1$ in Eqs. \eqref{eq:projVFFKqV} and \eqref{eq:projVFFKqAV},
\begin{align}
\label{eq:projVFFKlV}
    \Big(\mathbb{P}^{\hspace{.025 cm}\mu}_{\bar{\mathcal{V}}_\ell}\Big)_{n m} &= \frac{i}{4 e s(1-\ep)} \bar{v}_n (p_4)\gamma^\mu u_m (p_3)\,,
    \\
\label{eq:projVFFKlAV}
    \Big(\mathbb{P}^{\hspace{.025 cm} \mu}_{\bar{\mathcal{A}}_\ell}\Big)_{n m} &= \frac{i}{4 e s (1-\ep)} \bar{v}_n (p_4)\gamma^\mu \gamma_5 u_m (p_3)\,.
\end{align}
Note that the symmetric form of {\it e.g.} Eqs. \eqref{eq:projVFFKqV} and \eqref{eq:projVFFKqAV} is a reflection of the absence of chiral mismatch in Kreimer's $\gamma_5$ scheme (contrast with the asymmetric form of Eqs. \eqref{eq:projVFFHVBMV} and \eqref{eq:projVFFHVBMAV}).

We calculate our one- and two-loop scattering amplitudes in {\tt Mathematica}, using a code pipeline similar to that described in Section \ref{sec:HVBMcalcs}; as mentioned in Section \ref{sec:Kreimerdefs}, the algebraic form of the non-vanishing families of even and odd Dirac traces is the same in Kreimer's $\gamma_5$ scheme as in HVBM's $\gamma_5$ scheme. A key advantage of Kreimer's $\gamma_5$ scheme, however, is that, due to the standard anticommutation relation, Eq. \eqref{eq:anticommg5}, Dirac traces need not be calculated with split indices. Rather, one can first evaluate all traces and introduce scalar products of the form $\bar{g}_{\mu \nu} k_i^\mu k_j^\nu$ only at a later stage, when contractions and products of Levi-Civita tensors are eliminated using Eqs. \eqref{eq:epscontract4} - \eqref{eq:epscontract0}.
As a cross-check, we also calculated the interference of the one- and two-loop scattering amplitudes with the tree-level Drell-Yan amplitude in Kreimer's $\gamma_5$ scheme using an independent setup in {\tt FORM 4} \cite{Kuipers:2012rf} which, in particular, employed Passarino-Veltman reduction \cite{Passarino:1978jh} instead of the technology of Section \ref{sec:muterms} to simplify scalar products of the form $\bar{g}_{\mu \nu} k_i^\mu k_j^\nu$. Finding complete agreement between our independent implementations served as a crucial cross-check on our calculations.
\subsection{Infrared dipole singularity structure and subtraction functions}
\label{sec:IRsing}
The dipole formula \cite{Becher:2009cu,Gardi:2009qi,Becher:2009qa} provides a particularly concise and straightforward recipe for the generation of infrared subtraction terms for virtual gauge theory scattering amplitudes calculated in pure dimensional regularization. While, historically, the singularity structures of scattering amplitudes in particularly simple models such as massless QED, massless QCD, and $\mathcal{N} = 4$ super Yang-Mills theory were discussed most frequently in the literature, broader applications to theories with massive particles are certainly possible and have long been known (see {\it e.g.} \cite{Becher:2009kw}). Indeed, it was shown in \cite{Kilgore:2011pa} that with straightforward modifications, the dipole formula can be extended to describe the singularity structure of the two-loop mixed EW-QCD neutral-current Drell-Yan scattering amplitude of interest.

We begin by introducing the building blocks required to describe the infrared singularities of the order $\alpha \alpha_s$, order $\alpha^2$, and order $\alpha^2 \alpha_s$ neutral-current Drell-Yan scattering amplitudes. As has long been clear, the leading infrared singularities of gauge theory scattering amplitudes are governed by cusp anomalous dimensions \cite{Korchemsky:1985xj}. The one-loop quark cusp anomalous dimension of massless QCD may be extracted, for example, from the expression for the one-loop quark form factor of massless QCD given in Eq. \eqref{eq:oneloopQCDquarkFF}. In QCD, we have
\begin{align}
\label{eq:QCDquarkcusp}
    \Gamma_q^{(0,1)} = 4\,C_F\,,
\end{align}
whereas in QED we obtain the result from Eq. \eqref{eq:QCDquarkcusp} by replacing the quadratic Casimir invariant $C_F$ with the squared charge of the fermion flavor,
\begin{align}
\label{eq:QEDfermioncusp}
    \Gamma_f^{(1,0)} = 4\,Q_f^2\,.
\end{align}
However, as pointed out in \cite{Kilgore:2011pa}, the mixed quark cusp anomalous dimension vanishes:
\begin{align}
\label{eq:mixedquarkcusp}
    \Gamma_q^{(1,1)} = 0\,.
\end{align}

The next-to-leading infrared singularities are more complicated and receive contributions from both soft anomalous dimensions and resummation functions derived from the $\ep^{-1}$ poles of massless vertex form factors with an appropriate number of gluon and/or photon exchanges. The one-loop Drell-Yan soft anomalous dimensions are well-known and have been calculated many places in the literature, see {\it e.g.} \cite{Aybat:2006wq,Aybat:2006mz} for a thorough discussion.\footnote{Note that, for general QCD scattering processes, the soft anomalous dimension becomes a mixing matrix in color space. For processes with just two partons in the initial state such as the Drell-Yan process, the color space structure trivializes and all matrices in the dipole formula may be replaced with functions.} Rewriting the results to make all imaginary parts explicit in the physical kinematic region of interest, we find
\begin{align}
\label{eq:QCDDYsoftas}
    S^{(0,1)}_{\mathrm{DY}} &= \left(- 4 \ln\left(\frac{\mu^2}{s}\right) - 4 i \pi\right) C_F\\
    \label{eq:QCDDYsofta}
\mathrm{and}\qquad S^{(1,0)}_{\mathrm{DY}} &= \left(- 4 \ln\left(\frac{\mu^2}{s}\right) - 4 i \pi\right) \left(Q_\ell^2+Q_q^2\right)+8 Q_\ell Q_q\ln\left(\frac{t}{u}\right).
\end{align}
Once again, as shown in \cite{Kilgore:2011pa}, the two-loop mixed EW-QCD Drell-Yan soft anomalous dimension vanishes identically:
\begin{align}
\label{eq:mixedDYsoft}
    S^{(1,1)}_{\mathrm{DY}} = 0\,.
\end{align}

We define our massless QCD resummation functions in the framework of \cite{Moch:2005id}. In practice, following \cite{Kilgore:2011pa}, all results may be derived from the one- and two-loop quark resummation functions of massless QCD by making simple replacements. In what follows, $f[k]$ denotes the coefficient of $\ep^k$ in the Taylor series expansion of $f(\ep)$,
\begin{align}
    f(\ep) = \sum_{k = 0}^\infty f[k] \ep^k\,.
\end{align} 
To the relevant $\ep$ orders, we have 
\begin{align}
\label{eq:Gqalphasep0}
    G_q^{(0,1)}[0] &= 6\, C_F\,,\\
    G_q^{(0,1)}[1] &= \left(16-2\zeta_2\vphantom{\frac{28}{3}}\right)C_F\,,\\
    G_q^{(0,1)}[2] &= \left(32-3\zeta_2-\frac{28}{3}\zeta_3\right)C_F\,,\\
\label{eq:Gqalphasep3}
    G_q^{(0,1)}[3] &= \left(64-8\zeta_2-14\zeta_3-\frac{47}{10}\zeta_2^2\right)C_F\,,\\
\label{eq:Gqalphas2ep0}
    G_q^{(0,2)}[0] &= \left(3-24 \zeta_2+48\zeta_3\vphantom{\frac{28}{3}}\right)C_F^2
    \nonumber\\
    &+ \left(\frac{2545}{27}+\frac{44}{3}\zeta_2-52\zeta_3\right)C_A C_F\,,\\
\label{eq:Gqalphas2ep1}
  \mathrm{and}\qquad  G_q^{(0,2)}[1] &= \left(\frac{1}{2}-116\zeta_2+120\zeta_3+\frac{176}{5}\zeta_2^2\right)C_F^2
  \nonumber\\
  &+ \left(\frac{70165}{162}+\frac{575}{9}\zeta_2-\frac{520}{3}\zeta_3-\frac{176}{5}\zeta_2^2\right)C_A C_F
\end{align}
from Eqs. (3.10) and (3.11) of \cite{Moch:2005id} after discarding all contributions proportional to the number of light quarks. 

As with the cusp anomalous dimensions, the one-loop QED results may be obtained by making the replacement $C_F \rightarrow Q_f^2$ in Eqs. \eqref{eq:Gqalphasep0}-\eqref{eq:Gqalphasep3}. Explicitly, we have
\begin{align}
\label{eq:Galphaep0}
    G_f^{(1,0)}[0] &= 6\, Q_f^2\,,\\
    G_f^{(1,0)}[1] &= \left(16-2\zeta_2\vphantom{\frac{28}{3}}\right)Q_f^2\,,\\
    G_f^{(1,0)}[2] &= \left(32-3\zeta_2-\frac{28}{3}\zeta_3\right)Q_f^2\,,\\
\label{eq:Galphaep3}
\mathrm{and}\qquad    G_f^{(1,0)}[3] &= \left(64-8\zeta_2-14\zeta_3-\frac{47}{10}\zeta_2^2\right)Q_f^2\,.
\end{align}
Finally, the two-loop mixed EW-QCD quark resummation functions are obtained from Eqs. \eqref{eq:Gqalphas2ep0} and \eqref{eq:Gqalphas2ep1} by setting $C_A$ to zero and replacing $C_F^2$ with $Q_q^2 C_F$:
\begin{align}
\label{eq:Galphaalphasep0}
    G_q^{(1,1)}[0] &= \left(3-24 \zeta_2+48\zeta_3\vphantom{\frac{28}{3}}\right)Q_q^2 C_F \\
\label{eq:Galphaalphasep1}
  \mathrm{and}\qquad  G_q^{(1,1)}[1] &= \left(\frac{1}{2}-116\zeta_2+120\zeta_3+\frac{176}{5}\zeta_2^2\right)Q_q^2 C_F \,.
\end{align}

Given the ingredients discussed above, we are now in a position to present the predictions of the generalized dipole formula for the ultraviolet-renormalized, infrared-divergent scattering amplitudes $\mathcal{A}_{\mathrm{DY}}$ and $\bar{\mathcal{A}}_{\mathrm{DY}}$ in, respectively,  HVBM's $\gamma_5$ scheme and Kreimer's $\gamma_5$ scheme.
At tree level we have
\begin{align}
\label{eq:hardfuncnotationHVBM}
    \pbar{\mathcal{A}}_{\mathrm{DY}}
    &= 4\pi\alpha \pbar{\mathcal{H}}_{\rm DY}^{(0,0)}[0]  + \cdots,
\end{align}
where the dots stand for terms of higher order in the coupling constants.
 As the notation suggests, the tree-level hard functions, ${\mathcal{H}}_{\rm DY}^{(0,0)}[0]$ and $\bar{\mathcal{H}}_{\rm DY}^{(0,0)}[0]$, are defined to be independent of the coupling constants. We have
\begin{align}
\label{eq:expltreelevel}
\pbar{\mathcal{H}}_{\rm DY}^{(0,0)}[0] &= 
  \left(\frac{Q_q Q_\ell}{s}+\frac{v_q v_\ell}{s-m_z^2}\right)\pbar{\mathcal{T}}_{\mathrm{VV}}
- \frac{a_\ell v_q}{s-m_z^2}\pbar{\mathcal{T}}_{\mathrm{VA}}
- \frac{a_q v_\ell}{s-m_z^2}\pbar{\mathcal{T}}_{\mathrm{AV}}
+ \frac{a_q a_\ell}{s-m_z^2}\pbar{\mathcal{T}}_{\mathrm{AA}}
\end{align}
in the notation of \cite{Bohm:1986rj} for the couplings of the $Z$ boson to matter, see Eqs. \eqref{eq:zalias}.

As is clear from the form of Eqs. \eqref{eq:expltreelevel}, the tree-level amplitudes in the two $\gamma_5$ schemes we consider trivially coincide in the four-dimensional limit because all of the differences between the schemes disappear in four spacetime dimensions:
\begin{align}
\label{eq:treehardfuncprec}
    \lim_{\ep \to 0}\left\{ \mathcal{H}_{\rm DY}^{(0,0)}[0]\right\} = \lim_{\ep \to 0}\left\{ \bar{\mathcal{H}}_{\rm DY}^{(0,0)}[0]\right\}.
\end{align} 
For the remainder of this work, it will sometimes be convenient to view the hard scattering functions in the two schemes we consider as vectors of coefficients with respect to $\pbar{\mathcal{T}}_{\mathrm{VV}}$, $\pbar{\mathcal{T}}_{\mathrm{VA}}$, $\pbar{\mathcal{T}}_{\mathrm{AV}}$, and $\pbar{\mathcal{T}}_{\mathrm{AA}}$. This minor abuse of notation will allow us to rewrite {\it e.g.} Eq. \eqref{eq:treehardfuncprec} as
\begin{align}
    \mathcal{H}_{\rm DY}^{(0,0)}[0] = \bar{\mathcal{H}}_{\rm DY}^{(0,0)}[0]\,.
\end{align}

At higher perturbative orders, we have 
\begin{align}
\label{eq:subfuncs}
    &\pbar{\mathcal{A}}_{\mathrm{DY}}
    = 4\pi\alpha \Bigg( \coupling{\Big\{} \pbar{\mathcal{H}}_{\rm DY}^{(0,0)}[0] \coupling{\Big\}}
    \nonumber\\
&+ \coupling{\Bigg\{}\pole{\frac{1}{\ep^2}\Bigg(}-\frac{1}{2}\Gamma_q^{(0,1)}\pbar{\mathcal{H}}_{\rm DY}^{(0,0)}[0]\pole{\Bigg)}+\pole{\frac{1}{\ep}\Bigg(}\frac{1}{2}\Big[S_{\rm DY}^{(0,1)}-G_q^{(0,1)}[0]\Big]\pbar{\mathcal{H}}_{\rm DY}^{(0,0)}[0]\pole{\Bigg)}
\nonumber \\
&-\frac{1}{2}G_q^{(0,1)}[1]\pbar{\mathcal{H}}_{\rm DY}^{(0,0)}[0] + \pbar{\mathcal{H}}_{\rm DY}^{(0,1)}[0]+\pole{\ep\Bigg(}
    -\frac{1}{2}G_q^{(0,1)}[2]\pbar{\mathcal{H}}_{\rm DY}^{(0,0)}[0] + \pbar{\mathcal{H}}_{\rm DY}^{(0,1)}[1]\pole{\Bigg)}
\nonumber \\
    & 
    +\pole{\ep^2\Bigg(} -\frac{1}{2}G_q^{(0,1)}[3]\pbar{\mathcal{H}}_{\rm DY}^{(0,0)}[0] + \pbar{\mathcal{H}}_{\rm DY}^{(0,1)}[2]\pole{\Bigg)} + \cdots\coupling{\Bigg\}}\,
    \coupling{\left(\frac{\alpha_s}{4\pi}\right)}
    \nonumber\\
    &+ \coupling{\Bigg\{} \pole{\frac{1}{\ep^2}}\pole{\Bigg(}-\frac{1}{2}\left[ \Gamma_\ell^{(1,0)}+\Gamma_q^{(1,0)}\right]\pbar{\mathcal{H}}_{\rm DY}^{(0,0)}[0]\pole{\Bigg)}
    +\pole{\frac{1}{\ep}}\pole{\Bigg(}\frac{1}{2}\Big[ S_{\rm DY}^{(1,0)}-G_\ell^{(1,0)}[0]-G_q^{(1,0)}[0]\Big]\pbar{\mathcal{H}}_{\rm DY}^{(0,0)}[0]\pole{\Bigg)}
    \nonumber \\
    & -\frac{1}{2}\Big[G_\ell^{(1,0)}[1]+G_q^{(1,0)}[1]\Big]\pbar{\mathcal{H}}_{\rm DY}^{(0,0)}[0]+ \pbar{\mathcal{H}}_{\rm DY}^{(1,0)}[0]+\pole{\ep\Bigg(}
    -\frac{1}{2}\Big[G_\ell^{(1,0)}[2]+G_q^{(1,0)}[2]\Big]\pbar{\mathcal{H}}_{\rm DY}^{(0,0)}[0] 
    \nonumber \\
    &+ \pbar{\mathcal{H}}_{\rm DY}^{(1,0)}[1]\pole{\Bigg)}
   +\pole{\ep^2\Bigg(} -\frac{1}{2}\Big[G_\ell^{(1,0)}[3]+G_q^{(1,0)}[3]\Big]\pbar{\mathcal{H}}_{\rm DY}^{(0,0)}[0] + \pbar{\mathcal{H}}_{\rm DY}^{(1,0)}[2]\pole{\Bigg)} + \cdots\coupling{\Bigg\}}\,\coupling{\left(\frac{\alpha}{4\pi}\right)}
    \nonumber\\
    &+ \coupling{\Bigg\{}\pole{\frac{1}{\ep^4}\Bigg(}\frac{1}{4}\Gamma_q^{(0,1)}\left[\Gamma_\ell^{(1,0)}+\Gamma_q^{(1,0)}\right]\pbar{\mathcal{H}}_{\rm DY}^{(0,0)}[0]\pole{\Bigg)} + \pole{\frac{1}{\ep^3}\Bigg(}-\frac{1}{4}\bigg(\Gamma_q^{(0,1)}\left[S_{\rm DY}^{(1,0)}-G_\ell^{(1,0)}[0]-G_q^{(1,0)}[0]\right]
    \nonumber \\
    &+\left[\Gamma_\ell^{(1,0)}+\Gamma_q^{(1,0)}\right]\left[S_{\rm DY}^{(0,1)}-G_q^{(0,1)}[0]\right]\bigg)\pbar{\mathcal{H}}_{\rm DY}^{(0,0)}[0]\pole{\Bigg)}+\pole{\frac{1}{\ep^2}\Bigg(}\frac{1}{4}\bigg(\Gamma_q^{(0,1)}\left[G_\ell^{(1,0)}[1]+G_q^{(1,0)}[1]\right]
    \nonumber \\
    & +\left[\Gamma_\ell^{(1,0)}+\Gamma_q^{(1,0)}\right]G_q^{(0,1)}[1]+\left[S_{\rm DY}^{(0,1)}-G_q^{(0,1)}[0]\right]\left[S_{\rm DY}^{(1,0)}-G_\ell^{(1,0)}[0]-G_q^{(1,0)}[0]\right]\bigg)\pbar{\mathcal{H}}_{\rm DY}^{(0,0)}[0]
    \nonumber \\
    &-\frac{1}{2}\Gamma_q^{(0,1)}\pbar{\mathcal{H}}_{\rm DY}^{(1,0)}[0]-\frac{1}{2}\left[\Gamma_\ell^{(1,0)}+\Gamma_q^{(1,0)}\right]\pbar{\mathcal{H}}_{\rm DY}^{(0,1)}[0]\pole{\Bigg)}+ \pole{\frac{1}{\ep}\Bigg(}\frac{1}{4}\bigg(\Gamma_q^{(0,1)}\left[G_\ell^{(1,0)}[2]+G_q^{(1,0)}[2]\right]
    \nonumber \\
    &+\left[\Gamma_\ell^{(1,0)}+\Gamma_q^{(1,0)}\right]G_q^{(0,1)}[2]-G_q^{(0,1)}[1]\left[S_{\rm DY}^{(1,0)}-G_\ell^{(1,0)}[0]-G_q^{(1,0)}[0]\right]
    \nonumber \\
    &-\left[G_\ell^{(1,0)}[1]+G_q^{(1,0)}[1]\right]\left[S_{\rm DY}^{(0,1)}-G_q^{(0,1)}[0]\right]-2 G_q^{(1,1)}[0]\bigg)\pbar{\mathcal{H}}_{\rm DY}^{(0,0)}[0]-\frac{1}{2}\Gamma_q^{(0,1)}\pbar{\mathcal{H}}_{\rm DY}^{(1,0)}[1]
    \nonumber \\
    &+\frac{1}{2}\left[S_{\rm DY}^{(0,1)}-G_q^{(0,1)}[0]\right]\pbar{\mathcal{H}}_{\rm DY}^{(1,0)}[0]+\frac{1}{2}\left[S_{\rm DY}^{(1,0)}-G_\ell^{(1,0)}[0]-G_q^{(1,0)}[0]\right]\pbar{\mathcal{H}}_{\rm DY}^{(0,1)}[0] 
    \nonumber \\
    &-\frac{1}{2}\left[\Gamma_\ell^{(1,0)}+\Gamma_q^{(1,0)}\right]\pbar{\mathcal{H}}_{\rm DY}^{(0,1)}[1]\pole{\Bigg)}
    +\frac{1}{4}\bigg(\Gamma_q^{(0,1)}\left[G_\ell^{(1,0)}[3]+G_q^{(1,0)}[3]\right]+\left[\Gamma_\ell^{(1,0)}+\Gamma_q^{(1,0)}\right]G_q^{(0,1)}[3] 
    \nonumber \\
    &-G_q^{(0,1)}[2]\left[S_{\rm DY}^{(1,0)}-G_\ell^{(1,0)}[0] -G_q^{(1,0)}[0]\right]-\left[G_\ell^{(1,0)}[2]+G_q^{(1,0)}[2]\right]\left[S_{\rm DY}^{(0,1)}-G_q^{(0,1)}[0]\right]
    \nonumber \\
    &+G_q^{(0,1)}[1]\left[G_\ell^{(1,0)}[1]+G_q^{(1,0)}[1]\right]-2 G_q^{(1,1)}[1]\bigg)\pbar{\mathcal{H}}_{\rm DY}^{(0,0)}[0]-\frac{1}{2}G_q^{(0,1)}[1]\pbar{\mathcal{H}}_{\rm DY}^{(1,0)}[0]
    \nonumber \\
    &-\frac{1}{2}\left[G_\ell^{(1,0)}[1]+G_q^{(1,0)}[1]\right]\pbar{\mathcal{H}}_{\rm DY}^{(0,1)}[0]+\frac{1}{2}\left[S_{\rm DY}^{(0,1)}-G_q^{(0,1)}[0]\right]\pbar{\mathcal{H}}_{\rm DY}^{(1,0)}[1]
    \nonumber \\
    &+\frac{1}{2}\left[S_{\rm DY}^{(1,0)}-G_\ell^{(1,0)}[0]-G_q^{(1,0)}[0]\right]\pbar{\mathcal{H}}_{\rm DY}^{(0,1)}[1]-\frac{1}{2}\Gamma_q^{(0,1)}\pbar{\mathcal{H}}_{\rm DY}^{(1,0)}[2]-\frac{1}{2}\left[\Gamma_\ell^{(1,0)}+\Gamma_q^{(1,0)}\right]\pbar{\mathcal{H}}_{\rm DY}^{(0,1)}[2]
    \nonumber \\
    &+\pbar{\mathcal{H}}_{\rm DY}^{(1,1)}[0]+\cdots\coupling{\Bigg\}}\, \coupling{\left(\frac{\alpha}{4\pi}\right)\left(\frac{\alpha_s}{4\pi}\right)} + \cdots \Bigg)\,,
\end{align}
where the one- and two-loop hard scattering functions which appear above are determined by demanding equality between Eqs. \eqref{eq:subfuncs} and the corresponding explicit calculations at one and two loops.

All other things being equal, it is generally believed that HVBM's $\gamma_5$ scheme and Kreimer's $\gamma_5$ scheme produce identical results through to one loop and $\mathcal{O}\left(\ep^0\right)$ \cite{Denner:2019vbn}. This proposition was implicitly affirmed at order $\alpha \alpha_s$ in Section \ref{sec:finiteren}. At order $\alpha^2$, our explicit one-loop calculations, discussed in detail in the next section, do indeed show that
\begin{align}
\label{eq:hardscatalpha2}
    \mathcal{H}_{\rm DY}^{(1,0)}[0] = \bar{\mathcal{H}}_{\rm DY}^{(1,0)}[0]\,.
\end{align}
However, we also find that 
\begin{align}
    \mathcal{H}_{\rm DY}^{(1,0)}[k] \neq \bar{\mathcal{H}}_{\rm DY}^{(1,0)}[k]\qquad \qquad \mathrm{for}~k>0\,.
\end{align}
Thus, in light of the fact that the higher-order-in-$\ep$ one-loop hard functions first enter Eqs. \eqref{eq:subfuncs} at the level of the $\ep^{-1}$ poles of the relative order $\alpha \alpha_s$ expressions, it is not obvious that an analog of Eq. \eqref{eq:hardscatalpha2} continues to hold at higher orders in perturbation theory:
\begin{align}
\label{eq:hardscatalpha2alphas}
    \mathcal{H}_{\rm DY}^{(1,1)}[0] \stackrel{?}{=} \bar{\mathcal{H}}_{\rm DY}^{(1,1)}[0]\,.
\end{align}
In the Standard Model, it is expected that all terms of virtual and real radiative corrections sensitive to the $\gamma_5$ problem eventually cancel out of combinations which furnish complete fixed-order perturbative corrections to physical observables \cite{Korner:1985uj}. Therefore, it is expected that \eqref{eq:hardscatalpha2alphas} actually {\it is} an equality.
We will test this expectation with our explicit calculations below.

\section{One-loop scattering amplitudes}

\subsection{Diagrammatic structure}
\label{sec:1Ldiags}
In this section, we present the diagrammatic structure of the one-loop perturbative corrections to the neutral-current Drell-Yan process. Due to their familiarity, the renormalized self-energy corrections to the photon and $Z$ propagators,
\begin{align*}
    \includegraphics[valign=m,height=.5\linewidth,width=.5\linewidth,keepaspectratio]{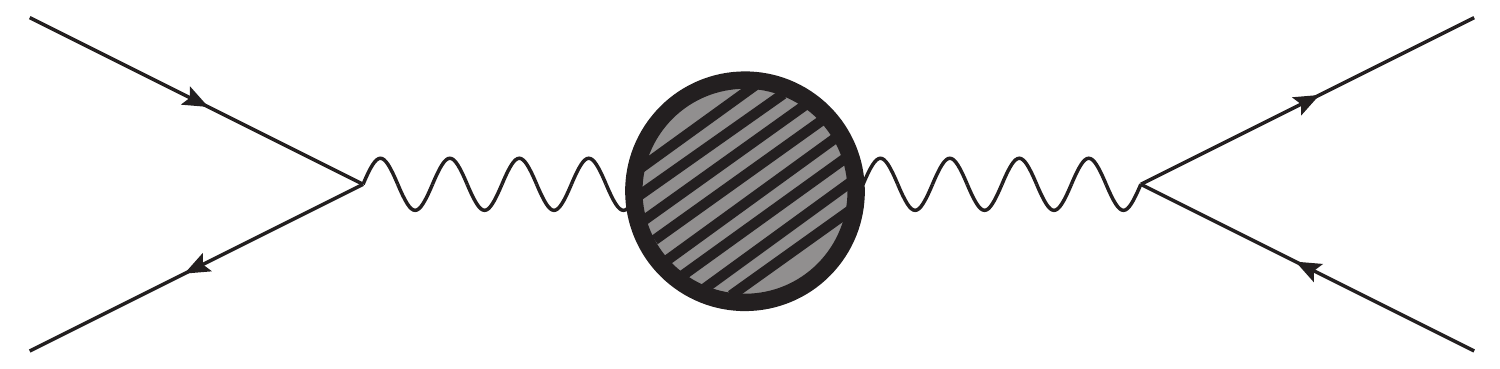}\,,
\end{align*}
will not be visualized in explicit detail. In what follows, we will also use a generic wavy line for both the photon and the $Z$, keeping in mind that some contributions involving {\it e.g.} a coupling to neutrinos can occur only for the $Z$. As is well-known, just a single class of diagram contributes to the one-loop corrections of relative order $\alpha_s$,
\begin{align*}
    \includegraphics[valign=m,height=.315\linewidth,width=.315\linewidth,keepaspectratio]{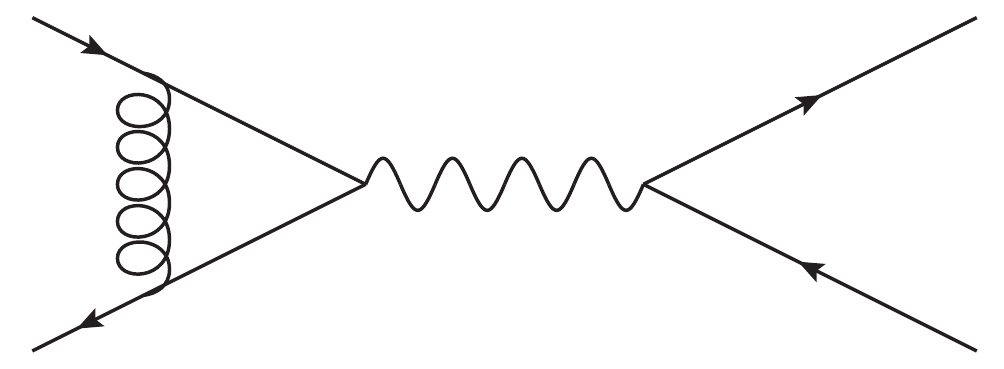}\,.
\end{align*}

The remaining one-loop corrections of relative order $\alpha$ fall into three categories: initial state vertex diagrams, final state vertex diagrams, and box diagrams. Both the initial state vertex diagrams
\begin{align*}
    \includegraphics[valign=m,height=.315\linewidth,width=.315\linewidth,keepaspectratio]{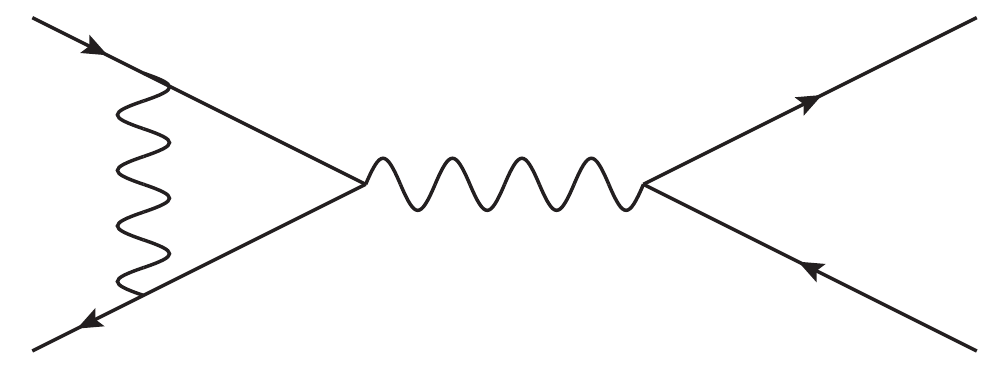}\quad \includegraphics[valign=m,height=.315\linewidth,width=.315\linewidth,keepaspectratio]{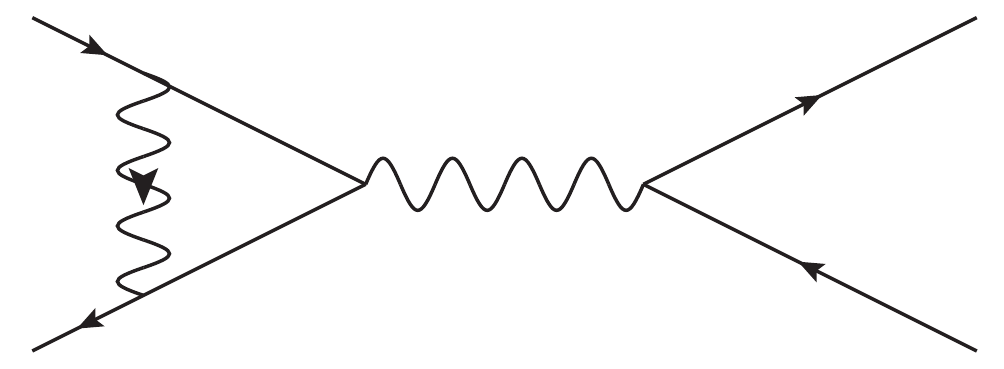}\quad
    \includegraphics[valign=m,height=.315\linewidth,width=.315\linewidth,keepaspectratio]{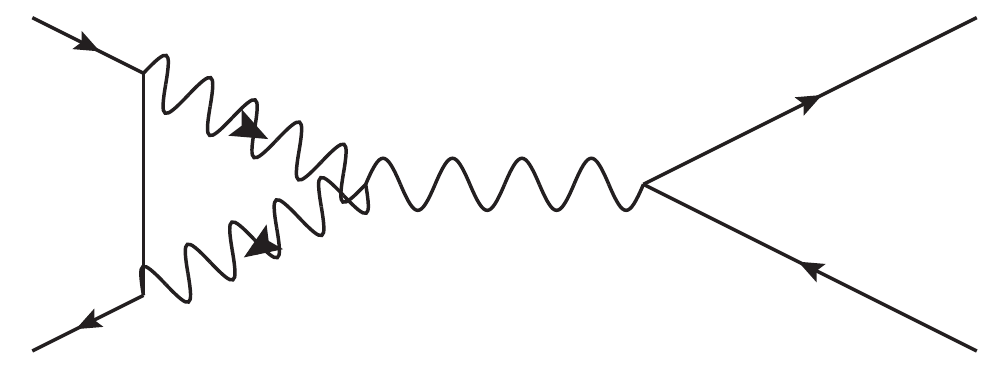}
\end{align*}
and the final state vertex diagrams
\begin{align*}
    \includegraphics[valign=m,height=.315\linewidth,width=.315\linewidth,keepaspectratio]{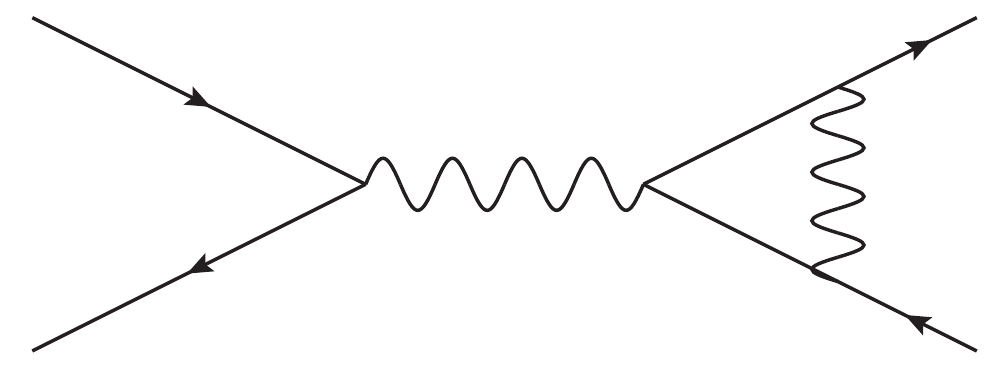}\quad \includegraphics[valign=m,height=.315\linewidth,width=.315\linewidth,keepaspectratio]{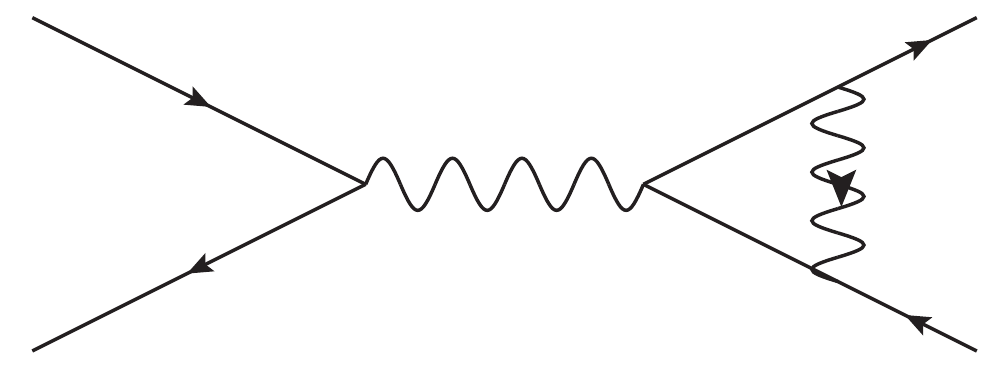}\quad
    \includegraphics[valign=m,height=.315\linewidth,width=.315\linewidth,keepaspectratio]{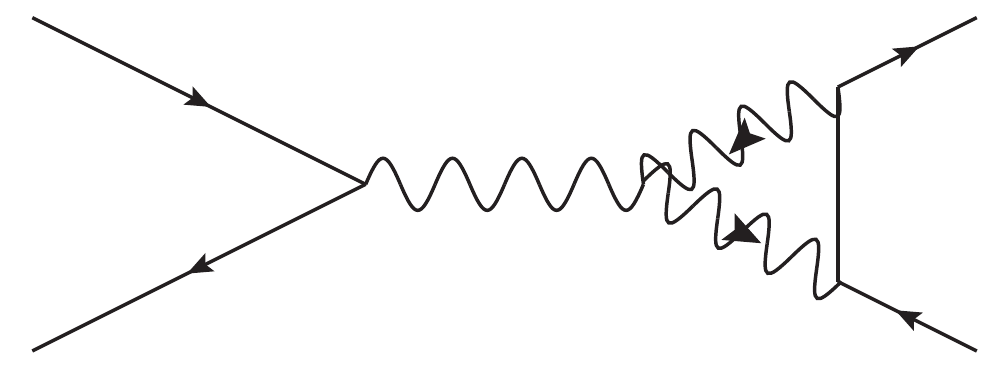}
\end{align*}
require renormalization,
\begin{align*}
    \includegraphics[valign=m,height=.315\linewidth,width=.315\linewidth,keepaspectratio]{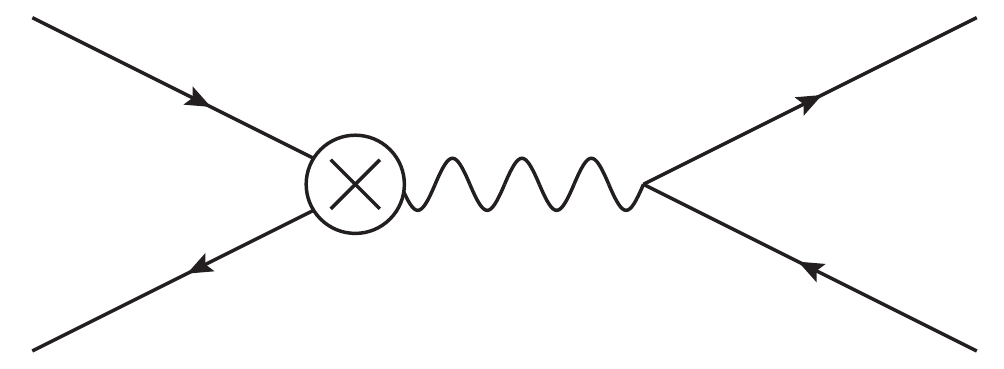}\quad
    \includegraphics[valign=m,height=.315\linewidth,width=.315\linewidth,keepaspectratio]{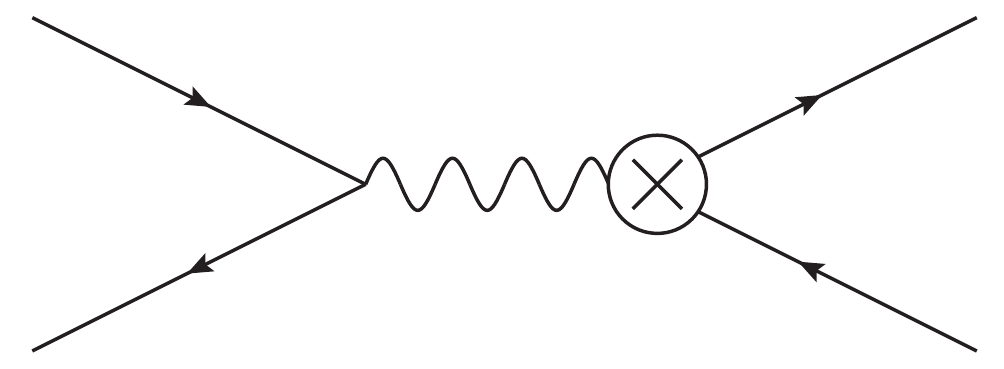}\,,
\end{align*}
due to the presence of ultraviolet divergences. The box diagrams
\begin{align*}
    \includegraphics[valign=m,height=.25\linewidth,width=.25\linewidth,keepaspectratio]{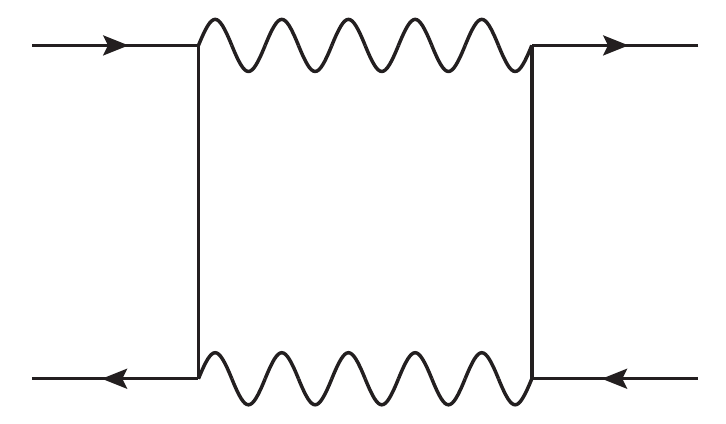}\quad
    \includegraphics[valign=m,height=.25\linewidth,width=.25\linewidth,keepaspectratio]{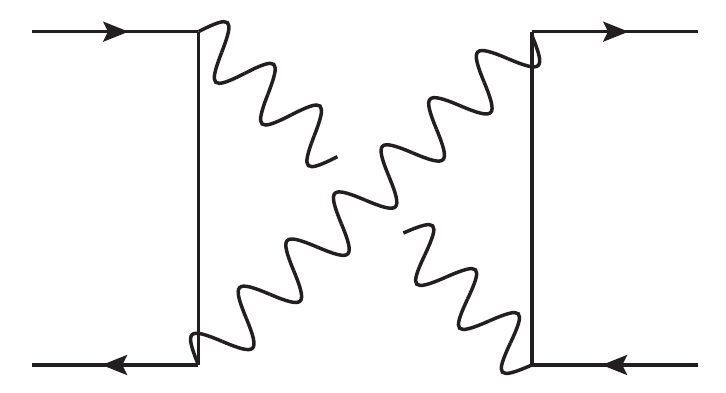}\quad
    \includegraphics[valign=m,height=.25\linewidth,width=.25\linewidth,keepaspectratio]{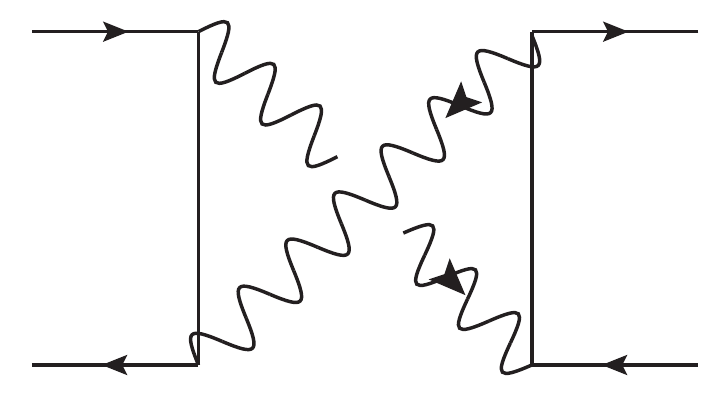}
\end{align*}
have infrared divergences only.
\subsection{One-loop integral definitions}
\label{sec:1Lints}
All of the one-loop integrals defined below are pure functions taken in the standard $\overline{\rm MS}$ normalization. Therefore, all of our one-loop Feynman integrals are understood to have a multiplicative factor of
\begin{equation}
    c_1\left(\epsilon,\mu^2\right) = \frac{e^{\gamma_E \epsilon} \left(\mu^2\right)^{\epsilon}}{i \pi^{2-\epsilon}}
\end{equation}
in the integration measure. In what follows, thin solid lines denote massless propagators or massless external momenta, thick, dotted lines denote propagators of mass $m_w$, thick, dashed lines denote propagators of mass $m_z$, and thick solid lines denote propagators of mass $m_h$. In total, 31 linearly-independent one-loop integrals appear in our calculations:
\begin{align}
    I_1 &= \epsilon ~~
    \includegraphics[valign=m,raise=.2cm,height=.15\linewidth,width=.15\linewidth,keepaspectratio]{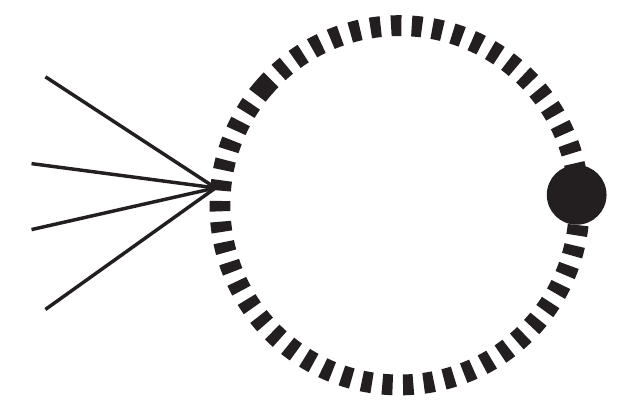}\left(m_w^2\right) \\
    I_2 &= \sqrt{s(s - 4\, m_w^2)}~ \epsilon ~~
    \includegraphics[valign=m,raise=.3cm,height=.15\linewidth,width=.15\linewidth,keepaspectratio]{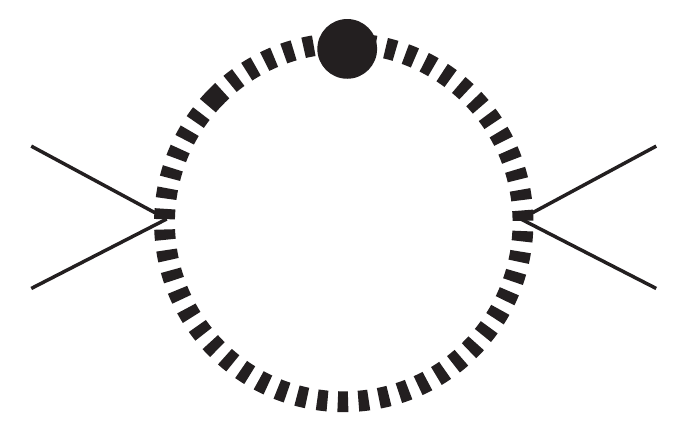}\left(s,m_w^2\right) \\
    I_3 &= \sqrt{m_z^2(4\, m_w^2 - m_z^2)}~ \epsilon ~~
    \includegraphics[valign=m,raise=.3cm,height=.18\linewidth,width=.18\linewidth,keepaspectratio]{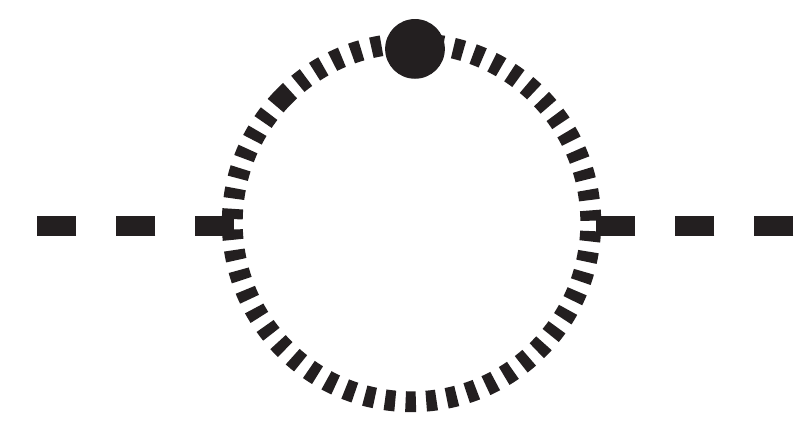}\left(m_w^2,m_z^2\right) \\
    I_4 &= \epsilon ~~
    \includegraphics[valign=m,raise=.2cm,height=.15\linewidth,width=.15\linewidth,keepaspectratio]{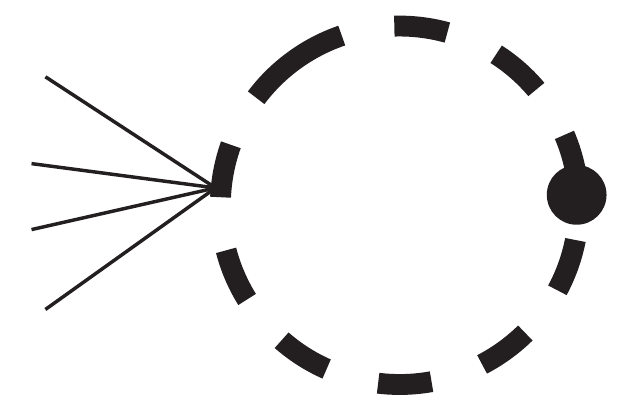}\left(m_z^2\right) \\
    I_5 &= \epsilon ~~
    \includegraphics[valign=m,raise=.2cm,height=.15\linewidth,width=.15\linewidth,keepaspectratio]{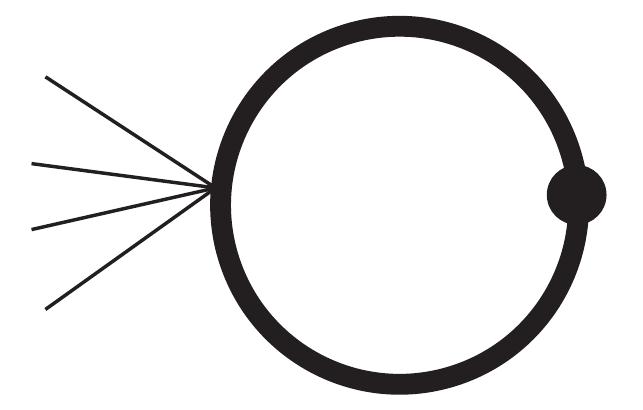}\left(m_h^2\right) \\
    I_6 &= \sqrt{\lambda\left(s, m_z^2, m_h^2\right)}~ \epsilon
    \left(\includegraphics[valign=m,raise=.3cm,height=.15\linewidth,width=.15\linewidth,keepaspectratio]{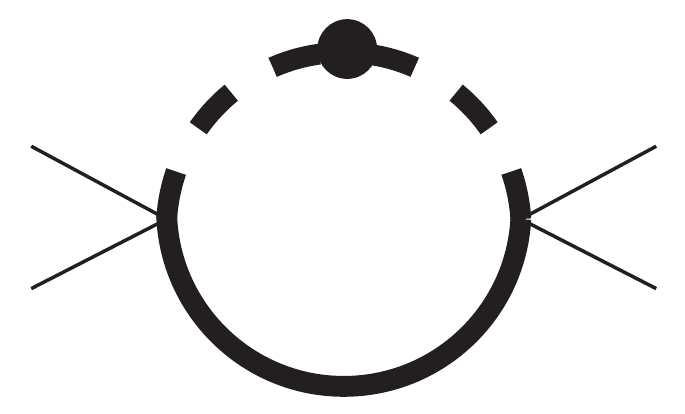}~+~ \includegraphics[valign=m,raise=.1cm,height=.15\linewidth,width=.15\linewidth,keepaspectratio]{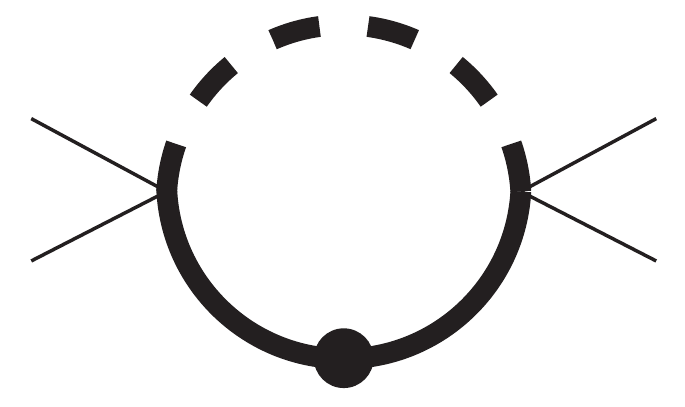}\right)\left(s, m_z^2, m_h^2\right) \\
    I_7 &= \sqrt{m_h^2(4\, m_z^2 - m_h^2)}~ \epsilon
    \left(\includegraphics[valign=m,raise=.3cm,height=.18\linewidth,width=.18\linewidth,keepaspectratio]{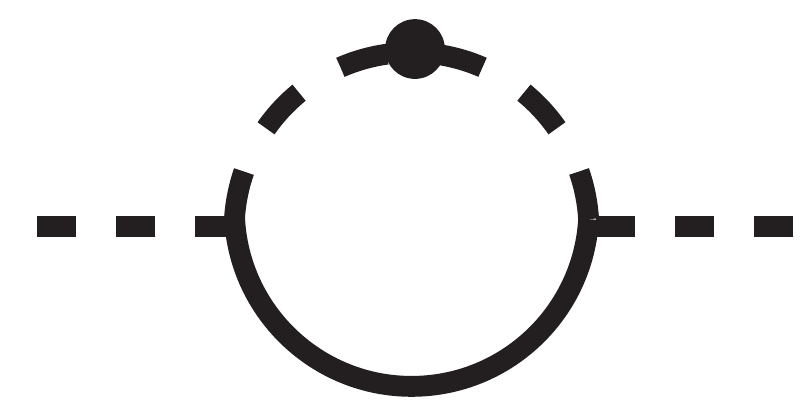}~+~ \includegraphics[valign=m,raise=.1cm,height=.18\linewidth,width=.18\linewidth,keepaspectratio]{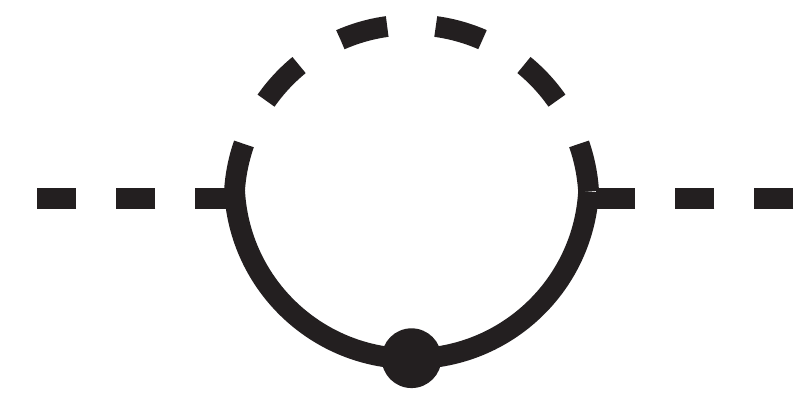}\right)\left(m_z^2,m_h^2\right) \\
    I_8 &= s\,\epsilon ~~
    \includegraphics[valign=m,raise=.3cm,height=.15\linewidth,width=.15\linewidth,keepaspectratio]{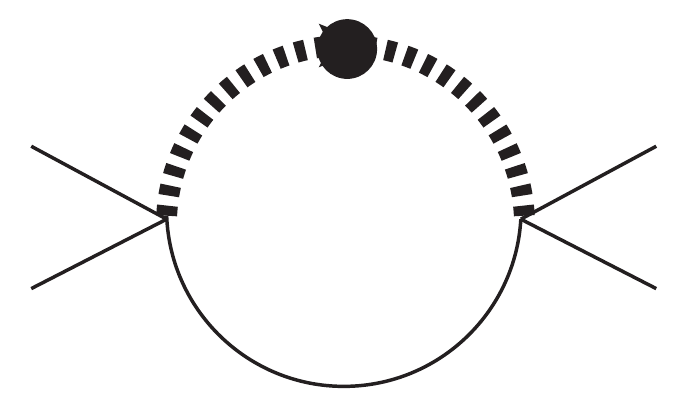}\left(s,m_w^2\right)\\
    I_9 &= \sqrt{\lambda\left(s, m_w^2, m_z^2\right)}~ \epsilon
    \left(\includegraphics[valign=m,raise=.3cm,height=.15\linewidth,width=.15\linewidth,keepaspectratio]{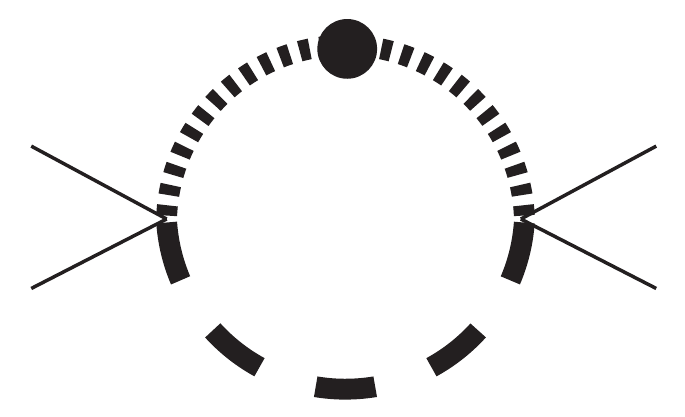}~+~ \includegraphics[valign=m,raise=.1cm,height=.15\linewidth,width=.15\linewidth,keepaspectratio]{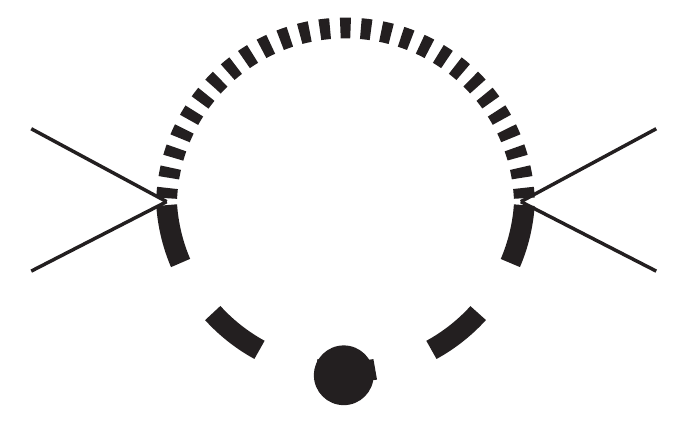}\right)\left(s, m_w^2, m_z^2\right) \\
    I_{10} &= \sqrt{\lambda\left(s, m_w^2, m_h^2\right)}~ \epsilon \left(\includegraphics[valign=m,raise=.3cm,height=.15\linewidth,width=.15\linewidth,keepaspectratio]{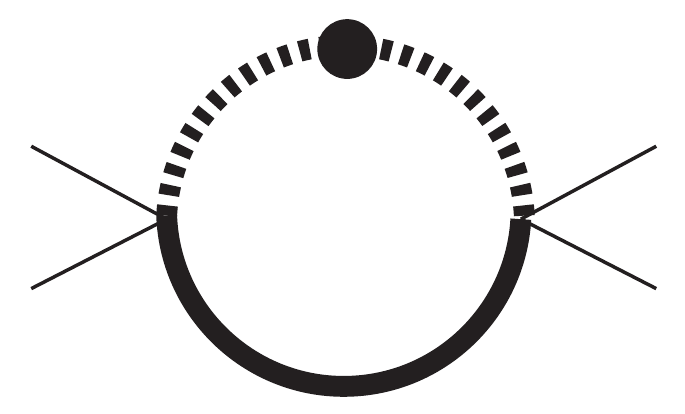}~+~ \includegraphics[valign=m,raise=.1cm,height=.15\linewidth,width=.15\linewidth,keepaspectratio]{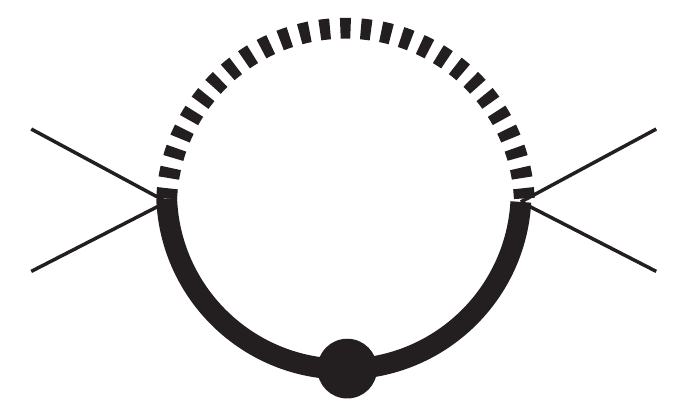}\right)\left(s, m_w^2, m_h^2\right) \\
    I_{11} &= \sqrt{m_z^2(4\, m_w^2 - m_z^2)}~ \epsilon
    \left(\includegraphics[valign=m,raise=.3cm,height=.18\linewidth,width=.18\linewidth,keepaspectratio]{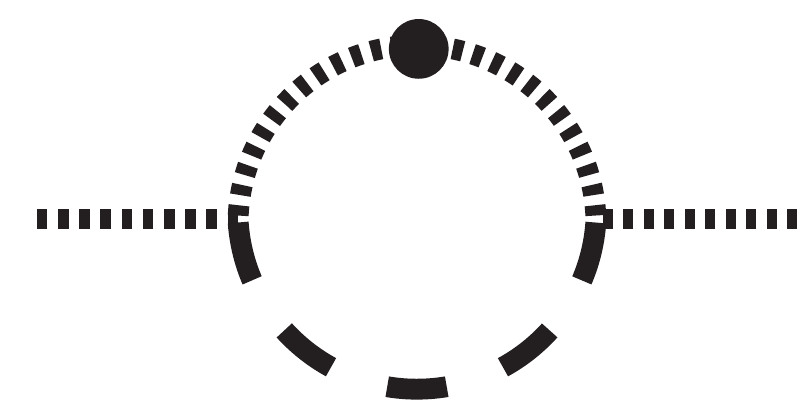}~+~ \includegraphics[valign=m,raise=.1cm,height=.18\linewidth,width=.18\linewidth,keepaspectratio]{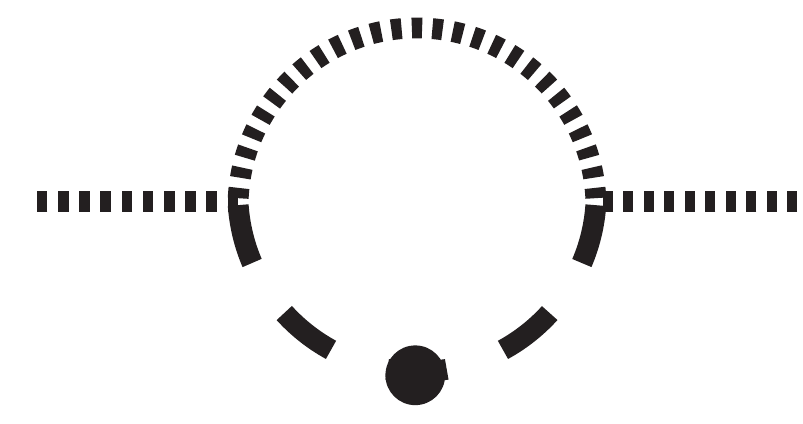}\right)\left(m_w^2,m_z^2\right) \\
    I_{12} &= \sqrt{m_h^2(4\, m_w^2 - m_h^2)}~ \epsilon
    \left(\includegraphics[valign=m,raise=.3cm,height=.18\linewidth,width=.18\linewidth,keepaspectratio]{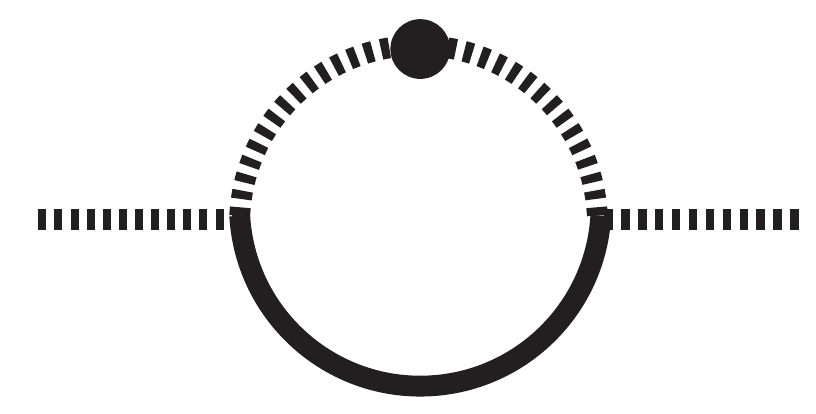}~+~ \includegraphics[valign=m,raise=.1cm,height=.18\linewidth,width=.18\linewidth,keepaspectratio]{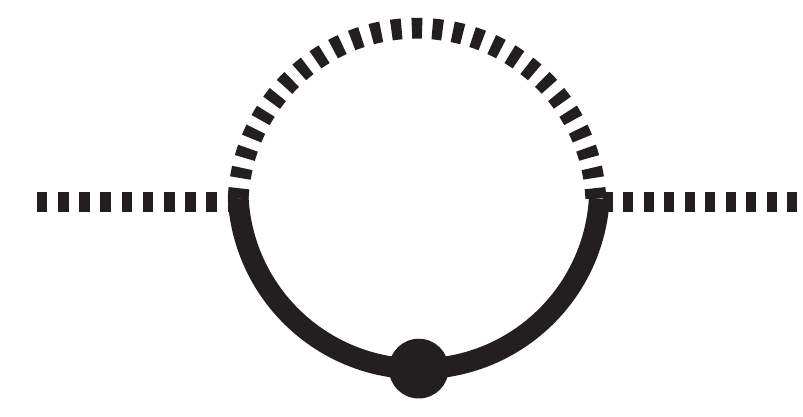}\right)\left(m_w^2,m_h^2\right) \\
    I_{13} &= s\, \epsilon ~~
    \includegraphics[valign=m,raise=.3cm,height=.15\linewidth,width=.15\linewidth,keepaspectratio]{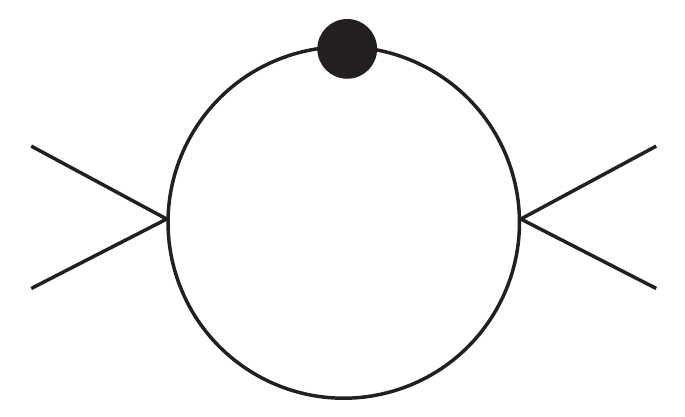}\left(s\right) \\
    I_{14} &= s\, \epsilon^2 ~~
    \includegraphics[valign=m,raise=.3cm,height=.15\linewidth,width=.15\linewidth,keepaspectratio]{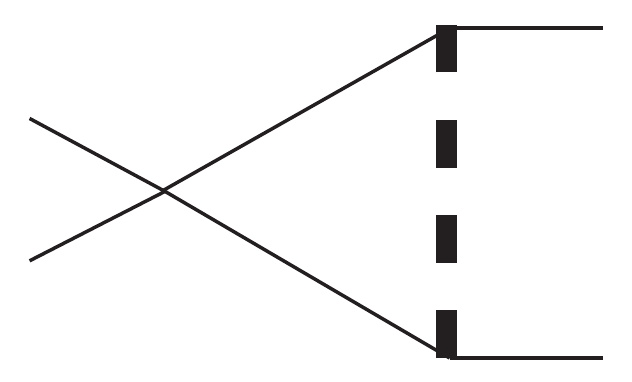}\left(s,m_z^2\right) \\
    I_{15} &= s\, \epsilon^2 ~~
    \includegraphics[valign=m,raise=.3cm,height=.15\linewidth,width=.15\linewidth,keepaspectratio]{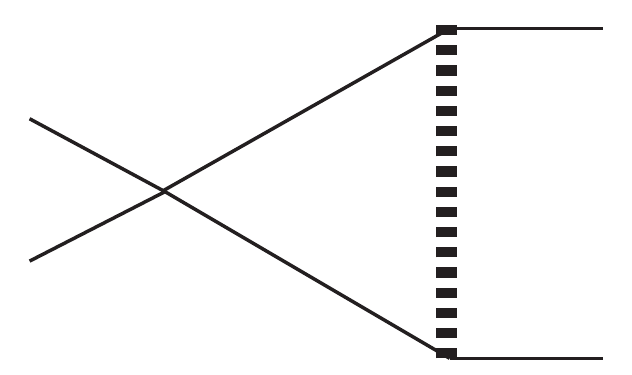}\left(s,m_w^2\right) \\
    I_{16} &= s\, \epsilon^2 ~~
    \includegraphics[valign=m,raise=.3cm,height=.15\linewidth,width=.15\linewidth,keepaspectratio]{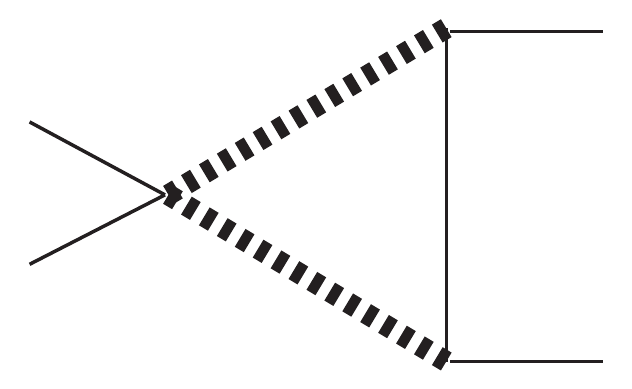}\left(s,m_w^2\right) \\
    I_{17} &= t\, \epsilon ~~
    \includegraphics[valign=m,raise=.3cm,height=.15\linewidth,width=.15\linewidth,keepaspectratio]{I131718}\left(t\right) \\
    I_{18} &= u\, \epsilon ~~
    \includegraphics[valign=m,raise=.3cm,height=.15\linewidth,width=.15\linewidth,keepaspectratio]{I131718}\left(u\right) \\
    I_{19} &= s \, t\, \epsilon^2~
    \includegraphics[valign=m,raise=.4cm,height=.15\linewidth,width=.15\linewidth,keepaspectratio]{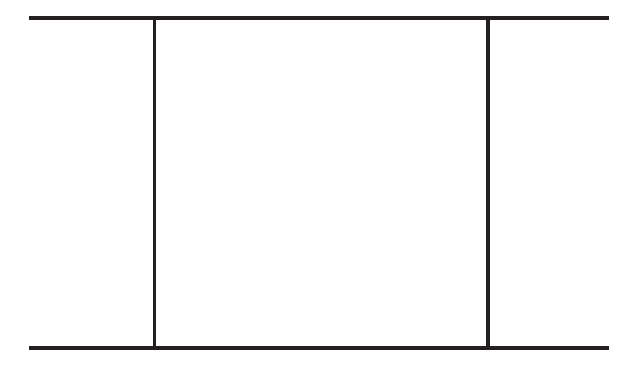}\left(s,t\right) \\
    I_{20} &= s \, u\, \epsilon^2~
    \includegraphics[valign=m,raise=.4cm,height=.15\linewidth,width=.15\linewidth,keepaspectratio]{I1920}\left(s,u\right) \\
    I_{21} &= s\,\epsilon ~~
    \includegraphics[valign=m,raise=.3cm,height=.15\linewidth,width=.15\linewidth,keepaspectratio]{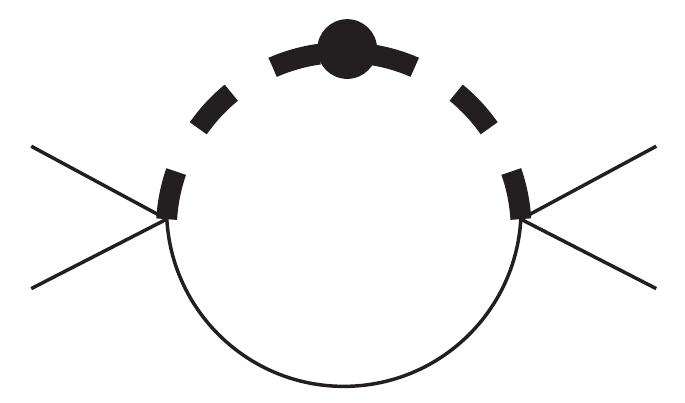}\left(s,m_z^2\right) \\
    I_{22} &= t\, \epsilon^2 ~~
    \includegraphics[valign=m,raise=.3cm,height=.15\linewidth,width=.15\linewidth,keepaspectratio]{I142223}\left(t,m_z^2\right) \\
    I_{23} &= u\, \epsilon^2 ~~
    \includegraphics[valign=m,raise=.3cm,height=.15\linewidth,width=.15\linewidth,keepaspectratio]{I142223}\left(u,m_z^2\right) \\
    I_{24} &= t \, \left(s-m_z^2\right)\, \epsilon^2~
    \includegraphics[valign=m,raise=.4cm,height=.15\linewidth,width=.15\linewidth,keepaspectratio]{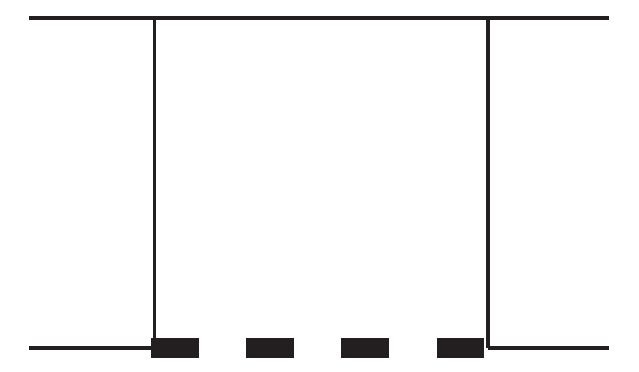}\left(s,t,m_z^2\right) \\
    I_{25} &= u \, \left(s-m_z^2\right)\, \epsilon^2~
    \includegraphics[valign=m,raise=.4cm,height=.15\linewidth,width=.15\linewidth,keepaspectratio]{I2425}\left(s,u,m_z^2\right) \\
    I_{26} &= \sqrt{s(s - 4\, m_z^2)}~ \epsilon ~~
    \includegraphics[valign=m,raise=.3cm,height=.15\linewidth,width=.15\linewidth,keepaspectratio]{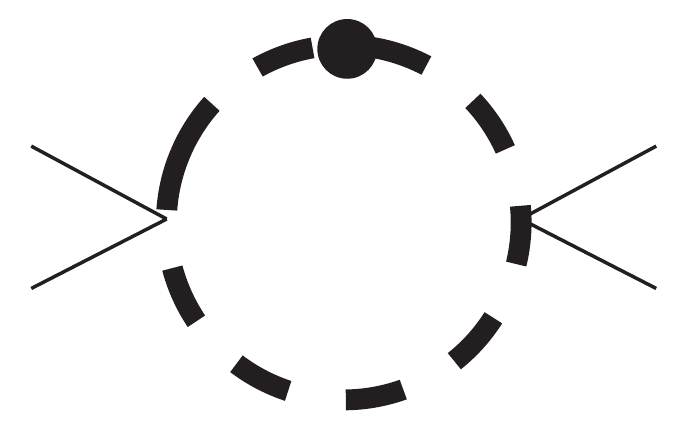}\left(s,m_z^2\right)\\
    I_{27} &= s\, \epsilon^2 ~~
    \includegraphics[valign=m,raise=.3cm,height=.15\linewidth,width=.15\linewidth,keepaspectratio]{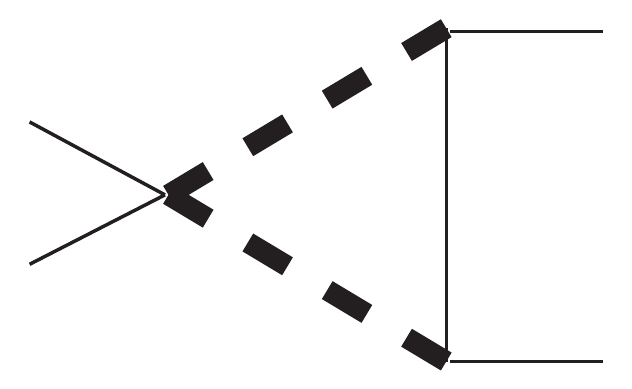}\left(s,m_z^2\right) \\
    I_{28} &= \sqrt{s\,t\left(s\,t - 4 m_z^2\left(t+m_z^2\right)\right)}~ \epsilon^2~
    \includegraphics[valign=m,raise=.4cm,height=.15\linewidth,width=.15\linewidth,keepaspectratio]{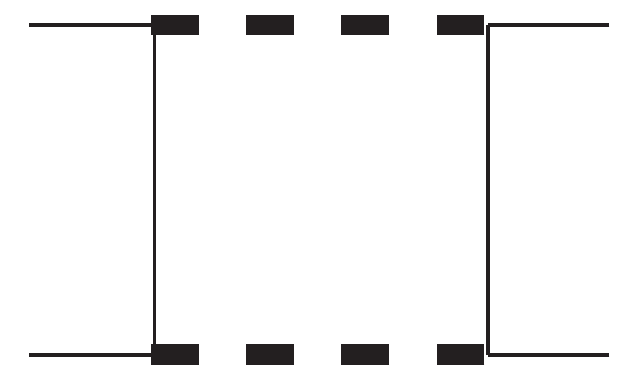}\left(s,t,m_z^2\right) \\
    I_{29} &= \sqrt{s\,u\left(s\,u - 4 m_z^2\left(u+m_z^2\right)\right)}~ \epsilon^2~
    \includegraphics[valign=m,raise=.4cm,height=.15\linewidth,width=.15\linewidth,keepaspectratio]{I2829}\left(s,u,m_z^2\right) \\
    I_{30} &= u\, \epsilon^2 ~~
    \includegraphics[valign=m,raise=.3cm,height=.15\linewidth,width=.15\linewidth,keepaspectratio]{I1530}\left(u,m_w^2\right) \\
    I_{31} &= \sqrt{s\,u\left(s\,u - 4 m_w^2\left(u+m_w^2\right)\right)}~ \epsilon^2~
    \includegraphics[valign=m,raise=.4cm,height=.15\linewidth,width=.15\linewidth,keepaspectratio]{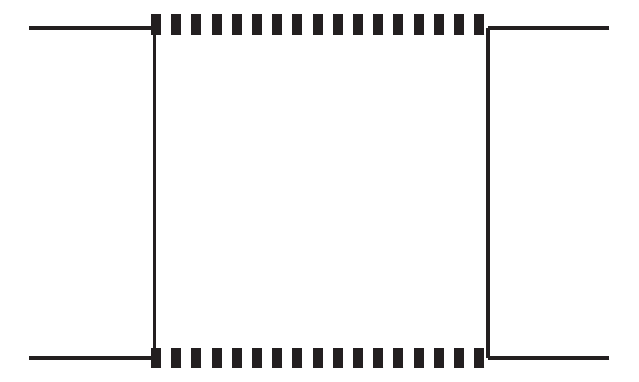}\left(s,u,m_w^2\right)\,,
\end{align}
where $\lambda(x,y,z)$ is the K{\"a}ll{\'e}n function,
\begin{equation}
    \lambda(x,y,z) = x^4 + y^4 + z^4 - 2 x^2 y^2 - 2 x^2 z^2 - 2 y^2 z^2.
\end{equation}
For clarity and later convenience, we have made all dependence on the kinematic variables $s$, $t$, $u = -s-t$, $m_w^2$, $m_z^2$, and $m_h^2$ completely explicit on the right-hand side. Moreover, we have included integrals for all relevant permutations of the kinematic variables separately. In order to actually calculate the master integrals, it is enough to find explicit expressions for $I_1$, $I_2$, $I_3$, $I_6$, $I_7$, $I_8$, $I_{13}$, $I_{14}$, $I_{16}$, $I_{17}$, $I_{19}$, $I_{22}$, $I_{24}$, and $I_{28}$, as all other integrals may be obtained by making simple replacements. For example, $I_{31}$ is obtained from $I_{28}$ by replacing $m_z^2$ with $m_w^2$ and $t$ with $u$. Due to the fact that we consider the physical region $s > (m_z + m_h)^2$ (see \eqref{eq:kinrestriction}), the same form of the analytic result can be used in a straightforward way.

Our calculation requires knowledge of the most complicated one-loop box-type master integrals ({\it i.e.} $I_{24}$ and $I_{28}$) expanded through to fourth order in $\epsilon$.
We could not find suitable analytic solutions for all integrals in the literature. Several of the explicit results we could locate in the {\tt Loopedia} database \cite{Bogner:2017xhp} were either not expanded to a sufficiently high order in $\epsilon$ for our purposes or not provided in a form convenient for numerical evaluations in the physical region of interest to us.

We therefore computed all of the integrals from scratch,
either by direct integration for generic $\epsilon$ followed by an expansion\footnote{For some of the more complicated expansions, we used the {\tt Mathematica} package {\tt HypExp} \cite{Huber:2005yg}.} in $\epsilon$ or by using the method of differential equations~\cite{Kotikov:1990kg,Kotikov:1991hm,Kotikov:1991pm,Bern:1992em,Bern:1993kr,Remiddi:1997ny,Gehrmann:1999as}.
The integral definitions given above lead to a $\epsilon$-decoupled form for the differential equations~\cite{Kotikov:2010gf,Henn:2013pwa}, which we integrate in terms of multiple polylogarithms.
To proceed, we construct real-valued multiple polylogarithms in the region of phase space of interest to us, employing the functional basis presented in~\cite{vonManteuffel:2017myy,Heller:2019gkq}, together with a few additional logarithms and polylogarithms required to integrate {\it e.g.} $I_6$ to a sufficiently high order in $\ep$. We provide our results in the ancillary file {\tt oneloopmasters.m} attached to the {\tt arXiv} submission of this paper.

\subsection{Assembly of one-loop results}
\label{sec:1Lfinalres}
Assembling the components provided in Appendix \ref{sec:1Lasres} and \ref{sec:1Lares}, we find
\begin{align}
\label{eq:1LCVVaas}
    {\bf C}^{(0,1)}_{\mathrm{VV}} &= -\frac{Q_\ell}{s}\bar{\mathcal{V}}^{(0,1)}_{\gamma \bar{q} q}(s) + \frac{v_\ell}{s-m_z^2}\bar{\mathcal{V}}^{(0,1)}_{Z \bar{q} q}(s)
    \,,
    \\
\label{eq:1LCVAVaas}
    {\bf C}^{(0,1)}_{\mathrm{VA}} &=-\frac{a_\ell}{s-m_z^2}\bar{\mathcal{V}}^{(0,1)}_{Z \bar{q} q}(s)
    \,,
    \\
\label{eq:1LCAVVaas}
    {\bf C}^{(0,1)}_{\mathrm{AV}} &=\frac{v_\ell}{s-m_z^2}\bar{\mathcal{A}}^{(0,1)}_{Z \bar{q} q}(s)
    \,,
    \\
\label{eq:1LCAVAVaas}
    {\bf C}^{(0,1)}_{\mathrm{AA}} &=-\frac{a_\ell}{s-m_z^2}\bar{\mathcal{A}}^{(0,1)}_{Z \bar{q} q}(s)
    \,,
    \\
\label{eq:1LCbarVVa2}
    {\bf C}^{(1,0)}_{\mathrm{VV}} &= -\frac{Q_q Q_\ell}{s^2}\left(\bar{\Sigma}^{(1,0)}_{\gamma \gamma}(s) - s \delta Z_{\gamma\gamma}^{(1,0)}\right)\notag\\
    &\quad+\frac{1}{{s(s-m_z^2)}}\left(Q_q v_\ell+Q_\ell v_q\right)\left(\bar{\Sigma}^{(1,0)}_{\gamma Z}(s) + \frac{1}{2}\left( s\left(\delta Z_{\gamma Z}^{(1,0)}+\delta Z_{Z \gamma}^{(1,0)}\right)-m_z^2 \delta Z_{Z \gamma}^{(1,0)}\right)\right)\notag\\
    &\quad-\frac{v_q v_\ell}{{(s-m_z^2)^2}}\left(\bar{\Sigma}^{(1,0)}_{Z Z}(s) + s\delta Z_{Z Z}^{(1,0)}-m_z^2\left(\delta Z_{Z Z}^{(1,0)} + \delta Z_{m_z^2}^{(1,0)}\right)\right)\notag\\
    &\quad-\frac{Q_q}{s}\left(\bar{\mathcal{V}}^{(1,0)}_{\gamma\bar{\ell}\ell}(s)-Q_\ell\left(\delta Z_e^{(1,0)}+\delta Z^{(1,0)}_{\mathrm{V},\,\ell} + \frac{1}{2}\delta Z_{\gamma \gamma}^{(1,0)}-\frac{v_\ell}{2 Q_\ell}\delta Z_{Z \gamma}^{(1,0)}\right)\right)\notag\\
    &\quad-\frac{Q_\ell}{s}\left(\bar{\mathcal{V}}^{(1,0)}_{\gamma\bar{q} q}(s)-Q_q\left(\delta Z_e^{(1,0)}+\delta Z^{(1,0)}_{\mathrm{V},\,q} + \frac{1}{2}\delta Z_{\gamma \gamma}^{(1,0)}-\frac{v_q}{2 Q_q}\delta Z_{Z \gamma}^{(1,0)}\right)\right)\notag\\
    &\quad+\frac{v_q}{s-m_z^2}\left(\bar{\mathcal{V}}^{(1,0)}_{Z\bar{\ell}\ell}(s) + v_\ell \left(\delta Z_e^{(1,0)}+\delta Z_{\mathrm{V},\,\ell}^{(1,0)}+\frac{1}{2}\delta Z_{Z Z}^{(1,0)}\right)-a_\ell \delta Z_{\mathrm{A},\,\ell}^{(1,0)}\notag\right.\\
    &\quad\left.\qquad-\frac{Q_\ell}{2}\delta Z_{\gamma Z}^{(1,0)}+\frac{m_z^2}{2(m_z^2-m_w^2)}\left(v_\ell - \frac{2 m_w^2}{m_z^2}a_\ell\right)\left(\delta Z_{m_z^2}^{(1,0)} - \delta Z_{m_w^2}^{(1,0)}\right)\right)\notag\\
    &\quad+\frac{v_\ell}{s-m_z^2}\left(\bar{\mathcal{V}}^{(1,0)}_{Z\bar{q} q}(s) + v_q \left(\delta Z_e^{(1,0)}+\delta Z_{\mathrm{V},\,q}^{(1,0)}+\frac{1}{2}\delta Z_{Z Z}^{(1,0)}\right)-a_q \delta Z_{\mathrm{A},\,q}^{(1,0)}\notag\right.\\
    &\quad\left.\qquad-\frac{Q_q}{2}\delta Z_{\gamma Z}^{(1,0)}+\frac{m_z^2}{2(m_z^2-m_w^2)}\left(v_q - \frac{2 m_w^2}{m_z^2}a_q\right)\left(\delta Z_{m_z^2}^{(1,0)} - \delta Z_{m_w^2}^{(1,0)}\right)\right)+\mathcal{B}_{\mathrm{VV}}^{(1,0)}\,,
    \\
\label{eq:1LCbarVAVa2}
    {\bf C}^{(1,0)}_{\mathrm{VA}} &= -\frac{Q_q a_\ell}{s(s-m_z^2)}\left(\bar{\Sigma}^{(1,0)}_{\gamma Z}(s) + \frac{1}{2}\left( s\left(\delta Z_{\gamma Z}^{(1,0)}+\delta Z_{Z \gamma}^{(1,0)}\right)-m_z^2 \delta Z_{Z \gamma}^{(1,0)}\right)\right)\notag\\
    &\quad+\frac{a_\ell v_q}{(s-m_z^2)^2}\left(\bar{\Sigma}^{(1,0)}_{Z Z}(s) + s\delta Z_{Z Z}^{(1,0)}-m_z^2\left(\delta Z_{Z Z}^{(1,0)} + \delta Z_{m_z^2}^{(1,0)}\right)\right)\notag\\
    &\quad-\frac{Q_q}{s}\left(\bar{\mathcal{A}}^{(1,0)}_{\gamma\bar{\ell}\ell}(s)-Q_\ell\left(\delta Z^{(1,0)}_{\mathrm{A},\,\ell} + \frac{a_\ell}{2 Q_\ell}\delta Z_{Z \gamma}^{(1,0)}\right)\right)\notag\\
    &\quad+\frac{v_q}{s-m_z^2}\left(\bar{\mathcal{A}}^{(1,0)}_{Z\bar{\ell}\ell}(s) + v_\ell \delta Z_{\mathrm{A},\,\ell}^{(1,0)}-a_\ell \left(\delta Z_e^{(1,0)}+\delta Z_{\mathrm{V},\,\ell}^{(1,0)}+\frac{1}{2}\delta Z_{Z Z}^{(1,0)}\right)\nonumber\right.\\
    &\quad\left.\qquad-a_\ell\frac{m_z^2}{2(m_z^2-m_w^2)}\left(1 - \frac{2 m_w^2}{m_z^2}\right)\left(\delta Z_{m_z^2}^{(1,0)} - \delta Z_{m_w^2}^{(1,0)}\right)\right)\notag\\
    &\quad-\frac{a_\ell}{s-m_z^2}\left(\bar{\mathcal{V}}^{(1,0)}_{Z\bar{q} q}(s) + v_q \left(\delta Z_e^{(1,0)}+\delta Z_{\mathrm{V},\,q}^{(1,0)}+\frac{1}{2}\delta Z_{Z Z}^{(1,0)}\right)-a_q \delta Z_{\mathrm{A},\,q}^{(1,0)}\notag\right.\\
    &\quad\left.\qquad-\frac{Q_q}{2}\delta Z_{\gamma Z}^{(1,0)}+\frac{m_z^2}{2(m_z^2-m_w^2)}\left(v_q - \frac{2 m_w^2}{m_z^2}a_q\right)\left(\delta Z_{m_z^2}^{(1,0)} - \delta Z_{m_w^2}^{(1,0)}\right)\right)+\mathcal{B}_{\mathrm{VA}}^{(1,0)}\,,
    \\
\label{eq:1LCbarAVVa2}
    {\bf C}^{(1,0)}_{\mathrm{AV}} &= -\frac{Q_\ell a_q}{s (s-m_z^2)}\left(\bar{\Sigma}^{(1,0)}_{\gamma Z}(s) + \frac{1}{2}\left( s\left(\delta Z_{\gamma Z}^{(1,0)}+\delta Z_{Z \gamma}^{(1,0)}\right)-m_z^2 \delta Z_{Z \gamma}^{(1,0)}\right)\right)\notag\\
    &\quad+\frac{a_q v_\ell}{(s-m_z^2)^2}\left(\bar{\Sigma}^{(1,0)}_{Z Z}(s) + s\delta Z_{Z Z}^{(1,0)}-m_z^2\left(\delta Z_{Z Z}^{(1,0)} + \delta Z_{m_z^2}^{(1,0)}\right)\right)\notag\\
    &\quad-\frac{Q_\ell}{s}\left(\bar{\mathcal{A}}^{(1,0)}_{\gamma\bar{q} q}(s)-Q_q\left(\delta Z^{(1,0)}_{\mathrm{A},\,q} + \frac{a_q}{2 Q_q}\delta Z_{Z \gamma}^{(1,0)}\right)\right)\notag\\
    &\quad-\frac{a_q}{s-m_z^2}\left(\bar{\mathcal{V}}^{(1,0)}_{Z\bar{\ell}\ell}(s) + v_\ell \left(\delta Z_e^{(1,0)}+\delta Z_{\mathrm{V},\,\ell}^{(1,0)}+\frac{1}{2}\delta Z_{Z Z}^{(1,0)}\right)-a_\ell \delta Z_{\mathrm{A},\,\ell}^{(1,0)}\notag\right.\\
    &\quad\left.\qquad-\frac{Q_\ell}{2}\delta Z_{\gamma Z}^{(1,0)}+\frac{m_z^2}{2(m_z^2-m_w^2)}\left(v_\ell - \frac{2 m_w^2}{m_z^2}a_\ell\right)\left(\delta Z_{m_z^2}^{(1,0)} - \delta Z_{m_w^2}^{(1,0)}\right)\right)\notag\\
    &\quad+\frac{v_\ell}{s-m_z^2}\left(\bar{\mathcal{A}}^{(1,0)}_{Z\bar{q}q}(s) + v_q \delta Z_{\mathrm{A},\,q}^{(1,0)}-a_q \left(\delta Z_e^{(1,0)}+\delta Z_{\mathrm{V},\,q}^{(1,0)}+\frac{1}{2}\delta Z_{Z Z}^{(1,0)}\right)\nonumber\right.\\
    &\quad\left.\qquad-a_q\frac{m_z^2}{2(m_z^2-m_w^2)}\left(1 - \frac{2 m_w^2}{m_z^2}\right)\left(\delta Z_{m_z^2}^{(1,0)} - \delta Z_{m_w^2}^{(1,0)}\right)\right)+\mathcal{B}_{\mathrm{AV}}^{(1,0)}\,,
    \\
\label{eq:1LCAVAVa2}
    {\bf C}^{(1,0)}_{\mathrm{AA}} &= -\frac{a_q a_\ell}{(s-m_z^2)^2}\left(\bar{\Sigma}^{(1,0)}_{Z Z}(s) + s\delta Z_{Z Z}^{(1,0)}-m_z^2\left(\delta Z_{Z Z}^{(1,0)} + \delta Z_{m_z^2}^{(1,0)}\right)\right)\notag\\
    &\quad-\frac{a_q}{s-m_z^2}\left(\bar{\mathcal{A}}^{(1,0)}_{Z\bar{\ell}\ell}(s) + v_\ell \delta Z_{\mathrm{A},\,\ell}^{(1,0)}-a_\ell \left(\delta Z_e^{(1,0)}+\delta Z_{\mathrm{V},\,\ell}^{(1,0)}+\frac{1}{2}\delta Z_{Z Z}^{(1,0)}\right)\nonumber\right.\\
    &\quad\left.\qquad-a_\ell\frac{m_z^2}{2(m_z^2-m_w^2)}\left(1 - \frac{2 m_w^2}{m_z^2}\right)\left(\delta Z_{m_z^2}^{(1,0)} - \delta Z_{m_w^2}^{(1,0)}\right)\right)\notag\\
    &\quad-\frac{a_\ell}{s-m_z^2}\left(\bar{\mathcal{A}}^{(1,0)}_{Z\bar{q} q}(s) + v_q \delta Z_{\mathrm{A},\,q}^{(1,0)}-a_q \left(\delta Z_e^{(1,0)}+\delta Z_{\mathrm{V},\,q}^{(1,0)}+\frac{1}{2}\delta Z_{Z Z}^{(1,0)}\right)\nonumber\right.\\
    &\quad\left.\qquad-a_q\frac{m_z^2}{2(m_z^2-m_w^2)}\left(1 - \frac{2 m_w^2}{m_z^2}\right)\left(\delta Z_{m_z^2}^{(1,0)} - \delta Z_{m_w^2}^{(1,0)}\right)\right) + \mathcal{B}_{\mathrm{AA}}^{(1,0)}
\end{align}
for the renormalized one-loop scattering amplitude coefficients in HVBM's $\gamma_5$ scheme. In writing the above expressions, we made use of the tabulated one-loop vertex counterterms collected in Appendix A of \cite{Denner:1991kt}.
Results in Kreimer's $\gamma_5$ scheme may be simply obtained from Eqs. \eqref{eq:1LCVVaas} - \eqref{eq:1LCAVAVa2}:
\begin{align}
\label{eq:1LCbarVVaas}
    {\bf \bar{C}}^{(0,1)}_{\mathrm{VV}} &= {\bf C}^{(0,1)}_{\mathrm{VV}}\,,\\
    {\bf \bar{C}}^{(0,1)}_{\mathrm{VA}} &= {\bf C}^{(0,1)}_{\mathrm{VA}}\,,\\
    {\bf \bar{C}}^{(0,1)}_{\mathrm{AV}} &= {\bf C}^{(0,1)}_{\mathrm{AV}}\,,\\
    {\bf \bar{C}}^{(0,1)}_{\mathrm{AA}} &= {\bf C}^{(0,1)}_{\mathrm{AA}}\,,\\
    {\bf \bar{C}}^{(1,0)}_{\mathrm{VV}} &= {\bf C}^{(1,0)}_{\mathrm{VV}}/.\, \mathcal{B}_{\mathrm{VV}}^{(1,0)}\rightarrow \bar{\mathcal{B}}_{\mathrm{VV}}^{(1,0)}\,,\\
    {\bf \bar{C}}^{(1,0)}_{\mathrm{VA}} &= {\bf C}^{(1,0)}_{\mathrm{VA}}/.\, \mathcal{B}_{\mathrm{VA}}^{(1,0)}\rightarrow \bar{\mathcal{B}}_{\mathrm{VA}}^{(1,0)}\,,\\
    {\bf \bar{C}}^{(1,0)}_{\mathrm{AV}} &= {\bf C}^{(1,0)}_{\mathrm{AV}}/.\, \mathcal{B}_{\mathrm{AV}}^{(1,0)}\rightarrow \bar{\mathcal{B}}_{\mathrm{AV}}^{(1,0)}\,,\\
\label{eq:1LCbarAVAVa2}
   \mathrm{and}\qquad {\bf \bar{C}}^{(1,0)}_{\mathrm{AA}} &= {\bf C}^{(1,0)}_{\mathrm{AA}}/.\, \mathcal{B}_{\mathrm{AA}}^{(1,0)}\rightarrow \bar{\mathcal{B}}_{\mathrm{AA}}^{(1,0)}\,.
\end{align}

Eqs. \eqref{eq:1LCVVaas} - \eqref{eq:1LCbarAVAVa2} may be directly matched onto Eqs. \eqref{eq:subfuncs} to determine $\mathcal{H}_{\rm DY}^{(0,1)}[0]$, $\mathcal{H}_{\rm DY}^{(0,1)}[1]$, $\mathcal{H}_{\rm DY}^{(0,1)}[2]$, $\bar{\mathcal{H}}_{\rm DY}^{(0,1)}[0]$, $\bar{\mathcal{H}}_{\rm DY}^{(0,1)}[1]$, $\bar{\mathcal{H}}_{\rm DY}^{(0,1)}[2]$, $\mathcal{H}_{\rm DY}^{(1,0)}[0]$, $\mathcal{H}_{\rm DY}^{(1,0)}[1]$, $\mathcal{H}_{\rm DY}^{(1,0)}[2]$, $\bar{\mathcal{H}}_{\rm DY}^{(1,0)}[0]$, \\$\bar{\mathcal{H}}_{\rm DY}^{(1,0)}[1]$, and $\bar{\mathcal{H}}_{\rm DY}^{(1,0)}[2]$. Let us emphasize that, upon substituting the explicit expressions for the master integrals discussed in the previous section (see also Appendix \ref{sec:1Lares}), we find
\begin{align}
    \mathcal{H}_{\rm DY}^{(1,0)}[0] = \bar{\mathcal{H}}_{\rm DY}^{(1,0)}[0]
\end{align}
but\footnote{Due to the fact that the relative order $\alpha_s$ one-loop corrections are comprised of simple vertex form factor diagrams with no sensitivity to the $\gamma_5$ problem, we have $\mathcal{H}_{\rm DY}^{(0,1)}[k] =\bar{\mathcal{H}}_{\rm DY}^{(0,1)}[k]$ for all $k$.}
\begin{align}
    \qquad\qquad\qquad\qquad~ \mathcal{H}_{\rm DY}^{(1,0)}[k] \neq \bar{\mathcal{H}}_{\rm DY}^{(1,0)}[k] \qquad \qquad \mathrm{for}~k>0\,.
\end{align}
In Section \ref{sec:helicities}, it is explained in detail how one-loop polarized hard scattering functions may be derived from $\mathcal{H}_{\rm DY}^{(0,1)}[0]$ and $\mathcal{H}_{\rm DY}^{(1,0)}[0]$ or, equivalently, from $\bar{\mathcal{H}}_{\rm DY}^{(0,1)}[0]$ and $\bar{\mathcal{H}}_{\rm DY}^{(1,0)}[0]$.

\section{Two-loop scattering amplitudes}
\subsection{Diagrammatic structure}
\label{sec:2Ldiags}
In this section, we present the diagrammatic structure of the two-loop perturbative corrections to the neutral-current Drell-Yan process of relative order $\alpha \alpha_s$. As at one loop, we do not explicitly identify the photon and $Z$.
Due to their familiarity, diagrams with one-loop renormalized self-energy insertions of the form
\begin{align*}
    \includegraphics[valign=m,height=.5\linewidth,width=.5\linewidth,keepaspectratio]{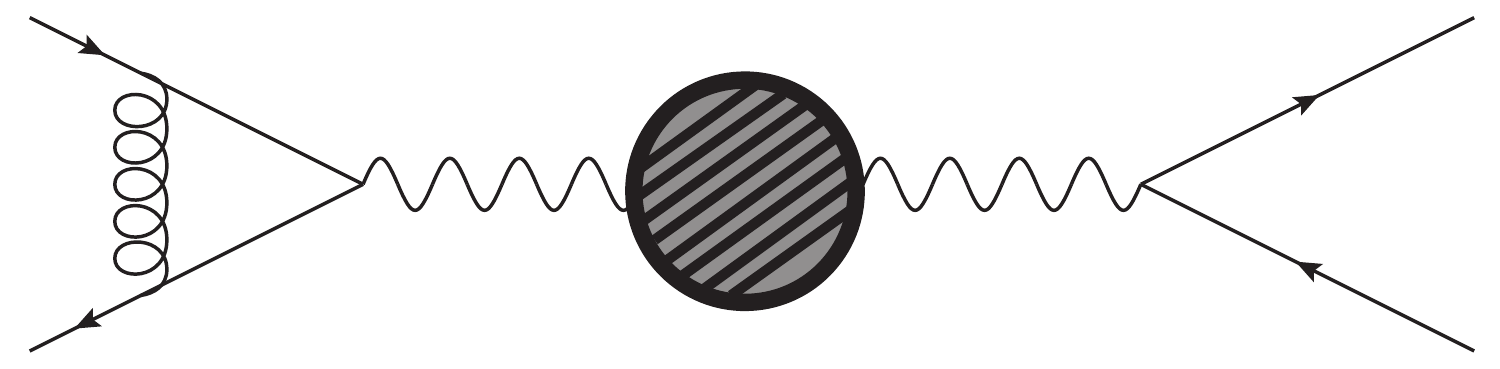}
\end{align*}
will not be visualized in explicit detail. Several other two-loop diagrams are also trivial, in that they essentially consist of simple products of one-loop contributions:
\begin{align*}
    &\includegraphics[valign=m,height=.315\linewidth,width=.315\linewidth,keepaspectratio]{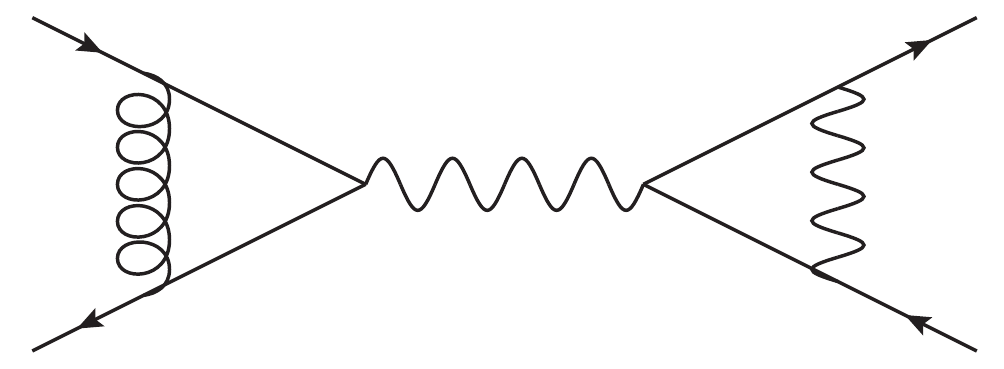}\quad \includegraphics[valign=m,height=.315\linewidth,width=.315\linewidth,keepaspectratio]{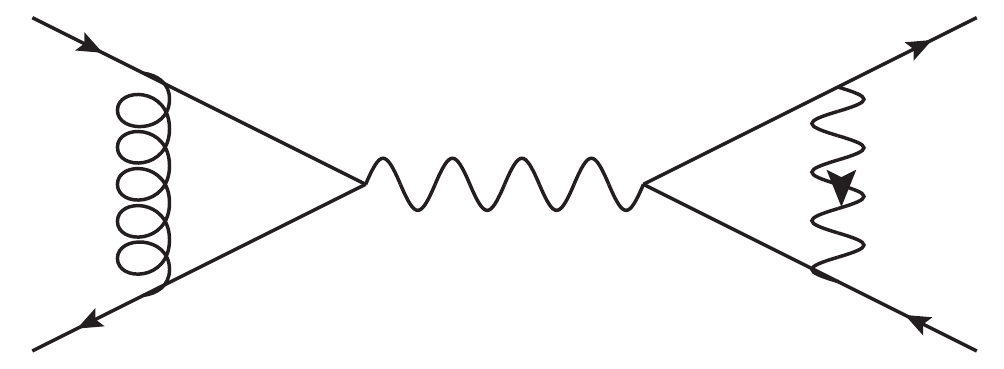}\quad
    \includegraphics[valign=m,height=.315\linewidth,width=.315\linewidth,keepaspectratio]{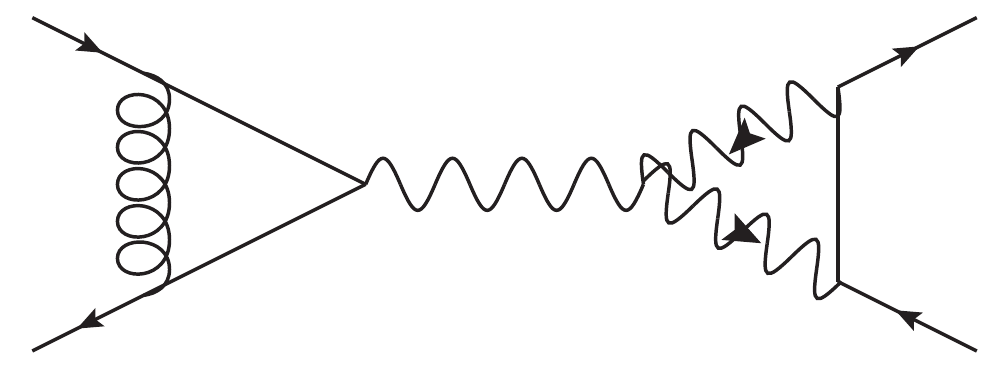}\\
    &\includegraphics[valign=m,height=.315\linewidth,width=.315\linewidth,keepaspectratio]{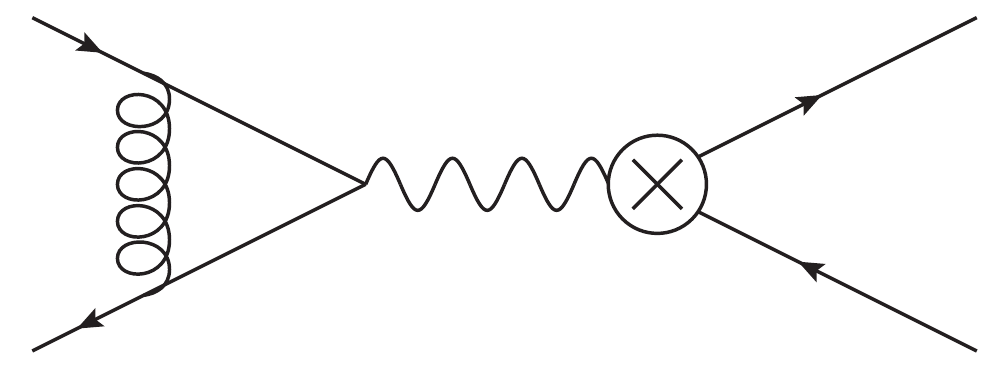}\quad \includegraphics[valign=m,height=.315\linewidth,width=.315\linewidth,keepaspectratio]{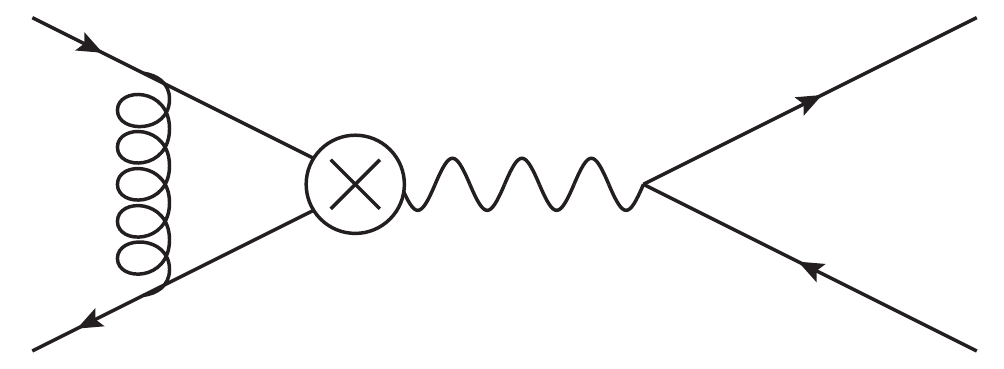}\quad
    \includegraphics[valign=m,height=.315\linewidth,width=.315\linewidth,keepaspectratio]{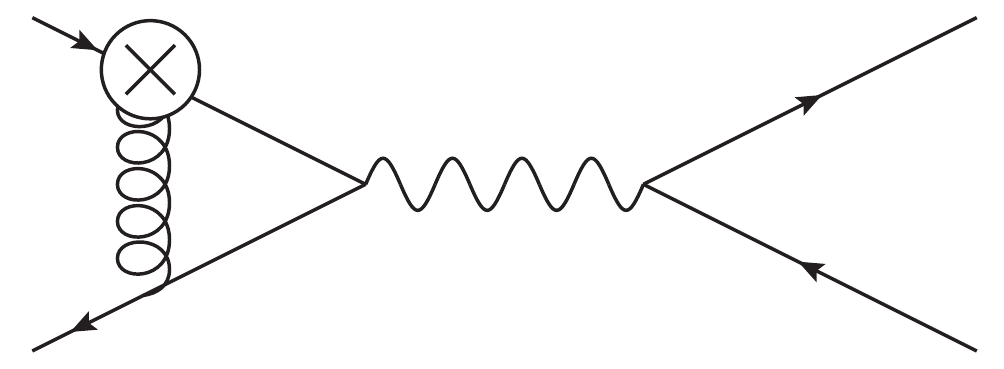}\\
    &\includegraphics[valign=m,height=.315\linewidth,width=.315\linewidth,keepaspectratio]{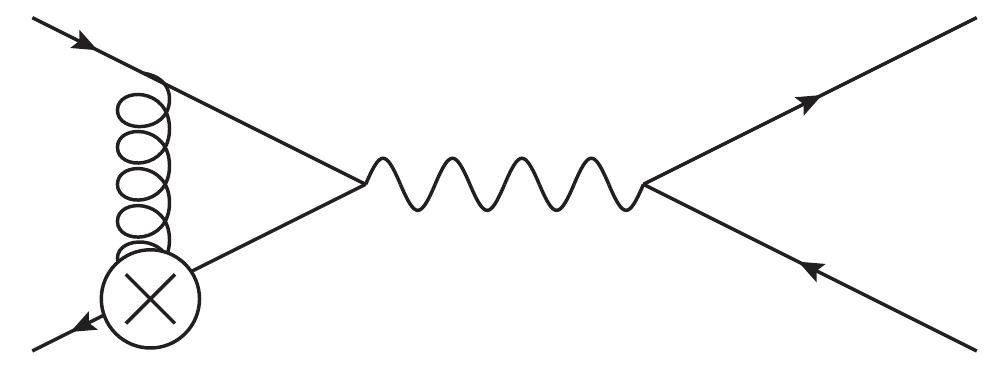}\quad \includegraphics[valign=m,height=.315\linewidth,width=.315\linewidth,keepaspectratio]{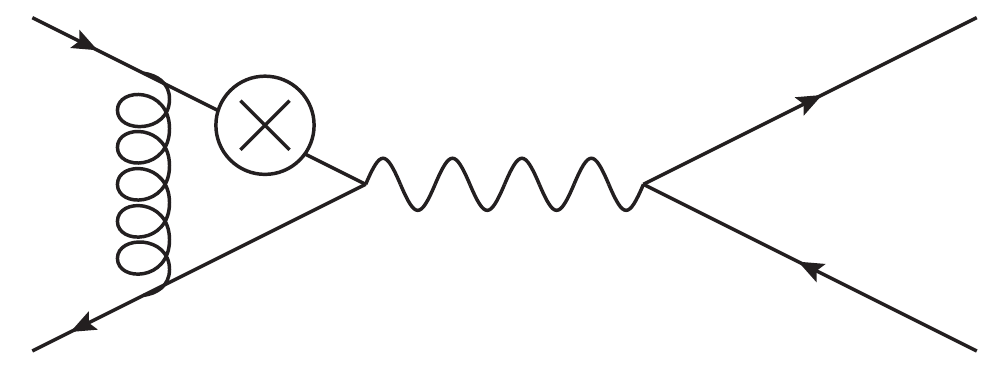}\quad
    \includegraphics[valign=m,height=.315\linewidth,width=.315\linewidth,keepaspectratio]{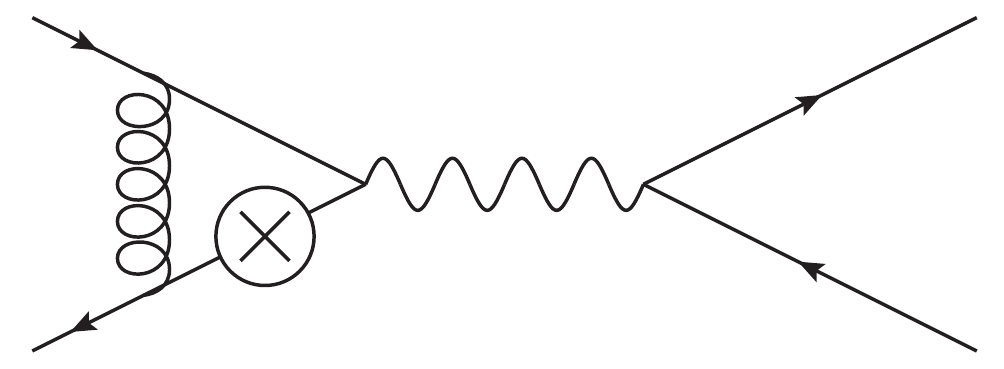}.
\end{align*}
Note that, in the above, the sum of the final four diagrams vanishes identically.

For the sake of brevity, we draw only one half of the two-loop vertex diagrams with a single massive vector boson stretched across the quark line:
\begin{align*}
    &\includegraphics[valign=m,height=.315\linewidth,width=.315\linewidth,keepaspectratio]{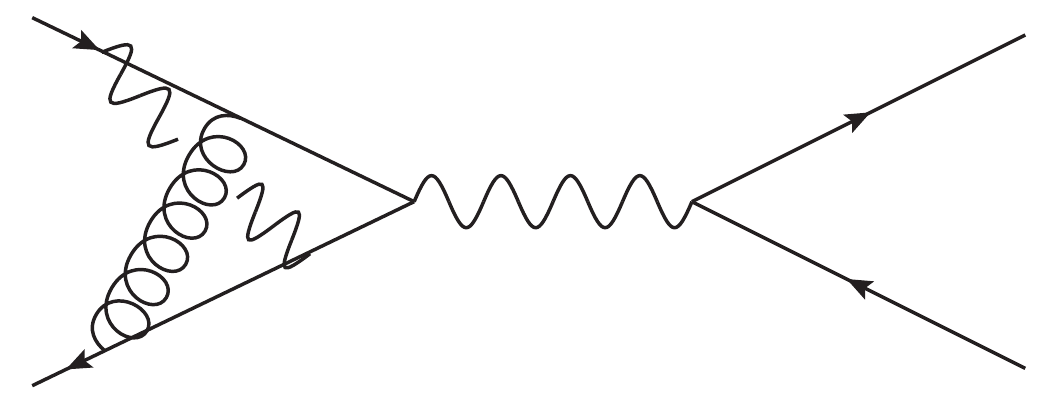}\quad \includegraphics[valign=m,height=.315\linewidth,width=.315\linewidth,keepaspectratio]{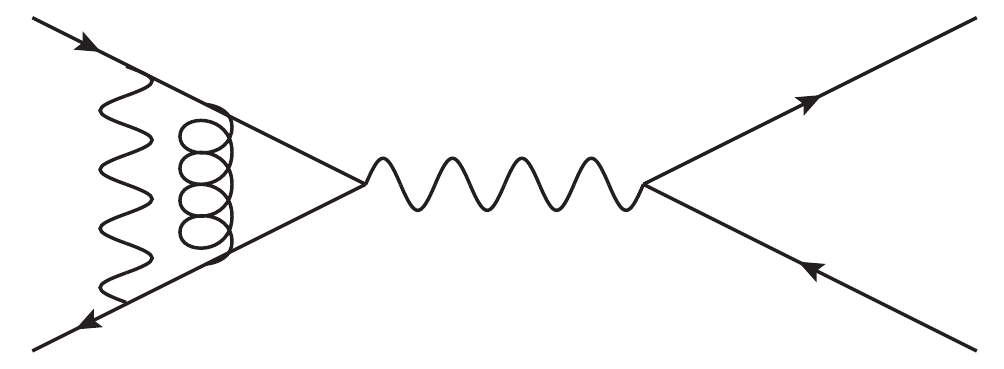}\quad
    \includegraphics[valign=m,height=.315\linewidth,width=.315\linewidth,keepaspectratio]{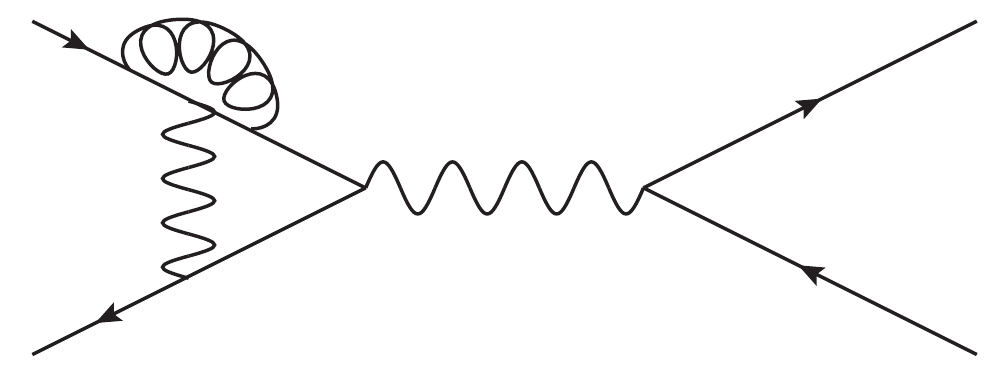}\\
    &\includegraphics[valign=m,height=.315\linewidth,width=.315\linewidth,keepaspectratio]{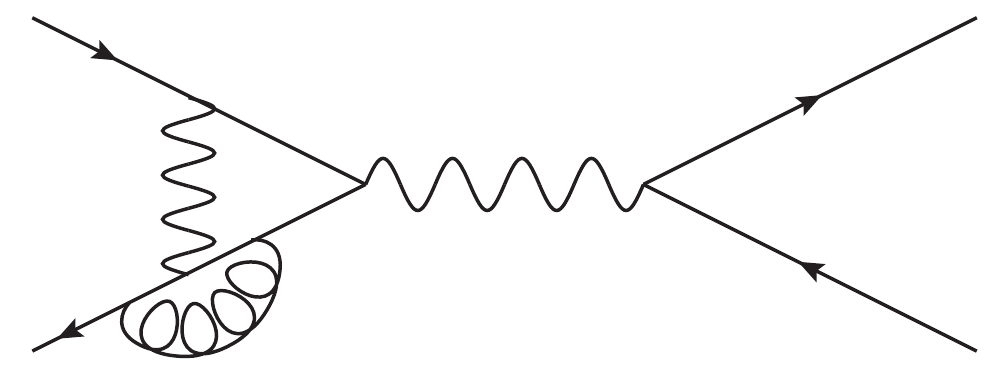}\quad \includegraphics[valign=m,height=.315\linewidth,width=.315\linewidth,keepaspectratio]{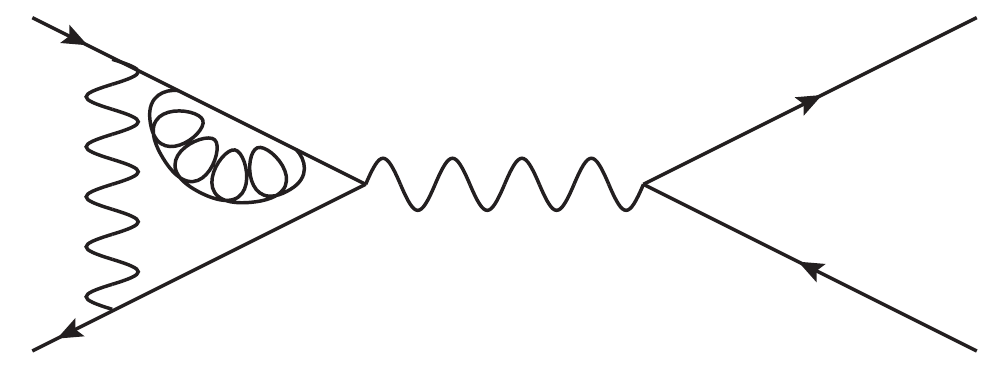}\quad
    \includegraphics[valign=m,height=.315\linewidth,width=.315\linewidth,keepaspectratio]{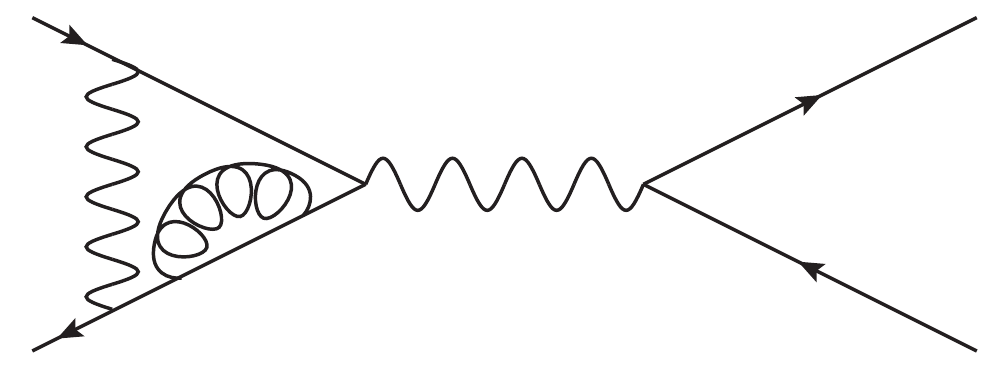}\\
    &\includegraphics[valign=m,height=.315\linewidth,width=.315\linewidth,keepaspectratio]{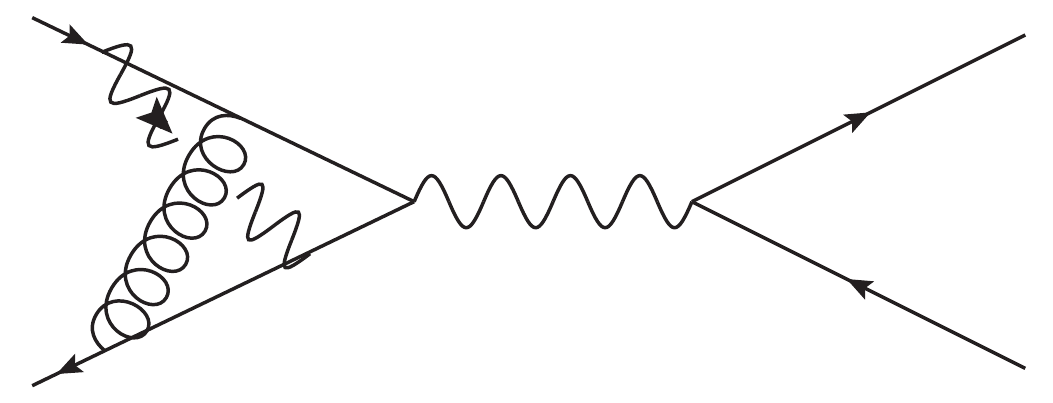}\quad \includegraphics[valign=m,height=.315\linewidth,width=.315\linewidth,keepaspectratio]{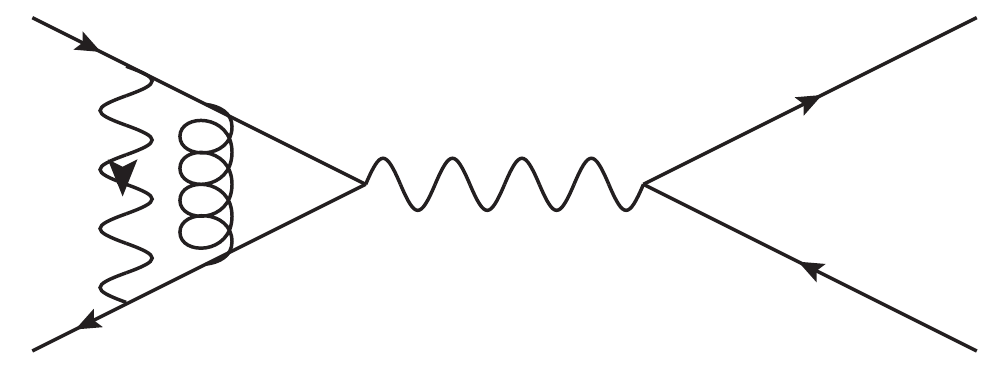}\quad
    \includegraphics[valign=m,height=.315\linewidth,width=.315\linewidth,keepaspectratio]{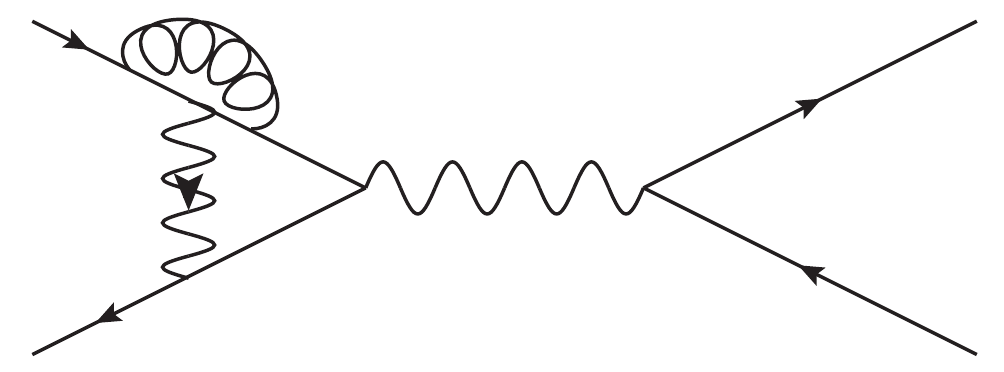}\\
    &\includegraphics[valign=m,height=.315\linewidth,width=.315\linewidth,keepaspectratio]{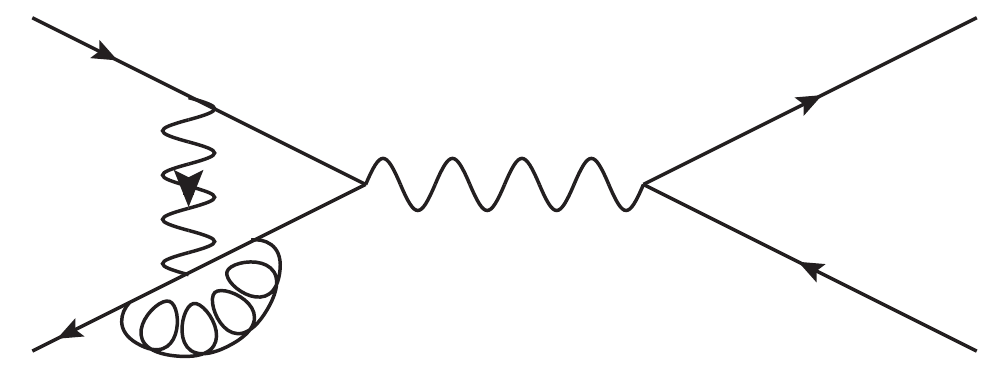}\quad \includegraphics[valign=m,height=.315\linewidth,width=.315\linewidth,keepaspectratio]{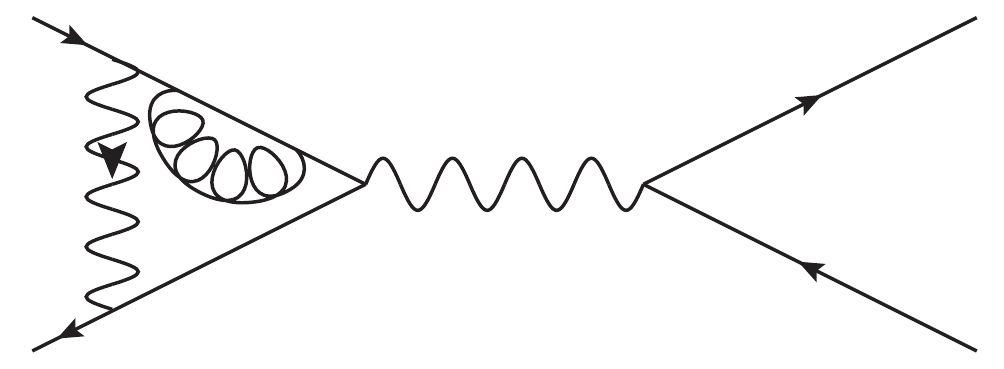}\quad
    \includegraphics[valign=m,height=.315\linewidth,width=.315\linewidth,keepaspectratio]{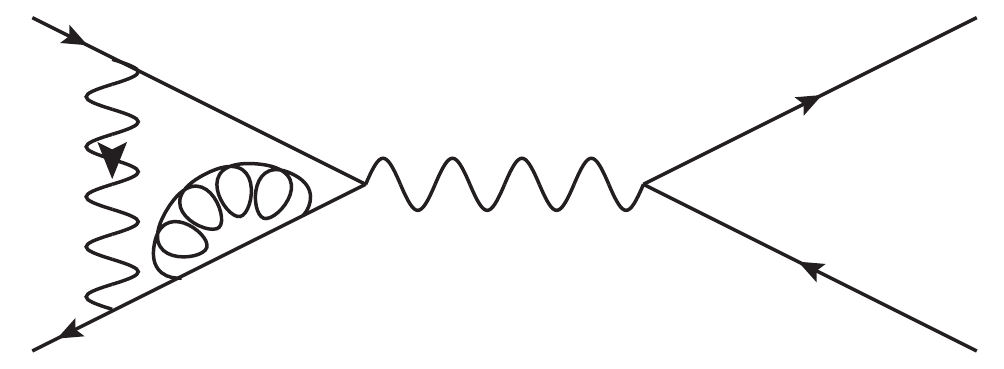}\,.
\end{align*}
The remaining diagrams of this type may be trivially recovered from the above by exchanging the gluon and the electroweak gauge boson attached to the quark line. 

Of course, there are also diagrams with $\gamma W^+  W^-$ and $Z W^+  W^-$ interactions:
\begin{align*}
    &\includegraphics[valign=m,height=.315\linewidth,width=.315\linewidth,keepaspectratio]{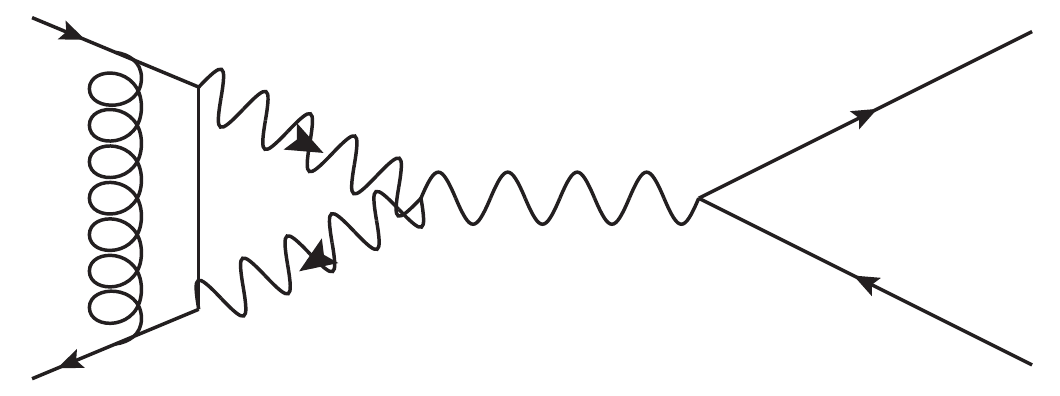}\quad \includegraphics[valign=m,height=.315\linewidth,width=.315\linewidth,keepaspectratio]{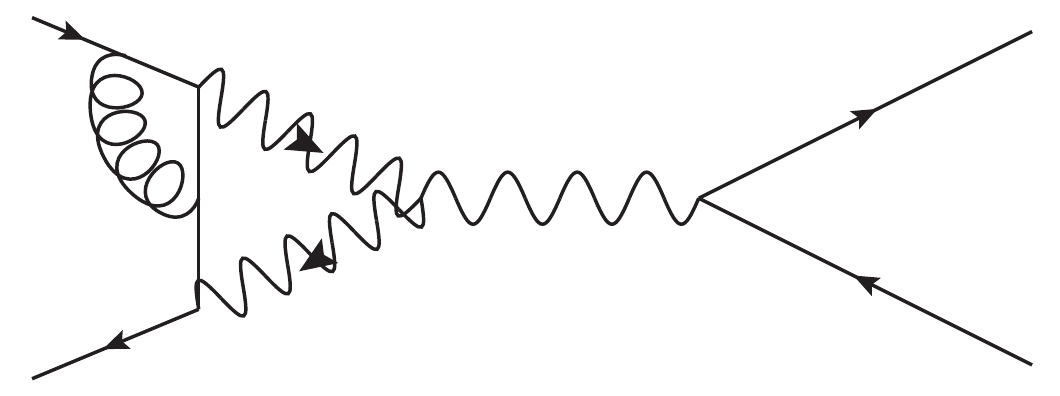}\\
    &\includegraphics[valign=m,height=.315\linewidth,width=.315\linewidth,keepaspectratio]{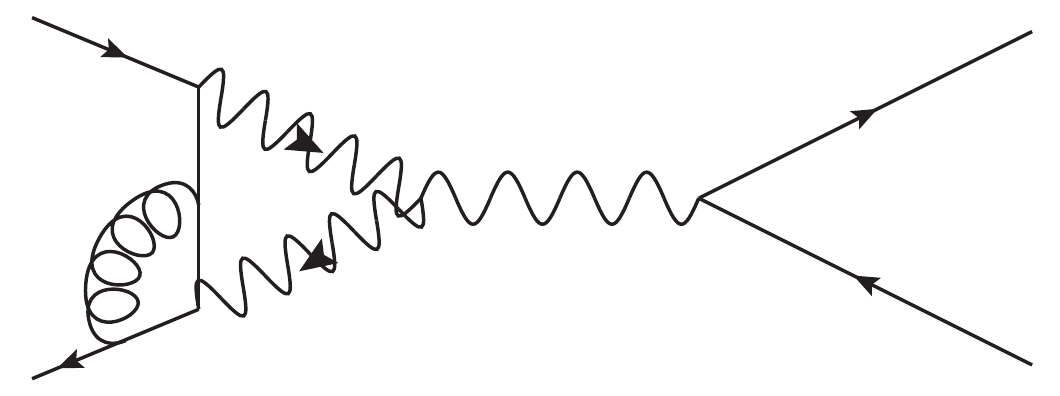}\quad
    \includegraphics[valign=m,height=.315\linewidth,width=.315\linewidth,keepaspectratio]{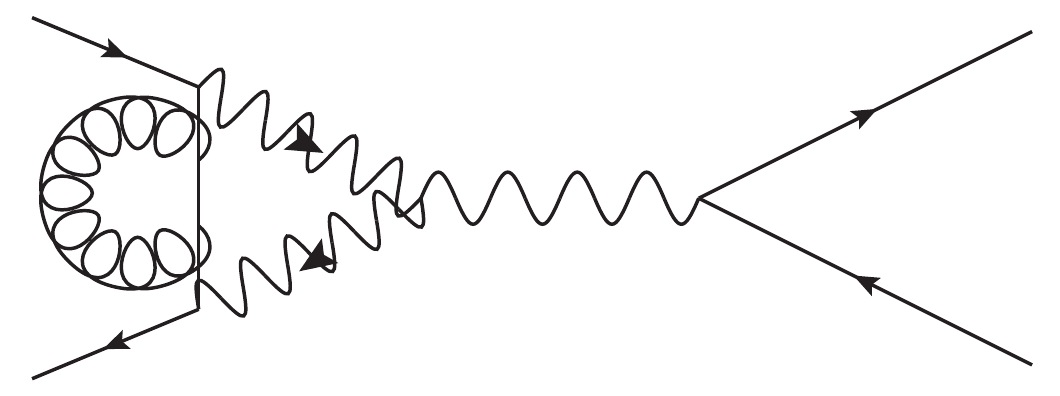}\,.
\end{align*}
Finally, the order $\alpha \alpha_s$ vertex receives a correction from a two-loop vertex counterterm insertion,
\begin{align*}
    &\includegraphics[valign=m,height=.315\linewidth,width=.315\linewidth,keepaspectratio]{vertex_CTi_1L}\,.
\end{align*}

The most complicated two-loop diagrams are those of box type:
\begin{align*}
    &\includegraphics[valign=m,height=.325\linewidth,width=.325\linewidth,keepaspectratio]{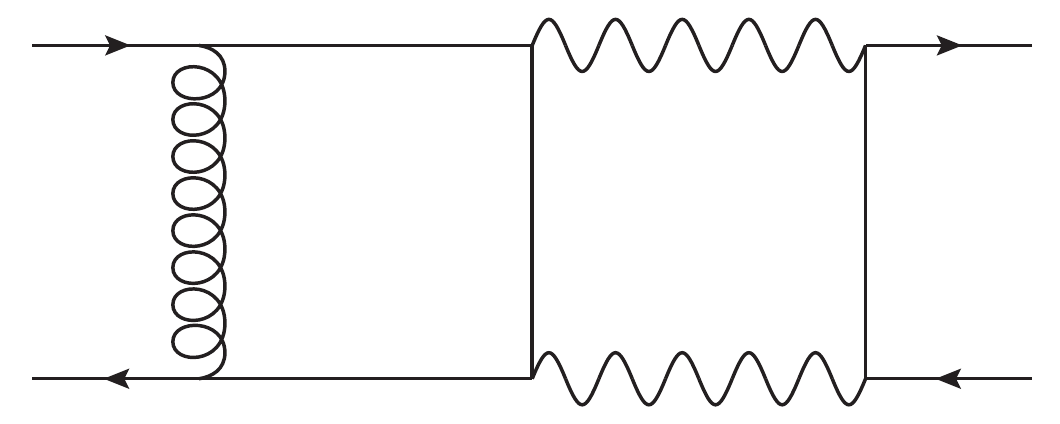}\quad \includegraphics[valign=m,height=.325\linewidth,width=.325\linewidth,keepaspectratio]{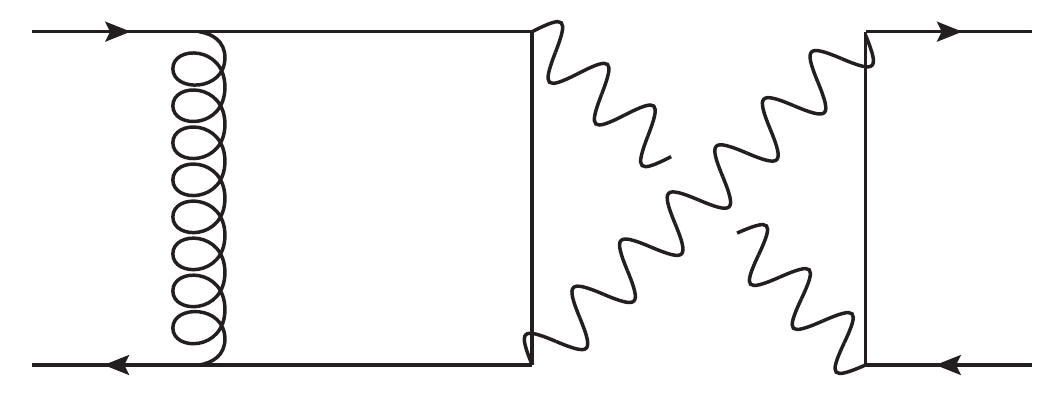}\quad
    \includegraphics[valign=m,height=.325\linewidth,width=.325\linewidth,keepaspectratio]{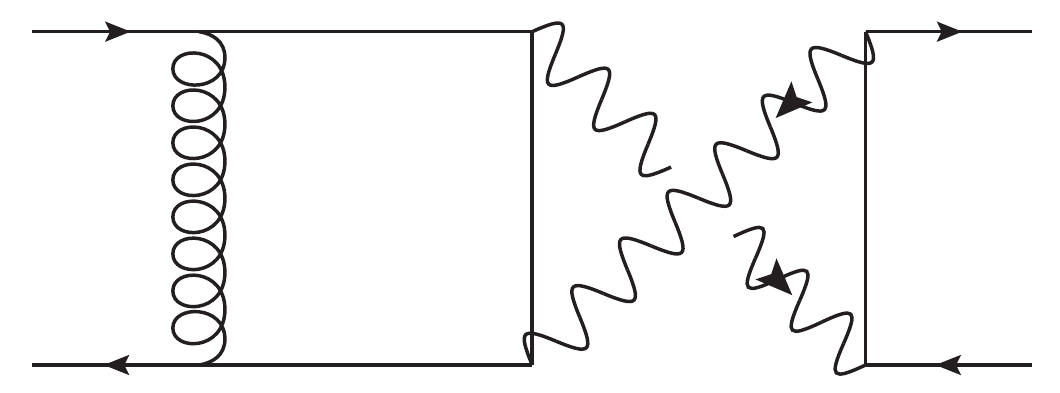}
    \\
    &\qquad\qquad \includegraphics[valign=m,height=.25\linewidth,width=.25\linewidth,keepaspectratio]{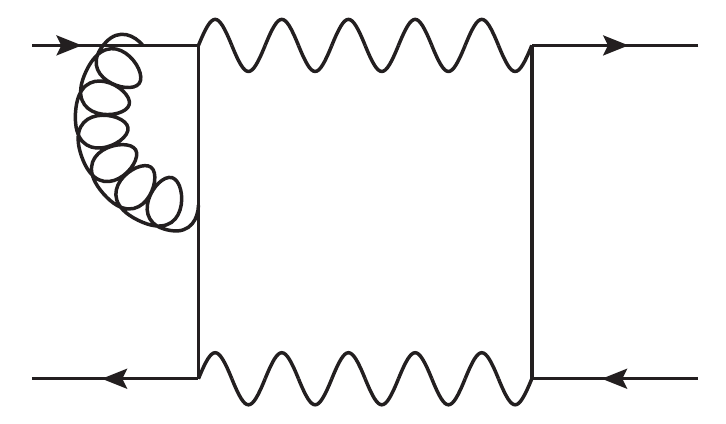}\quad \includegraphics[valign=m,height=.25\linewidth,width=.25\linewidth,keepaspectratio]{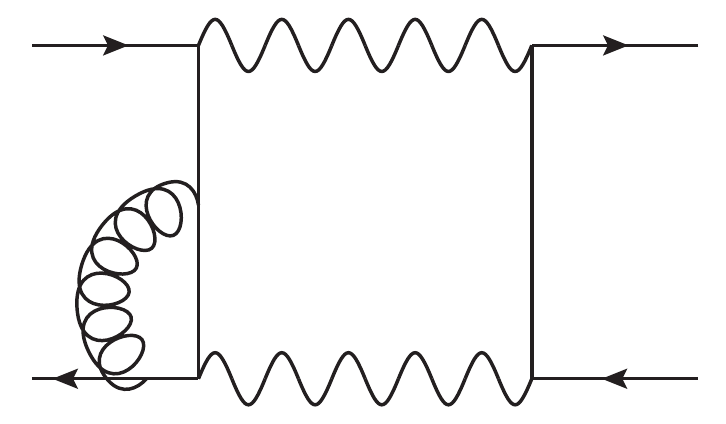}\quad
    \includegraphics[valign=m,height=.25\linewidth,width=.25\linewidth,keepaspectratio]{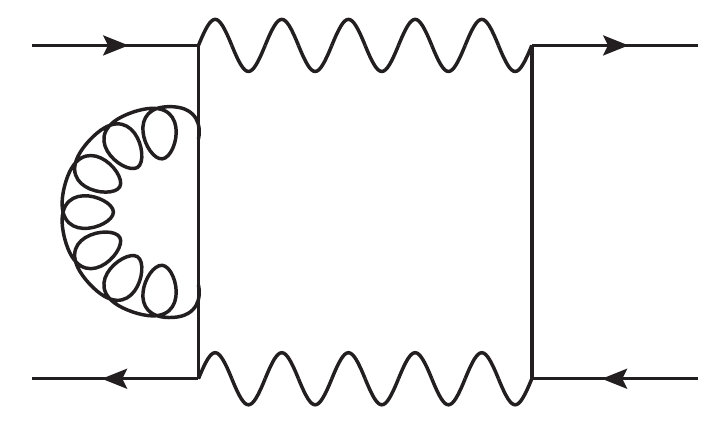}
    \\
    &\qquad\qquad \includegraphics[valign=m,height=.25\linewidth,width=.25\linewidth,keepaspectratio]{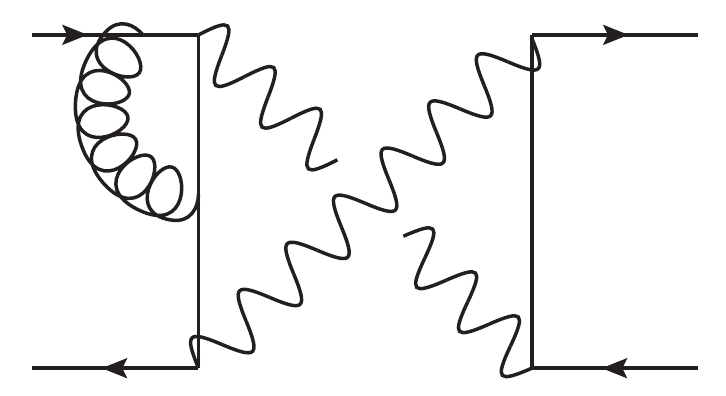}\quad
    \includegraphics[valign=m,height=.25\linewidth,width=.25\linewidth,keepaspectratio]{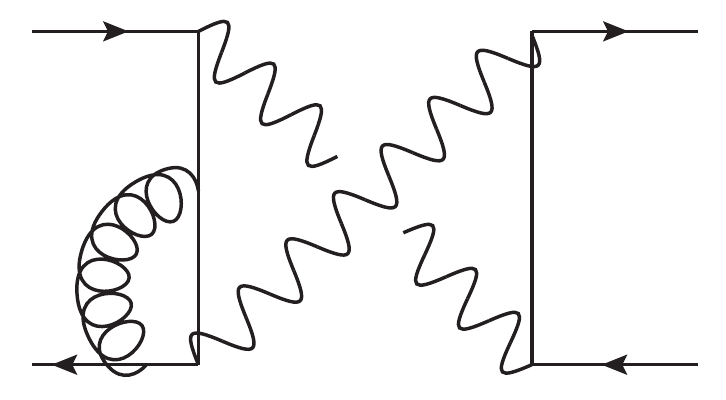}\quad
    \includegraphics[valign=m,height=.25\linewidth,width=.25\linewidth,keepaspectratio]{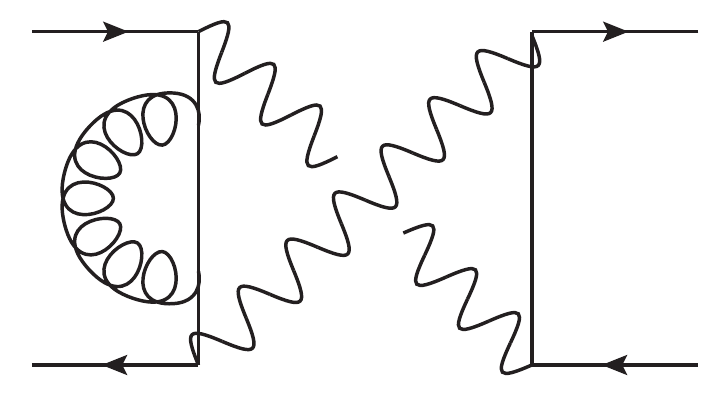}\quad
    \\
    &\qquad\qquad \includegraphics[valign=m,height=.25\linewidth,width=.25\linewidth,keepaspectratio]{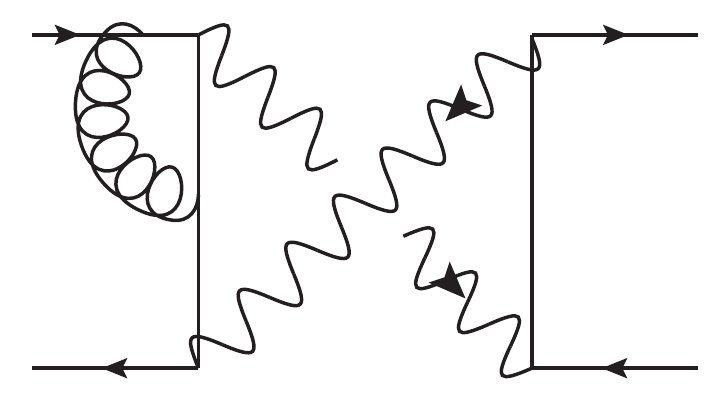}\quad \includegraphics[valign=m,height=.25\linewidth,width=.25\linewidth,keepaspectratio]{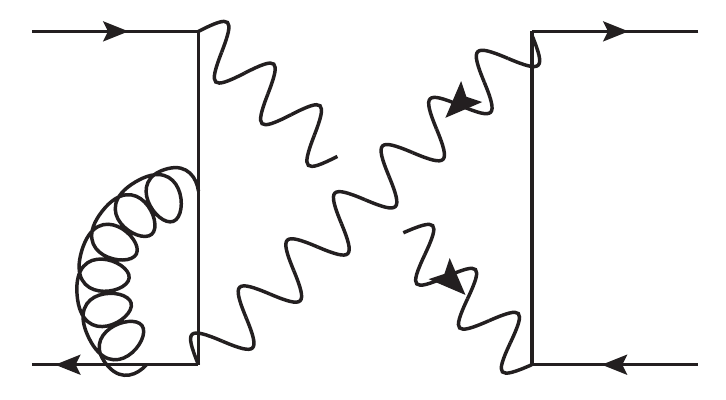}\quad
    \includegraphics[valign=m,height=.25\linewidth,width=.25\linewidth,keepaspectratio]{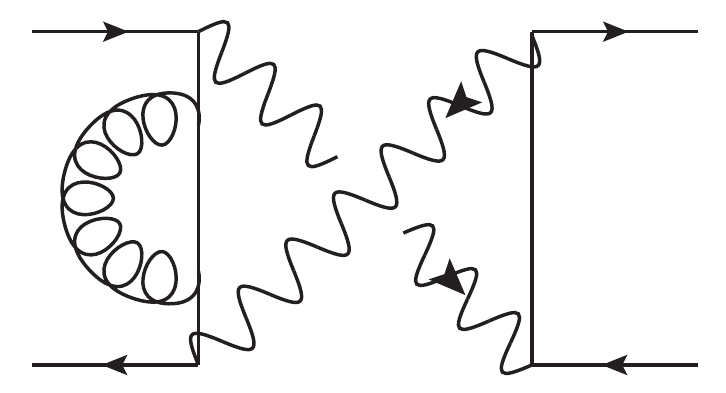}\,.
\end{align*}
Our evaluation of these box-type diagrams is one of the most important new results of this work. In HVBM's $\gamma_5$ scheme, further diagrams,
\begin{align*}
    &\includegraphics[valign=m,height=.25\linewidth,width=.25\linewidth,keepaspectratio]{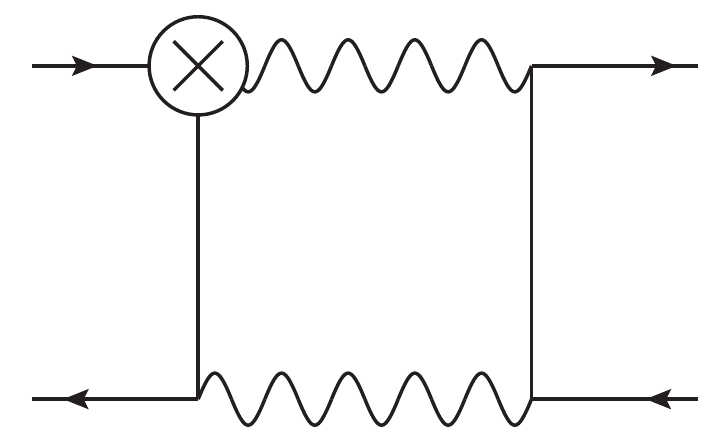}\quad \includegraphics[valign=m,height=.25\linewidth,width=.25\linewidth,keepaspectratio]{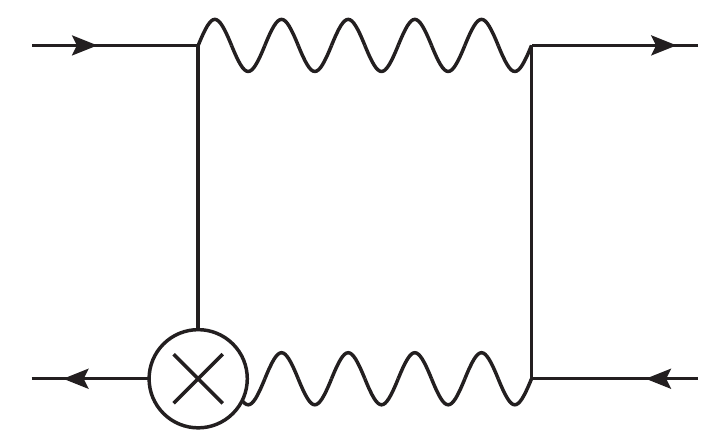}\quad
    \includegraphics[valign=m,height=.25\linewidth,width=.25\linewidth,keepaspectratio]{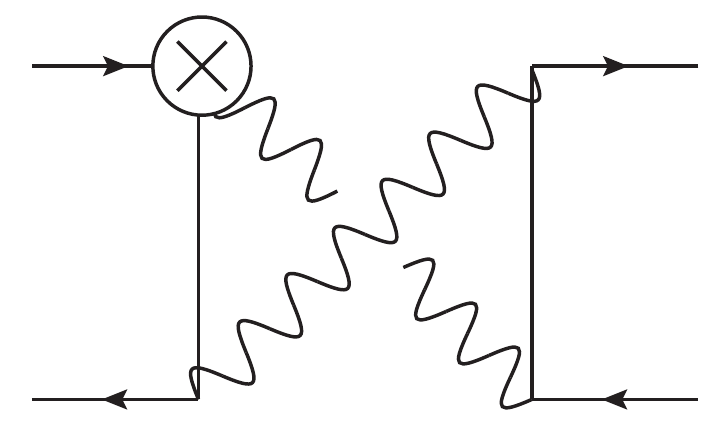}\\
    &\includegraphics[valign=m,height=.25\linewidth,width=.25\linewidth,keepaspectratio]{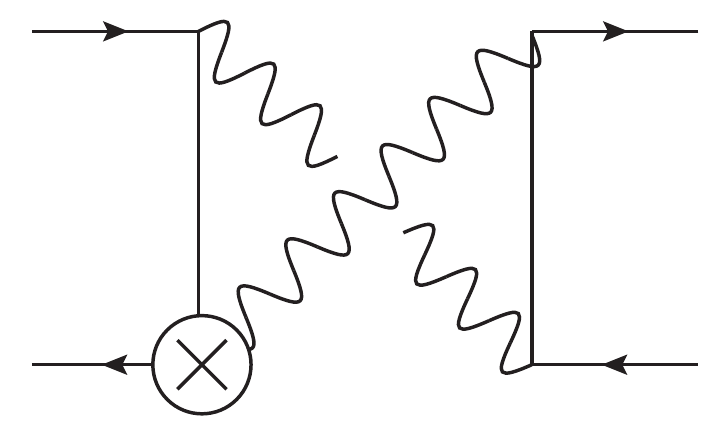}\quad
    \includegraphics[valign=m,height=.25\linewidth,width=.25\linewidth,keepaspectratio]{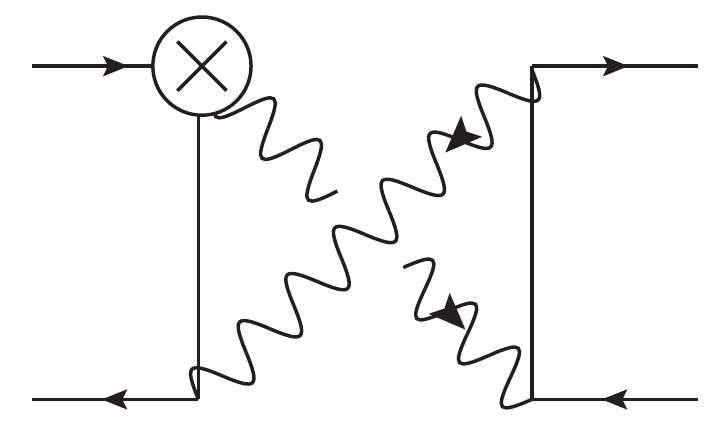}\quad
    \includegraphics[valign=m,height=.25\linewidth,width=.25\linewidth,keepaspectratio]{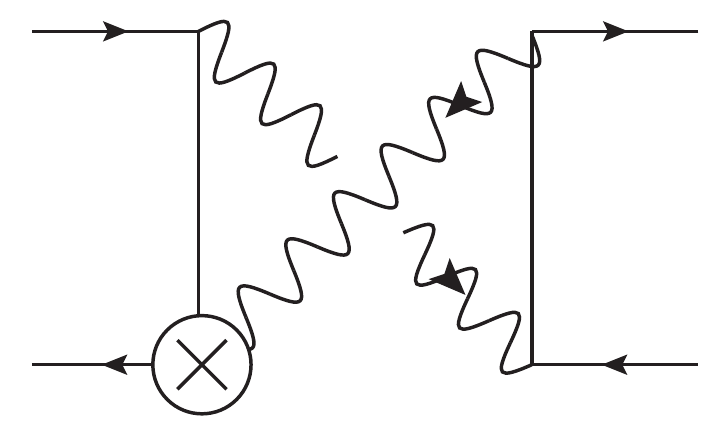}\,,
\end{align*}
need to be included to ensure that the order $\alpha^2 \alpha_s$ hard scattering function in HVBM's $\gamma_5$ scheme respects the chiral symmetry of the Standard Model (see Section \ref{sec:finiteren} for more details).
\subsection{Two-loop integral definitions}
\label{sec:2Lints}
All of the two-loop integrals defined below are pure functions, converted from the idiosyncratic notation of \cite{vonManteuffel:2017myy,Heller:2019gkq}, where they were originally evaluated in the physical region above all two-particle thresholds, to standard $\overline{\rm MS}$ normalization. To be explicit, the $m_i$ integrals from \cite{vonManteuffel:2017myy} and the $\mathbf{m}_i$ integrals from \cite{Heller:2019gkq} must be multiplied by
\begin{equation}
    c_2\left(\epsilon,\mu^2,s\right) = \frac{e^{2 \gamma_E \epsilon}}{\Gamma^2(1-\epsilon)}\left(\frac{\mu^2}{s}\right)^{2 \epsilon}.
\end{equation}
That is to say, the integral measure of our two-loop integrals is exactly:
\begin{equation}
    \left(\frac{e^{\gamma_E \epsilon}\left(\mu^2\right)^\epsilon}{i \pi^{2-\epsilon}}\right)^2\int \mathrm{d}^d k_1\int \mathrm{d}^d k_2\,.
\end{equation}
For brevity, we suppress the factors of $c_2\left(\epsilon,\mu^2,s\right)$ which belong in the definitions of all of the non-factorizable two-loop integrals given below. In total, 133 linearly-independent two-loop integrals appear in our calculations:
\begin{align}
    J_1 &= m_{17}\left(m_z^2\right) &
    J_2 &= m_{17}\left(m_w^2\right) &
    J_3 &= m_1\left(s\right) \notag\\
    J_4 &= \Big(I_{13}\Big)^2 &
    J_5 &= m_5\left(s\right) &
    J_6 &= m_{12}\left(s\right) \notag\\
    J_7 &= I_4 I_{13} &
    J_8 &= m_{18}\left(s,m_z^2\right) &
    J_9 &= m_{19}\left(s,m_z^2\right) \notag\\
    J_{10} &= m_{28}\left(s,m_z^2\right) &
    J_{11} &= m_{31}\left(s,m_z^2\right) &
    J_{12} &= m_{32}\left(s,m_z^2\right) \notag\\
    J_{13} &= I_{13} I_{14} &
    J_{14} &= m_{37}\left(s,m_z^2\right) &
    J_{15} &= m_{39}\left(s,m_z^2\right) \notag\\
    J_{16} &= m_{42}\left(s,m_z^2\right) &
    J_{17} &= m_{43}\left(s,m_z^2\right) &
    J_{18} &= m_{44}\left(s,m_z^2\right) \notag\\
    J_{19} &= m_{56}\left(s,m_z^2\right) &
    J_{20} &= m_{57}\left(s,m_z^2\right) &
    J_{21} &= m_{58}\left(s,m_z^2\right) \notag\\
    J_{22} &= m_{15}\left(s,m_w^2\right) &
    J_{23} &= m_{18}\left(s,m_w^2\right) &
    J_{24} &= m_{19}\left(s,m_w^2\right) \notag\\
    J_{25} &= m_{28}\left(s,m_w^2\right) &
    J_{26} &= m_{31}\left(s,m_w^2\right) &
    J_{27} &= m_{32}\left(s,m_w^2\right) \notag\\
    J_{28} &= I_{13} I_{15} &
    J_{29} &= m_{37}\left(s,m_z^2\right) &
    J_{30} &= m_{39}\left(s,m_w^2\right) \notag\\
    J_{31} &= m_{42}\left(s,m_w^2\right) &
    J_{32} &= m_{43}\left(s,m_w^2\right) &
    J_{33} &= m_{44}\left(s,m_w^2\right) \notag\\
    J_{34} &= m_{56}\left(s,m_w^2\right) &
    J_{35} &= m_{57}\left(s,m_w^2\right) &
    J_{36} &= m_{58}\left(s,m_w^2\right) \notag\\
    J_{37} &= I_2 I_{13} &
    J_{38} &= \mathbf{m}_{10}\left(s,m_w^2\right) &
    J_{39} &= \mathbf{m}_{11}\left(s,m_w^2\right) \notag\\
    J_{40} &= \mathbf{m}_{12}\left(s,m_w^2\right) &
    J_{41} &= \mathbf{m}_{23}\left(s,m_w^2\right) &
    J_{42} &= \mathbf{m}_{24}\left(s,m_w^2\right) \notag\\
    J_{43} &= \mathbf{m}_{25}\left(s,m_w^2\right) &
    J_{44} &= m_2\left(t\right) &
    J_{45} &= m_2\left(u\right) \notag\\
    J_{46} &= m_6\left(t\right) &
    J_{47} &= m_6\left(u\right) &
    J_{48} &= m_7\left(s,t\right) \notag\\
    J_{49} &= m_7\left(s,u\right) &
    J_{50} &= m_8\left(s,t\right) &
    J_{51} &= m_8\left(s,u\right) \notag\\
    J_{52} &= m_9\left(s,t\right) &
    J_{53} &= m_9\left(s,u\right) &
    J_{54} &= m_{10}\left(s,t\right) \notag\\
    J_{55} &= m_{10}\left(s,u\right) &
    J_{56} &= m_{11}\left(s,t\right) &
    J_{57} &= m_{11}\left(s,u\right) \notag\\
    J_{58} &= I_{13}I_{21} &
    J_{59} &= m_{25}\left(s,m_z^2\right) &
    J_{60} &= m_{26}\left(s,m_z^2\right) \notag\\
    J_{61} &= m_{27}\left(t,m_z^2\right) &
    J_{62} &= m_{37}\left(t,m_z^2\right) &
    J_{63} &= m_{38}\left(t,m_z^2\right) \notag\\
    J_{64} &= m_{40}\left(s,m_z^2\right) &
    J_{65} &= m_{45}\left(s,t,m_z^2\right) &
    J_{66} &= m_{46}\left(s,t,m_z^2\right) \notag\\
    J_{67} &= m_{47}\left(s,t,m_z^2\right)&
    J_{68} &= m_{48}\left(s,t,m_z^2\right) &
    J_{69} &= m_{49}\left(s,t,m_z^2\right) \notag\\
    J_{70} &= m_{50}\left(s,t,m_z^2\right) &
    J_{71} &= m_{51}\left(s,t,m_z^2\right) &
    J_{72} &= m_{52}\left(s,t,m_z^2\right) \notag\\
    J_{73} &= m_{53}\left(s,t,m_z^2\right) &
    J_{74} &= m_{54}\left(s,t,m_z^2\right) &
    J_{75} &= m_{55}\left(s,t,m_z^2\right) \notag\\
    J_{76} &= m_{27}\left(u,m_z^2\right) &
    J_{77} &= m_{37}\left(u,m_z^2\right) &
    J_{78} &= m_{38}\left(u,m_z^2\right) \notag\\
    J_{79} &= m_{45}\left(s,u,m_z^2\right) &
    J_{80} &= m_{46}\left(s,u,m_z^2\right) &
    J_{81} &= m_{47}\left(s,u,m_z^2\right) \notag\\
    J_{82} &= m_{48}\left(s,u,m_z^2\right) &
    J_{83} &= m_{49}\left(s,u,m_z^2\right) &
    J_{84} &= m_{50}\left(s,u,m_z^2\right) \notag\\
    J_{85} &= m_{51}\left(s,u,m_z^2\right) &
    J_{86} &= m_{52}\left(s,u,m_z^2\right) &
    J_{87} &= m_{53}\left(s,u,m_z^2\right) \notag\\
    J_{88} &= m_{54}\left(s,u,m_z^2\right) &
    J_{89} &= m_{55}\left(s,u,m_z^2\right) &
    J_{90} &= I_{13}I_{26} \notag\\
    J_{91} &= \mathbf{m}_{10}\left(s,m_z^2\right) &
    J_{92} &= \mathbf{m}_{11}\left(s,m_z^2\right) &
    J_{93} &= \mathbf{m}_{12}\left(s,m_z^2\right) \notag\\
    J_{94} &= \mathbf{m}_{23}\left(s,m_z^2\right) &
    J_{95} &= \mathbf{m}_{24}\left(s,m_z^2\right) &
    J_{96} &= \mathbf{m}_{25}\left(s,m_z^2\right) \notag\\
    J_{97} &= I_{13}I_{27} &
    J_{98} &= \mathbf{m}_{26}\left(s,t,m_z^2\right) &
    J_{99} &= \mathbf{m}_{27}\left(s,t,m_z^2\right) \notag\\
    J_{100} &= \mathbf{m}_{30}\left(s,m_z^2\right) &
    J_{101} &= \mathbf{m}_{31}\left(s,t,m_z^2\right) &
    J_{102} &= \mathbf{m}_{32}\left(s,t,m_z^2\right) \notag\\
    J_{103} &= \mathbf{m}_{33}\left(s,t,m_z^2\right) &
    J_{104} &= \mathbf{m}_{34}\left(s,t,m_z^2\right) &
    J_{105} &= \mathbf{m}_{35}\left(s,t,m_z^2\right) \notag\\
    J_{106} &= \mathbf{m}_{36}\left(s,t,m_z^2\right) &
    J_{107} &= \mathbf{m}_{26}\left(s,u,m_z^2\right) &
    J_{108} &= \mathbf{m}_{27}\left(s,u,m_z^2\right) \notag\\
    J_{109} &= \mathbf{m}_{31}\left(s,u,m_z^2\right) &
    J_{110} &= \mathbf{m}_{32}\left(s,u,m_z^2\right) &
    J_{111} &= \mathbf{m}_{33}\left(s,u,m_z^2\right) \notag\\
    J_{112} &= \mathbf{m}_{34}\left(s,u,m_z^2\right) &
    J_{113} &= \mathbf{m}_{35}\left(s,u,m_z^2\right) &
    J_{114} &= \mathbf{m}_{36}\left(s,u,m_z^2\right) \notag\\
    J_{115} &= m_{27}\left(u,m_w^2\right) &
    J_{116} &= m_{37}\left(u,m_w^2\right) &
    J_{117} &= m_{38}\left(u,m_w^2\right) \notag\\
    J_{118} &= m_{46}\left(s,u,m_w^2\right) &
    J_{119} &= m_{47}\left(s,u,m_w^2\right) &
    J_{120} &= m_{48}\left(s,u,m_w^2\right) \notag\\
    J_{121} &= m_{49}\left(s,u,m_w^2\right) &
    J_{122} &= m_{52}\left(s,u,m_w^2\right) &
    J_{123} &= m_{53}\left(s,u,m_w^2\right) \notag\\
    J_{124} &= I_{13}I_{16} &
    J_{125} &= \mathbf{m}_{26}\left(s,u,m_w^2\right) &
    J_{126} &= \mathbf{m}_{27}\left(s,u,m_w^2\right) \notag\\
    J_{127} &= \mathbf{m}_{30}\left(s,m_w^2\right) &
    J_{128} &= \mathbf{m}_{31}\left(s,u,m_w^2\right) &
    J_{129} &= \mathbf{m}_{32}\left(s,u,m_w^2\right) \notag\\
    J_{130} &= \mathbf{m}_{33}\left(s,u,m_w^2\right) &
    J_{131} &= \mathbf{m}_{34}\left(s,u,m_w^2\right) &
    J_{132} &= \mathbf{m}_{35}\left(s,u,m_w^2\right) \notag\\
    J_{133} &= \mathbf{m}_{36}\left(s,u,m_w^2\right).
\end{align}

In writing the above definitions, we have made all dependence on the kinematic variables $s$, $t$, $u = -s-t$, $m_w^2$, $m_z^2$, and $m_h^2$ explicit on the right-hand side for all functions not expressed as a simple product of the one-loop integrals from Section \ref{sec:1Lints}. The attentive reader will note that, out of 124 non-factorizable integrals, only 57 actually need to be evaluated due to the fact that the rest may subsequently be accessed by considering simple permutations of the kinematic invariants.

One noteworthy fact is that, apart from $J_{29}$, all non-factorizable crossed two-loop integrals not treated explicitly in references \cite{vonManteuffel:2017myy} or \cite{Heller:2019gkq} may be directly obtained through to weight four from the integral evaluations carried out in \cite{vonManteuffel:2017myy,Heller:2019gkq} via the replacement $t \rightarrow u$. For the sake of completeness, we present a simple expression for $J_{29}$ in the physical region above its one-mass threshold. Indeed, the integral admits a compact evaluation in Feynman/Schwinger parameters to all orders in $\epsilon$:
\begin{align}
\label{eq:J29alt}
    J_{29} &= \frac{\epsilon\, \Gamma^2(1-\epsilon)\Gamma(2+2\epsilon)e^{2\gamma_E \epsilon}}{\Gamma(1-2\epsilon)}\left(\frac{\mu^2}{s}\right)^{2 \epsilon}\notag\\
    &\quad \left\{\left(\frac{m_z^2}{s}\right)^{-1-2\epsilon}\frac{\Gamma(-1-2\epsilon)\Gamma(1+\epsilon)}{(1-\epsilon)\Gamma(-\epsilon)}{}_2 F_1\left(1,1;2-\epsilon;-\frac{s}{m_z^2}\right) \nonumber \right. \notag\\
    &\quad \left.- \frac{\Gamma(1-2\epsilon)\Gamma(-2-2\epsilon)\Gamma(2+\epsilon)e^{2 i \pi \epsilon}}{\Gamma(1-3\epsilon)}{}_2 F_1\left(1,-2\epsilon;1-3\epsilon;-\frac{s}{m_z^2}\right)\right\}.
\end{align}
Expanding Eq. \eqref{eq:J29alt} in $\epsilon$, we find results equivalent to those obtained from $J_{62}$ by performing a crossing from $t$ to $s$ and then carefully rewriting the result in terms of the basis of multiple polylogarithms suitable for the $\epsilon$ expansion of $J_{28}$ in the physical region above the one-particle threshold.

\subsection{Assembly of two-loop results}
\label{sec:2Lassem}
In this section, we provide further details relevant to the assembly of our renormalized two-loop scattering amplitudes for the mixed EW-QCD corrections to $q \bar{q} \rightarrow \ell^+ \ell^-$ and discuss some implications of our results. As alluded to in Section \ref{sec:couplingren}, we have
\begin{align}
\label{eq:2LEWrenorm}
    \delta Z_{V V^\prime}^{(1,1)} = 0 \qquad \mathrm{and} \qquad \delta Z_{m_v^2}^{(1,1)} = 0\,,
\end{align}
due to our neglect of all contributions proportional to the number of light fermion flavors or involving the top quark. Together with Eq. \eqref{eq:2Lchargerenorm}, Eqs. \eqref{eq:2LEWrenorm} imply exceptionally-simple forms for the order $\alpha \alpha_s$ $\gamma \bar{q}q$ and $Z \bar{q}q$ vertex counterterms:
\begin{align}
\label{eq:2Lcounterterm1}
    \delta Z_{\mathrm{V}, \gamma \bar{q} q}^{(1,1)} &= -Q_q \delta Z_{\mathrm{V},\,q}^{(1,1)}\,,\\
\label{eq:2Lcounterterm2}
    \delta Z_{\mathrm{A}, \gamma \bar{q} q}^{(1,1)} &= -Q_q \delta Z_{\mathrm{A},\,q}^{(1,1)}\,,\\
\label{eq:2Lcounterterm3}
    \delta Z_{\mathrm{V}, Z \bar{q} q}^{(1,1)} &= v_q \delta Z_{\mathrm{V},\,q}^{(1,1)}- a_q \delta Z_{\mathrm{A},\,q}^{(1,1)}\,,\\
\label{eq:2Lcounterterm4}
  \mathrm{and} \qquad \delta Z_{\mathrm{A}, Z \bar{q} q}^{(1,1)} &= -a_q \delta Z_{\mathrm{V},\,q}^{(1,1)} +v_q \delta Z_{\mathrm{A},\,q}^{(1,1)}\,,
\end{align}
where $\delta Z_{\mathrm{V},\,q}^{(1,1)}$ and $\delta Z_{\mathrm{A},\,q}^{(1,1)}$ are, respectively, the vector and axial vector components of the order $\alpha \alpha_s$ quark wavefunction counterterm.\footnote{We remind the reader that explicit all-orders-in-$\ep$ results for the components of the quark wavefunction counterterm at order $\alpha \alpha_s$ are provided in Eqs. \eqref{eq:wavefuncaasV} and \eqref{eq:wavefuncaasAV}.} Eqs. \eqref{eq:2Lcounterterm1} - \eqref{eq:2Lcounterterm4} are in fact the {\it only} non-zero counterterms of order $\alpha \alpha_s$ which directly contribute to the renormalization of the scattering amplitudes we calculate.

Of course, it is crucial to bear in mind that the renormalization program in HVBM's $\gamma_5$ scheme is complicated by the fact that insertions of the finite counterterms\footnote{Expressions for the finite counterterms relevant to this work are provided in unintegrated form in Eqs. \eqref{eq:fincounterZ} and \eqref{eq:fincounterW} and, due to the trivial ratio of $\delta Z_{Z \bar{q} q}^{(0,1)}$ and $\delta Z_{W^\pm \bar{q} q}^{(0,1)}$, in integrated form in Eq. \eqref{eq:qbarZqfiniteCT}.} $\delta Z_{Z \bar{q} q}^{(0,1)}$ and $\delta Z_{W^\pm \bar{q} q}^{(0,1)}$ into one-loop box diagrams are required on top of all of the conventional renormalizations required to remove ultraviolet divergences. As emphasized in Section \ref{sec:finiteren}, $\delta Z_{Z \bar{q} q}^{(0,1)}$ and $\delta Z_{W^\pm \bar{q} q}^{(0,1)}$ do {\it not} contribute to the renormalization of the two-loop scattering amplitudes in Kreimer's $\gamma_5$ scheme. This fundamental difference between the systematics of renormalized Standard Model perturbation theory in the two $\gamma_5$ schemes we consider indicates that the two-loop analog of Eq. \eqref{eq:hardscatalpha2},
\begin{align}
\label{eq:finalhardscatalpha2alphas}
    \mathcal{H}_{\rm DY}^{(1,1)}[0] = \bar{\mathcal{H}}_{\rm DY}^{(1,1)}[0]\,,
\end{align}
is a highly non-trivial relation. 

As the equality sign in \eqref{eq:finalhardscatalpha2alphas} suggests, after applying the computational methods and master integral evaluations described above to the Feynman diagrams of Section \ref{sec:2Ldiags}, our explicit two-loop calculations confirm that \eqref{eq:hardscatalpha2alphas} does in fact hold: the hard scattering functions for the two-loop mixed EW-QCD corrections to the neutral-current Drell-Yan process we consider in HVBM's $\gamma_5$ scheme and in Kreimer's $\gamma_5$ scheme coincide. 
In our view, the rigorous establishment of Eq. \eqref{eq:finalhardscatalpha2alphas} constitutes a very important cross-check on our entire formulation and a significant theoretical result. In the future, our detailed comparison of HVBM and Kreimer at the two-loop level should benefit further calculations of still-unknown multi-loop perturbative corrections to more complicated Standard Model processes in pure dimensional regularization.

In future studies of Standard Model scattering processes, it would certainly be possible to perform two parallel calculations for each process in the two $\gamma_5$ schemes considered in this work, but there may be a more practical way to rigorously cross-check perturbative calculations going forward. During the course of our investigation of $q \bar{q} \rightarrow \ell^+ \ell^-$ in Kreimer's $\gamma_5$ scheme, we discovered that we could effectively check the correctness of the two-loop hard scattering functions within Kreimer's $\gamma_5$ scheme itself by setting up our code pipeline to allow the user to make different choices of reading point prescription for the Dirac traces. For example, we found that the fully-symmetrized reading point prescription mentioned in Section \ref{sec:Kreimerdefs} led to different higher-order-in-$\ep$ one-loop hard scattering functions, but the same order $\alpha^2 \alpha_s$ hard scattering function at $\mathcal{O}\left(\ep^0\right)$.

The two-loop hard scattering functions calculated in this work are too long to be included in the article.
Once we have implemented support for the remaining kinematic regions of the physical phase space, we plan to provide public access on {\tt HEPForge} to an efficient {\tt C++} program for the evaluation of the polarized hard scattering functions discussed in the next section.

\section{Numerical results}
\subsection{Helicity amplitudes}
\label{sec:helicities}

Having arrived at finite quantities in four dimensions, we can directly cast our results into expressions for the scattering of fermions with definite helicities using explicit four-dimensional representations.

Due to the form of the massless fermion propagator and the manner in which the electroweak gauge bosons couple to matter, the helicity quantum number is conserved along the initial and final fermion lines (see {\it e.g.} \cite{Dixon:2013uaa} for a review). This implies that there are just four non-vanishing helicity amplitudes, those where both the quark and the antiquark have opposite helicity and the lepton and antilepton have opposite helicity.
We parametrize the center-of-momentum frame according to
\begin{align}
\label{eq:explmomenta}
p_1 &= \frac{\sqrt{s}}{2} (1, 0, 0, 1), &
p_3 &= \frac{\sqrt{s}}{2} (1, \sin\theta, 0, \cos\theta ),\notag\\
p_2 &= \frac{\sqrt{s}}{2} (1, 0, 0, -1), &
p_4 &= \frac{\sqrt{s}}{2} (1, -\sin\theta, 0, -\cos\theta ),
\end{align}
with
\begin{equation}
\label{eq:tofcostheta}
-\frac{t}{s} = \frac{1 - \cos\theta}{2},
\end{equation}
and use an explicit basis of spinor states in the chiral representation \cite{Peskin:1995ev}
\begin{align}
\label{eq:explicitspinors}
  u_+(p_1) &= s^{1/4}\begin{pmatrix} 0,& 0,& 1,& 0\end{pmatrix}, &
  u_+(p_3) &= s^{-1/4}\begin{pmatrix} 0,& 0,&  \sqrt{-u}, &  \sqrt{-t}\end{pmatrix}, \nonumber\\
  u_-(p_1) &= s^{1/4}\begin{pmatrix}0,& 1,& 0,& 0\end{pmatrix}, &
  u_-(p_3) &= s^{-1/4}\begin{pmatrix} -\sqrt{-t}, &  \sqrt{-u}, & 0,& 0\end{pmatrix}, \nonumber\\
  v_+(p_2) &= s^{1/4}\begin{pmatrix}-1,& 0,& 0,& 0\end{pmatrix}, &
  v_+(p_4) &= s^{-1/4}\begin{pmatrix}-\sqrt{-u},& -\sqrt{-t},& 0,& 0\end{pmatrix}, \nonumber\\
  v_-(p_2) &= s^{1/4}\begin{pmatrix}0,& 0,& 0,& 1\end{pmatrix},\qquad \mathrm{and} &
  v_-(p_4) &= s^{-1/4}\begin{pmatrix}0,& 0,& -\sqrt{-t},&  \sqrt{-u}\end{pmatrix}.
\end{align}
Plugging these representations\footnote{The phase conventions of \eqref{eq:explicitspinors} above are in line with the spinor helicity formalism, as in \cite{Dixon:1996wi} where $u_\pm (p_i) = v_\mp (p_i)$ for all $i$. Let us stress, however, that we do {\it not} adopt the all-outgoing convention for the external four-momenta.
We made use of Eqs. \eqref{eq:explmomenta} and \eqref{eq:tofcostheta} to derive Eqs. \eqref{eq:explicitspinors}.}
 into our Lorentz structures directly gives us results for the helicity amplitudes.
Decomposing the hard functions (or finite remainders) according to
\begin{equation}
\label{eq:hardfuncnot}
\pbar{\mathcal{H}}^{(m,n)}_{\mathrm{DY}}[0] = \pbar{\mathbf{C}}^{(m,n),\,\text{fin}}_\mathrm{VV}\, \pbar{\mathcal{T}}_\mathrm{VV} + \pbar{\mathbf{C}}^{(m,n),\,\text{fin}}_\mathrm{AA}\, \pbar{\mathcal{T}}_\mathrm{AA} + \pbar{\mathbf{C}}^{(m,n),\,\text{fin}}_\mathrm{VA}\, \pbar{\mathcal{T}}_\mathrm{VA} + \pbar{\mathbf{C}}^{(m,n),\,\text{fin}}_\mathrm{AV}\, \pbar{\mathcal{T}}_\mathrm{AV}\,,
\end{equation}
we obtain for the polarized hard scattering functions $\mathcal{H}^{(m,n)}_{\lambda_1\lambda_2\lambda_3\lambda_4}$:
\begin{align}
\label{eq:helampnot1}
\mathcal{H}^{(m,n)}_{+-+-} &=  -2 (s + t) \left( \mathbf{C}^{(m,n),\,\text{fin}}_{\mathrm{VV}} + \mathbf{C}^{(m,n),\,\text{fin}}_{\mathrm{AA}} + \mathbf{C}^{(m,n),\,\text{fin}}_{\mathrm{VA}} + \mathbf{C}^{(m,n),\,\text{fin}}_{\mathrm{AV}} \right),\\
\label{eq:helampnot2}
\mathcal{H}^{(m,n)}_{-+-+} &= -2 (s + t) \left( \mathbf{C}^{(m,n),\,\text{fin}}_{\mathrm{VV}} + \mathbf{C}^{(m,n),\,\text{fin}}_{\mathrm{AA}} - \mathbf{C}^{(m,n),\,\text{fin}}_{\mathrm{VA}} - \mathbf{C}^{(m,n),\,\text{fin}}_{\mathrm{AV}} \right),\\
\label{eq:helampnot3}
\mathcal{H}^{(m,n)}_{+--+} &= -2 t \left( \mathbf{C}^{(m,n),\,\text{fin}}_{\mathrm{VV}} - \mathbf{C}^{(m,n),\,\text{fin}}_{\mathrm{AA}} - \mathbf{C}^{(m,n),\,\text{fin}}_{\mathrm{VA}} + \mathbf{C}^{(m,n),\,\text{fin}}_{\mathrm{AV}} \right),\\
\label{eq:helampnot4}
\mathrm{and}\qquad\mathcal{H}^{(m,n)}_{-++-} &= -2 t \left( \mathbf{C}^{(m,n),\,\text{fin}}_{\mathrm{VV}} - \mathbf{C}^{(m,n),\,\text{fin}}_{\mathrm{AA}} + \mathbf{C}^{(m,n),\,\text{fin}}_{\mathrm{VA}} - \mathbf{C}^{(m,n),\,\text{fin}}_{\mathrm{AV}} \right).
\end{align}
for coupling orders $m$ and $n$. In Eqs. \eqref{eq:helampnot1} - \eqref{eq:helampnot4}, due to the observed $\gamma_5$ scheme-independence of the hard scattering functions at zeroth order in the $\epsilon$ expansion (see Section \ref{sec:2Lassem}), we  drop the dual notation employed in Eq. \eqref{eq:hardfuncnot}.
\subsection{Final results}
\label{sec:2Lfinalres}

In this section we present visualizations of the polarized hard scattering functions $\mathcal{H}^{(0,0)}_{\lambda_1\lambda_2\lambda_3\lambda_4}$, $\mathcal{H}^{(0,1)}_{\lambda_1\lambda_2\lambda_3\lambda_4}$, $\mathcal{H}^{(1,0)}_{\lambda_1\lambda_2\lambda_3\lambda_4}$, and $\mathcal{H}^{(1,1)}_{\lambda_1\lambda_2\lambda_3\lambda_4}$ defined in Section \ref{sec:helicities} for all four non-trivial helicity configurations.
In the numerical analysis, we set the gauge boson and Higgs masses to their on-shell values as listed in~\cite{Zyla:2020zbs},
\begin{align}
    m_w &= 80.379~\mathrm{GeV}, & m_z &= 91.1876~\mathrm{GeV}, & m_h &= 125.10~\mathrm{GeV},
\end{align}
and the renormalization scale $\mu_R=\sqrt{s}$.
Since we keep powers of $\alpha$ and $\alpha_s$ factored out of the expressions we plot, we do not need to specify their values here.
For illustrative purposes, we found it sufficient to consider up-type quarks in the initial state. Our figures do look rather different for down-type quarks, but our impression is that, on the whole, they do not introduce completely new features which would be of paramount importance to discuss here. We elected to plot only the real parts of our final results, as the real parts of $\mathcal{H}^{(1,1)}_{\lambda_1\lambda_2\lambda_3\lambda_4}$ contain the most complicated weight four multiple polylogarithms appearing in our calculations. 

In our numerical analysis, we focus on larger values of $\sqrt{s}$ where the previously unknown non-factorizable two-loop box-type Feynman diagrams of Section \ref{sec:2Ldiags} are expected to be important. In order to compare the different orders in $\alpha$ and $\alpha_s$, we find it convenient to include in our plots the $4 \pi$ suppression factors taken out of each higher-order term in Eq. \eqref{eq:subfuncs}; that is, we consider $\mathcal{H}^{(m,n)}_{\lambda_1\lambda_2\lambda_3\lambda_4}/(4\pi)^{m+n}$, where the relative orders in the electroweak and strong couplings are given by $(m,n)$ (see also Eq. \eqref{eq:relativenotation}).

\begin{figure}
    \centering
    \includegraphics*[height=53mm]{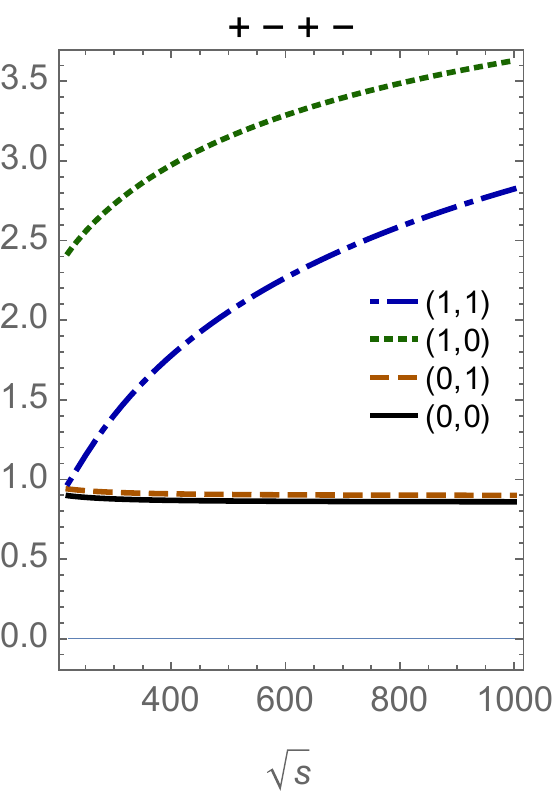}\hspace{-1.5mm}
    \includegraphics*[height=53mm]{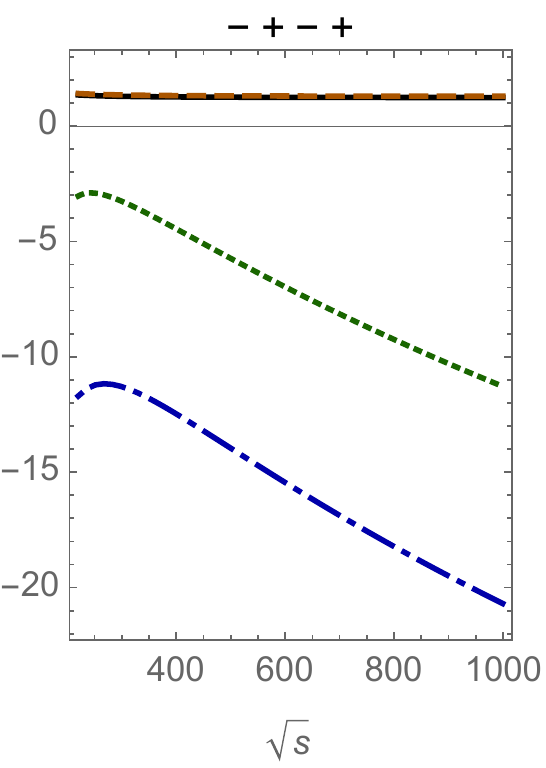}\hspace{-1.5mm}
    \includegraphics*[height=53mm]{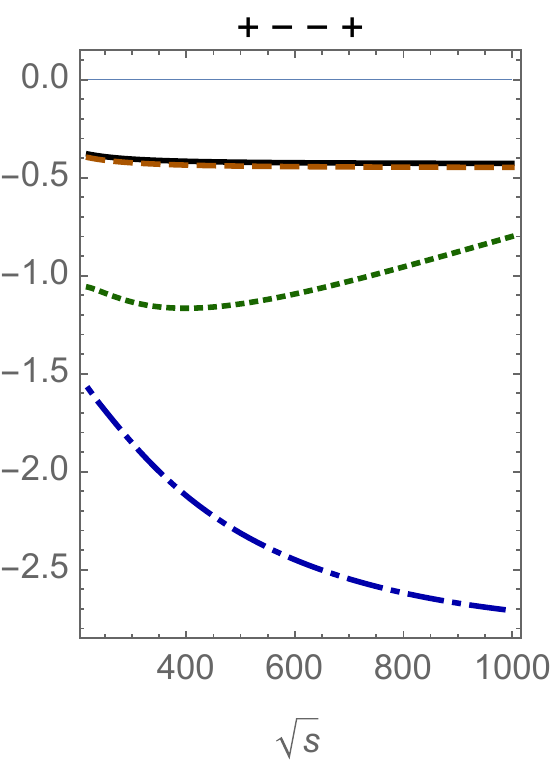}\hspace{-1.5mm}
    \includegraphics*[height=53mm]{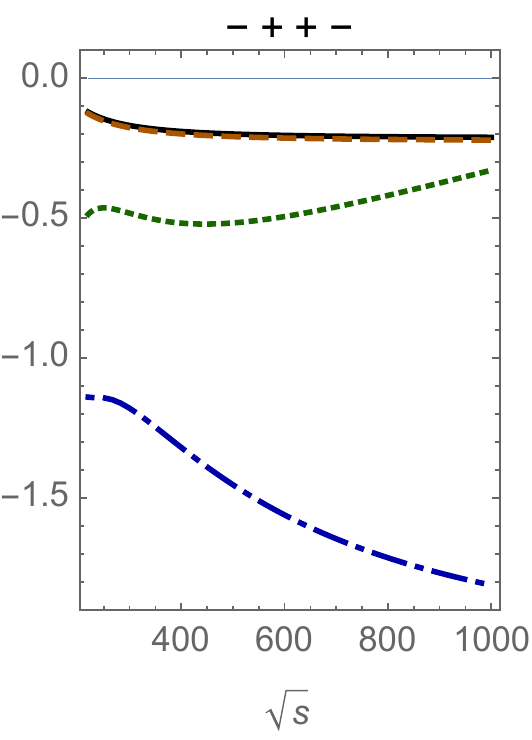}
    \caption{Helicity amplitudes for $u\bar{u}\to \ell^+ \ell^-$ in dependence on the center-of-momentum energy for central scattering, $\cos\theta=0$, and $\mu_\text{R}=\sqrt{s}$.
    The figure shows the real parts of the polarized hard scattering functions (finite remainders) $\mathrm{Re}\,\mathcal{H}^{(m,n)}_{\lambda_1\lambda_2\lambda_3\lambda_4}/(4\pi)^{m+n}$, see Eqs.~\eqref{eq:helampnot1}-\eqref{eq:helampnot4}, where
    the relative orders in the electroweak and strong couplings are denoted by $(m,n)$.
    }
    \label{fig:helampus}
\end{figure}

In all plots, we include the tree-level and relative order $\alpha_s$ results for reference, they show a rather simple behavior. The one-loop QCD corrections lie almost on top of the tree level results. Indeed, they fully factorize from the tree results; we have
\begin{equation}
 \mathcal{H}^{(0,1)}_{\lambda_1\lambda_2\lambda_3\lambda_4}/(4\pi) = 
\left(\frac{\pi}{3}-i\right)\mathcal{H}^{(0,0)}_{\lambda_1\lambda_2\lambda_3\lambda_4} 
\end{equation}
for all helicity configurations and both up- and down-type quarks in the initial state. For the real part we see that $\pi/3\approx 1.05$, thus explaining the observed similarity.

Figure~\ref{fig:helampus} shows the dependence of the hard functions on the center-of-mass energy $\sqrt{s}$ for fixed central scattering angle.
We observe that the absolute values of all plotted real parts of relative order $\alpha \alpha_s$ increase as a function of $\sqrt{s}$, justifying the expectation that the calculation of the mixed two-loop EW-QCD corrections for the leptonic final state is well-motivated in the kinematic regime depicted in the figure.
We note that the real parts of the one-loop EW and the two-loop EW-QCD corrections are not aligned for all helicity configurations.

\begin{figure}
    \centering
    \includegraphics*[height=55mm]{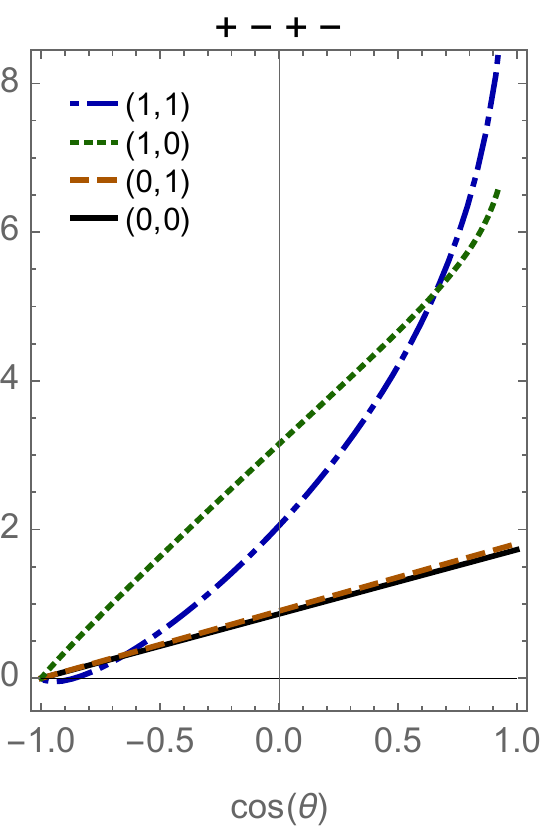}\hspace{-1.5mm}
    \includegraphics*[height=55mm]{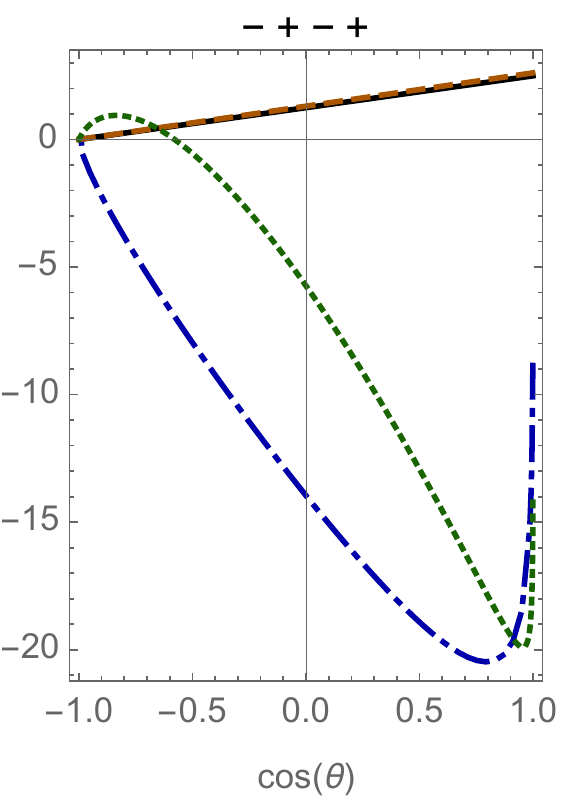}\hspace{-1.5mm}
    \includegraphics*[height=55mm]{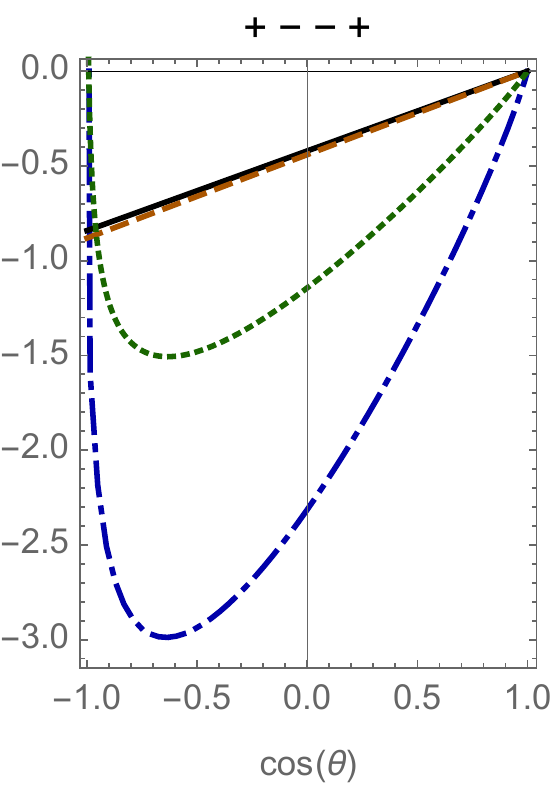}\hspace{-1.5mm}
    \includegraphics*[height=55mm]{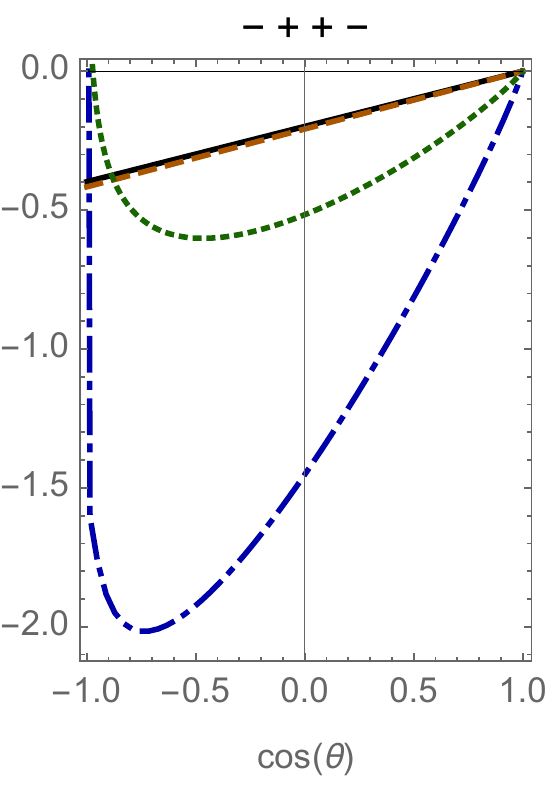}
    \caption{Helicity amplitudes for $u\bar{u}\to \ell^+ \ell^-$ in dependence on the cosine of the scattering angle for $\sqrt{s}=500~\text{GeV}$ and $\mu_\text{R}=\sqrt{s}$.
    The figure shows the real parts of the polarized hard scattering functions (finite remainders) $\mathrm{Re}\,\mathcal{H}^{(m,n)}_{\lambda_1\lambda_2\lambda_3\lambda_4}/(4\pi)^{m+n}$, see Eqs.~\eqref{eq:helampnot1}-\eqref{eq:helampnot4}, where
    the relative orders in the electroweak and strong couplings are denoted by $(m,n)$.
    }
    \label{fig:helampuct}
\end{figure}

Figure~\ref{fig:helampuct} shows the dependence of the hard functions on the cosine of the scattering angle for fixed center-of-mass energy $\sqrt{s}$.
The electroweak corrections show a rather complex angular dependence.
While the one-loop EW and the two-loop EW-QCD corrections show a similarity in their angular dependence, they differ in the details.

We also compared the curves of Figure \ref{fig:helampuct} to analogous ones for the pure QED-QCD model studied in \cite{Kilgore:2011pa}.\footnote{To our knowledge, the authors of \cite{Kilgore:2011pa} calculated unpolarized hard scattering functions only.} While maybe not instructive from the phenomenological point of view, it was interesting to see a somewhat similar qualitative angular dependence emerge when comparing the two-loop polarized hard scattering functions for QED-QCD and EW-QCD normalized by their respective tree level contributions.
\section{Summary and outlook}
\label{sec:outlook}
In this article, we calculated the relative order $\alpha \alpha_s$ mixed EW-QCD corrections to Drell-Yan lepton pair production, $q\bar{q}\to\ell^+ \ell^-$.
We performed the calculation in both the 't\,Hooft-Veltman-Breitenlohner-Maison $\gamma_5$ scheme and in the Kreimer $\gamma_5$ scheme using the projector method for chiral fermions.
While our two-loop scattering amplitudes were found to be scheme dependent starting at $\mathcal{O}\left(\ep^{-1}\right)$, unique polarized hard scattering functions in $d=4$ were obtained after infrared subtraction. In the 't\,Hooft-Veltman-Breitenlohner-Maison scheme, we restored chiral symmetry using local counterterms, where we found it essential to consistently calculate our renormalization constants to higher orders in $\ep$. To the best of our knowledge, our application of Kreimer's $\gamma_5$ scheme to the calculation of genuine two-loop $2\to 2$ scattering amplitudes with non-trivial electroweak effects is the first of its kind.

Our calculation provides a major building block for the calculation of the relative order $\alpha \alpha_s$ corrections to off-shell Drell-Yan production in the high energy region, which is particularly relevant to new physics searches and the establishment of constraints on possible extensions of the Standard Model.

\acknowledgments
We acknowledge helpful discussions with Ayres Freitas, Dominik St\"ockinger, Lorenzo Tancredi, and C.-P.\ Yuan. MH was supported in part by the German Research Foundation (DFG), through the Collaborative Research Center, Project ID 204404729, SFB 1044, and the Cluster of Excellence PRISMA$^+$, Project ID 39083149, EXC 2118/1.
AvM was supported in part by the National Science Foundation under Grants No.\ 1719863
and 2013859.
The authors would like to express a special thanks to the Mainz Institute for Theoretical Physics (MITP) of the Cluster of Excellence PRISMA$^+$ for its hospitality and support.
Our figures were generated using {\tt Jaxodraw} \cite{Binosi:2003yf}, based on {\tt AxoDraw} \cite{Vermaseren:1994je}.

\appendix
\section{Explicit one-loop results}
\subsection{Relative order $\alpha_s$ results}
\label{sec:1Lasres}
Corrections to the neutral-current Drell-Yan process of relative order $\alpha_s$ are produced exclusively via gluon exchange across the initial quark line. As should be clear from the diagrams of Section \ref{sec:1Ldiags}, the corresponding order $\alpha_s$ vertex form factors,
\begin{align}
    \bar{\mathcal{V}}_{\gamma \bar{q} q}^{(0,1)}(s) &= -Q_q C_F \mathbf{t}_1 I_{13}\,,\\
    \bar{\mathcal{V}}_{Z \bar{q} q}^{(0,1)}(s) &= v_q C_F \mathbf{t}_1 I_{13}\,,\\
    \mathrm{and}\qquad\bar{\mathcal{A}}_{Z \bar{q} q}^{(0,1)}(s) &= -a_q C_F \mathbf{t}_1 I_{13}\,,
\end{align}
where
\begin{align}
    \mathbf{t}_1 = \frac{2-\ep+2\ep^2}{(1-2\ep)\ep^2}\,,
\end{align}
are nearly trivial.

In HVBM's $\gamma_5$ scheme, the finite counterterms relevant to the renormalization of the order $\alpha^2 \alpha_s$ scattering amplitude,
\begin{align}
\label{eq:fincounterZ}
    \delta Z_{Z \bar{q} q}^{(0,1)} &= -a_q C_F\mathbf{t}_2 I_{13}\\
\label{eq:fincounterW}
    \mathrm{and}\qquad \delta Z_{W^\pm \bar{q} q}^{(0,1)} &= -a_w C_F \mathbf{t}_2 I_{13}\,,
\end{align}
where
\begin{align}
    \mathbf{t}_2 = \frac{2 (2-\ep)}{(1-\ep)(1-2\ep)}\,,
\end{align}
are also of order $\alpha_s$ (see Section \ref{sec:finiteren} for more details).
\subsection{Relative order $\alpha$ results}
\label{sec:1Lares}
Corrections to the neutral-current Drell-Yan process of relative order $\alpha$ are generated by electroweak gauge boson self-energy diagrams, vertex diagrams, and box diagrams. For the sake of completeness, we also present expressions for the counterterms of order $\alpha$ discussed in Sections \ref{sec:EWren} and \ref{sec:fermren}.

The fermion wavefunction counterterms, calculated to all orders in $\ep$ in Section \ref{sec:fermren}, are
\begin{align}
    \delta Z^{(1,0)}_{\mathrm{V},\,f} &= \Big(a_w^2+v_w^2\Big) \mathbf{p}\, I_1+\Big(a_f^2+v_f^2\Big)\mathbf{p}\,I_4 \\
    \mathrm{and}\qquad \delta Z^{(1,0)}_{\mathrm{A},\,f} &= -2 a_w v_w \mathbf{p}\, I_1 - 2 a_f v_f \mathbf{p}\, I_4\,,
\end{align}
where
\begin{align}
    \mathbf{p} = -\frac{2 (1-\epsilon )}{(2-\epsilon ) \epsilon }\,.
\end{align}

Due to their excessive length, we abbreviate the integral coefficients of the electroweak gauge boson self-energies and associated counterterms as $\mathbf{r}_i$ and present explicit expressions for them in the ancillary file \texttt{twopointcoeffs.m} attached to the \texttt{arXiv} submission of this paper. In the notation of Section \ref{sec:EWren}, we have
\begin{align}
\label{eq:explSigmagammagamma}
    \bar{\Sigma}_{\gamma \gamma}^{(1,0)}(s) &= \mathbf{r}_1 I_1 + \mathbf{r}_2 I_2
   \,,
   \\
   \bar{\Sigma}_{\gamma Z}^{(1,0)}(s) &= \mathbf{r}_3 I_1 + \mathbf{r}_4 I_2
   \,,
   \\
   \bar{\Sigma}_{Z Z}^{(1,0)}(s) &= \mathbf{r}_5 I_1+\mathbf{r}_6 I_2+\mathbf{r}_7 I_4 + \mathbf{r}_8 I_5 + \mathbf{r}_9 I_6
   \,,
   \\
   \mathrm{and}\qquad\bar{\Sigma}_{W^+ W^-}^{(1,0)}(s) &= \mathbf{r}_{10} I_1+\mathbf{r}_{11} I_4+\mathbf{r}_{12} I_5 + \mathbf{r}_{13} I_8 + \mathbf{r}_{14} I_9 + \mathbf{r}_{15} I_{10}
\end{align}
for the transverse parts of the bare electroweak gauge boson self-energies of relative order $\alpha$. Inserting explicit expressions for our master integrals, we find complete agreement through to $\mathcal{O}\left(\ep^0\right)$ with the results of \cite{Bohm:1986rj,Denner:1991kt} after setting contributions proportional to the number of fermion flavors to zero. 

Through Eqs. \eqref{eq:countertermdefs}, we subsequently find
\begin{align}
    \delta Z_{\gamma \gamma}^{(1,0)} &= \mathbf{r}_{16} I_1\,,
    \\
    \delta Z_{Z \gamma}^{(1,0)} &= \mathbf{r}_{17}I_1\,,
    \\
    \delta Z_{\gamma Z}^{(1,0)} &= \mathbf{r}_{18}I_1+\mathbf{r}_{19}I_3
   \,,
   \\
    \delta Z_{Z Z}^{(1,0)} &=  \mathbf{r}_{20}I_1+\mathbf{r}_{21}I_3+\mathbf{r}_{22}I_4+\mathbf{r}_{23}I_5+\mathbf{r}_{24}I_7 
   \,,
   \\
    \delta Z_{m_z^2}^{(1,0)} &= \mathbf{r}_{25}I_1+\mathbf{r}_{26}I_3+\mathbf{r}_{27}I_4+\mathbf{r}_{28}I_5+\mathbf{r}_{29}I_7  
   \,,
   \\
\label{eq:expldelZmw}
    \mathrm{and} \qquad 
   \delta Z_{m_w^2}^{(1,0)} &= \mathbf{r}_{30}I_1+\mathbf{r}_{31}I_4+\mathbf{r}_{32}I_5+\mathbf{r}_{33}I_{11}+\mathbf{r}_{34}I_{12}
\end{align}
for the relative order $\alpha$ electroweak gauge boson mass and wavefunction counterterms appearing in our calculations. As it was not possible for us to find Eqs. \eqref{eq:explSigmagammagamma}-\eqref{eq:expldelZmw} in the literature verbatim, it may be that we present explicit all-orders-in-$\ep$ results for the above quantities for the first time.

Although the vertex form factor integral coefficients are all reasonably compact, we nevertheless abbreviate them as $\mathbf{v}_i$ in what follows in order to highlight the strong constraints imposed on their structure by the chiral symmetry of the Standard Model. We present explicit expressions for the $\mathbf{v}_i$ in the ancillary file \texttt{vertexcoeffs.m} attached to the \texttt{arXiv} submission of this paper. For the order $\alpha$ vertex form factors, we find
\begin{align}
    \bar{\mathcal{V}}^{(1,0)}_{\gamma \bar{q} q}(s) &= -Q_q^3 \mathbf{v}_1 I_{13} -Q_q\left(a_q^2+v_q^2\right)\left(\mathbf{v}_2 I_4+\mathbf{v}_3 I_{13}+ \mathbf{v}_4 I_{14}\right)\\
   &-Q_{q^\prime}\left(a_w^2+v_w^2\right) \left(\mathbf{v}_5 I_1+\mathbf{v}_6 I_{13}+\mathbf{v}_7 I_{15}\right)
   -I_q^3 \left(a_w^2+v_w^2\right)\left(\mathbf{v}_8 I_1+\mathbf{v}_9 I_2 + \mathbf{v}_{10} I_{16}\right)
   \,,
   \nonumber
   \\
    \bar{\mathcal{A}}^{(1,0)}_{\gamma \bar{q} q}(s) &= 2 Q_q a_q v_q \left(\mathbf{v}_2 I_4+\mathbf{v}_3 I_{13}+\mathbf{v}_4 I_{14}\right)+2 Q_{q^\prime} a_w v_w\left(\mathbf{v}_5 I_1 +\mathbf{v}_6 I_{13} +\mathbf{v}_7 I_{15}\right)\\
   &+2 I_q^3 a_w v_w\left(\mathbf{v}_8 I_1 +\mathbf{v}_9 I_2+\mathbf{v}_{10} I_{16}\right)
    \,,
    \nonumber
    \\
    \bar{\mathcal{V}}^{(1,0)}_{Z\bar{q} q}(s) &= Q_q^2 v_q \mathbf{v}_1 I_{13}+v_q \left(3 a_q^2+v_q^2\right)\left(\mathbf{v}_2 I_4+\mathbf{v}_3 I_{13}+\mathbf{v}_4 I_{14}\right)\\
    &+\left(2 a_w v_w a_{q^\prime}+\left(a_w^2+v_w^2\right)v_{q^\prime}\right)\left(\mathbf{v}_5 I_1+\mathbf{v}_6 I_{13}+\mathbf{v}_7 I_{15}\right)\nonumber\\
   &+I_q^3\left(a_w^2+v_w^2\right)\frac{m_w}{\sqrt{m_z^2-m_w^2}}\left(\mathbf{v}_8 I_1+\mathbf{v}_9 I_2+\mathbf{v}_{10} I_{16} \right)
    \,,
    \nonumber
    \\
    \bar{\mathcal{A}}^{(1,0)}_{Z\bar{q} q}(s) &= -Q_q^2 a_q \mathbf{v}_1 I_{13}-a_q \left(a_q^2+3 v_q^2\right)\left(\mathbf{v}_2 I_4+\mathbf{v}_3 I_{13}+\mathbf{v}_4 I_{14}\right)\\
    &-\left(2 a_w v_w v_{q^\prime}+\left(a_w^2+v_w^2\right)a_{q^\prime}\right)\left(\mathbf{v}_5 I_1+\mathbf{v}_6 I_{13}+\mathbf{v}_7 I_{15}\right)\nonumber\\
    &-2 I_q^3 a_w v_w \frac{m_w}{\sqrt{m_z^2-m_w^2}} \left(\mathbf{v}_8 I_1+\mathbf{v}_9 I_2+\mathbf{v}_{10} I_{16}\right)
    \,,
    \nonumber
    \\
    \bar{\mathcal{V}}^{(1,0)}_{\gamma \bar{\ell} \ell}(s) &= -Q_\ell^3 \mathbf{v}_1 I_{13} -Q_\ell\left(a_\ell^2+v_\ell^2\right)\left(\mathbf{v}_2 I_4+\mathbf{v}_3 I_{13}+ \mathbf{v}_4 I_{14}\right)\\
   &-Q_{\ell^\prime}\left(a_w^2+v_w^2\right) \left(\mathbf{v}_5 I_1+\mathbf{v}_6 I_{13}+\mathbf{v}_7 I_{15}\right)
   -I_\ell^3 \left(a_w^2+v_w^2\right)\left(\mathbf{v}_8 I_1+\mathbf{v}_9 I_2 + \mathbf{v}_{10} I_{16}\right)
   \,,
   \nonumber
   \\
    \bar{\mathcal{A}}^{(1,0)}_{\gamma \bar{\ell} \ell}(s) &= 2 Q_\ell a_\ell v_\ell \left(\mathbf{v}_2 I_4+\mathbf{v}_3 I_{13}+\mathbf{v}_4 I_{14}\right)+2 Q_{\ell^\prime} a_w v_w\left(\mathbf{v}_5 I_1 +\mathbf{v}_6 I_{13} +\mathbf{v}_7 I_{15}\right)\\
   &+2 I_\ell^3 a_w v_w\left(\mathbf{v}_8 I_1 +\mathbf{v}_9 I_2+\mathbf{v}_{10} I_{16}\right)
    \,,
    \nonumber
    \\
    \bar{\mathcal{V}}^{(1,0)}_{Z\bar{\ell} \ell}(s) &= Q_\ell^2 v_\ell \mathbf{v}_1 I_{13}+v_\ell \left(3 a_\ell^2+v_\ell^2\right)\left(\mathbf{v}_2 I_4+\mathbf{v}_3 I_{13}+\mathbf{v}_4 I_{14}\right)\\
    &+\left(2 a_w v_w a_{\ell^\prime}+\left(a_w^2+v_w^2\right)v_{\ell^\prime}\right)\left(\mathbf{v}_5 I_1+\mathbf{v}_6 I_{13}+\mathbf{v}_7 I_{15}\right)\nonumber\\
   &+I_\ell^3\left(a_w^2+v_w^2\right)\frac{m_w}{\sqrt{m_z^2-m_w^2}}\left(\mathbf{v}_8 I_1+\mathbf{v}_9 I_2+\mathbf{v}_{10} I_{16} \right)
    \,,
    \qquad
    \mathrm{and}
    \nonumber
    \\
    \bar{\mathcal{A}}^{(1,0)}_{Z\bar{\ell} \ell}(s) &= -Q_\ell^2 a_\ell \mathbf{v}_1 I_{13}-a_\ell \left(a_\ell^2+3 v_\ell^2\right)\left(\mathbf{v}_2 I_4+\mathbf{v}_3 I_{13}+\mathbf{v}_4 I_{14}\right)\\
    &-\left(2 a_w v_w v_{\ell^\prime}+\left(a_w^2+v_w^2\right)a_{\ell^\prime}\right)\left(\mathbf{v}_5 I_1+\mathbf{v}_6 I_{13}+\mathbf{v}_7 I_{15}\right)\nonumber\\
    &-2 I_\ell^3 a_w v_w \frac{m_w}{\sqrt{m_z^2-m_w^2}} \left(\mathbf{v}_8 I_1+\mathbf{v}_9 I_2+\mathbf{v}_{10} I_{16}\right) \,,\nonumber
\end{align}
where $f^\prime$ denotes the isospin partner of fermion $f$. It is worth pointing out that, in the above, $\mathbf{v}_5$, $\mathbf{v}_6$, and $\mathbf{v}_7$ may be trivially obtained from $\mathbf{v}_2$, $\mathbf{v}_3$, and $\mathbf{v}_4$ respectively by replacing $m_z^2$ with $m_w^2$. For the infrared-finite contributions featuring massive electroweak vector bosons in the loop, our results for the one-loop vertex form factors are directly comparable to the original calculation of \cite{Bohm:1986rj}. Indeed, upon inserting explicit expressions for our master integrals, we find complete agreement through to $\mathcal{O}\left(\ep^0\right)$ with the relevant terms of Eqs. (5.28) and (5.30) of \cite{Bohm:1986rj} modulo minor typos.\footnote{While comparing to \cite{Bohm:1986rj}, we noticed two typos in Eq. (B.6), the definition of the form factor $\Lambda_3(k^2,M)$ appearing in Eqs. (5.28) and (5.30): the sixth and seventh lines of Eq. (B.6) should read
\begin{align}
   + \frac{2}{3}w (w+2) \left[\ln^2\left(\frac{1+\sqrt{1-4 w}}{1-\sqrt{1-4 w}}\right) - \pi^2\right]
    -i \pi \left[\frac{2 w + 1}{3}\sqrt{1 - 4 w} + \frac{4}{3}w(w+2)\ln\left(\frac{1+\sqrt{1-4 w}}{1-\sqrt{1-4 w}}\right)\right]. \nonumber
\end{align}}

Finally, for HVBM's $\gamma_5$ scheme, we let $\mathcal{B}_{\mathrm{VV}}^{(1,0)}$, $\mathcal{B}_{\mathrm{VA}}^{(1,0)}$, $\mathcal{B}_{\mathrm{AV}}^{(1,0)}$, and $\mathcal{B}_{\mathrm{AA}}^{(1,0)}$ denote the contributions from order $\alpha^2$ box diagrams to ${\bf C}^{(1,0)}_{\mathrm{VV}}$, ${\bf C}^{(1,0)}_{\mathrm{VA}}$, ${\bf C}^{(1,0)}_{\mathrm{AV}}$, and ${\bf C}^{(1,0)}_{\mathrm{AA}}$ and, for Kreimer's $\gamma_5$ scheme, $\bar{\mathcal{B}}_{\mathrm{VV}}^{(1,0)}$, $\bar{\mathcal{B}}_{\mathrm{VA}}^{(1,0)}$, $\bar{\mathcal{B}}_{\mathrm{AV}}^{(1,0)}$, and $\bar{\mathcal{B}}_{\mathrm{AA}}^{(1,0)}$ denote the contributions from order $\alpha^2$ box diagrams to ${\bf \bar{C}}^{(1,0)}_{\mathrm{VV}}$, ${\bf \bar{C}}^{(1,0)}_{\mathrm{VA}}$, ${\bf \bar{C}}^{(1,0)}_{\mathrm{AV}}$, and ${\bf \bar{C}}^{(1,0)}_{\mathrm{AA}}$. We find
\begin{align}
    \mathcal{B}_{\mathrm{VV}}^{(1,0)} &= Q_q^2 Q_\ell^2 \left({\bf a}_1 I_{13} + {\bf a}_2 I_{17}+{\bf a}_3 I_{18} + {\bf a}_4 I_{19}+{\bf a}_5 I_{20}\right)\nonumber\\
    &\hspace{-.1 cm}+ Q_q Q_\ell a_q a_\ell\left({\bf a}_6 I_{4} + {\bf a}_7 I_{17}+{\bf a}_8 I_{18} + {\bf a}_9 I_{21}+{\bf a}_{10} I_{22}+ {\bf a}_{11} I_{23}+{\bf a}_{12} I_{24}+ {\bf a}_{13} I_{25}\right)\nonumber\\
    &\hspace{-.1 cm}+ Q_q Q_\ell v_q v_\ell \left({\bf a}_{14} I_{4} + {\bf a}_{15} I_{17}+{\bf a}_{16} I_{18} + {\bf a}_{17} I_{21}+{\bf a}_{18} I_{22}+ {\bf a}_{19} I_{23}+{\bf a}_{20} I_{24}+ {\bf a}_{21} I_{25}\right)\nonumber\\
    &\hspace{-.1 cm}+a_q^2 a_\ell^2 \left({\bf a}_{22} I_{4} + {\bf a}_{23} I_{17}+{\bf a}_{24} I_{18} +{\bf a}_{25} I_{22}+ {\bf a}_{26} I_{23}+{\bf a}_{27} I_{26}+ {\bf a}_{28} I_{27}+{\bf a}_{29} I_{28}+ {\bf a}_{30} I_{29}\right)\nonumber\\
    &\hspace{-.1 cm}+a_q^2 v_\ell^2 \left({\bf a}_{31} I_{4} + {\bf a}_{32} I_{17}+{\bf a}_{33} I_{18} +{\bf a}_{34} I_{22}+ {\bf a}_{35} I_{23}+{\bf a}_{36} I_{26}+ {\bf a}_{37} I_{27}+{\bf a}_{38} I_{28}+ {\bf a}_{39} I_{29}\right)\nonumber\\
    &\hspace{-.1 cm}+a_\ell^2 v_q^2 \left({\bf a}_{40} I_{4} + {\bf a}_{41} I_{17}+{\bf a}_{42} I_{18} +{\bf a}_{43} I_{22}+ {\bf a}_{44} I_{23}+{\bf a}_{45} I_{26}+ {\bf a}_{46} I_{27}+{\bf a}_{47} I_{28}+ {\bf a}_{48} I_{29}\right)\nonumber\\
    &\hspace{-.1 cm}+v_q^2 v_\ell^2 \left({\bf a}_{49} I_{4} + {\bf a}_{50} I_{17}+{\bf a}_{51} I_{18} +{\bf a}_{52} I_{22}+ {\bf a}_{53} I_{23}+{\bf a}_{54} I_{26}+ {\bf a}_{55} I_{27}+{\bf a}_{56} I_{28}+ {\bf a}_{57} I_{29}\right)\nonumber\\
    &\hspace{-.75 cm}+ a_q a_\ell v_q v_\ell \left({\bf a}_{58} I_{4} + {\bf a}_{59} I_{17}+{\bf a}_{60} I_{18} +{\bf a}_{61} I_{22}+ {\bf a}_{62} I_{23}+{\bf a}_{63} I_{26}+ {\bf a}_{64} I_{27}+{\bf a}_{65} I_{28}+ {\bf a}_{66} I_{29}\right)\nonumber\\
    &+a_w^4 \left({\bf a}_{67} I_{1} + {\bf a}_{68} I_{2}+{\bf a}_{69} I_{16} +{\bf a}_{70} I_{18}+ {\bf a}_{71} I_{30}+{\bf a}_{72} I_{31}\right)\nonumber\\
    &+v_w^4 \left({\bf a}_{73} I_{1} + {\bf a}_{74} I_{2}+{\bf a}_{75} I_{16} +{\bf a}_{76} I_{18}+ {\bf a}_{77} I_{30}+{\bf a}_{78} I_{31}\right)\nonumber\\
\label{eq:BVV}
    &+a_w^2 v_w^2 \left({\bf a}_{79} I_{1} + {\bf a}_{80} I_{2}+{\bf a}_{81} I_{16} +{\bf a}_{82} I_{18}+ {\bf a}_{83} I_{30}+{\bf a}_{84} I_{31}\right)\,,\\
    \mathcal{B}_{\mathrm{VA}}^{(1,0)} &= Q_q Q_\ell a_q v_\ell\left({\bf b}_{1} I_{4} + {\bf b}_{2} I_{17}+{\bf b}_{3} I_{18} + {\bf b}_{4} I_{21}+{\bf b}_{5} I_{22}+ {\bf b}_{6} I_{23}+{\bf b}_{7} I_{24}+ {\bf b}_{8} I_{25}\right)\nonumber\\
    &+ Q_q Q_\ell a_\ell v_q \left({\bf b}_{9} I_{4} + {\bf b}_{10} I_{17}+{\bf b}_{11} I_{18} + {\bf b}_{12} I_{21}+{\bf b}_{13} I_{22}+ {\bf b}_{14} I_{23}+{\bf b}_{15} I_{24}+ {\bf b}_{16} I_{25}\right)\nonumber\\
    &\hspace{-.8 cm}+a_q^2 a_\ell v_\ell \left({\bf b}_{17} I_{4} + {\bf b}_{18} I_{17}+{\bf b}_{19} I_{18} +{\bf b}_{20} I_{22}+ {\bf b}_{21} I_{23}+{\bf b}_{22} I_{26}+ {\bf b}_{23} I_{27}+{\bf b}_{24} I_{28}+ {\bf b}_{25} I_{29}\right)\nonumber\\
    &\hspace{-.8 cm}+a_q a_\ell^2 v_q \left({\bf b}_{26} I_{4} + {\bf b}_{27} I_{17}+{\bf b}_{28} I_{18} +{\bf b}_{29} I_{22}+ {\bf b}_{30} I_{23}+{\bf b}_{31} I_{26}+ {\bf b}_{32} I_{27}+{\bf b}_{33} I_{28}+ {\bf b}_{34} I_{29}\right)\nonumber\\
    &\hspace{-.8 cm}+a_q v_q v_\ell^2 \left({\bf b}_{35} I_{4} + {\bf b}_{36} I_{17}+{\bf b}_{37} I_{18} +{\bf b}_{38} I_{22}+ {\bf b}_{39} I_{23}+{\bf b}_{40} I_{26}+ {\bf b}_{41} I_{27}+{\bf b}_{42} I_{28}+ {\bf b}_{43} I_{29}\right)\nonumber\\
    &\hspace{-.8 cm}+a_\ell v_q^2 v_\ell \left({\bf b}_{44} I_{4} + {\bf b}_{45} I_{17}+{\bf b}_{46} I_{18} +{\bf b}_{47} I_{22}+ {\bf b}_{48} I_{23}+{\bf b}_{49} I_{26}+ {\bf b}_{50} I_{27}+{\bf b}_{51} I_{28}+ {\bf b}_{52} I_{29}\right)\nonumber\\
    &+a_w^3 v_w \left({\bf b}_{53} I_{16} + {\bf b}_{54} I_{30} + {\bf b}_{55} I_{31}\right)\nonumber\\
\label{eq:BVAV}
    &+a_w v_w^3 \left({\bf b}_{56} I_{16} +{\bf b}_{57} I_{30}+{\bf b}_{58} I_{31}\right)
    \,,
    \\
    \mathcal{B}_{\mathrm{AV}}^{(1,0)} &= Q_q Q_\ell a_q v_\ell\left({\bf c}_{1} I_{4} + {\bf c}_{2} I_{17}+{\bf c}_{3} I_{18} + {\bf c}_{4} I_{21}+{\bf c}_{5} I_{22}+ {\bf c}_{6} I_{23}+{\bf c}_{7} I_{24}+ {\bf c}_{8} I_{25}\right)\nonumber\\
    &+ Q_q Q_\ell a_\ell v_q \left({\bf c}_{9} I_{4} + {\bf c}_{10} I_{17}+{\bf c}_{11} I_{18} + {\bf c}_{12} I_{21}+{\bf c}_{13} I_{22}+ {\bf c}_{14} I_{23}+{\bf c}_{15} I_{24}+ {\bf c}_{16} I_{25}\right)\nonumber\\
    &\hspace{-.8 cm}+a_q^2 a_\ell v_\ell \left({\bf c}_{17} I_{4} + {\bf c}_{18} I_{17}+{\bf c}_{19} I_{18} +{\bf c}_{20} I_{22}+ {\bf c}_{21} I_{23}+{\bf c}_{22} I_{26}+ {\bf c}_{23} I_{27}+{\bf c}_{24} I_{28}+ {\bf c}_{25} I_{29}\right)\nonumber\\
    &\hspace{-.8 cm}+a_q a_\ell^2 v_q \left({\bf c}_{26} I_{4} + {\bf c}_{27} I_{17}+{\bf c}_{28} I_{18} +{\bf c}_{29} I_{22}+ {\bf c}_{30} I_{23}+{\bf c}_{31} I_{26}+ {\bf c}_{32} I_{27}+{\bf c}_{33} I_{28}+ {\bf c}_{34} I_{29}\right)\nonumber\\
    &\hspace{-.8 cm}+a_q v_q v_\ell^2 \left({\bf c}_{35} I_{4} + {\bf c}_{36} I_{17}+{\bf c}_{37} I_{18} +{\bf c}_{38} I_{22}+ {\bf c}_{39} I_{23}+{\bf c}_{40} I_{26}+ {\bf c}_{41} I_{27}+{\bf c}_{42} I_{28}+ {\bf c}_{43} I_{29}\right)\nonumber\\
    &\hspace{-.8 cm}+a_\ell v_q^2 v_\ell \left({\bf c}_{44} I_{4} + {\bf c}_{45} I_{17}+{\bf c}_{46} I_{18} +{\bf c}_{47} I_{22}+ {\bf c}_{48} I_{23}+{\bf c}_{49} I_{26}+ {\bf c}_{50} I_{27}+{\bf c}_{51} I_{28}+ {\bf c}_{52} I_{29}\right)\nonumber\\
    &+a_w^3 v_w \left({\bf c}_{53} I_{16} + {\bf c}_{54} I_{30}+{\bf c}_{55} I_{31}\right)\nonumber\\
\label{eq:BAVV}
    &+a_w v_w^3 \left({\bf c}_{56} I_{16} + {\bf c}_{57} I_{30}+{\bf c}_{58} I_{31}\right)
    \,,
    \\
   \mathcal{B}_{\mathrm{AA}}^{(1,0)} &= 
   Q_q^2 Q_\ell^2 \left({\bf d}_1 I_{13} + {\bf d}_2 I_{17}+{\bf d}_3 I_{18} + {\bf d}_4 I_{19}+{\bf d}_5 I_{20}\right)\nonumber\\
    &\hspace{-.2 cm}+ Q_q Q_\ell a_q a_\ell\left({\bf d}_6 I_{4} + {\bf d}_7 I_{17}+{\bf d}_8 I_{18} + {\bf d}_9 I_{21}+{\bf d}_{10} I_{22}+ {\bf d}_{11} I_{23}+{\bf d}_{12} I_{24}+ {\bf d}_{13} I_{25}\right)\nonumber\\
    &\hspace{-.2 cm}+ Q_q Q_\ell v_q v_\ell \left({\bf d}_{14} I_{4} + {\bf d}_{15} I_{17}+{\bf d}_{16} I_{18} + {\bf d}_{17} I_{21}+{\bf d}_{18} I_{22}+ {\bf d}_{19} I_{23}+{\bf d}_{20} I_{24}+ {\bf d}_{21} I_{25}\right)\nonumber\\
    &\hspace{-.2 cm}+a_q^2 a_\ell^2 \left({\bf d}_{22} I_{4} + {\bf d}_{23} I_{17}+{\bf d}_{24} I_{18} +{\bf d}_{25} I_{22}+ {\bf d}_{26} I_{23}+{\bf d}_{27} I_{26}+ {\bf d}_{28} I_{27}+{\bf d}_{29} I_{28}+ {\bf d}_{30} I_{29}\right)\nonumber\\
    &\hspace{-.2 cm}+a_q^2 v_\ell^2 \left({\bf d}_{31} I_{4} + {\bf d}_{32} I_{17}+{\bf d}_{33} I_{18} +{\bf d}_{34} I_{22}+ {\bf d}_{35} I_{23}+{\bf d}_{36} I_{26}+ {\bf d}_{37} I_{27}+{\bf d}_{38} I_{28}+ {\bf d}_{39} I_{29}\right)\nonumber\\
    &\hspace{-.2 cm}+a_\ell^2 v_q^2 \left({\bf d}_{40} I_{4} + {\bf d}_{41} I_{17}+{\bf d}_{42} I_{18} +{\bf d}_{43} I_{22}+ {\bf d}_{44} I_{23}+{\bf d}_{45} I_{26}+ {\bf d}_{46} I_{27}+{\bf d}_{47} I_{28}+ {\bf d}_{48} I_{29}\right)\nonumber\\
    &\hspace{-.2 cm}+v_q^2 v_\ell^2 \left({\bf d}_{49} I_{4} + {\bf d}_{50} I_{17}+{\bf d}_{51} I_{18} +{\bf d}_{52} I_{22}+ {\bf d}_{53} I_{23}+{\bf d}_{54} I_{26}+ {\bf d}_{55} I_{27}+{\bf d}_{56} I_{28}+ {\bf d}_{57} I_{29}\right)\nonumber\\
    &\hspace{-.85 cm}+ a_q a_\ell v_q v_\ell \left({\bf d}_{58} I_{4} + {\bf d}_{59} I_{17}+{\bf d}_{60} I_{18} +{\bf d}_{61} I_{22}+ {\bf d}_{62} I_{23}+{\bf d}_{63} I_{26}+ {\bf d}_{64} I_{27}+{\bf d}_{65} I_{28}+ {\bf d}_{66} I_{29}\right)\nonumber\\
    &+a_w^4 \left({\bf d}_{67} I_{1} + {\bf d}_{68} I_{2}+{\bf d}_{69} I_{16} +{\bf d}_{70} I_{18}+ {\bf d}_{71} I_{30}+{\bf d}_{72} I_{31}\right)\nonumber\\
    &+v_w^4 \left({\bf d}_{73} I_{1} + {\bf d}_{74} I_{2}+{\bf d}_{75} I_{16} +{\bf d}_{76} I_{18}+ {\bf d}_{77} I_{30}+{\bf d}_{78} I_{31}\right)\nonumber\\
\label{eq:BAVAV}
    &+a_w^2 v_w^2 \left({\bf d}_{79} I_{1} + {\bf d}_{80} I_{2}+{\bf d}_{81} I_{16} +{\bf d}_{82} I_{18}+ {\bf d}_{83} I_{30}+{\bf d}_{84} I_{31}\right)
    \,,
    \\
    \bar{\mathcal{B}}_{\mathrm{VV}}^{(1,0)} &= Q_q^2 Q_\ell^2 \left({\bf \bar{a}}_1 I_{13} + {\bf \bar{a}}_2 I_{17}+{\bf \bar{a}}_3 I_{18} + {\bf \bar{a}}_4 I_{19}+{\bf \bar{a}}_5 I_{20}\right)\nonumber\\
    &\hspace{-.1 cm}+ Q_q Q_\ell a_q a_\ell\left({\bf \bar{a}}_6 I_{4} + {\bf \bar{a}}_7 I_{17}+{\bf \bar{a}}_8 I_{18} + {\bf \bar{a}}_9 I_{21}+{\bf \bar{a}}_{10} I_{22}+ {\bf \bar{a}}_{11} I_{23}+{\bf \bar{a}}_{12} I_{24}+ {\bf \bar{a}}_{13} I_{25}\right)\nonumber\\
    &\hspace{-.1 cm}+ Q_q Q_\ell v_q v_\ell \left({\bf \bar{a}}_{14} I_{4} + {\bf \bar{a}}_{15} I_{17}+{\bf \bar{a}}_{16} I_{18} + {\bf \bar{a}}_{17} I_{21}+{\bf \bar{a}}_{18} I_{22}+ {\bf \bar{a}}_{19} I_{23}+{\bf \bar{a}}_{20} I_{24}+ {\bf \bar{a}}_{21} I_{25}\right)\nonumber\\
    &\hspace{-.1 cm}+a_q^2 a_\ell^2 \left({\bf \bar{a}}_{22} I_{4} + {\bf \bar{a}}_{23} I_{17}+{\bf \bar{a}}_{24} I_{18} +{\bf \bar{a}}_{25} I_{22}+ {\bf \bar{a}}_{26} I_{23}+{\bf \bar{a}}_{27} I_{26}+ {\bf \bar{a}}_{28} I_{27}+{\bf \bar{a}}_{29} I_{28}+ {\bf \bar{a}}_{30} I_{29}\right)\nonumber\\
    &\hspace{-.1 cm}+a_q^2 v_\ell^2 \left({\bf \bar{a}}_{31} I_{4} + {\bf \bar{a}}_{32} I_{17}+{\bf \bar{a}}_{33} I_{18} +{\bf \bar{a}}_{34} I_{22}+ {\bf \bar{a}}_{35} I_{23}+{\bf \bar{a}}_{36} I_{26}+ {\bf \bar{a}}_{37} I_{27}+{\bf \bar{a}}_{38} I_{28}+ {\bf \bar{a}}_{39} I_{29}\right)\nonumber\\
    &\hspace{-.1 cm}+a_\ell^2 v_q^2 \left({\bf \bar{a}}_{40} I_{4} + {\bf \bar{a}}_{41} I_{17}+{\bf \bar{a}}_{42} I_{18} +{\bf \bar{a}}_{43} I_{22}+ {\bf \bar{a}}_{44} I_{23}+{\bf \bar{a}}_{45} I_{26}+ {\bf \bar{a}}_{46} I_{27}+{\bf \bar{a}}_{47} I_{28}+ {\bf \bar{a}}_{48} I_{29}\right)\nonumber\\
    &\hspace{-.1 cm}+v_q^2 v_\ell^2 \left({\bf \bar{a}}_{49} I_{4} + {\bf \bar{a}}_{50} I_{17}+{\bf \bar{a}}_{51} I_{18} +{\bf \bar{a}}_{52} I_{22}+ {\bf \bar{a}}_{53} I_{23}+{\bf \bar{a}}_{54} I_{26}+ {\bf \bar{a}}_{55} I_{27}+{\bf \bar{a}}_{56} I_{28}+ {\bf \bar{a}}_{57} I_{29}\right)\nonumber\\
    &\hspace{-.75 cm}+ a_q a_\ell v_q v_\ell \left({\bf \bar{a}}_{58} I_{4} + {\bf \bar{a}}_{59} I_{17}+{\bf \bar{a}}_{60} I_{18} +{\bf \bar{a}}_{61} I_{22}+ {\bf \bar{a}}_{62} I_{23}+{\bf \bar{a}}_{63} I_{26}+ {\bf \bar{a}}_{64} I_{27}+{\bf \bar{a}}_{65} I_{28}+ {\bf \bar{a}}_{66} I_{29}\right)\nonumber\\
    &+a_w^4 \left({\bf \bar{a}}_{67} I_{1} + {\bf \bar{a}}_{68} I_{2}+{\bf \bar{a}}_{69} I_{16} +{\bf \bar{a}}_{70} I_{18}+ {\bf \bar{a}}_{71} I_{30}+{\bf \bar{a}}_{72} I_{31}\right)\nonumber\\
    &+v_w^4 \left({\bf \bar{a}}_{73} I_{1} + {\bf \bar{a}}_{74} I_{2}+{\bf \bar{a}}_{75} I_{16} +{\bf \bar{a}}_{76} I_{18}+ {\bf \bar{a}}_{77} I_{30}+{\bf \bar{a}}_{78} I_{31}\right)\nonumber\\
\label{eq:BbarVV}
    &+a_w^2 v_w^2 \left({\bf \bar{a}}_{79} I_{1} + {\bf \bar{a}}_{80} I_{2}+{\bf \bar{a}}_{81} I_{16} +{\bf \bar{a}}_{82} I_{18}+ {\bf \bar{a}}_{83} I_{30}+{\bf \bar{a}}_{84} I_{31}\right)
    \,,
    \\
    \bar{\mathcal{B}}_{\mathrm{VA}}^{(1,0)} &=Q_q Q_\ell a_q v_\ell\left({\bf \bar{b}}_{1} I_{4} + {\bf \bar{b}}_{2} I_{17}+{\bf \bar{b}}_{3} I_{18} + {\bf \bar{b}}_{4} I_{21}+{\bf \bar{b}}_{5} I_{22}+ {\bf \bar{b}}_{6} I_{23}+{\bf \bar{b}}_{7} I_{24}+ {\bf \bar{b}}_{8} I_{25}\right)\nonumber\\
    &+ Q_q Q_\ell a_\ell v_q \left({\bf \bar{b}}_{9} I_{4} + {\bf \bar{b}}_{10} I_{17}+{\bf \bar{b}}_{11} I_{18} + {\bf \bar{b}}_{12} I_{21}+{\bf \bar{b}}_{13} I_{22}+ {\bf \bar{b}}_{14} I_{23}+{\bf \bar{b}}_{15} I_{24}+ {\bf \bar{b}}_{16} I_{25}\right)\nonumber\\
    &\hspace{-.775 cm}+a_q^2 a_\ell v_\ell \left({\bf \bar{b}}_{17} I_{4} + {\bf \bar{b}}_{18} I_{17}+{\bf \bar{b}}_{19} I_{18} +{\bf \bar{b}}_{20} I_{22}+ {\bf \bar{b}}_{21} I_{23}+{\bf \bar{b}}_{22} I_{26}+ {\bf \bar{b}}_{23} I_{27}+{\bf \bar{b}}_{24} I_{28}+ {\bf \bar{b}}_{25} I_{29}\right)\nonumber\\
    &\hspace{-.775 cm}+a_q a_\ell^2 v_q \left({\bf \bar{b}}_{26} I_{4} + {\bf \bar{b}}_{27} I_{17}+{\bf \bar{b}}_{28} I_{18} +{\bf \bar{b}}_{29} I_{22}+ {\bf \bar{b}}_{30} I_{23}+{\bf \bar{b}}_{31} I_{26}+ {\bf \bar{b}}_{32} I_{27}+{\bf \bar{b}}_{33} I_{28}+ {\bf \bar{b}}_{34} I_{29}\right)\nonumber\\
    &\hspace{-.775 cm}+a_q v_q v_\ell^2 \left({\bf \bar{b}}_{35} I_{4} + {\bf \bar{b}}_{36} I_{17}+{\bf \bar{b}}_{37} I_{18} +{\bf \bar{b}}_{38} I_{22}+ {\bf \bar{b}}_{39} I_{23}+{\bf \bar{b}}_{40} I_{26}+ {\bf \bar{b}}_{41} I_{27}+{\bf \bar{b}}_{42} I_{28}+ {\bf \bar{b}}_{43} I_{29}\right)\nonumber\\
    &\hspace{-.775 cm}+a_\ell v_q^2 v_\ell \left({\bf \bar{b}}_{44} I_{4} + {\bf \bar{b}}_{45} I_{17}+{\bf \bar{b}}_{46} I_{18} +{\bf \bar{b}}_{47} I_{22}+ {\bf \bar{b}}_{48} I_{23}+{\bf \bar{b}}_{49} I_{26}+ {\bf \bar{b}}_{50} I_{27}+{\bf \bar{b}}_{51} I_{28}+ {\bf \bar{b}}_{52} I_{29}\right)\nonumber\\
    &+a_w^3 v_w \left({\bf \bar{b}}_{53} I_{1} + {\bf \bar{b}}_{54} I_{2}+{\bf \bar{b}}_{55} I_{16} +{\bf \bar{b}}_{56} I_{18}+ {\bf \bar{b}}_{57} I_{30}+{\bf \bar{b}}_{58} I_{31}\right)\nonumber\\
\label{eq:BbarVAV}
    &+a_w v_w^3 \left({\bf \bar{b}}_{59} I_{1} + {\bf \bar{b}}_{60} I_{2}+{\bf \bar{b}}_{61} I_{16} +{\bf \bar{b}}_{62} I_{18}+ {\bf \bar{b}}_{63} I_{30}+{\bf \bar{b}}_{64} I_{31}\right)
    \,,
    \\
    \bar{\mathcal{B}}_{\mathrm{AV}}^{(1,0)} &= Q_q Q_\ell a_q v_\ell\left({\bf \bar{c}}_{1} I_{4} + {\bf \bar{c}}_{2} I_{17}+{\bf \bar{c}}_{3} I_{18} + {\bf \bar{c}}_{4} I_{21}+{\bf \bar{c}}_{5} I_{22}+ {\bf \bar{c}}_{6} I_{23}+{\bf \bar{c}}_{7} I_{24}+ {\bf \bar{c}}_{8} I_{25}\right)\nonumber\\
    &+ Q_q Q_\ell a_\ell v_q \left({\bf \bar{c}}_{9} I_{4} + {\bf \bar{c}}_{10} I_{17}+{\bf \bar{c}}_{11} I_{18} + {\bf \bar{c}}_{12} I_{21}+{\bf \bar{c}}_{13} I_{22}+ {\bf \bar{c}}_{14} I_{23}+{\bf \bar{c}}_{15} I_{24}+ {\bf \bar{c}}_{16} I_{25}\right)\nonumber\\
    &\hspace{-.775 cm}+a_q^2 a_\ell v_\ell \left({\bf \bar{c}}_{17} I_{4} + {\bf \bar{c}}_{18} I_{17}+{\bf \bar{c}}_{19} I_{18} +{\bf \bar{c}}_{20} I_{22}+ {\bf \bar{c}}_{21} I_{23}+{\bf \bar{c}}_{22} I_{26}+ {\bf \bar{c}}_{23} I_{27}+{\bf \bar{c}}_{24} I_{28}+ {\bf \bar{c}}_{25} I_{29}\right)\nonumber\\
    &\hspace{-.775 cm}+a_q a_\ell^2 v_q \left({\bf \bar{c}}_{26} I_{4} + {\bf \bar{c}}_{27} I_{17}+{\bf \bar{c}}_{28} I_{18} +{\bf \bar{c}}_{29} I_{22}+ {\bf \bar{c}}_{30} I_{23}+{\bf \bar{c}}_{31} I_{26}+ {\bf \bar{c}}_{32} I_{27}+{\bf \bar{c}}_{33} I_{28}+ {\bf \bar{c}}_{34} I_{29}\right)\nonumber\\
    &\hspace{-.775 cm}+a_q v_q v_\ell^2 \left({\bf \bar{c}}_{35} I_{4} + {\bf \bar{c}}_{36} I_{17}+{\bf \bar{c}}_{37} I_{18} +{\bf \bar{c}}_{38} I_{22}+ {\bf \bar{c}}_{39} I_{23}+{\bf \bar{c}}_{40} I_{26}+ {\bf \bar{c}}_{41} I_{27}+{\bf \bar{c}}_{42} I_{28}+ {\bf \bar{c}}_{43} I_{29}\right)\nonumber\\
    &\hspace{-.775 cm}+a_\ell v_q^2 v_\ell \left({\bf \bar{c}}_{44} I_{4} + {\bf \bar{c}}_{45} I_{17}+{\bf \bar{c}}_{46} I_{18} +{\bf \bar{c}}_{47} I_{22}+ {\bf \bar{c}}_{48} I_{23}+{\bf \bar{c}}_{49} I_{26}+ {\bf \bar{c}}_{50} I_{27}+{\bf \bar{c}}_{51} I_{28}+ {\bf \bar{c}}_{52} I_{29}\right)\nonumber\\
    &+a_w^3 v_w \left({\bf \bar{c}}_{53} I_{1} + {\bf \bar{c}}_{54} I_{2}+{\bf \bar{c}}_{55} I_{16} +{\bf \bar{c}}_{56} I_{18}+ {\bf \bar{c}}_{57} I_{30}+{\bf \bar{c}}_{58} I_{31}\right)\nonumber\\
\label{eq:BbarAVV}
    &+a_w v_w^3 \left({\bf \bar{c}}_{59} I_{1} + {\bf \bar{c}}_{60} I_{2}+{\bf \bar{c}}_{61} I_{16} +{\bf \bar{c}}_{62} I_{18}+ {\bf \bar{c}}_{63} I_{30}+{\bf \bar{c}}_{64} I_{31}\right)
    \,,
    \qquad
    \mathrm{and}
    \\
    \bar{\mathcal{B}}_{\mathrm{AA}}^{(1,0)} &= Q_q^2 Q_\ell^2 \left({\bf \bar{d}}_1 I_{13} + {\bf \bar{d}}_2 I_{17}+{\bf \bar{d}}_3 I_{18} + {\bf \bar{d}}_4 I_{19}+{\bf \bar{d}}_5 I_{20}\right)\nonumber\\
    &\hspace{-.25 cm}+ Q_q Q_\ell a_q a_\ell\left({\bf \bar{d}}_6 I_{4} + {\bf \bar{d}}_7 I_{17}+{\bf \bar{d}}_8 I_{18} + {\bf \bar{d}}_9 I_{21}+{\bf \bar{d}}_{10} I_{22}+ {\bf \bar{d}}_{11} I_{23}+{\bf \bar{d}}_{12} I_{24}+ {\bf \bar{d}}_{13} I_{25}\right)\nonumber\\
    &\hspace{-.25 cm}+ Q_q Q_\ell v_q v_\ell \left({\bf \bar{d}}_{14} I_{4} + {\bf \bar{d}}_{15} I_{17}+{\bf \bar{d}}_{16} I_{18} + {\bf \bar{d}}_{17} I_{21}+{\bf \bar{d}}_{18} I_{22}+ {\bf \bar{d}}_{19} I_{23}+{\bf \bar{d}}_{20} I_{24}+ {\bf \bar{d}}_{21} I_{25}\right)\nonumber\\
    &\hspace{-.25 cm}+a_q^2 a_\ell^2 \left({\bf \bar{d}}_{22} I_{4} + {\bf \bar{d}}_{23} I_{17}+{\bf \bar{d}}_{24} I_{18} +{\bf \bar{d}}_{25} I_{22}+ {\bf \bar{d}}_{26} I_{23}+{\bf \bar{d}}_{27} I_{26}+ {\bf \bar{d}}_{28} I_{27}+{\bf \bar{d}}_{29} I_{28}+ {\bf \bar{d}}_{30} I_{29}\right)\nonumber\\
    &\hspace{-.25 cm}+a_q^2 v_\ell^2 \left({\bf \bar{d}}_{31} I_{4} + {\bf \bar{d}}_{32} I_{17}+{\bf \bar{d}}_{33} I_{18} +{\bf \bar{d}}_{34} I_{22}+ {\bf \bar{d}}_{35} I_{23}+{\bf \bar{d}}_{36} I_{26}+ {\bf \bar{d}}_{37} I_{27}+{\bf \bar{d}}_{38} I_{28}+ {\bf \bar{d}}_{39} I_{29}\right)\nonumber\\
    &\hspace{-.25 cm}+a_\ell^2 v_q^2 \left({\bf \bar{d}}_{40} I_{4} + {\bf \bar{d}}_{41} I_{17}+{\bf \bar{d}}_{42} I_{18} +{\bf \bar{d}}_{43} I_{22}+ {\bf \bar{d}}_{44} I_{23}+{\bf \bar{d}}_{45} I_{26}+ {\bf \bar{d}}_{46} I_{27}+{\bf \bar{d}}_{47} I_{28}+ {\bf \bar{d}}_{48} I_{29}\right)\nonumber\\
    &\hspace{-.25 cm}+v_q^2 v_\ell^2 \left({\bf \bar{d}}_{49} I_{4} + {\bf \bar{d}}_{50} I_{17}+{\bf \bar{d}}_{51} I_{18} +{\bf \bar{d}}_{52} I_{22}+ {\bf \bar{d}}_{53} I_{23}+{\bf \bar{d}}_{54} I_{26}+ {\bf \bar{d}}_{55} I_{27}+{\bf \bar{d}}_{56} I_{28}+ {\bf \bar{d}}_{57} I_{29}\right)\nonumber\\
    &\hspace{-.9 cm}+ a_q a_\ell v_q v_\ell \left({\bf \bar{d}}_{58} I_{4} + {\bf \bar{d}}_{59} I_{17}+{\bf \bar{d}}_{60} I_{18} +{\bf \bar{d}}_{61} I_{22}+ {\bf \bar{d}}_{62} I_{23}+{\bf \bar{d}}_{63} I_{26}+ {\bf \bar{d}}_{64} I_{27}+{\bf \bar{d}}_{65} I_{28}+ {\bf \bar{d}}_{66} I_{29}\right)\nonumber\\
    &+a_w^4 \left({\bf \bar{d}}_{67} I_{1} + {\bf \bar{d}}_{68} I_{2}+{\bf \bar{d}}_{69} I_{16} +{\bf \bar{d}}_{70} I_{18}+ {\bf \bar{d}}_{71} I_{30}+{\bf \bar{d}}_{72} I_{31}\right)\nonumber\\
    &+v_w^4 \left({\bf \bar{d}}_{73} I_{1} + {\bf \bar{d}}_{74} I_{2}+{\bf \bar{d}}_{75} I_{16} +{\bf \bar{d}}_{76} I_{18}+ {\bf \bar{d}}_{77} I_{30}+{\bf \bar{d}}_{78} I_{31}\right)\nonumber\\
\label{eq:BbarAVAV}
    &+a_w^2 v_w^2 \left({\bf \bar{d}}_{79} I_{1} + {\bf \bar{d}}_{80} I_{2}+{\bf \bar{d}}_{81} I_{16} +{\bf \bar{d}}_{82} I_{18}+ {\bf \bar{d}}_{83} I_{30}+{\bf \bar{d}}_{84} I_{31}\right)\,,
\end{align}
where, due to their excessive length, the ${\bf a}_i$, ${\bf b}_i$, ${\bf c}_i$, ${\bf d}_i$, ${\bf \bar{a}}_i$, ${\bf \bar{b}}_i$, ${\bf \bar{c}}_i$, and ${\bf \bar{d}}_i$
are provided in the ancillary files {\tt HVBMboxcoeffs.m} and {\tt Kreimerboxcoeffs.m} attached to the {\tt arXiv} submission of this paper.

In fact, it is possible to compare our results for the one-loop box diagrams to the original calculation of \cite{Bohm:1986rj}, where a photon mass $\lambda$ was employed to regulate infrared singularities, by making the identification $\ln\left(\lambda^2/s\right) \rightarrow 1/\ep$. Inserting explicit expressions for our master integrals into Eqs. \eqref{eq:BVV}-\eqref{eq:BbarAVAV}, we find, modulo typos, complete agreement through to $\mathcal{O}\left(\ep^0\right)$ with Eqs. (B.10) and (B.12) - (B.14) of \cite{Bohm:1986rj}. To find agreement with \cite{Bohm:1986rj}, we first set the factor of $\alpha/(2\pi s)$ in Eq. (B.10) to one, the factors of $\alpha/(2\pi)$ in Eq. (B.12) to $2/s$, the factors of $\alpha/(2\pi)$ in Eq. (B.13) to $2/\left(s-M_Z^2\right)$, and the factors of $\alpha/(2\pi)$ in Eq. (B.14) to $2/s$. In addition, we find that the term inside the brackets on the second line of Eq. (B.12) should read
\begin{align}
    -2 \left[\ln\left(\frac{-t}{s}\right)-\ln\left(\frac{-u}{s}\right)\right]\ln\left(\frac{-s-i\varepsilon}{\lambda^2}\right)\,,\nonumber
\end{align}
that the term inside the parentheses on the right-hand side of the second equality of Eq. (B.13) should read\footnote{The Spence function of \cite{Bohm:1986rj}, $\mathrm{Sp}(x)$, is exactly the dilogarithm function, $\mathrm{Li}_2(x)$.}
\begin{align}
    &\mathrm{Sp}\left(\frac{M_Z^2 + t}{t}\right) - \mathrm{Sp}\left(\frac{M_Z^2 + u}{u}\right) + \frac{1}{2}\left[\ln\left(\frac{t^2}{s^2}\right)-\ln\left(\frac{u^2}{s^2}\right)\right]\ln\left(\frac{M_Z^2}{M_Z^2-s-i\varepsilon}\right)\nonumber\\
    &\qquad\qquad- \left[\ln\left(\frac{-t}{s}\right)-\ln\left(\frac{-u}{s}\right)\right]\ln\left(\frac{M_Z^2-s-i\varepsilon}{\lambda^2}\right)\,,\nonumber
\end{align}
and that the factor of $x_2 + x_1$ in the denominator on the third line of Eq. (B.14) should be replaced by $x_2 - x_1$.

From the ingredients presented above, we distill complete results for ${\bf C}^{(0,1)}_{\mathrm{VV}}$, ${\bf C}^{(0,1)}_{\mathrm{VA}}$, ${\bf C}^{(0,1)}_{\mathrm{AV}}$, ${\bf C}^{(0,1)}_{\mathrm{AA}}$, ${\bf \bar{C}}^{(0,1)}_{\mathrm{VV}}$, ${\bf \bar{C}}^{(0,1)}_{\mathrm{VA}}$, ${\bf \bar{C}}^{(0,1)}_{\mathrm{AV}}$, ${\bf \bar{C}}^{(0,1)}_{\mathrm{AA}}$, ${\bf C}^{(1,0)}_{\mathrm{VV}}$, ${\bf C}^{(1,0)}_{\mathrm{VA}}$, ${\bf C}^{(1,0)}_{\mathrm{AV}}$, ${\bf C}^{(1,0)}_{\mathrm{AA}}$, ${\bf \bar{C}}^{(1,0)}_{\mathrm{VV}}$, ${\bf \bar{C}}^{(1,0)}_{\mathrm{VA}}$, ${\bf \bar{C}}^{(1,0)}_{\mathrm{AV}}$, and ${\bf \bar{C}}^{(1,0)}_{\mathrm{AA}}$ in Section \ref{sec:1Lfinalres}.
\bibliographystyle{JHEP}
\bibliography{dyamp}

\end{document}